\documentclass{article}

\usepackage{amsmath}
\usepackage{amssymb}
\usepackage{physics}
\usepackage{graphicx}
\usepackage{tikz-cd}
\usepackage{authblk}
\usepackage{hyperref}
\usepackage{setspace}
\usepackage{fullpage}

\title{Axiomatic Quantum Field Theory in Discrete Spacetime via Multiway Causal Structure: The Case of Entanglement Entropies}
\author[1,2]{Jonathan Gorard\footnote{\href{mailto:gorardj@cardiff.ac.uk}{gorardj@cardiff.ac.uk}, \href{mailto:jg865@cantab.ac.uk}{jg865@cantab.ac.uk}}}
\author[3]{Julia Dannemann-Freitag\footnote{\href{mailto:jad318@ic.ac.uk}{jad318@ic.ac.uk}}}
\affil[1]{Cardiff University, Cardiff, UK}
\affil[2]{University of Cambridge, Cambridge, UK}
\affil[3]{Imperial College London, London, UK}

\begin{document}
\maketitle

\begin{abstract}
The causal set and Wolfram model approaches to discrete quantum gravity both permit the formulation of a manifestly covariant notion of entanglement entropy for quantum fields. In the causal set case, this is given by a construction (due to Sorkin and Johnston) of a 2-point correlation function for a Gaussian scalar field from causal set Feynman propagators and Pauli-Jordan functions, from which an eigendecomposition, and hence an entanglement entropy, can be computed. In the Wolfram model case, it is given instead in terms of the Fubini-Study metric on branchial graphs, whose tensor product structure is inherited functorially from that of finite-dimensional Hilbert spaces. In both cases, the entanglement entropies in question are most naturally defined over an extended spacetime region (hence the manifest covariance), in contrast to the generically non-covariant definitions over single spacelike hypersurfaces common to most continuum quantum field theories. In this article, we show how an axiomatic field theory for a free, massless scalar field (obeying the appropriate bosonic commutation relations) may be rigorously constructed over multiway causal graphs: a combinatorial structure sufficiently general as to encompass both causal sets and Wolfram model evolutions as special cases. We proceed to show numerically that the entanglement entropies computed using both the Sorkin-Johnston approach and the branchial graph approach are monotonically related for a large class of Wolfram model evolution rules. We also prove a special case of this monotonic relationship using a recent geometrical entanglement monotone proposed by Cocchiarella et al. The resulting construction is non-trivial, since the evolution of an arbitrary Wolfram model rule will not, in general, result in an integer-dimensional causal graph, and so the definition of scalar field Green's functions (and hence Feynman propagators) on causal sets must be analytically continued to accommodate the non-integer-dimensional case. Finally, we propose potential extensions of the approaches developed herein to more general spacetime geometries, to discrete spacelike hypersurfaces/Cauchy surfaces in the form of hypergraphs and to fully-interacting, non-Gaussian scalar field theories involving higher correlators.
\end{abstract}

\newpage

\section{Introduction}

Just as the von Neumann entropy of a mixed quantum system may be thought of as quantifying the deviation of that system from being a \textit{pure} superposition of eigenstates (by analogy to the Gibbs entropy from classical statistical mechanics), the entanglement entropy of a multipartite quantum system may be thought of as quantifying the deviation of that system from being in a fully-separable tensor product state (or, equivalently, as quantifying the deviation of that state's tensor product structure away from being purely Cartesian). Concretely, the entanglement entropy for a pair of entangled subsystems may therefore be computed by taking a partial trace over one of the two subsystems, and then evaluating the standard von Neumann entropy of the resulting (reduced) density matrix corresponding to the other subsystem\cite{popescu}\cite{vedral}. In the context of a quantum gravity theory, it is natural to think of the quantum state of spacetime in its entirety as being composed of a large tensor product of smaller quantum states associated with individual spacetime subregions, where the nature of this tensor product structure, and thus the rule for how the quantum states of smaller spacetime regions may be ``glued'' together, remains more-or-less mysterious (indeed, one can think of this question regarding the compositional structure of the tensor product as being at the heart of the mystery of quantum gravity). It is, consequently, equally natural to ask how the entanglement entropy of spacetime itself may be calculated within such a theory, wherein one may calculate the degree to which a single spacetime subregion is entangled with the remainder of spacetime\cite{sorkin}\cite{bombelli}. In a continuum quantum gravity theory (or indeed even a conventional quantum field theory in a continuous curved spacetime), such an entanglement entropy exhibits a very direct and appealing physical intuition underlying it: it quantifies the extent to which quantum information is apparently ``lost'' as a consequence of the geometry of the background spacetime (for instance, certain quantum degrees of freedom may be rendered inaccessible due to the presence of black hole event horizons, cosmic event horizons, etc., whenever the $n$-point correlations of the field theory are allowed to involve points lying on both sides of the horizon simultaneously)\cite{ryu}.

On the other hand, both causal set theory and the Wolfram model are examples of \textit{discrete} quantum gravity models. Causal set theory represents spacetime as a partially-ordered set\cite{bombelli2}\cite{bombelli3}, with the partial order relation determining, in some appropriate continuum limit, the conformal structure of a Lorentzian manifold\cite{sorkin2}\cite{sorkin3} (and hence, due to the classic theorems of Hawking, King and McCarthy, and later David Malament, the full topology of continuous timelike curves in spacetime\cite{hawking}\cite{malament}), and the cardinalities of subsets of the causal set determining the volumes of the corresponding subregions in the continuum. The Wolfram model also represents spacetime as a partially-ordered set (known as a \textit{causal graph}, due to its conventional representation as a directed, acyclic graph)\cite{wolfram}\cite{wolfram2}\cite{gorard}\cite{gorard2} with the same limiting properties as a causal set, but where the partial order is determined by the causal interactions of abstract rewrites in a hypergraph rewriting system. Indeed, the \textit{transitive reduction} of a causal graph corresponds precisely to the Hasse diagram of the corresponding causal set. In this way, the Wolfram model may be thought of as endowing causal set theory (which is otherwise a largely kinematic formalism) with an explicit algorithmic dynamics\cite{bolognesi}\cite{bolognesi2}\cite{gorard3}. However, a Wolfram model evolution possesses strictly more mathematical structure than a causal set does, since it also encompasses a time-ordered sequence of hypergraphs (corresponding to a sequence of spacelike hypersurfaces representing the evolution of the Einstein field equations from Cauchy initial data, where this sequence is dependent upon a choice of hypergraph rewriting order, corresponding in the continuum to a global choice of gauge conditions\cite{gorard4}), as well as a \textit{multiway} causal structure (encoding the fact that Wolfram model evolution is inherently non-deterministic, since there is no canonical choice of rewriting order, and hence no preferred choice of gauge). The resulting \textit{multiway system}, which effectively parametrizes all possible evolution histories for the Wolfram model, as well as the \textit{branchial graphs} of which it is composed, exhibits certain quantum mechanical properties (and, in particular, is equipped with a tensor product structure which provably satisfies the axioms of a dagger-symmetric, compact-closed monoidal category, and thus constitutes an appropriate setting for a categorical theory of quantum mechanics\cite{gorard5}\cite{gorard6}), a fact which has been exploited to great effect within previous work in the study of quantum circuits via the ZX-calculus formalism of Coecke and Duncan\cite{coecke}\cite{coecke2} and generalized multiway hypergraph rewriting\cite{gorard7}.

Discrete formulations of spacetime offer several key advantages over continuous ones with regards to the computation of spacetime entanglement entropies. For one, the existence of a countable set of degrees of gravitational freedom potentially enables a very directly statistical-mechanical intuition for what the entropy associated with a particular spacetime region actually means, based purely on combinatorics and very much in line with the original ``microstate vs. macrostate'' spirit of Boltzmann. Specifically, if the underlying discrete structure of spacetime is represented by (for example) a causal graph, then one can imagine simply calculating the logarithm of the number of distinct (non-isomorphic) causal graphs that are consistent with a given continuum spacetime geometry directly, and using this as the fundamental definition of spacetime entropy. Indeed, one can interpret the results presented within this article as being part of a much larger and more ambitious research program, intended to determine whether this rather tempting intuition regarding gravitational microstates and macrostates can be made mathematically precise. Less speculatively, one pathology that is common to all standard definitions of entanglement entropy in quantum field theories defined over continuous spacetimes is the presence of unwelcome ultraviolet divergences: for instance, a scalar field theory defined over a black hole geometry will generically exhibit an infinite number of high-frequency modes in the region surrounding the event horizon, and therefore computations of the entanglement entropy between regions of the scalar field separated by the horizon will not be well-defined. However, the existence of a finite underlying discretization scale in discrete quantum gravity models (such as causal set theory) imposes a natural ultraviolet cutoff on the quantum field theory, in a manner akin to lattice regularization in lattice gauge theories, thus rendering the theory ultraviolet-finite and the resulting entanglement entropies sensible, without the need to construct an explicit ultraviolet completion. Moreover, entropies in continuum quantum field theories are traditionally formulated in terms of a density matrix ${\rho \left( \Sigma \right)}$ for the field localized to a particular spacelike hypersurface ${\Sigma}$. This certainly appears to break the spirit, if not the letter, of the principle of general covariance, and, more pragmatically, many quantum fields (especially those that appear in quantum gravity contexts) are conjectured to be much too singular to permit any meaningful embedding onto lower-dimensional hypersurfaces in this way. The requisite ultraviolet cutoffs, necessary to render the entanglement entropies finite, are then defined relative to these hypersurfaces, in a manner which is not possible to reproduce within an inherently non-local formalism such as causal set theory, where such localizations simply cannot be constructed (although it may be possible in the more general case of Wolfram model evolution, where localizations onto spacelike hypersurfaces can be constructed for appropriate choices of gauge, as we shall discuss subsequently). This limitation ultimately motivated Sorkin's eventual introduction\cite{sorkin4} of a manifestly covariant definition of spacetime entropy, including entanglement entropies defined over extended regions of spacetime as a special case, in a manner directly analogous to Peierls' introduction of a Lorentz-invariant formulation of quantum field theory based on spacetime commutators and advanced/retarded Green's functions\cite{peierls} (indeed, Sorkin's construction makes explicit use of the so-called ``Peierls bracket'', more commonly known as the Pauli-Jordan function\cite{jordan}). Thus, this apparent ``limitation'', stemming from the inherent spacetime non-locality of causal set theory, also ensures that the natural definition of an ultraviolet-finite notion of entanglement entropy is necessarily manifestly covariant, in stark contrast to the continuum case.

In the causal set case, one may begin by constructing a discrete d'Alembertian operator (which plays the role of the continuum Klein-Gordon operator) for a causal set sprinkled into a flat, integer-dimensional spacetime using the ansatz given by Dowker and Glaser\cite{dowker}. Using methods due to Sorkin\cite{sorkin5}, this construction can then be extended to causal sets sprinkled into a Riemann normal neighborhood of any curved (integer-dimensional) Lorentzian manifold, and the resulting d'Alembertian operator may subsequently be inverted in order to derive discrete advanced and retarded Green's function for a massive free (Gaussian) scalar field; in the general (continuum) case, the performance of such an inversion necessitates the introduction of certain Fourier-analytic methods. As shown by Johnston\cite{johnston}, these advanced and retarded Green's functions may then be used to construct discrete propagators, and in particular the discrete Feynman propagator\cite{johnston2}, by means of an extremely elegant construction (known as the ``hops and stops'' formalism\cite{johnston3}) that makes manifest the intuitive correspondence between the ``sum over paths'' approach to path integrals in continuum quantum field theories and the ``sum over chains''/``sum over paths'' approaches to propagators in causal set quantum field theories. The discrete (and manifestly covariant) Peierls bracket/Pauli-Jordan operator, given by the difference between the retarded and advanced Green's functions, now permits an eigendecomposition with a natural splitting into positive and negative eigenvalue pairs (more precisely, this is an eigendecomposition of the \textit{imaginary variant} of the Pauli-Jordan operator), corresponding in the continuum case to a mode decomposition of the solutions to the Klein-Gordon equation into positive and negative frequency classes. This eigendecomposition defines (in both the discrete and continuum cases) a distinguished vacuum state, known as the \textit{Sorkin-Johnston}\cite{sorkin6} (or SJ) vacuum, with respect to which one may then define a two-point correlation function (or ``Wightman'' function) by considering only the positive part of the eigenspectrum. For a free (Gaussian) scalar field, this Wightman function is sufficient to determine the structure of the resulting quantum field theory in its entirety. The Sorkin spacetime entanglement entropy may then be defined purely in terms of a generalized eigenvalue problem for the (discrete) Wightman and Pauli-Jordan operators, although truncations of the eigenspectrum are required in order to reproduce the expected ``area'' scaling law for the entanglement entropy, as opposed to the ``volume'' law that appears to emerge more naturally within causal set models of this kind\cite{sorkin7}. On the other hand, a multiway system describing the evolution of a generic Wolfram model rule may be decomposed (subject to certain gauge conditions) into a time-ordered sequence of combinatorial structures known as ``branchial graphs'', each of which effectively represents the tensor product structure of eigenstates at each instant of time. In cases where an appropriate continuum limit exists, the discrete metric on branchial graphs converges to the Fubini-Study metric on projective Hilbert spaces, which, when restricted to pure states only, reduces to the quantum Bures metric: a standard measure of pure state entanglement in quantum information theory\cite{marian}. Although this definition appears on the surface to break general covariance, manifest covariance may nevertheless be reintroduced by instead considering \textit{causal} multiway systems (and hence \textit{causal} branchial graphs), in which each vertex corresponds not to the instantaneous state of a hypergraph (i.e. a single spacelike hypersurface), but rather to a complete causal graph (i.e. the complete causal history of an extended region of spacetime). The central objective of this article is to investigate the correspondence between these two, apparently distinct, covariant definitions of entanglement entropy for discrete spacetimes.

In Section \ref{sec:Section1}, we begin by reviewing the standard Dowker-Glaser ansatz for discrete d'Alembertian operators in flat, integer-dimensional spacetimes, outline how these operators may be extended to Riemann normal neighborhoods of more general curved spacetimes, and illustrate how the corresponding solutions to the Klein-Gordon equation in integer-dimensions (and hence the corresponding advanced and retarded Green's functions) may be derived by means of Fourier analysis. We also describe how Johnston's ``hops and stops'' formalism for summing over causal set chains/paths may be used to derive a highly intuitive form of the (massless) causal set propagator, and show once again how this analysis may be extended to more general Riemann normal neighborhoods in discrete spacetimes via the Ollivier-Ricci curvature construction\cite{ollivier}\cite{eidi} for causal graphs. However, for causal sets that have been constructed algorithmically (for instance via Wolfram model evolution), there is no guarantee that the limiting causal graph will exhibit integer dimensionality. As we show in Section \ref{sec:Section2}, it is therefore necessary to ``analytically continue'' the contributions to the massless causal set Green's functions derived in Section \ref{sec:Section1} as meromorphic functions of a complex parameter $d$, interpreted as the analytic continuation of the number of spacetime dimensions (through a procedure that is directly analogous to the dimensional regularization of Feynman integrals in quantum field theory, as developed by 't Hooft and Veltman\cite{hooft}). In the process, we recapitulate the general definition of Hausdorff dimensionality for arbitrary causal graphs, and argue in favor of its greater suitability for quantifying the limiting dimension of algorithmically-grown causal sets, as compared to other standard dimension estimators for causal sets such as the (generalized) Myrheim-Meyer estimator\cite{myrheim}\cite{meyer}. We also show how the uniqueness of this ``analytic continuation'' of the massless Green's functions follows immediately by virtue of Carlson's theorem. In Section \ref{sec:Section3}, we provide an overview of the axioms that must be obeyed by a family of free, bosonic scalar field operators acting on an arbitrary (bosonic/symmetric) Fock space, and illustrate how the causal set advanced and retarded Green's functions derived within the preceding sections may be used to construct covariant Peierls brackets/Pauli-Jordan operators, whose eigendecompositions can then be used to introduce the Sorkin-Johnston/SJ vacuum states, along with the corresponding Wightman/2-point correlation functions. The resulting causal set field operators provably satisfy the aforementioned axioms, and can be used to construct much of the mathematical apparatus of a typical algebraic quantum field theory, including Feynman propagators and creation and annihilation operators. We show how the resulting algebraic quantum field theory may be used to construct a rigorous notion of entanglement entropy on spacetimes with countable degrees of freedom, via an inductive construction based on the techniques of Sorkin. We validate this construction numerically by applying the resulting algorithm(s) to 2000 randomly-generated causal sets (sprinkled into diamond-shaped regions of 1+1-dimensional Minkowski space), and show that we reproduce the expected spatial ``area'' and ``volume'' laws obtained previously by Sorkin and Yazdi\cite{sorkin7}, depending upon whether or not we impose the requisite truncations on the spectrum of the discrete Pauli-Jordan operator.

In Section \ref{sec:Section4}, we show how the conventionally non-covariant definition of entanglement entropy in Wolfram model/hypergraph rewriting systems, based on the Fubini-Study metric on branchial graphs, may be modified to yield a manifestly covariant extension based on \textit{causal multiway systems} (thus yielding branchial graphs whose constituent eigenstates correspond not to instantaneous states of spacelike hypersurfaces, as in the conventional case, but rather to states of extended spacetime regions). We prove that the resulting branchial metric satisfies the expected axioms of an entanglement monotone on spacetime (in particular, it is monotonically-decreasing under local unitary transformations, is equal to zero for fully-separable states and attains its maximum value for maximally-entangled states), and in fact corresponds to a special case of the entanglement measure for finite-dimensional hybrid quantum systems recently proposed by Cocchiarella et al\cite{cocchiarella}. We then present substantive numerical evidence for the expected monotonic relationship between the entanglement entropy computed via SJ Wightman functions and the entanglement monotone computed via branchial distances, for a large class of algorithmic causal sets generated via Wolfram model evolution, suggesting that the correspondence may continue to hold even beyond the aforementioned cases for which analytic proofs currently exist. Finally, in Section \ref{sec:Section5}, we outline some of the broader implications of these results, including the possibility of a purely combinatorial definition of spacetime entanglement entropy that generalizes both approaches studied within this article, and the potential interpretations of the ``dimensional regularization'' procedure required to extend the causal set Green's functions to non-integer dimensions. We also discuss future directions for investigation, including extensions to the case of black hole spacetimes (in which it is tempting to conjecture that the spatial ``area'' law obeyed by spacetime entropies under Sorkin's prescription may be the fundamental origin of the semiclassical black hole entropy law of Bekenstein and Hawking\cite{bekenstein}\cite{hawking2}) and other non-trivial geometries, potential implications for ER=EPR\cite{vanraamsdonk}\cite{maldacena} and other conjectural relationships between entanglement entropy and spacetime geometry (especially in discrete spacetimes\cite{shah}), possible non-covariant reformulations of ``localized'' entanglement entropies on hypergraphs (i.e. discrete spacelike hypersurfaces) within the Wolfram model context, and possible reformulations of the largely algebraic constructions presented within this article in the more mathematically elegant language of functorial quantum field theory (noting that, in general, homotopic\cite{arsiwalla}\cite{arsiwalla2} and functorial\cite{gorard8} descriptions of arbitrary multiway systems appear to be more natural than purely algebraic ones). Although the majority of this article is dedicated to the case of massless free scalar field theories, we also discuss some preliminary thoughts regarding the generalization of these methods to the case of massive interacting scalar field theories (in which the state is no longer Gaussian, Wick's theorem is no longer applicable and contributions from higher-order correlators must be considered), for instance by means of perturbation theory.

Note that all of the code necessary to reproduce the results presented within this article is open source, and much of it is fully-documented and freely available through the \textit{Wolfram Function Repository}. For instance, the functions \href{https://resources.wolframcloud.com/FunctionRepository/resources/CausalGraphEntanglementEntropyNaive/}{CausalGraphEntanglementEntropyNaive} and \href{https://resources.wolframcloud.com/FunctionRepository/resources/CausalGraphEntanglementEntropyGeneralized/}{CausalGraphEntanglementEntropyGeneralized} for computing causal set entanglement entropies through either the naive or generalized eigenvalue approaches, the function \href{https://resources.wolframcloud.com/FunctionRepository/resources/MultiwaySystem}{MultiwaySystem} for simulating arbitrary Wolfram model evolutions (and their corresponding multiway evolution graphs, causal graphs, branchial graphs, etc.), the functions \href{https://resources.wolframcloud.com/FunctionRepository/resources/WolframHausdorffDimension/}{WolframHausdorffDimension}, \href{https://resources.wolframcloud.com/FunctionRepository/resources/WolframRicciCurvatureScalar/}{WolframRicciCurvatureScalar}, \href{https://resources.wolframcloud.com/FunctionRepository/resources/WolframRicciCurvatureTensor/}{WolframRicciCurvatureTensor}, etc., for computing discrete dimension and curvature estimates for arbitrary causal graphs, are all exposed in this way. More up-to-date (though not yet fully-documented) code for performing many of the same functions is exposed through the open source \href{https://github.com/JonathanGorard/Gravitas}{\textsc{Gravitas} package} on GitHub. With regards to the structure of the present article, Section \ref{sec:Section1} is partly expository, recapitulating the core aspects of the Dowker-Glaser construction and Johnston's derivations of causal set propagators, albeit with more explicit discussion of the underlying Fourier-analytic techniques, as well as a more complete explanation of the extension to Riemann normal neighborhoods in more general curved spacetimes (via the Ollivier-Ricci curvature construction). The unique ``analytic continuation'' of the causal set Green's functions to non-integer-dimensional discrete spacetimes presented within Section \ref{sec:Section2} is, to the best of our knowledge, original. Much of the material in Section \ref{sec:Section3} is derivative, since this section is largely concerned with the validation of our algorithmic implementation against the prior results of Sorkin and Yazdi for causal sets sprinkled into diamond-shaped regions of ${1 + 1}$-dimensional Minkowski space, using algebraic field theory and eigendecomposition methods due to Johnston. In the process, however, we present a definition of the discrete spacetime entanglement entropy that is more mathematically (and computationally) explicit than any that we have been able to find elsewhere in the literature thus far. The entirety of Section \ref{sec:Section4} is, to the best of our knowledge, original, since it builds upon prior work regarding the formalism of \textit{causal multiway systems} in order to produce a manifestly covariant definition of spacetime entanglement entropy in arbitrary Wolfram model systems, and shows, via a combination of mathematical analysis and explicit numerical simulation, that the resulting definition is monotonically related to the notion of entanglement entropy in discrete spacetimes resulting from the Sorkin-Johnston approach (at least for algorithmically-generated causal sets).

\section{Scalar Field Green's Functions in Discrete, Integer-Dimensional Spacetimes}
\label{sec:Section1}

\begin{figure}[ht]
\centering
\includegraphics[width=0.295\textwidth]{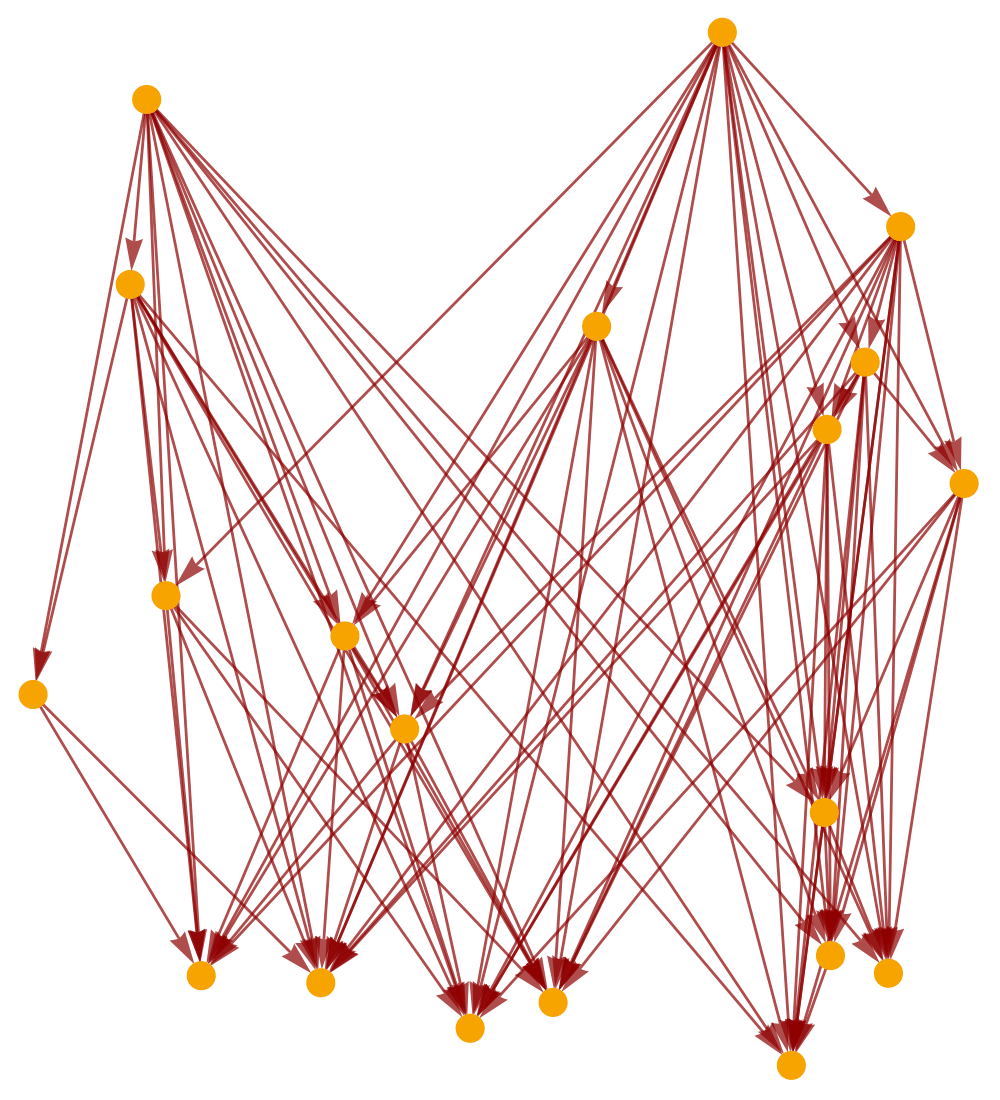}\hspace{0.2\textwidth}
\includegraphics[width=0.295\textwidth]{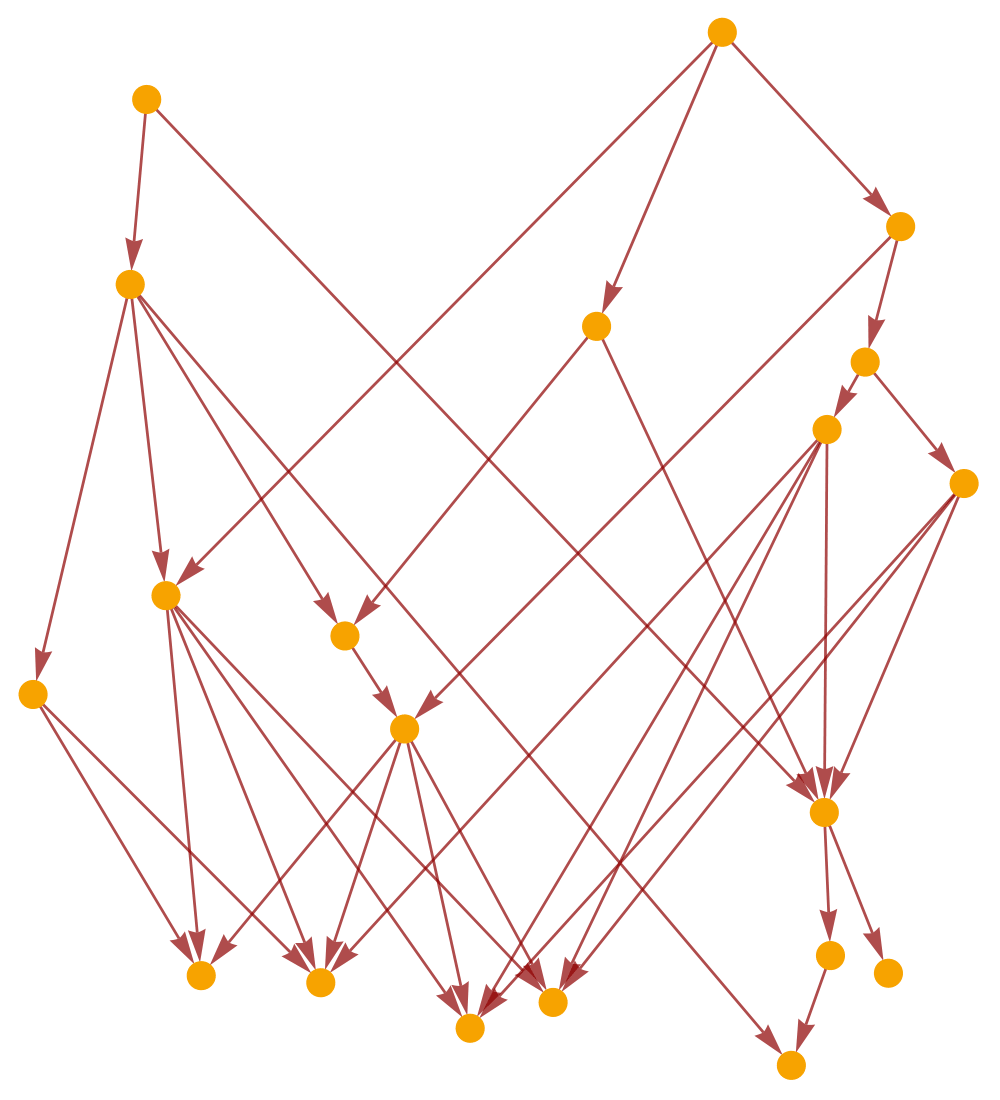}
\caption{On the left, a directed graph generated by connecting all pairs of 20 (uniformly-sprinkled) points that are related by the causal partial order relation ${\prec_{\mathcal{M}}}$ on the rectangular region of ${1 + 1}$-dimensional flat (Minkowski) spacetime. On the right, the transitive reduction (i.e. the Hasse diagram) of this same causal partial order graph.}
\label{fig:Figure1}
\end{figure}

\begin{figure}[ht]
\centering
\includegraphics[width=0.395\textwidth]{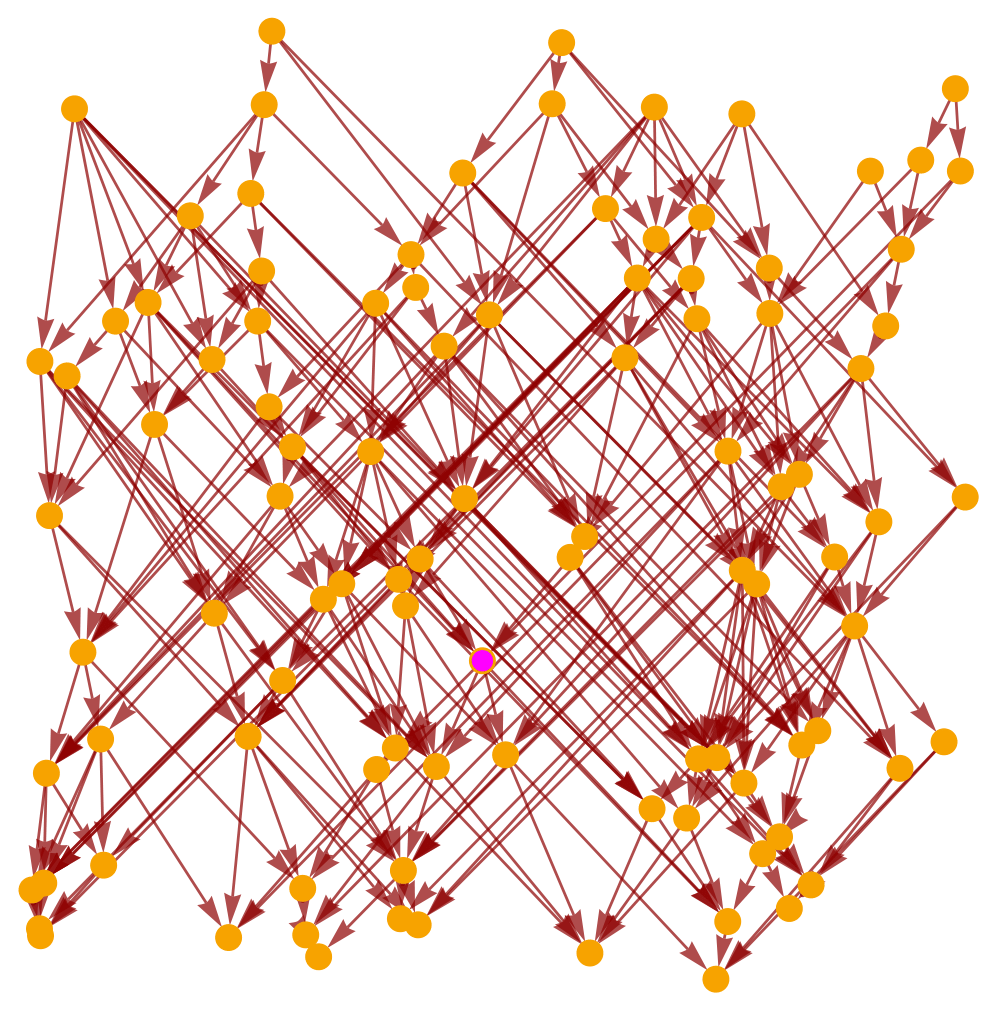}\hspace{0.1\textwidth}
\includegraphics[width=0.395\textwidth]{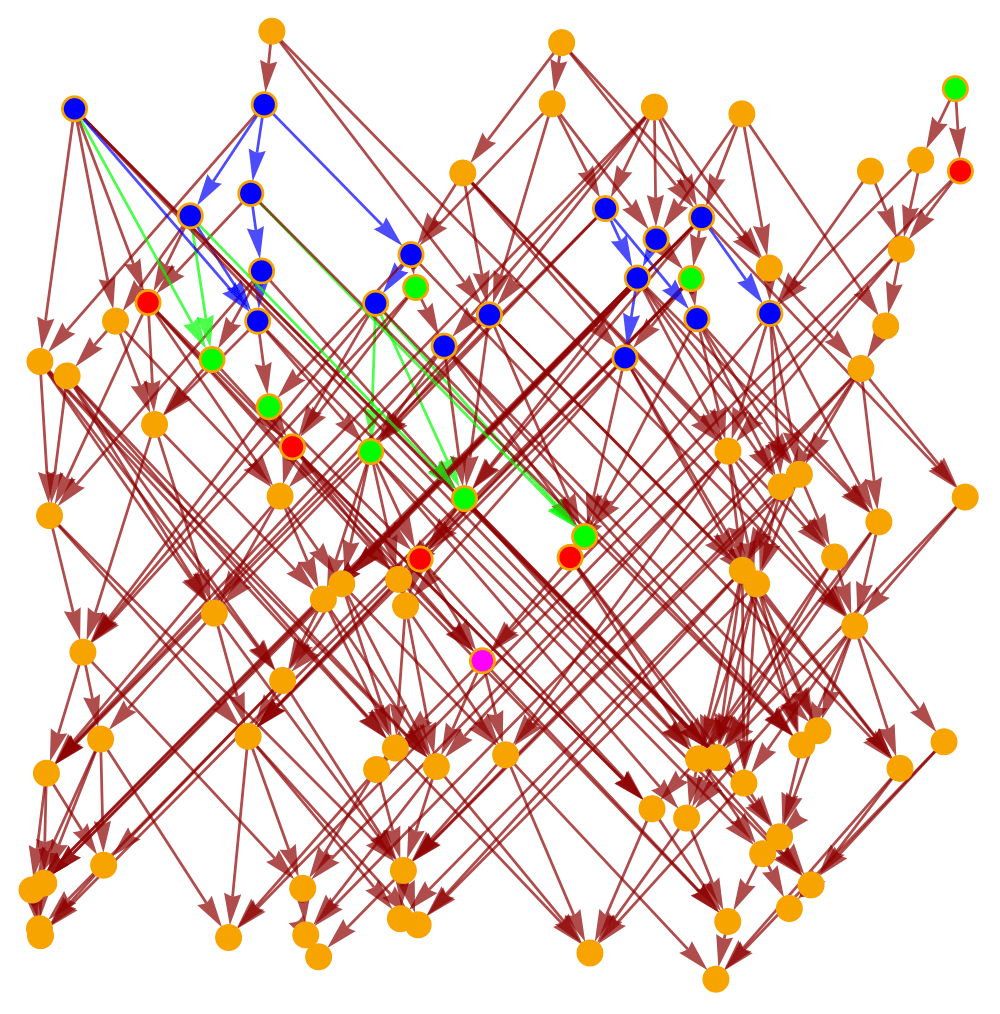}
\caption{On the left, the transitive reduction (i.e. the Hasse diagram) of the directed graph generated by connecting all pairs of 100 (uniformly-sprinkled) points that are related by the causal partial order relation ${\prec_{\mathcal{M}}}$ on the rectangular region of ${1 + 1}$-dimensional flat (Minkowski) spacetime, with the discrete d'Alembertian operator being computed for the point highlighted in magenta. On the right, the ``layer sets'' ${L_0}$, ${L_1}$ and ${L_2}$ of 0-nearest, 1-nearest and 2-nearest neighbors, over which the values of the scalar field ${\phi}$ are summed, are highlighted in red, green and blue, respectively.}
\label{fig:Figure2}
\end{figure}

Recall that, if a causal set ${\mathcal{C}}$ is equipped with a free (real) scalar field ${\phi : \mathcal{C} \to \mathbb{R}}$, then the discrete d'Alembertian operator ${B^{\left( 2 \right)}}$ for causal sets generated via Poisson sprinklings into 2-dimensional flat (Minkowski) spacetimes ${\mathbb{M}^2 = \mathbb{R}^{1, 1}}$ was given by Sorkin\cite{sorkin5} and Henson\cite{henson} to be:

\begin{equation}
B^{\left( 2 \right)} \phi \left( e \right) = \frac{1}{l^2} \left[ - 2 \phi \left( e \right) + 4 \left( \sum_{e^{\prime} \in L_0 \left( e \right)} \phi \left( e^{\prime} \right) - 2 \sum_{e^{\prime} \in L_1 \left( e \right)} \phi \left( e^{\prime} \right) + \sum_{e^{\prime} \in L_2 \left( e \right)} \phi \left( e^{\prime} \right) \right) \right],
\end{equation}
and was later extended to 4-dimensional flat (Minkowski) spacetimes ${\mathbb{M}^4 = \mathbb{R}^{1, 3}}$ by Benincasa and Dowker\cite{benincasa} as:

\begin{multline}
B^{\left( 4 \right)} \phi \left( e \right) = \frac{1}{l^2} \left[ - \frac{4}{\sqrt{6}} \phi \left( e \right) + \frac{4}{\sqrt{6}} \left( \sum_{e^{\prime} \in L_0 \left( e \right)} \phi \left( e^{\prime} \right) - 9 \sum_{e^{\prime} \in L_1 \left( e \right)} \phi \left( e^{\prime} \right) + 16 \sum_{e^{\prime} \in L_2 \left( e \right)} \phi \left( e^{\prime} \right) \right. \right.\\
\left. \left. - 8 \sum_{e^{\prime} \in L_3 \left( e \right)} \phi \left( e^{\prime} \right) \right) \right],
\end{multline}
where $l$ designates the characteristic discretization scale (or ``lattice spacing'') of the sprinkled causal set ${\mathcal{C}}$, given for instance by ${\rho_c = l^{-d}}$ for a causal set with sprinkling density ${\rho_c}$ and dimension $d$, and the ``layer'' sets ${L_k \left( e \right)}$ over which the sums are evaluated correspond to the sets of $k$-nearest neighbors in the causal past of a given event ${e \in \mathcal{C}}$:

\begin{equation}
L_k \left( e \right) = \left\lbrace e^{\prime} \prec e : n \left[ e, e^{\prime} \right] = k \right\rbrace,
\end{equation}
with ${n \left[ p, q \right]}$ being related to the cardinality of the discrete Alexandrov interval ${\mathbf{I} \left[ p, q \right]}$ (i.e. the spacetime order interval) up to an additive constant:

\begin{equation}
n \left[ p, q \right] = \left\lvert \mathbf{I} \left[ p, q \right] \right\rvert - 2, \qquad \text{ where } \qquad \mathbf{I} \left[ p, q \right] = \left\lbrace r \in \mathcal{C} : q \prec r \prec p \right\rbrace.
\end{equation}
The sprinkling density ${\rho_c}$ effectively determines the expectation value of the number of causal set elements ${\hat{n}}$ lying within a spacetime region of volume $v$, namely ${\left\langle \hat{n} \right\rangle = \rho_c v}$, since the sprinkling procedure itself is enacted by performing a Poisson point process in which the probability of $n$ causal set elements lying within a spacetime region of volume $v$ is defined to be precisely:

\begin{equation}
P_v \left( n \right) = \frac{\left( \rho_c v \right)^n}{n!} \exp \left( - \rho_c v \right);
\end{equation}
elements are then connected pairwise in accordance with the partial order relation ${\prec_{\mathcal{M}}}$ in the continuous manifold. An example of a causal set generated via a Poisson sprinkling of 20 points into a rectangular region of ${1 + 1}$-dimensional flat (Minkowski) spacetime is shown in Figure \ref{fig:Figure1}. An example of how the discrete d'Alembertian operator ${B^{\left( d \right)}}$ is computed within a causal set constructed via a Poisson sprinkling of 100 points into a rectangular region of ${1 + 1}$-dimensional flat (Minkowski) spacetime is shown in Figure \ref{fig:Figure2}. As per Dowker and Glaser\cite{dowker}, the following ansatz:

\begin{equation}
B^{\left( d \right)} \phi \left( e \right) = \frac{1}{l^2} \left( \alpha_d \phi \left( e \right) + \beta_d \sum_{i = 0}^{n_d - 1} C_{i}^{\left( d \right)} \sum_{y \in L_i \left( e \right)} \phi \left( y \right) \right),
\end{equation}
may be employed to construct discrete d'Alembertian operators ${B^{\left( d \right)}}$ for more general flat (Minkowksi) spacetimes ${\mathbb{M}^d = \mathbb{R}^{1, d - 1}}$ in arbitrary (integer) numbers of dimensions $d$, given undetermined and dimension-dependent constants ${\alpha_d, \beta_d}$ and ${C_{i}^{\left( d \right)}}$ (for ${i = 0, \dots, n_d - 1}$), with fixed initial coefficient ${C_{0}^{\left( d \right)} = 1}$, and where ${n_d}$ designates the total number of $k$-nearest neighbor layers to sum over.

\begin{figure}[ht]
\centering
\includegraphics[width=0.495\textwidth]{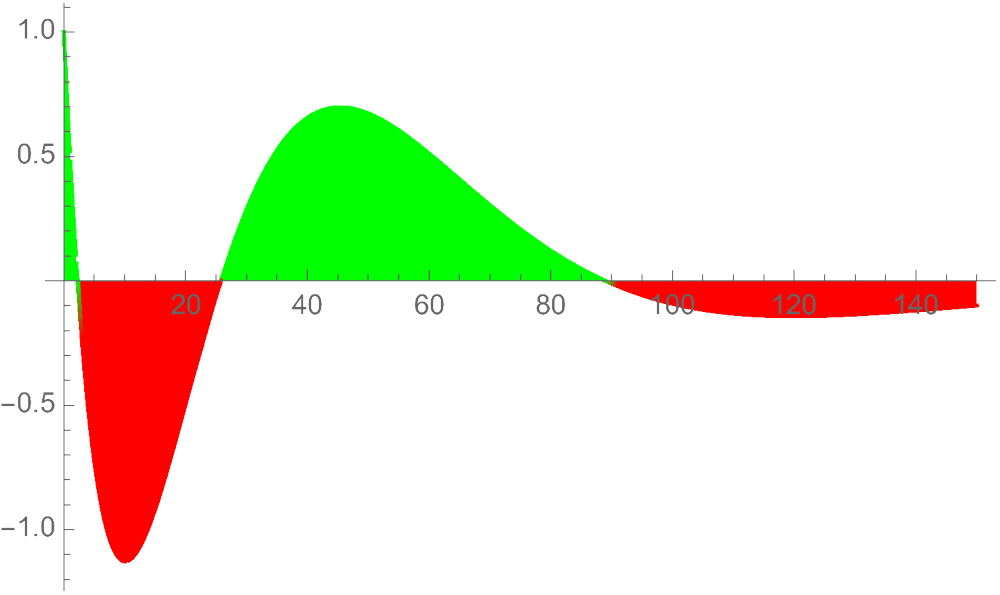}
\caption{The behavior of the 4-dimensional ``smearing'' function ${f \left( n, \varepsilon \right)}$ for ${\varepsilon = 0.05}$ over the interval ${0 \leq n \leq 150}$, clearly exhibiting the four regions of alternating sign (with the two positive regions highlighted in green and the two negative regions highlighted in red) over which the four layers are ``smeared out'' in the computation of the expectation value for the discrete d'Alembertian operator.}
\label{fig:Figure3}
\end{figure}

In fact, we may extend these constructions to any Riemann normal neighborhood ${U_p}$ of a point $p$ in an arbitrarily curved Lorentzian manifold ${\left( \mathcal{M}, g \right)}$ by using the methods of Sorkin\cite{sorkin5}, in which one introduces a ``mesoscopic'' (intermediate) length scale parameter ${l_k > l}$, which has the effect of ``damping'' the microscopic fluctuations in the d'Alembertian operator ${B^{\left( d \right)} \phi \left( x \right)}$, yielding instead a ``smoothing'' discrete operator ${B_k}$, whose expectation value ${\left\langle \hat{B}_{k}^{\left( d \right)} \phi \left( x \right) \right\rangle}$ may be computed by means of the following spacetime volume integral, evaluated over the causal past ${J^{-} \left( x \right)}$ of the element ${x \in \mathcal{C}}$:

\begin{multline}
\left\langle \hat{B}_{k}^{\left( d \right)} \phi \left( x \right) \right\rangle = \alpha_d l_{k}^{-2} \phi \left( x \right)\\
+ \beta_d l_{k}^{- \left( d + 2 \right)} \int_{J^{-} \left( x \right)} \sqrt{-g \left( y \right)} \phi \left( y \right) \sum_{i = 0}^{n_d - 1} C_{i}^{\left( d \right)} \frac{\left( \mathrm{Vol} \left( \mathbf{I} \left( y, x \right] \right) l_{k}^{-d} \right)^{i - 2}}{\Gamma \left( i - 1 \right)} \exp \left( - \mathrm{Vol} \left( \mathbf{I} \left[ y, x \right] \right) l_{k}^{-d} \right) d^d y,
\end{multline}
where ${\mathrm{Vol} \left( \mathbf{I} \left[ x, y \right] \right)}$ denotes the volume of the causal interval ${\mathbf{I} \left[ x, y \right]}$. Here, we have assumed that the scalar field ${\phi : \mathcal{M} \to \mathbb{R}}$ with which the manifold ${\left( \mathcal{M}, g \right)}$ is equipped is of compact support. As an instructive example, consider the ${d = 4}$ case; here, we can introduce a ``smearing'' function ${f \left( n, \epsilon \right)}$ of the form:

\begin{equation}
f \left( n, \varepsilon \right) = \left( 1 - \varepsilon \right)^n \left[ 1 - \frac{9 \varepsilon n}{1 - \varepsilon} + \frac{8 \varepsilon^2 \Gamma \left( n + 1 \right)}{\Gamma \left( n - 1 \right) \left( 1 - \varepsilon \right)^2} - \frac{4 \varepsilon^3 \Gamma \left( n + 1 \right)}{3 \Gamma \left( n - 2 \right) \left( 1 - \varepsilon^3 \right)} \right],
\end{equation}
where ${\varepsilon = \left( \frac{l}{l_k} \right)^4}$ is a constant parameter quantifying the degree of nonlocality present within the causal set. This function $f$ can be generalized naturally to the $d$-dimensional case ${f_d}$ as:

\begin{equation}
f_d \left( n, \varepsilon \right) = \left( 1 - \varepsilon \right)^n \sum_{i = 0}^{n_d} C_{i}^{\left( d \right)} {n \choose i - 1} \left( \frac{\varepsilon}{1 - \epsilon} \right)^{i - 1};
\end{equation}
intuitively, the ``smearing'' operation performed by the function ${f_d}$ works by sampling values of the scalar field ${\phi}$ over ${n_d}$ $k$-nearest neighbor layers (e.g. four, in the ${d = 4}$ case) with alternating sign, and each with depth approximately equal to the characteristic mesoscale length ${l_k}$, as illustrated by the plot of the ${d = 4}$ case of ${f \left( n, \varepsilon \right)}$ shown in Figure \ref{fig:Figure3}. In terms of the smearing function, the expectation value ${\left\langle \hat{B}_{k}^{\left( 4 \right)} \phi \left( x \right) \right\rangle}$ for the discrete d'Alembertian operator ${B_k}$ may be written directly as:

\begin{equation}
\left\langle \hat{B}_{k}^{\left( 4 \right)} \phi \left( x \right) \right\rangle = \frac{4}{\sqrt{6} l_{k}^{2}} \left[ - \phi \left( x \right) + \varepsilon \sum_{y \prec x} f \left( n \left[ x, y \right] + 2, \varepsilon \right) \phi \left( y \right) \right],
\end{equation}
with ${n \left[ x, y \right]}$ being related to the cardinality of the discrete spacetime interval, as described above. More explicitly, the expectation values in the ${d = 2}$ and ${d = 4}$ cases may be written, assuming the presence of non-zero spacetime curvature, in the form of the spacetime volume integrals over causal pasts as:

\begin{equation}
\left\langle \hat{B}_{k}^{\left( 2 \right)} \phi \left( x \right) \right\rangle = \frac{2}{l_{k}^{2}} \left[ - \phi \left( x \right) + \frac{2}{l_{k}^{2}} \int_{y \in J^{-} \left( x \right)} \sqrt{-g} \exp \left( - \xi_2 \right) \left( 1 - 2 \xi_2 + \frac{1}{2} \xi_{2}^{2} \right) \phi \left( y \right) d^2 y \right],
\end{equation}
and:

\begin{equation}
\left\langle \hat{B}_{k}^{\left( 4 \right)} \phi \left( x \right) \right\rangle = \frac{4}{\sqrt{6} l_{k}^{2}} \left[ - \phi \left( x \right) + \frac{1}{l_{k}^{2}} \int_{y \in J^{-} \left( x \right)} \sqrt{-g} \exp \left( - \xi_4 \right) \left( 1 - 9 \xi_4 + 8 \xi_{4}^{2} - \frac{4}{3} \xi_{4}^{3} \right) \phi \left( y \right) d^4 y \right],
\end{equation}
respectively, with the parameters ${\xi_2}$ and ${\xi_4}$ being defined in terms of the 2- and 4-dimensional volumes of the causal intervals ${\mathbf{I} \left[ x, y \right]}$ as:

\begin{equation}
\xi_2 = \mathrm{Vol} \left( \mathbf{I} \left[ x, y \right] \right) l_{k}^{-2}, \qquad \text{ and } \qquad \xi_4 = \mathrm{Vol} \left( \mathbf{I} \left[ x, y \right] \right) l_{k}^{-4},
\end{equation}
respectively.

As we shall now proceed to illustrate, in much the same way as one can derive a continuum Green's function $G$ for the Klein-Gordon equation by simply inverting the continuum d'Alembertian operator ${\Box}$ to obtain ${\Box^{-1}}$, one can equally construct a discrete Green's function by inverting the discrete (smoothed) d'Alembertian operator ${B_k}$ to obtain ${B_{k}^{-1}}$. Recall that, in the continuum case, the d'Alembertian operator ${\Box}$ may be written (assuming natural units ${\hbar = c = 1}$) as a sum of partial derivative operators:

\begin{equation}
\Box = \frac{\partial^2}{\partial \left( x^0 \right)^2} - \frac{\partial^2}{\partial \left( x^1 \right)^2} - \frac{\partial^2}{\partial \left( x^2 \right)^2} - \dots - \frac{\partial^2}{\partial \left( x^{d - 1} \right)^2},
\end{equation}
which is, in turn, related to the massive $d$-dimensional Green's function ${G_{m}^{\left( d \right)} \left( \mathbf{x} \right)}$ (for mass $m$) by means of the Klein-Gordon equation:

\begin{equation}
\left( \Box + m^2 \right) G_{m}^{\left( d \right)} \left( \mathbf{x} \right) = \delta^d \left( \mathbf{x} \right),
\end{equation}
where ${\delta^d}$ designates the Dirac delta function in $d$-dimensional space. For the sake of simplicity, we consider first the case of a flat spacetime ${\mathbb{M}^d = \mathbb{R}^{1, d - 1}}$ (in which one has translation invariance, and hence in which ${G_{m}^{\left( d \right)} \left( \mathbf{y} - \mathbf{x} \right)}$ may consequently be considered to correspond to the transition amplitude between events ${\mathbf{x}}$ and ${\mathbf{y}}$), before extending the analysis to more general spacetimes in due course. Much of the preliminary analysis will follow techniques outlined in Johnston's PhD thesis\cite{johnston3}. Using the modified Fourier transform and inverse Fourier transform operations (assuming a general multivariate function ${f \left( \mathbf{x} \right)}$ and its Fourier-transformed analog ${\tilde{f} \left( \mathbf{p} \right)}$) defined by:

\begin{equation}
\tilde{f} \left( \mathbf{p} \right) = \int_{- \infty}^{\infty} f \left( \mathbf{x} \right) e^{i \mathbf{p} \cdot \mathbf{x}} d^d \mathbf{x}, \qquad \text{ and } \qquad f \left( \mathbf{x} \right) = \frac{1}{\left( 2 \pi \right)^d} \int_{- \infty}^{\infty} \mathbf{p} \tilde{f} \left( \mathbf{p} \right) e^{- i \mathbf{p} \cdot \mathbf{x}} d^d \mathbf{p},
\end{equation}
with the dot product ${\mathbf{p} \cdot \mathbf{x}}$ given by:

\begin{equation}
\mathbf{p} \cdot \mathbf{x} = p_0 x^0 - p_1 x^1 - p_2 x^2 - \dots - p_{d - 1} x^{d - 1},
\end{equation}
we are able trivially to obtain the following solution to the massive Klein-Gordon equation:

\begin{equation}
\tilde{G}_{m}^{\left( d \right)} \left( \mathbf{p} \right) = - \frac{1}{p_{0}^{2} - p_{1}^{2} - p_{2}^{2} - \dots - p_{d - 1}^{2} - m^2},
\end{equation}
such that:

\begin{equation}
G_{m}^{\left( d \right)} \left( \mathbf{x} \right) = - \frac{1}{\left( 2 \pi \right)^d} \int_{- \infty}^{\infty} \frac{e^{- i \mathbf{p} \cdot \mathbf{x}}}{p_{0}^{2} - p_{1}^{2} - p_{2}^{2} - \dots - p_{d - 1}^{2} - m^2} d^d \mathbf{p}.
\end{equation}
The poles that appear in the integrand, namely at:

\begin{equation}
p_0 = \pm \sqrt{p_{1}^{2} + p_{2}^{2} + \dots + p_{d - 1}^{2} + m^2},
\end{equation}
may be avoided if one simply chooses an integration contour such that the Green's function ${G_{m}^{\left( d \right)}}$ is constrained to be non-zero only inside the future light cone of the event ${\mathbf{x}}$, thus yielding the \textit{retarded propagator} ${\left( G_R \right)_{m}^{\left( d \right)} \left( \mathbf{x} \right)}$ by means of the following (distributional) limit:

\begin{equation}
\left( G_R \right)_{m}^{\left( d \right)} \left( \mathbf{x} \right) = \lim_{\epsilon \to 0^{+}} \left[ - \frac{1}{\left( 2 \pi \right)^d} \int_{- \infty}^{\infty} \frac{e^{- i \mathbf{p} \cdot \mathbf{x}}}{\left( p_0 + i \epsilon \right)^2 - p_{1}^{2} - p_{2}^{2} - \dots - p_{d -1}^{2} - m^2} d^d \mathbf{p} \right];
\end{equation}
the corresponding \textit{advanced propagator} ${\left( G_A \right)_{m}^{\left( d \right)} \left( \mathbf{x} \right)}$ may, dually, be obtained by constraining the Green's function ${G_{m}^{\left( d \right)}}$ to be non-zero only inside the \text{past} light cone of the event ${\mathbf{x}}$.

This integral may be evaluated straightforwardly for general $d$ using elementary Fourier-analytic techniques (the general mathematical theory can be found in the work of Gel'fand and Shilov\cite{gelfand} and Egorov and Shubin\cite{egorov}, with the 4-dimensional/physical case analyzed rigorously by Bogoliubov and Shirkov\cite{bogoliubov} and de Jager\cite{dejager}); for instance, the cases ${d = 1}$, ${d = 2}$, ${d = 3}$ and ${d = 4}$ yield retarded Green's functions of the form:

\begin{equation}
\left( G_R \right)_{m}^{\left( 1 \right)} \left( x \right) = \theta \left( x \right) \frac{\sin \left( m x \right)}{m}, \qquad \left( G_R \right)_{m}^{\left( 2 \right)} \left( \mathbf{x} \right) = \theta \left( x^0 \right) \theta \left( \tau^2 \right) \frac{1}{2} J_0 \left( m \tau \right),
\end{equation}
\begin{equation}
\left( G_R \right)_{m}^{\left( 3 \right)} \left( \mathbf{x} \right) = \theta \left( x^0 \right) \theta \left( \tau^2 \right) \frac{1}{2 \pi} \frac{\cos \left( m \tau \right)}{\tau}, \qquad \left( G_R \right)_{m}^{\left( 4 \right)} \left( \mathbf{x} \right) = \theta \left( x^0 \right) \theta \left( \tau^2 \right) \left( \frac{1}{2 \pi} \delta \left( \tau^2 \right) - \frac{m}{4 \pi} \frac{J_1 \left( m \tau \right)}{\tau} \right),
\end{equation}
respectively, where ${\theta \left( \alpha \right)}$ denotes the usual Heaviside step function:

\begin{equation}
\theta \left( \alpha \right) = \begin{cases}
1, \qquad & \text{ if } \alpha \geq 0,\\
0, \qquad & \text{ if } \alpha < 0,
\end{cases}
\end{equation}
${\tau}$ denotes the proper time length of the vector ${\mathbf{x}}$:

\begin{equation}
\tau = \sqrt{\left( x^0 \right)^2 - \left( x^1 \right)^2 - \left( x^2 \right)^2 - \dots - \left( x^{d - 1} \right)^2},
\end{equation}
the ${J_{\alpha} \left( x \right)}$ are Bessel functions of the first kind of order ${\alpha}$, defined traditionally in terms of their series expansions around ${x = 0}$:

\begin{equation}
J_{\alpha} \left( x \right) = \sum_{m = 0}^{\infty} \frac{\left( -1 \right)^m}{m! \Gamma \left( m + \alpha + 1 \right)} \left( \frac{x}{2} \right)^{2m + \alpha},
\end{equation}
and ${\delta}$ is the conventional (1-dimensional) Dirac delta function. Consider, as an illustrative example of how the Fourier analysis goes, the calculation for the familiar ${d = 4}$ case (following the approach of de Jager\cite{dejager}). We begin by noting that, in this case, the modified Fourier transform operator defined above, henceforth denoted ${F^{*}}$:

\begin{equation}
F^{*} \left[ f \left( \mathbf{x} \right) \right] = \tilde{f} \left( \mathbf{p} \right) = \int_{- \infty}^{\infty} f \left( \mathbf{x} \right) e^{i \mathbf{p} \cdot \mathbf{x}} d^d \mathbf{x},
\end{equation}
is related to the standard Fourier transform operator $F$:

\begin{equation}
F \left[ f \left( \mathbf{p} \right) \right] = \int_{- \infty}^{\infty} f \left( \mathbf{p} \right) e^{i \left( x_0 p_0 + x_1 p_1 + x_2 p_2 + x_3 p_3 \right)} d^d \mathbf{p},
\end{equation}
by the identity:

\begin{equation}
F F^{*} \left[ f \left( x_0, x_1, x_2, x_3 \right) \right] = \left( 2 \pi \right)^4 f \left( - x_0, x_1, x_2, x_3 \right),
\end{equation}
which holds for all integral functions \textit{and} for all distributions. Applying the modified Fourier operator ${F^{*}}$ to our original Klein-Gordon equation:

\begin{equation}
\left( \Box + m^2 \right) G_{m}^{\left( 4 \right)} \left( \mathbf{x} \right) = \delta^4 \left( \mathbf{x} \right),
\end{equation}
we obtain:

\begin{equation}
\left( \mathbf{p}^2 - m^2 \right) \tilde{G}_{m}^{\left( 4 \right)} \left( \mathbf{p} \right) = 1.
\end{equation}
Following de Jager, since the modified Fourier operator ${F^{*}}$ and its inverse ${\left( F^{*} \right)^{-1}}$ are guaranteed to preserve Lorentz invariance, it suffices to determine all of the solutions of this equation which satisfy Lorentz invariance, and then to transform the results back into configuration space.

A particular solution ${\tilde{\left( G_p \right)}_{m}^{\left( 4 \right)} \left( \mathbf{p} \right)}$ to this inhomogeneous equation which is known to satisfy Lorentz invariance is given by:

\begin{equation}
\tilde{\left( G_p \right)}_{m}^{\left( 4 \right)} \left( \mathbf{p} \right) = \frac{1}{\mathbf{p}^2 - m^2},
\end{equation}
where, for our present purposes, ${\left( \mathbf{p}^2 - m^2 \right)^{-1}}$ is a distribution defined in terms of the Cauchy principal value of the following improper integral:

\begin{equation}
\left\langle \frac{1}{\mathbf{p}^2 - m^2}, \Phi \left( \mathbf{p} \right) \right\rangle = \lim_{\epsilon \to 0^{+}} \left[ \int_{\left\lvert \mathbf{p}^2 - m^2 \right\rvert > \epsilon} \frac{\tilde{\Phi} \left( \mathbf{p} \right)}{\mathbf{p}^2 - m^2} d^d \mathbf{p} \right],
\end{equation}
where ${\Phi \left( \mathbf{p} \right)}$ here denotes an arbitrary test function. On the other hand, the general solution ${\tilde{\left( G_H \right)}_{m}^{\left( 4 \right)} \left( \mathbf{p} \right)}$ to the corresponding homogeneous equation:

\begin{equation}
\left( \mathbf{p}^2 - m^2 \right) \tilde{\left( G_H \right)}_{m}^{\left( 4 \right)} \left( \mathbf{p} \right) = 0,
\end{equation}
which lies strictly on the surface of the hyperboloid ${\mathbf{p}^2 - m^2 = 0}$, is given by:

\begin{equation}
\tilde{\left( G_H \right)}_{m}^{\left( 4 \right)} \left( \mathbf{p} \right) = c_{+} \delta_{+} \left( \mathbf{p}^2 - m^2 \right) + c_{-} \delta_{-} \left( \mathbf{p}^2 - m^2 \right),
\end{equation}
where ${c_{+}}$ and ${c_{-}}$ are arbitrary numerical constants, and ${\delta_{\pm} \left( \mathbf{p}^2 - m^2 \right)}$ are distributions defined over the upper and lower sheets of the hyperboloid ${\mathbf{p}^2 - m^2 = 0}$, respectively, defined by means of the following nested integral:

\begin{equation}
\left\langle \delta_{\pm} \left( \mathbf{p}^2 - m^2 \right), \hat{\Phi} \left( \mathbf{p} \right) \right\rangle = \frac{1}{2} \int_{0}^{\infty} \int_{\Omega} \frac{1}{\sqrt{\kappa^2 + m^2}} \kappa^2 \hat{\Phi} \left( \kappa \omega_1, \kappa \omega_2, \kappa \omega_3, \pm \sqrt{\kappa^2 + m^2} \right) d \Omega d \kappa,
\end{equation}
in which we have introduced the parameter ${\kappa = p_{1}^{2} + p_{2}^{2} + p_{3}^{2}}$, and we are integrating over the unit sphere ${\Omega}$ in ${\left( p_1, p_2, p_3 \right)}$-space (with surface measure ${d \Omega}$). By introducing a new function ${\bar{\hat{\Phi}} \left( \kappa, p_0 \right)}$ and defining it to be equal to the mean value of the test function ${\hat{\Phi} \left( \mathbf{p} \right)}$ over a sphere in ${\mathbb{R}^3}$ with radius ${\kappa}$ (up to some additive constant), we can rewrite this nested integral as:

\begin{multline}
\frac{1}{2} \int_{0}^{\infty} \int_{\Omega} \frac{1}{\sqrt{ \kappa^2 + m^2}} \kappa^2 \hat{\Phi} \left( \kappa \omega_1, \kappa \omega_2, \kappa \omega_3, \pm \sqrt{\kappa^2 + m^2} \right) d \Omega d \kappa\\
= \frac{1}{2} \int_{0}^{\infty} \frac{1}{\sqrt{\kappa^2 + m^2}} \kappa^2 \bar{\hat{\Phi}} \left( \kappa, \pm \sqrt{\kappa^2 + m^2} \right) d \kappa.
\end{multline}
The sum of the two distributions ${\delta_{\pm} \left( \mathbf{p}^2 - m^2 \right)}$ defined on the two sheets of the hyperboloid yields the usual Dirac delta function:

\begin{equation}
\delta \left( \mathbf{p}^2 - m^2 \right) = \delta_{+} \left( \mathbf{p}^2 - m^2 \right) + \delta_{-} \left( \mathbf{p}^2 - m^2 \right).
\end{equation}
Therefore, by taking a sum of the particular inhomogeneous solution and the general homogeneous solution, we obtain the following general solution to the inhomogeneous (Fourier-transformed) Klein-Gordon equation:

\begin{equation}
\tilde{G}_{m}^{\left( 4 \right)} \left( \mathbf{p} \right) = \frac{1}{\mathbf{p}^2 - m^2} + c_{+} \delta_{+} \left( \mathbf{p}^2 - m^2 \right) + c_{-} \delta_{-} \left( \mathbf{p}^2 - m^2 \right),
\end{equation}
and hence, to find the general solution ${G_{m}^{\left( 4 \right)} \left( \mathbf{p} \right)}$ to the Klein-Gordon equation in configuration space, it suffices to compute the inverse Fourier transforms ${\left( F^{*} \right)^{-1}}$ of the distributions ${\left( \mathbf{p}^2 - m^2 \right)^{-1}}$ and ${\delta_{\pm} \left( \mathbf{p^2} - m^2 \right)}$.

We also know, as a consequence of the aforementioned relation:

\begin{equation}
F F^{*} \left[ f \left( x_0, x_1, x_2, x_3 \right) \right] = \left( 2 \pi \right)^4 f \left( - x_0, x_1, x_2, x_3 \right),
\end{equation}
that computing the inverse modified Fourier transform ${\left( F^{*} \right)^{-1}}$ is formally equivalent to computing the ordinary Fourier transform $F$, then applying the (time) coordinate transformation ${x_0 \to - x_0}$, and finally dividing the resulting function by ${\left( 2 \pi \right)^4}$. The ordinary Fourier transform of the distribution ${\left( \mathbf{p}^2 - m^2 \right)^{-1}}$ is given by:

\begin{equation}
F \left[ \frac{1}{\mathbf{p}^2 - m^2} \right] = 2 \pi^2 i m \left[ \frac{K_1 \left( m \sqrt{- \tau^2 + 0 i} \right)}{\sqrt{- \tau^2 + 0 i}} - \frac{K_1 \left( m \sqrt{- \tau^2 - 0 i} \right)}{\sqrt{- \tau^2 - 0 i}} \right],
\end{equation}
whereas for the usual Dirac delta distribution ${\delta \left( \mathbf{p}^2 - m^2 \right)}$ it is given by:

\begin{equation}
F \left[ \delta \left( \mathbf{p}^2 - m^2 \right) \right] = 2 \pi m \left[ \frac{K_1 \left( m \sqrt{- \tau^2 + 0 i} \right)}{\sqrt{- \tau^2 + 0 i}} + \frac{K_1 \left( m \sqrt{- \tau^2 - 0 i} \right)}{\sqrt{- \tau^2 - 0 i}} \right],
\end{equation}
where, in the above, ${\tau^2 = x_{0}^{2} - x_{1}^{2} - x_{2}^{2} - x_{3}^{2}}$ as previously, and ${K_1 \left( x \right)}$ corresponds to a modified Bessel function of the second kind (of order 1), defined for general order ${\alpha}$ by:

\begin{equation}
K_{\alpha} \left( x \right) = \frac{\pi}{2} \frac{I_{- \alpha} \left( x \right) - I_{\alpha} \left( x \right)}{\sin \left( \alpha \pi \right)},
\end{equation}
with ${I_{\alpha} \left( x \right)}$ being a modified Bessel function of the first kind (of order ${\alpha}$):

\begin{equation}
I_{\alpha} \left( x \right) = i^{- \alpha} J_{\alpha} \left( i x \right) = \sum_{m = 0}^{\infty} \frac{1}{m! \Gamma \left( m + \alpha + 1 \right)} \left( \frac{x}{2} \right)^{2m + \alpha}.
\end{equation}
Moreover, the ordinary Fourier transforms of the distributions ${\delta_{\pm} \left( \mathbf{p}^2 - m^2 \right)}$ are given by:

\begin{multline}
F \left[ \delta_{+} \left( \mathbf{p}^2 - m^2 \right) \right] = 2 i \pi^2 \left[ \delta_{+} \left( \tau^2 \right) - \delta_{-} \left( \tau^2 \right) \right] + 2 \pi m \theta \left( - \tau^2 \right) \frac{K_1 \left( m \sqrt{- \tau^2} \right)}{\sqrt{- \tau^2}}\\
- i \pi^2 m \theta \left( \tau^2 \right) \left[ \theta \left( x_0 \right) \frac{H_{1}^{\left( 1 \right)} \left( m \sqrt{\tau^2} \right)}{\sqrt{\tau^2}} - \theta \left( - x_0 \right) \frac{H_{1}^{\left( 2 \right)} \left( m \sqrt{\tau^2} \right)}{\sqrt{\tau^2}} \right],
\end{multline}
and:

\begin{multline}
F \left[ \delta_{-} \left( \mathbf{p}^2 - m^2 \right) \right] = - 2 i \pi^2 \left[ \delta_{+} \left( \tau^2 \right) - \delta_{-} \left( \tau^2 \right) \right] + 2 \pi m \theta \left( - \tau^2 \right) \frac{K_1 \left( m \sqrt{- \tau^2} \right)}{\sqrt{- \tau^2}}\\
+ i \pi^2 m \theta \left( \tau^2 \right) \left[ \theta \left( x_0 \right) \frac{H_{1}^{\left( 2 \right)} \left( m \sqrt{\tau^2} \right)}{\sqrt{\tau^2}} - \theta \left( - x_0 \right) \frac{H_{1}^{\left( 1 \right)} \left( m \sqrt{\tau^2} \right)}{\sqrt{\tau^2}} \right],
\end{multline}
respectively, where, in the above, ${H_{1}^{\left( 1 \right)} \left( x \right)}$ and ${H_{1}^{\left( 2 \right)} \left( x \right)}$ correspond to Hankel functions of the first and second kind (of order 1), defined for general order ${\alpha}$ by:

\begin{equation}
H_{\alpha}^{\left( 1 \right)} \left( x \right) = J_{\alpha} \left( x \right) + i Y_{\alpha} \left( x \right) = \frac{J_{- \alpha} \left( x \right) - e^{- \alpha \pi i} J_{\alpha} \left( x \right)}{i \sin \left( \alpha \pi \right)},
\end{equation}
and:

\begin{equation}
H_{\alpha}^{\left( 2 \right)} \left( x \right) = J_{\alpha} \left( x \right) - i Y_{\alpha} \left( x \right) = \frac{J_{- \alpha} \left( x \right) - e^{\alpha \pi i} J_{\alpha} \left( x \right)}{-i \sin \left( \alpha \pi \right)},
\end{equation}
respectively, with ${Y_{\alpha} \left( x \right)}$ being a Bessel function of the second kind (of order ${\alpha}$):

\begin{equation}
Y_{\alpha} \left( x \right) = \frac{J_{\alpha} \left( x \right) \cos \left( \alpha \pi \right) - J_{- \alpha} \left( x \right)}{\sin \left( \alpha \pi \right)}.
\end{equation}

From here, we are finally in a position to complete the calculation of ${\left( G_R \right)_{m}^{\left( 4 \right)} \left( \mathbf{x} \right)}$, by first setting the constants ${c_{+} = c_{-} = \pi i}$ and ${c_{+} = c_{-} = - \pi i}$ to obtain the causal Green's function ${\left( G_C \right)_{m}^{\left( 4 \right)} \left( \mathbf{x} \right)}$:

\begin{equation}
\left( G_C \right)_{m}^{\left( 4 \right)} \left( \mathbf{x} \right) = \frac{\delta \left( \tau^2 \right)}{4 \pi} - \frac{m}{8 \pi} \theta \left( \tau^2 \right) \frac{H_{1}^{\left( 2 \right)} \left( m \sqrt{\tau^2} \right)}{\sqrt{\tau^2}} + \frac{i m}{4 \pi^2} \theta \left( - \tau^2 \right) \frac{K_1 \left( m \sqrt{- \tau^2} \right)}{\sqrt{- \tau^2}},
\end{equation}
and the anticausal Green's function ${\left( G_{AC} \right)_{m}^{\left( 4 \right)} \left( \mathbf{x} \right)}$:

\begin{equation}
\left( G_{AC} \right)_{m}^{\left( 4 \right)} \left( \mathbf{x} \right) = \frac{\delta \left( \tau^2 \right)}{4 \pi} - \frac{m}{8 \pi} \theta \left( \tau^2 \right) \frac{H_{1}^{\left( 1 \right)} \left( m \sqrt{\tau^2} \right)}{\sqrt{\tau^2}} - \frac{i m}{4 \pi^2} \theta \left( - \tau^2 \right) \frac{K_1 \left( m \sqrt{- \tau^2} \right)}{\sqrt{- \tau^2}},
\end{equation}
respectively. Conversely, by setting the constants ${c_{+} = - c_{-} = \pi i}$ and ${c_{=} = - c_{-} = - \pi i}$, one obtains instead the retarded Green's function ${\left( G_R \right)_{m}^{\left( 4 \right)} \left( \mathbf{x} \right)}$:

\begin{equation}
\left( G_R \right)_{m}^{\left( 4 \right)} \left( \mathbf{x} \right) = \begin{cases}
\frac{1}{2 \pi} \delta_{+} \left( \tau^2 \right) - \frac{m}{4 \pi} \theta \left( \tau^2 \right) \frac{J_1 \left( m \sqrt{\tau^2} \right)}{\sqrt{\tau^2}}, \qquad & \text{ if } x_0 > 0,\\
0, \qquad & \text{ if } x_0 < 0,
\end{cases}
\end{equation}
as well as, for the sake of completeness, the advanced Green's function ${\left( G_A \right)_{m}^{\left( 4 \right)} \left( \mathbf{x} \right)}$:

\begin{equation}
\left( G_A \right)_{m}^{\left( 4 \right)} \left( \mathbf{x} \right) = \begin{cases}
0, \qquad & \text{ if } x_0 > 0,\\
\frac{1}{2 \pi} \delta_{-} \left( \tau^2 \right) - \frac{m}{4 \pi} \theta \left( \tau^2 \right) \frac{J_1 \left( m \sqrt{\tau^2} \right)}{\sqrt{\tau^2}}, \qquad & \text{ if } x_0 < 0,
\end{cases}
\end{equation}
respectively, as required. In general, the advanced and retarded propagators differ only in terms of which boundary conditions are enforced (i.e. whether they are constrained to be non-zero only in the future light cone or only in the past light cone), meaning that they can be related by means of the following identity:

\begin{equation}
\left( G_A \right)_{m}^{\left( d \right)} \left( \mathbf{x} \right) = \left( G_R \right)_{m}^{\left( d \right)} \left( - \mathbf{x} \right).
\end{equation}
The calculation for more general values of $d$ may be performed analogously (at least for the case of massless propagators), with only slight modifications depending upon whether $d$ is even or odd, as outlined below.

The massless retarded Green's functions ${\left( G_R \right)_{0}^{\left( d \right)} \left( \mathbf{x} \right)}$ may be constructed by simply evaluating the ${m \to 0}$ limit of the massive functions ${\left( G_R \right)_{m}^{\left( d \right)} \left( \mathbf{x} \right)}$, yielding (in the ${d = 1}$, ${d = 2}$, ${d = 3}$ and ${d = 4}$ cases considered previously):

\begin{equation}
\left( G_R \right)_{0}^{\left( 1 \right)} \left( x \right) = \lim_{m \to 0} \left[ \left( G_R \right)_{m}^{\left( 1 \right)} \left( x \right) \right] = \theta \left( x \right) x, \qquad \left( G_R \right)_{0}^{\left( 2 \right)} \left( \mathbf{x} \right) = \lim_{m \to 0} \left[ \left( G_R \right)_{m}^{\left( 2 \right)} \left( \mathbf{x} \right) \right] = \theta \left( x^0 \right) \theta \left( \tau^2 \right) \frac{1}{2},
\end{equation}
\begin{equation}
\left( G_R \right)_{0}^{\left( 3 \right)} \left( \mathbf{x} \right) = \lim_{m \to 0} \left[ \left( G_R \right)_{m}^{\left( 3 \right)} \left( \mathbf{x} \right) \right] = \theta \left( x^0 \right) \theta \left( \tau^2 \right) \frac{1}{2 \pi \tau},
\end{equation}
and:

\begin{equation}
\left( G_R \right)_{0}^{\left( 4 \right)} \left( \mathbf{x} \right) = \lim_{m \to 0} \left[ \left( G_R \right)_{m}^{\left( 4 \right)} \left( \mathbf{x} \right) \right] = \theta \left( x^0 \right) \theta \left( \tau^2 \right) \frac{1}{2 \pi} \delta \left( \tau^2 \right),
\end{equation}
respectively. As evidenced by the analysis presented above, every massless retarded Green's function, in any number of spacetime dimensions $d$, is given by a product of arbitrary-order distributional derivatives of either ${\delta \left( \tau^2 \right)}$ (for the case in which $d$ is even) or ${\frac{1}{\tau}}$ (for the case in which $d$ is odd), where the arbitrary-order distributional derivatives of ${\delta \left( \tau^2 \right)}$ may themselves be written purely as products of ${\delta \left( \tau^2 \right)}$ and ${\frac{1}{\tau}}$. Thus, the extension of the analysis into arbitrary (integer) numbers of spacetime dimensions involves merely ascertaining appropriate values to assign to the weights that appear within these products.

\begin{figure}[ht]
\centering
\includegraphics[width=0.325\textwidth]{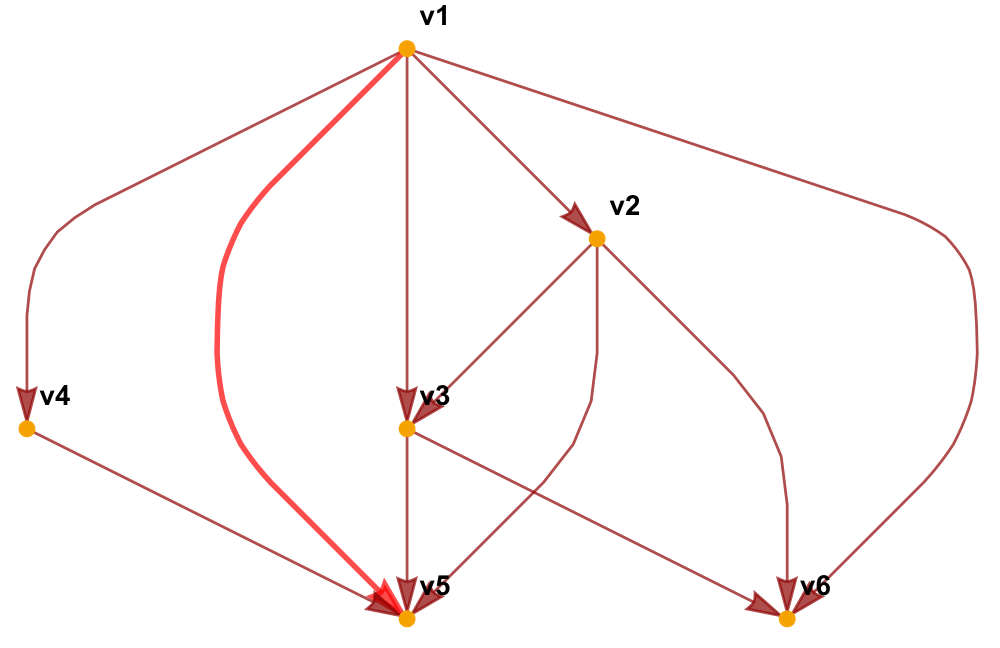}
\includegraphics[width=0.325\textwidth]{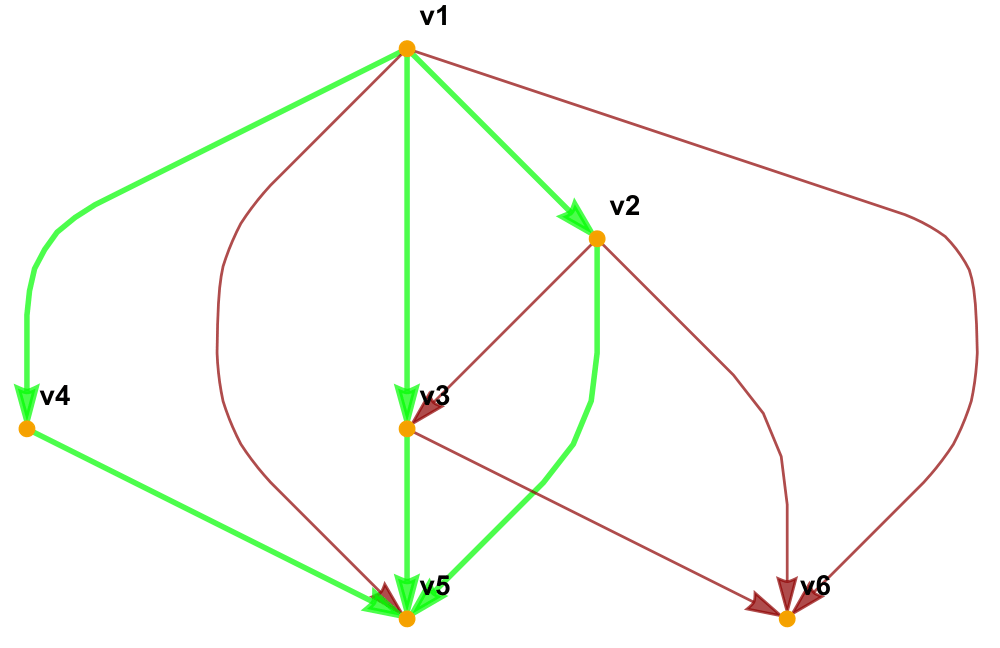}
\includegraphics[width=0.325\textwidth]{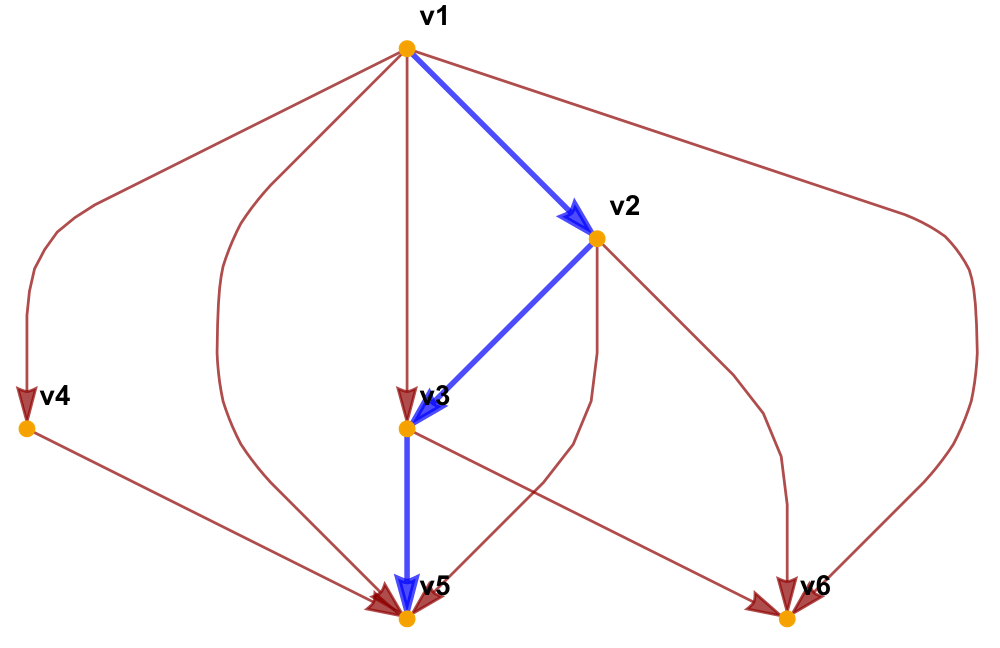}
\caption{The calculation of the overall amplitude for chains connecting elements ${v_1}$ and ${v_5}$ in a 6-element causal set, by means of the discrete propagator matrix ${K \left( 1, 5 \right)}$. On the left, the contribution to the amplitude from the 1 chain of length 1 (highlighted in red) is $a$. In the middle, the contribution to the amplitude from the 3 chains of length 2 (highlighted in green) is ${3 a^2 b}$. On the right, the contribution to the amplitude from the 1 chain of length 3 (highlighted in blue) is ${a^3 b^2}$.}
\label{fig:Figure4}
\end{figure}

Henceforth, we choose to adopt Johnston's ``hops and stops'' formalism\cite{johnston3} for constructing and evaluating discrete path integrals over causal sets, in which there is freedom regarding whether one wishes to compute the path integral as a sum over \textit{chains} or over \textit{paths}. Recall that a \textit{chain} of length $k$ is simply a sequence of causal set elements of the form ${e_1 \prec e_2 \prec \cdots \prec e_{k - 1} \prec e_k}$, for some arbitrary $k$, while a \textit{path} of length $k$ is a sequence of causal set elements related by \textit{links} of the form ${e_1 \prec^{*} e_2 \prec^{*} \cdots \prec^{*} e_{k - 1} \prec^{*} e_k}$, where ${\prec^{*}}$ denotes the link relation, and where a \textit{link} ${e \prec^{*} e^{\prime}}$ in the causal set ${\mathcal{C}}$ is a pair of elements ${e, e^{\prime} \in \mathcal{C}}$ such that ${e \prec e^{\prime}}$ and:

\begin{equation}
\nexists e^{\prime \prime} \in \mathcal{C}, \qquad \text{ such that } \qquad e^{\prime \prime} \neq e, e^{\prime \prime} \neq e^{\prime} \text{ and } e \prec e^{\prime \prime} \prec e^{\prime},
\end{equation}
i.e. the links ${e \prec^{*} e^{\prime}}$ form precisely the directed edges in the corresponding Hasse diagram for ${\mathcal{C}}$. As an initial approximation, we proceed on the assumption that each ``hop'' (i.e. each traversal from one element to another) is associated with a constant amplitude $a$, and each ``stop'' (i.e. each intermediate element reached within the chain/path) is also associated with a constant amplitude $b$, such that the overall amplitude for a chain or path of length $n$ is given by the product ${a^n b^{n - 1}}$ (since there will exist exactly $n$ ``hops'' and ${n - 1}$ ``stops'' along this chain/path). In order to define the discrete propagator for a finite causal set ${\mathcal{C}}$ of cardinality ${\left\lvert \mathcal{C} \right\rvert = p}$, one must introduce a ${p \times p}$ matrix ${\Phi \left( x, y \right)}$ which is related either to the ${p \times p}$ \textit{causal matrix} ${C_0 \left( x, y \right)}$ on ${\mathcal{C}}$ (for the case in which one is summing over chains):

\begin{equation}
\Phi \left( x, y \right) = a C_0 \left( x, y \right), \qquad \text{ with } \qquad C_0 \left( x, y \right) = \begin{cases}
1, \qquad & \text{ if } y \prec x,\\
0, \qquad & \text{ otherwise},
\end{cases}
\end{equation}
or to the ${p \times p}$ \textit{link matrix} ${L_0 \left( x, y \right)}$ on ${\mathcal{C}}$ (for the case in which one is summing over paths):

\begin{equation}
\Phi \left( x, y \right) = a L_0 \left( x, y \right), \qquad \text{ with } \qquad L_0 \left( x, y \right) = \begin{cases}
1, \qquad & \text{ if } y \prec^{*} x,\\
0, \qquad & \text{ otherwise}.
\end{cases}
\end{equation}
For chains/paths of arbitrary length, connecting causal set elements ${v_x}$ and ${v_y}$, the overall amplitude is therefore given by a ${p \times p}$ discrete propagator matrix ${K \left( x, y \right)}$ of the general form:

\begin{equation}
K \left( x, y \right) = \Phi \left( x, y \right) + b \Phi^2 \left( x, y \right) + b^2 \Phi^3 \left( x, y \right) + \dots = \sum_{n = 1}^{\infty} b^{n - 1} \Phi^n \left( x, y \right),
\end{equation}
where the $n$-th term in this sum corresponds to the individual contribution to the overall amplitude of chains/paths of length $n$; for instance, for the ${n = 2}$ contributions, one has the term:

\begin{equation}
b \Phi^2 \left( x, y \right) = \sum_{z = 1}^{p} b \Phi \left( x, z \right) \Phi \left( z, y \right),
\end{equation}
designating a sum (evaluated over all intermediate causal set elements ${v_z}$) of the individual amplitudes for all chains/paths of length 2 connecting element ${v_x}$ to element ${v_z}$, and element ${v_z}$ to element ${v_y}$. An illuminating example (due to Johnston) of the computation of the overall amplitude for chains connecting elements ${v_1}$ and ${v_5}$ in a 6-element causal set, yielding the discrete propagator matrix:

\begin{equation}
K \left( 1, 5 \right) = a + 3 a^2 b + a^3 b^2,
\end{equation}
is shown in Figure \ref{fig:Figure4} (note that there is a small error in the presentation of this example given within Johnston's PhD thesis, which we correct here). The sum within the expression for the discrete propagator ${K \left( x, y \right)}$ is guaranteed to truncate, since by hypothesis the causal set ${\mathcal{C}}$ has finite cardinality $p$, so every element ${x \in \mathcal{C}}$ has only a finite set of elements in its causal future, allowing us to evaluate ${K \left( x, y \right)}$ pragmatically via the following straightforward matrix inversion:

\begin{equation}
K \left( x, y \right) = \Phi \left( x, y \right) \left( I \left( x, y \right) - b \Phi \left( x, y \right) \right)^{-1},
\end{equation}
for the ${p \times p}$ identity matrix ${I \left( x, y \right)}$.

In order to evaluate the discrete path integral as a sum over \textit{chains}, we must first introduce a procedure for calculating the number (or \textit{abundance}) ${C_n}$ of chains of length $n$ separating the elements $x$ and $y$ in the causal set ${\mathcal{C}}$:

\begin{multline}
C_n \left( x, y \right) = \left\lvert \left\lbrace \left( x, z_1, z_2, \dots, z_{n - 1}, y \right) : \right. \right. \\
\left. \left. x, z_1, z_2, \dots, z_{n - 1}, y \in \mathcal{C} \text{ such that } x \prec z_1 \prec z_2 \prec \cdots \prec z_{n - 1} \prec y \right\rbrace \right\rvert.
\end{multline}
Following Meyer\cite{meyer}, when considered as a random variable ${\hat{C_n} \left( x, y \right)}$, we can compute the expectation value ${\left\langle \hat{C_1} \left( x, y \right) \right\rangle}$ of chains of length 1 simply using the continuum analog of the causal matrix ${C_0 \left( x, y \right)}$, namely the function ${C \left( x, y \right)}$ defined over $d$-dimensional flat (Minkowski) spacetime ${\mathbb{M}^d = \mathbb{R}^{1, d - 1}}$ (with the causal partial order relation on the continuum being denoted ${\prec_{\mathcal{M}}}$):

\begin{equation}
C \left( x, y \right) = \begin{cases}
1, \qquad & \text{ if } x \prec_{\mathcal{M}} y,\\
0, \qquad & \text{ otherwise},
\end{cases} \qquad \text{ such that } \qquad
\left\langle \hat{C_1} \left( x, y \right) \right\rangle = C \left( x, y \right).
\end{equation}
Using this construction, the expected abundance ${\left\langle \hat{C_n} \left( x, y \right) \right\rangle}$ of chains ${x \prec z_1 \prec z_2 \prec \cdots \prec z_{n - 1} \prec y}$ of arbitrary length ${n > 1}$ may be computed by virtue of a nested integral, evaluated over the following nested sequence of Alexandrov intervals ${\mathbf{A}}$ in the continuum spacetime:

\begin{equation}
\mathcal{M} \supset \mathbf{A} \left[ x, y \right] \supset \left( J^{+} \left( z_1 \right) \cap \mathbf{A} \left[ x, y \right] \right) \supset \left( J^{+} \left( z_2 \right) \cap \mathbf{A} \left[ x, y \right] \right) \supset \cdots \supset \left( J^{+} \left( z_{n - 2} \right) \cap \mathbf{A} \left[ x, y \right] \right),
\end{equation}
namely:

\begin{multline}
\left\langle \hat{C_n} \left( x, y \right) \right\rangle = \rho^{n - 1} \int_{\mathbf{A} \left[ x, y \right]} \int_{J^{+} \left( z_1 \right) \cap \mathbf{A} \left[ x, y \right]} \int_{J^{+} \left( z_2 \right) \cap \mathbf{A} \left[ x, y \right]} \cdots\\
\int_{J^{+} \left( z_{n - 2} \right) \cap \mathbf{A} \left[ x, y \right]} C \left( z_{n - 1}, y \right) C \left( z_{n - 2}, z_{n - 1} \right) \cdots C \left( x, z_1 \right) d^d z_{n - 1} \cdots d^d z_2 d^d z_1,
\end{multline}
whose closed form solution is given by:

\begin{equation}
\left\langle \hat{C_n} \left( x, y \right) \right\rangle = C \left( x, y \right) \frac{\left( \rho \mathrm{Vol} \left( \mathbf{A} \left[ y, x \right] \right) \right)^{n - 1}}{n - 1} \left( \frac{\Gamma \left( d + 1 \right)}{2} \right)^{n - 2} \frac{\Gamma \left( \frac{d}{2} \right) \Gamma \left( d \right)}{\Gamma \left( \left( n - 1 \right) \frac{d}{2} \right) \Gamma \left( \frac{n d}{2} \right)}.
\end{equation}
As per Rideout and Zohren\cite{rideout}, the volume factor contribution ${\mathrm{Vol} \left( \mathbf{A} \left[ y, x \right] \right)}$ may be determined by integrating over the proper time interval ${\tau}$ separating events $x$ and $y$:

\begin{equation}
\mathrm{Vol} \left( \mathbf{A} \left[ y, x \right] \right) = 2 \int_{0}^{\frac{\tau}{2}} \mathrm{Vol} \left( S_{d - 1} \left( t \right) \right) dt,
\end{equation}
where we use ${S_{d - 1} \left( t \right)}$ to denote the volume of a ${\left( d - 1 
\right)}$-dimensional ball of radius $t$, such that this integral evaluates to yield:

\begin{equation}
2 \int_{0}^{\frac{\tau}{2}} \mathrm{Vol} \left( S_{d - 1} \left( t \right) \right) dt = 2 \int_{0}^{\frac{\tau}{2}} \frac{\pi^{\frac{d}{2}} r^d}{\Gamma \left( \frac{d}{2} + 1 \right)} dt = \frac{\pi^{\frac{d - 1}{2}}}{2^{d - 1} d \Gamma \left( \frac{d + 1}{2} \right)} \tau^d.
\end{equation}
From here, the expectation value ${\left\langle \hat{K_C} \left( x, y \right) \right\rangle}$ for the variant of the discrete propagator ${K_C \left( x, y \right)}$ evaluated by summing over chains (when considered as a random variable ${\hat{K_C} \left( x, y \right)}$) can now be written in the form of a sum over expectation values for abundances of chains:

\begin{equation}
\left\langle \hat{K_C} \left( x, y \right) \right\rangle = \sum_{n = 1}^{\infty} a^n b^{n - 1} \left\langle \hat{C_n} \left( x, y \right) \right\rangle,
\end{equation}
thus satisfying the elegant integral equation:

\begin{equation}
\left\langle \hat{K_C} \left( x, y \right) \right\rangle = a C \left( x, y \right) + a b \rho \int_{- \infty}^{\infty} C \left( z, y \right) \left\langle \hat{K_C} \left( x, z \right) \right\rangle d^d z.
\end{equation}

As an illustration of how Fourier-analytic techniques, when combined with this general \textit{summing over chains} philosophy, may once again be used in a relatively straightforward manner to obtain an explicit form for the discrete (massless) propagator matrix ${\left( K_C \right)_{0} \left( x, y \right)}$, we consider the computation of the discrete propagator ${\left( K_{C} \right)_{0}^{\left( 2 \right)} \left( x, y \right)}$ in ${1 + 1}$-dimensional flat (Minkowski) spacetime\cite{johnston3}. Since the integral equation for the expectation value ${\left\langle \hat{K_C} \left( x, y \right) \right\rangle}$ of the discrete propagator presented above takes the form of a convolution, its Fourier transform ${\left\langle \tilde{\hat{K_C}} \left( \mathbf{p} \right) \right\rangle}$ simply reduces to a product:

\begin{equation}
\left\langle \tilde{\hat{K_C}} \left( \mathbf{p} \right) \right\rangle = a \tilde{C} \left( \mathbf{p} \right) + a b \rho \tilde{C} \left( \mathbf{p} \right) \left\langle \tilde{\hat{K_C}} \left( \mathbf{p} \right) \right\rangle = \frac{a \tilde{C} \left( \mathbf{p} \right)}{1 - a b \rho \tilde{C} \left( \mathbf{p} \right)},
\end{equation}
and, moreover, since the continuum analog of the causal matrix ${C \left( x, y \right)}$ is related directly to the massless retarded Green's function ${\left( G_R \right)_{0}^{\left( 2 \right)} \left( x, y \right)}$ by a factor of 2:

\begin{equation}
C \left( x, y \right) = 2 \left( G_R \right)_{0}^{\left( 2 \right)} \left( x, y \right),
\end{equation}
it follows, by taking the aforementioned Fourier transform of the massive Green's function ${G_{m}^{\left( 2 \right)} \left( \mathbf{x} \right)}$:

\begin{equation}
\tilde{G}_{m}^{\left( 2 \right)} \left( \mathbf{p} \right) = - \frac{1}{p_{0}^{2} - p_{1}^{2} - m^2},
\end{equation}
and setting ${m = 0}$, that the Fourier transform ${\tilde{C} \left( \mathbf{p} \right)}$ is given simply by:

\begin{equation}
\tilde{C} \left( \mathbf{p} \right) = - \frac{2}{\left( p_0 + i \epsilon \right)^2 - p_{1}^{2}}.
\end{equation}
When substituted into the expression for the expectation value ${\left\langle \tilde{\hat{K_C}} \left( \mathbf{p} \right) \right\rangle}$, this yields (following simplification):

\begin{equation}
\left\langle \tilde{\hat{K_C}} \left( \mathbf{p} \right) \right\rangle = \frac{- \frac{2 a}{\left( p_0 + i \epsilon \right)^2 - p_{1}^{2}}}{1 + \frac{2 a b \rho}{\left( p_0 + i \epsilon \right)^2 - p_{1}^{2}}} = - \frac{2 a}{\left( p_0 + i \epsilon \right)^2 - p_{1}^{2} + 2 a b \rho},
\end{equation}
implying that the amplitudes $a$ and $b$ must be given by ${a = \frac{1}{2}}$ and ${b = - \frac{m^2}{\rho}}$, respectively, in order for the expectation value of the (Fourier-transformed) discrete propagator ${\left\langle \tilde{\hat{K_C}} \left( \mathbf{p} \right) \right\rangle}$ to match the known value of the (Fourier-transformed) continuum massive Green's function ${\tilde{G}_{m}^{\left( 2 \right)} \left( \mathbf{p} \right)}$, yielding the following explicit form of the massless discrete propagator in ${1 + 1}$-dimensional flat (Minkowski) spacetime:

\begin{equation}
\left( K_C \right)_{0}^{\left( 2 \right)} \left( x, y \right) = \frac{1}{2} C_0 \left( x, y \right),
\end{equation}
where ${C_0 \left( x, y \right)}$ designates, as usual, the causal matrix on the causal set ${\mathcal{C}}$.

In much the same way, evaluating the discrete path integral over \textit{paths} necessitates introducing a technique for calculating the number/abundance ${P_n}$ of paths of length $n$ separating the elements $x$ and $y$ in the causal set ${\mathcal{C}}$:

\begin{multline}
P_n \left( x, y \right) = \left\lvert \left\lbrace \left( x, z_1, z_2, \dots, z_{n - 1}, y \right) : \right. \right.\\
\left. \left. x, z_1, z_2, \dots, z_{n - 1}, y \in \mathcal{C} \text{ such that } x \prec^{*} z_1 \prec^{*} z_2 \prec^{*} \cdots \prec^{*} z_{n - 1} \prec^{*} y \right\rbrace \right\rvert.
\end{multline}
Just as before, we follow the approach of Bombelli\cite{bombelli4} and consider the path abundance as a random variable ${\hat{P}_n \left( x, y \right)}$, such that the expectation value ${\left\langle \hat{P}_1 \left( x, y \right) \right\rangle}$ of paths of length 1 is given now by the continuum analog of the link matrix ${L_0 \left( x, y \right)}$, namely the function ${P \left( x, y \right)}$ defined over $d$-dimensional flat (Minkowski) spacetime ${\mathbb{M}^d = \mathbb{R}^{1, d - 1}}$ (with the causal partial order relation on the continuum being denoted, as before, by ${\prec_{\mathcal{M}}}$):

\begin{equation}
P \left( x, y \right) = \begin{cases}
e^{- \rho \mathrm{Vol} \left( \mathbf{I} \left[ y, x \right] \right)}, \qquad & \text{ if } x \prec_{\mathcal{M}} y,\\
0, \qquad & \text{ otherwise},
\end{cases} \qquad \text{ such that } \qquad
\left\langle \hat{P}_1 \left( x, y \right) \right\rangle = P \left( x, y \right).
\end{equation}
As previously, the expected abundance ${\left\langle \hat{P}_n \left( x, y \right) \right\rangle}$ of paths ${x \prec^{*} z_1 \prec^{*} z_2 \prec^{*} \cdots \prec^{*} z_{n - 1} \prec^{*} y}$ of arbitrary length ${n > 1}$ now becomes a nested integral over the usual sequence of Alexandrov intervals ${\mathbf{A}}$:

\begin{equation}
\mathcal{M} \supset \mathbf{A} \left[ x, y \right] \supset \left( J^{+} \left( z_1 \right) \cap \mathbf{A} \left[ x, y \right] \right) \supset \left( J^{+} \left( z_2 \right) \cap \mathbf{A} \left[ x, y \right] \right) \supset \cdots \supset \left( J^{+} \left( z_{n - 2} \right) \cap \mathbf{A} \left[ x, y \right] \right),
\end{equation}
i.e:

\begin{multline}
\left\langle \hat{P}_n \left( x, y \right) \right\rangle = \rho^{n - 1} \int_{\mathbf{A} \left[ x, y \right]} \int_{J^{+} \left( z_1 \right) \cap \mathbf{A} \left[ x, y \right]} \int_{J^{+} \left( z_2 \right) \cap \mathbf{A} \left[ x, y \right]} \cdots\\
\int_{J^{+} \left( z_{n - 2} \right) \cap \mathbf{A} \left[ x, y \right]} P \left( z_{n - 1}, y \right) P \left( z_{n - 2}, z_{n - 1} \right) \cdots P \left( x, z_1 \right) d^d z_{n - 1} \cdots d^d z_2 d^d z_1,
\end{multline}
for which no known closed-form solution exists. However, following the approach of Johnston\cite{johnston3}, we use the existing closed-form solution for the chain abundance expectation value ${\left\langle \hat{C}_n \left( x, y \right) \right\rangle}$, namely:

\begin{equation}
\left\langle \hat{C}_n \left( x, y \right) \right\rangle = C \left( x, y \right) \frac{\left( \rho \mathrm{Vol} \left( \mathbf{A} \left[ y, x \right] \right) \right)^{n - 1}}{n - 1} \left( \frac{\Gamma \left( d + 1 \right)}{2} \right)^{n - 2} \frac{\Gamma \left( \frac{d}{2} \right) \Gamma \left( d \right)}{\Gamma \left( \left( n - 1 \right) \frac{d}{2} \right) \Gamma \left( \frac{n d}{2} \right)},
\end{equation}
in order to expand the continuum analog of the link matrix function ${P \left( x, y \right)}$ as a formal power series in terms of the expectation values of the chain abundance functions ${\left\langle \hat{C}_n \left( x, y \right) \right\rangle}$, using:

\begin{equation}
P \left( x, y \right) = C \left( x, y \right) e^{- \rho \mathrm{Vol} \left( \mathbf{I} \left[ y, x \right] \right)} = C \left( x, y \right) \sum_{n = 0}^{\infty} \frac{\left( - \rho \mathrm{Vol} \left( \mathbf{I} \left[ y, x \right] \right) \right)^n}{n!},
\end{equation}
where we know that:

\begin{equation}
C \left( x, y \right) \sum_{n = 0}^{\infty} \frac{\left( - \rho \mathrm{Vol} \left( \mathbf{I} \left[ y, x \right] \right) \right)^n}{n!} = \sum_{n = 0}^{\infty} \frac{\left( -1 \right)^n \left\langle \hat{C}_{n + 1} \left( x, y \right) \right\rangle}{\left( n - 1 \right)! \left( \frac{\Gamma \left( d + 1 \right)}{2} \right)^{n - 1} \frac{\Gamma \left( \frac{d}{2} \right) \Gamma \left( d \right)}{\Gamma \left( \frac{n d}{2} \right) \Gamma \left( \left( n + 1 \right) \frac{d}{2} \right)}}.
\end{equation}
Since any formal power series in a single variable $x$ may be exponentiated via the following rule:

\begin{equation}
\left( \sum_{k = 0}^{\infty} a_k x^k \right)^n = \sum_{k = 0}^{\infty} c_k x^k, \qquad \text{ with } \qquad c_0 = a_{0}^{n},
\end{equation}
where the new coefficients ${c_m}$ are defined by the recurrence relation:

\begin{equation}
c_m = \frac{1}{m a_0} \sum_{k = 1}^{m} \left( k n - m + k \right) a_k c_{n - k},
\end{equation}
we can consequently employ the formal power series expansion for the continuum analog of the link matrix function ${P \left( x, y \right)}$ to construct a corresponding formal power series expansion for the expectation value of the path abundance ${\left\langle \hat{P}_n \left( x, y \right) \right\rangle}$ as follows:

\begin{equation}
\left\langle \hat{P}_n \left( x, y \right) \right\rangle = \sum_{m = 0}^{\infty} g_m \left\langle \hat{C}_{m + n} \left( x, y \right) \right\rangle, \qquad \text{ with } \qquad g_0 = 1,
\end{equation}
where we also know that:

\begin{multline}
\sum_{m = 0}^{\infty} g_m \left\langle \hat{C}_{m + n} \left( x, y \right) \right\rangle\\
= \sum_{m = 0}^{\infty} \frac{g_m}{m + n - 1} \left( \frac{ \Gamma \left( d + 1 \right)}{2} \right)^{m + n - 2} \frac{\Gamma \left( \frac{d}{2} \right) \Gamma \left( d \right)}{\Gamma \left( \left( m + n - 1 \right) \frac{d}{2} \right) \Gamma \left( \left( m + n \right) \frac{d}{2} \right)} C \left( x, y \right) \left( \rho \mathrm{Vol} \left( \mathbf{I} \left[ y, x \right] \right) \right)^{n + m - 1},
\end{multline}
with new coefficients ${g_m}$ defined by the recurrence relation:

\begin{equation}
g_m = \frac{1}{m} \sum_{k = 1}^{m} \left( k \left( n + 1 \right) - m \right) \frac{\left( -1 \right)^k}{\left( k - 1 \right)! \left( \frac{\Gamma \left( d + 1 \right)}{2} \right)^{k - 1} \frac{\Gamma \left( \frac{d}{2} \right) \Gamma \left( d \right)}{\Gamma \left( \frac{k d}{2} \right) \Gamma \left( \left( k + 1 \right) \frac{d}{2} \right)}} g_{m - k}.
\end{equation}
Much as in the previously-described \textit{summing over chains} case, the expectation value ${\left\langle \hat{K_P} \left( x, y \right) \right\rangle}$ for the variant of the discrete propagator ${K_P \left( x, y \right)}$ evaluated by summing over paths (when considered as a random variable ${\hat{K_P} \left( x, y \right)}$) may resultingly be written as a sum over expectation values for abundances of paths:

\begin{equation}
\left\langle \hat{K_P} \left( x, y \right) \right\rangle = \sum_{n = 1}^{\infty} a^n b^{n - 1} \left\langle \hat{P}_n \left( x, y \right) \right\rangle,
\end{equation}
satisfying an analogous integral equation:

\begin{equation}
\left\langle \hat{K_P} \left( x, y \right) \right\rangle = a P \left( x, y \right) + a b \rho \int_{- \infty}^{\infty} P \left( z, y \right) \left\langle \hat{K_P} \left( x, z \right) \right\rangle d^d z.
\end{equation}

The same Fourier-analytic techniques outlined previously may also be used in conjunction with this \textit{summing over paths} philosophy, so as to yield explicit forms for the discrete (massless) propagator matrix ${\left( K_P \right)_0 \left( x, y \right)}$, which we shall now proceed to illustrate by considering the computation of the discrete propagator ${\left( K_P \right)_{0}^{\left( 4 \right)} \left( x, y \right)}$ in ${3 + 1}$-dimensional flat (Minkowski) spacetime\cite{johnston3}. Just as in the \textit{sum over chains} case, since the integral equation for the expectation value ${\left\langle \hat{K_P} \left( x, y \right) \right\rangle}$ of the discrete propagator is a convolution, its Fourier transform ${\left\langle \tilde{\hat{K_P}} \left( \mathbf{p} \right) \right\rangle}$ also reduces to a product of the same underlying form:

\begin{equation}
\left\langle \tilde{\hat{K_P}} \left( \mathbf{p} \right) \right\rangle = a \tilde{P} \left( \mathbf{p} \right) + a b \rho \tilde{P} \left( \mathbf{p} \right) \left\langle \tilde{\hat{K_P}} \left( \mathbf{p} \right) \right\rangle = \frac{a \tilde{P} \left( \mathbf{p} \right)}{1 - a b \rho \tilde{P} \left( \mathbf{p} \right)},
\end{equation}
and, furthermore, since we know that the volume contribution ${\mathrm{Vol} \left( \mathbf{I} \left[ y, x \right] \right)}$ to the continuum analog of the link matrix ${P \left( x, y \right)}$ in the ${3 + 1}$-dimensional Minkowski spacetime ${\mathbb{M}^4 = \mathbb{R}^{1, 3}}$ is given explicitly by:

\begin{equation}
\mathrm{Vol} \left( \mathbf{I} \left[ y, x \right] \right) = \frac{\pi}{24} \tau_{x y}^{4},
\end{equation}
with ${\tau_{x y}}$ denoting, as usual, the proper time distance between events $x$ and $y$, we are able to write the continuum analog of the link matrix ${P \left( x, y \right)}$ in an altogether more concrete fashion:

\begin{equation}
P \left( x, y \right) = \begin{cases}
e^{- \rho \mathrm{Vol} \left( \mathbf{I} \left[ y, x \right] \right)} = e^{- \frac{\pi}{24} \rho \tau_{x y}^{4}}, \qquad & \text{ if } x \prec_{\mathcal{M}} y,\\
0, \qquad & \text{ otherwise}.
\end{cases}
\end{equation}
Taking the (distributional) limit of an infinite sprinkling density (${\rho \to \infty}$), we therefore find that:

\begin{equation}
\lim_{\rho \to \infty} \left[ \sqrt{\rho} P \left( x, y \right) \right] = \lim_{\rho \to \infty} \left[ \begin{cases}
f_{\rho} \left( \tau_{x y}^{2}, \frac{1}{24} \right), \qquad & \text{ if } x^0 \leq y^0,\\
0, \qquad & \text{ otherwise},
\end{cases} \right] = \begin{cases}
\frac{\sqrt{24}}{2} \delta \left( \tau_{x y}^{2} \right), \qquad & \text{ if } x \prec_{\mathcal{M}} y,\\
0, \qquad & \text{ otherwise},
\end{cases}
\end{equation}
where, for the sake of notational convenience, we have introduced the function ${f_{\rho} \left( z, c \right)}$ defined by:

\begin{equation}
f_{\rho} \left( z, c \right) = \begin{cases}
\sqrt{\rho} e^{- \pi c \rho z^2}, \qquad & \text{ if } z \geq 0,\\
0, \qquad & \text{ if } z < 0,
\end{cases} \qquad \text{ for some positive real constant } \qquad c \in \mathbb{R}^{+},
\end{equation}
and have exploited the fact that its (distributional) limit yields a Dirac delta function ${\delta \left( z \right)}$:

\begin{equation}
\lim_{\rho \to \infty} \left[ f_{\rho} \left( z, c \right) \right] = \frac{1}{2 \sqrt{c}} \delta \left( z \right).
\end{equation}
From here, we may employ the following theorem\cite{champeney} regarding limits of Fourier transforms of arbitrary functions:

\begin{equation}
\lim_{\rho \to \infty} \left[ \sqrt{\rho} P \left( \mathbf{p} \right) \right] = \lim_{\rho \to \infty} \left[ f \left( \mathbf{p} \right) \right] \qquad \iff \qquad \lim_{\rho \to \infty} \left[ \tilde{\left( \sqrt{\rho} P \left( \mathbf{p} \right) \right)} \right] = \lim_{\rho \to \infty} \left[ \tilde{f} \left( \mathbf{p} \right) \right],
\end{equation}
for an arbitrary function ${f \left( \mathbf{p} \right)}$ with Fourier transform ${\tilde{f} \left( x, y \right)}$, in order to establish the following relationship between the (Fourier-transformed) continuum analog of the link matrix ${\tilde{P} \left( \mathbf{p} \right)}$ and the (Fourier-transformed) massless retarded Green's function ${\left( \tilde{G_R} \right)_{0}^{\left( 4 \right)} \left( \mathbf{p} \right)}$ in the infinite sprinkling density limit:

\begin{equation}
\lim_{\rho \to \infty} \left[ \sqrt{\rho} \tilde{P} \left( \mathbf{p} \right) \right] = 2 \pi \frac{\sqrt{24}}{2} \left( \tilde{G_R} \right)_{0}^{\left( 4 \right)} \left( \mathbf{p} \right) = - 2 \pi \sqrt{6} \frac{1}{\left( p_0 + i \epsilon \right)^2 - p_1 - p_2 - p_3},
\end{equation}
since the previously-derived form of the massless retarded Green's function in ${3 + 1}$-dimensional Minkowski space ${\left( G_R \right)_{0}^{\left( 4 \right)} \left( \mathbf{x} \right)}$ implies that:

\begin{equation}
\left( G_R \right)_{0}^{\left( 4 \right)} \left( \mathbf{x} \right) = \theta \left( x^0 \right) \theta \left( \tau_{x y}^{2} \right) \frac{1}{2 \pi} \delta \left( \tau^2 \right) = \begin{cases}
\frac{1}{2 \pi} \delta \left( \tau_{x y}^{2} \right), \qquad & \text{ if } x \prec_{\mathcal{M}} y,\\
0, \qquad & \text{ otherwise},
\end{cases}
\end{equation}
and, moreover, setting ${m = 0}$ in the expression for the Fourier transform of the massive Green's function ${G_{m}^{\left( 4 \right)} \left( \mathbf{x} \right)}$:

\begin{equation}
\tilde{G}_{m}^{\left( 4 \right)} \left( \mathbf{p} \right) = - \frac{1}{p_{0}^{2} - p_{1}^{2} - p_{2}^{2} - p_{3}^{2} - m^2},
\end{equation}
consequently yields the following form for the Fourier transform of the massless retarded Green's function ${\left( G_R \right)_{0}^{\left( 4 \right)} \left( \mathbf{x} \right)}$:

\begin{equation}
\left( \tilde{G_R} \right)_{0}^{\left( 4 \right)} \left( \mathbf{p} \right) = - \frac{1}{\left( p_0 + i \epsilon \right)^2 - p_{1}^{2} - p_{2}^{2} - p_{3}^{2}}.
\end{equation}

By introducing an ansatz for the constant amplitudes $a$ and $b$ given by ${a = A \sqrt{\rho}}$ and ${b = \frac{B}{\rho}}$, for a pair of additional constants $A$ and $B$ (taken, by hypothesis, to be independent of the density ${\rho}$), the prior expression for the expectation value ${\left\langle \tilde{\hat{K}}_P \left( \mathbf{p} \right) \right\rangle}$ becomes instead:

\begin{equation}
\left\langle \tilde{\hat{K}}_P \left( \mathbf{p} \right) \right\rangle = \frac{A \sqrt{\rho} \tilde{P} \left( \mathbf{p} \right)}{1 - A B \sqrt{\rho} \tilde{P} \left( \mathbf{p} \right)},
\end{equation}
and so, by combining this with the expression for the (Fourier-transformed) density-weighted continuum analog of the link matrix ${\sqrt{\rho} \tilde{P} \left( \mathbf{p} \right)}$ in the limit of infinite sprinkling density, one obtains (following simplification):

\begin{equation}
\lim_{\rho \to \infty} \left[ \left\langle \tilde{\hat{K}}_P \left( \mathbf{p} \right) \right\rangle \right] = \frac{- \frac{2 \pi A \sqrt{6}}{\left( p_0 + i \epsilon \right)^2 - p_{1}^{2} - p_{2}^{2} - p_{3}^{2}}}{1 + \frac{2 \pi A B \sqrt{6}}{\left( p_0 + i \epsilon \right)^2 - p_{1}^{2} - p_{2}^{2} - p_{3}^{2}}} = - \frac{2 \pi A \sqrt{6}}{\left( p_0 + i \epsilon \right)^2 - p_{1}^{2} - p_{2}^{2} - p_{3}^{2} + 2 \pi A B \sqrt{6}},
\end{equation}
implying that the constants $A$ and $B$ must be given by ${2 \pi A \sqrt{6} = 1}$ and ${B = - m^2}$, respectively, in order for the expectation value of the (Fourier-transformed) discrete propagator ${\left\langle \tilde{\hat{K}}_P \left( \mathbf{p} \right) \right\rangle}$ to match the known value of the (Fourier-transformed) continuum massive Green's function ${\tilde{G}_{m}^{\left( 4 \right)} \left( \mathbf{p} \right)}$:

\begin{equation}
\tilde{G}_{m}^{\left( 4 \right)} \left( \mathbf{p} \right) = - \frac{1}{p_{0}^{2} - p_{1}^{2} - p_{2}^{2} - p_{3}^{2} - m^2},
\end{equation}
in the infinite sprinkling density limit. This, in turn, implies that the amplitudes $a$ and $b$ must be given by ${a = \sqrt{\rho}{2 \pi \sqrt{6}}}$ and ${b = - \frac{m^2}{\rho}}$, respectively, hence allowing us to deduce the following explicit form of the massless discrete propagator in ${3 + 1}$-dimensional flat (Minkowski) spacetime:

\begin{equation}
\left( K_P \right)_{0}^{\left( 4 \right)} \left( x, y \right) = \frac{1}{2 \pi} \sqrt{\frac{1}{6}} L_0 \left( x, y \right),
\end{equation}
where ${L_0 \left( x, y \right)}$ designates the usual link matrix on the causal set ${\mathcal{C}}$.

\begin{figure}[ht]
\centering
\includegraphics[width=0.345\textwidth]{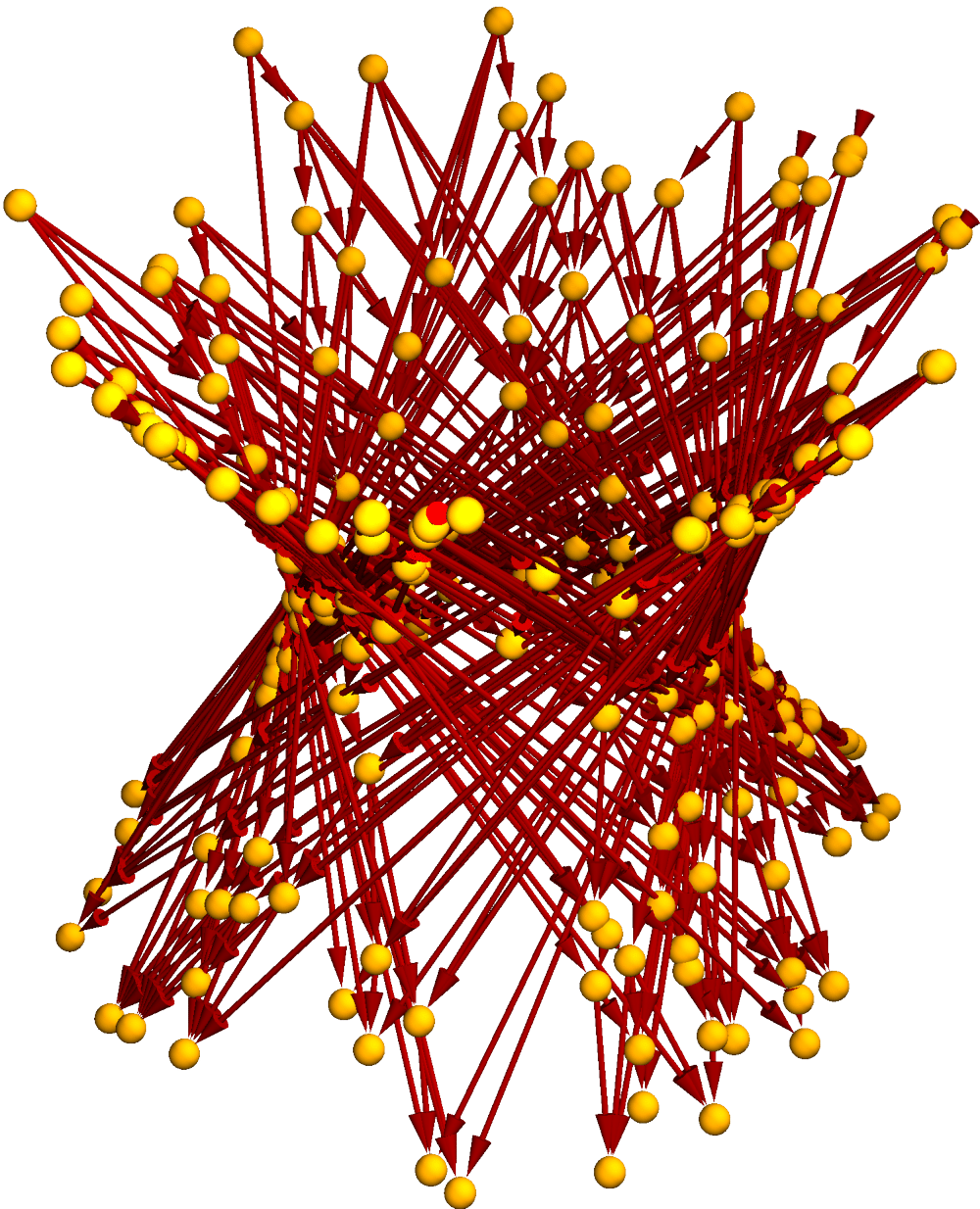}\hspace{0.1\textwidth}
\includegraphics[width=0.395\textwidth]{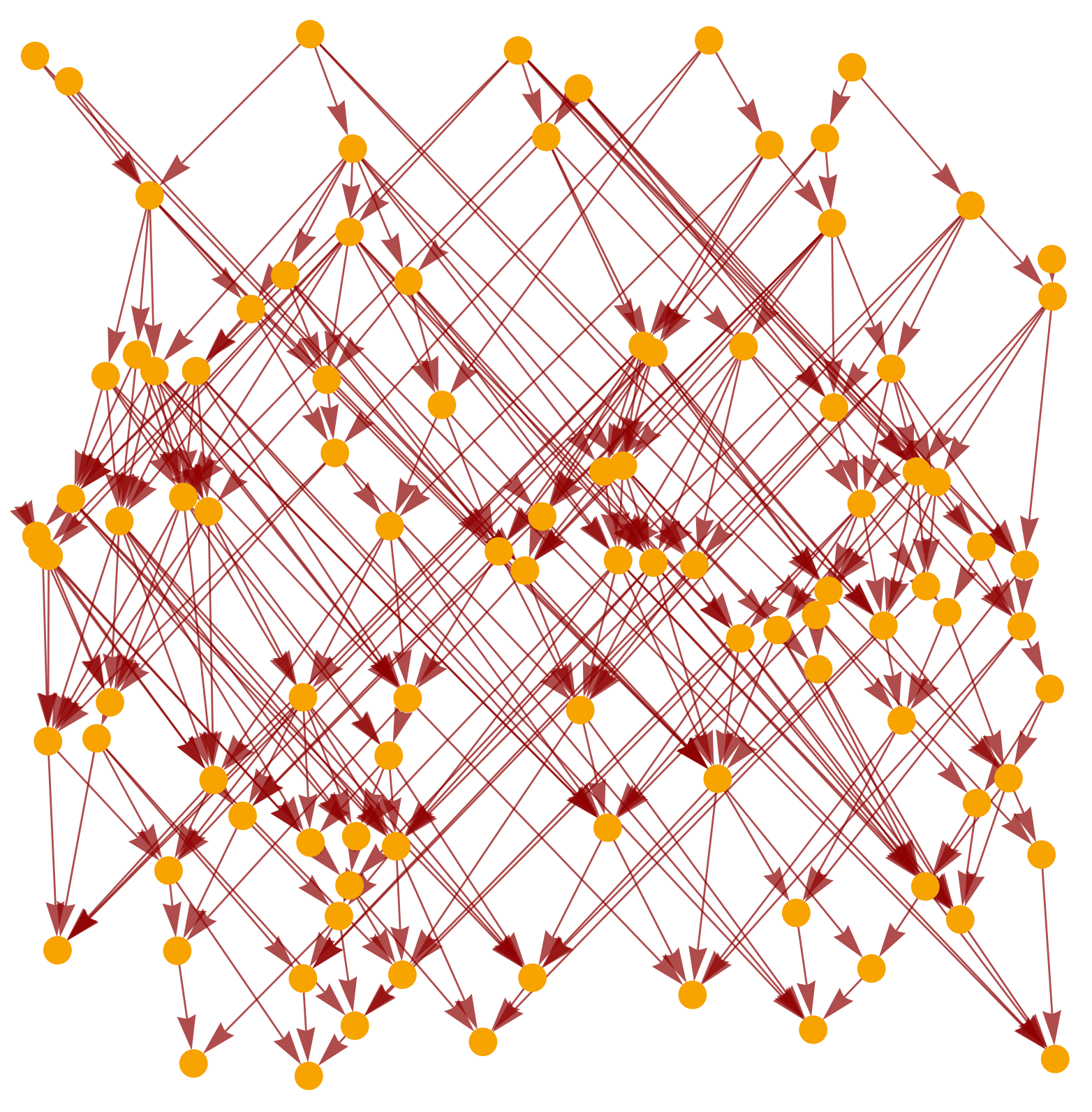}
\caption{On the left, the transitive reduction (i.e. the Hasse diagram) of the directed graph generated by connecting all pairs of 200 (uniformly-sprinkled) points that are related by the causal partial order relation ${\prec_{\mathcal{M}}}$ on a hyperboloidal region of a ${1 + 1}$-dimensional de Sitter spacetime embedded within a ${2 + 1}$-dimensional background spacetime. On the right, a comparison against a similar transitive reduction graph generated via a uniform sprinkling of 100 points into a rectangular region of ${1 + 1}$-dimensional flat (Minkowski) spacetime.}
\label{fig:Figure5}
\end{figure}

For the sake of simplicity, the analyses presented so far have assumed flat (Minkowski) spacetimes of the form ${\mathbb{M}^d = \mathbb{R}^{1, d - 1}}$; however, the mathematical methods outlined within this section are sufficiently general that they may be extended to causal sets produced by sprinklings into curved spacetimes (as illustrated for the case of a ${1 + 1}$-dimensional de Sitter spacetime in Figure \ref{fig:Figure5}) with only very minimal modification, following the approach of Birrell and Davies\cite{birrell} and Fulling\cite{fulling}. First, one must relax the assumption that translation invariance holds (and therefore that, for instance, a simple massive Green's function of the form ${G_{m}^{\left( d \right)} \left( \mathbf{y} - \mathbf{x} \right)}$ may be used to compute the transition amplitude between events ${\mathbf{x}}$ and ${\mathbf{y}}$; one must instead use a two-argument function of the form ${G_{m}^{\left( d \right)} \left( \mathbf{x}, \mathbf{y} \right)}$). Somewhat more substantially, the form of the Klein-Gordon equation for the massive $d$-dimensional Green's function ${G_{m}^{\left( d \right)} \left( \mathbf{x}, \mathbf{y} \right)}$ itself must be modified, so that it now reads:

\begin{equation}
\left( \Box + m^2 + \zeta R \left( \mathbf{x} \right) \right) G_{m}^{\left( d \right)} \left( \mathbf{x}, \mathbf{y} \right) = \frac{\delta \left( \mathbf{x} - \mathbf{y} \right)}{\sqrt{g}}, \qquad \text{ with } \qquad g = \sqrt{\det \left( g^{\mu \nu} \right)},
\end{equation}
for some real constant ${\zeta \in \mathbb{R}}$, where ${\Box}$ is now the generalization of the d'Alembertian operator to arbitrary curved spacetimes, defined in terms of its action on an arbitrary scalar field ${\phi \left( \mathbf{x} \right)}$

\begin{equation}
\Box \phi \left( \mathbf{x} \right) = \frac{1}{\sqrt{g}} \partial_{\mu} \left[ g^{\mu \nu} \sqrt{g} \partial_{\nu} \phi \left( \mathbf{x} \right) \right],
\end{equation}
and where ${R \left( \mathbf{x} \right)}$ designates the Ricci scalar curvature of the underlying Lorentzian manifold at point ${\mathbf{x}}$. Thus, in order to extend the flat spacetime construction of discrete (massless) propagators to more general curved spacetimes, all that remains is to introduce a systematic procedure for computing the Ricci scalar curvature ${R \left( x \right)}$ of an element ${x \in \mathcal{C}}$ in the causal set ${\mathcal{C}}$. Note that the formalism proposed by Sorkin\cite{sorkin5}, as well as Benincasa and Dowker\cite{benincasa}, provides a more-or-less tautological method for computing ${R \left( x \right)}$ in terms of the expectation value ${\left\langle \hat{B}_{k}^{\left( d \right)} \phi \left( x \right) \right\rangle}$ of the action of the discrete $d$-dimensional d'Alembertian ${B_{k}^{\left( d \right)}}$ on the scalar field ${\phi \left( x \right)}$ in the infinite sprinkling density limit, namely:

\begin{equation}
\lim_{\rho_c \to \infty} \left[ \frac{1}{\sqrt{\rho_c}} \left. \left\langle \hat{B}_{k}^{\left( d \right)} \phi \left( x \right) \right\rangle \right\rvert_{\mathcal{W}_1} \right] = \Box \phi \left( x \right) - \frac{1}{2} R \left( x \right) \phi \left( x \right),
\end{equation}
where we have opted to subdivide the causal past ${J^{-} \left( x \right)}$ of element $x$ into non-overlapping subregions ${\mathcal{W}_1}$, ${\mathcal{W}_2}$ and ${\mathcal{W}_3}$:

\begin{equation}
J^{-} \left( x \right) = \mathcal{W}_1 \cup \mathcal{W}_2 \cup \mathcal{W}_3, \qquad \text{ such that } \qquad \left( \mathcal{W}_1 \cap \mathcal{W}_2 \right) = \left( \mathcal{W}_1 \cap \mathcal{W}_3 \right) = \left( \mathcal{W}_2 \cap \mathcal{W}_3 \right) = \varnothing,
\end{equation}
where ${\mathcal{W}_1}$ is the Riemann normal neighborhood of event $x$, ${\mathcal{W}_2}$ is the Riemann normal neighborhood of the boundary ${\partial J^{-} \left( x \right)}$ of the past region (namely the region ``down the light cone'', bounded away from the origin), and ${\mathcal{W}_3}$ is bounded away from the boundary ${\partial J^{-} \left( x \right)}$ of the past region (namely the ``deep chronological past'' region).

However, a non-tautological method for computing the discrete Ricci scalar curvature\cite{forman} for an arbitrary causal set ${\mathcal{C}}$ is furnished by the construction of the \textit{Ollivier-Ricci} scalar curvature\cite{ollivier}\cite{ollivier2}\cite{ollivier3} for arbitrary (potentially discrete) metric-measure spaces (including hypergraphs and causal graphs as special cases\cite{eidi}). One starts from the standard geometrical intuition for the Ricci scalar curvature ${R \left( p \right)}$ evaluated at a point ${p \in \mathcal{M}}$ in a (pseudo-)Riemannian manifold ${\left( \mathcal{M}, g \right)}$ of dimension $d$, in which ${R \left( p \right)}$ effectively quantifies the discrepancy between the volume of a ball ${B_{\varepsilon} \left( p \right)}$ of radius ${\varepsilon}$ in the manifold ${\mathcal{M}}$, and the volume of a ball ${B_{\varepsilon} \left( 0 \right)}$ of the same radius in (flat) Euclidean space ${\mathbb{R}^d}$:

\begin{equation}
\frac{\mathrm{Vol} \left( B_{\varepsilon} \left( p \right) \subset \mathcal{M} \right)}{\mathrm{Vol} \left( B_{\varepsilon} \left( 0 \right) \subset \mathbb{R}^d \right)} = 1 - \frac{R \left( p \right)}{6 \left( d + 2 \right)} \varepsilon^2 + O \left( \varepsilon^4 \right),
\end{equation}
evaluated in the limit ${\varepsilon \to 0}$. Equivalently, the Ricci scalar curvature ${R \left( p \right)}$ quantifies the discrepancy between the distance ${\delta}$ between the centers of two balls ${B_{\varepsilon} \left( p \right)}$ and ${B_{\varepsilon} \left( q \right)}$ of radius ${\varepsilon}$ (centered at points $p$ and $q$) in the manifold ${\mathcal{M}}$, in which the latter is obtained via parallel transport of the former, and the average distance $W$ between the corresponding points on the surfaces of the balls:

\begin{equation}
W \left( B_{\varepsilon} \left( p \right), B_{\varepsilon} \left( q \right) \right) = \delta \left( 1 - \frac{\varepsilon^2}{2 \left( d + 2 \right)} R \left( p \right) + O \left( \varepsilon^3 + \varepsilon^2 \delta \right) \right),
\end{equation}
evaluated now in the limit ${\varepsilon, \delta \to 0}$, where ${\delta = d \left( p, q \right)}$ designates the distance between the centers of the balls. For a general (Polish) metric-measure space ${\left( X, d \right)}$, equipped with a Borel ${\sigma}$-algebra along with a random walk $m$, i.e. a family of probability measures of the form:

\begin{equation}
m = \left\lbrace m_x : x \in X \right\rbrace,
\end{equation}
in which each ${m_x}$ has finite first moment, and the map ${x \to m_x}$ is measurable, the \textit{Ollivier-Ricci} scalar curvature ${\kappa \left( x, y \right)}$ between distinct points ${x, y \in X}$ may be defined as:

\begin{equation}
\kappa \left( x, y \right) = 1 - \frac{W_1 \left( m_x, m_y \right)}{d \left( x, y \right)}.
\end{equation}
In the above, ${W_1 \left( m_x, m_y \right)}$ denotes the \textit{1-Wasserstein} distance between the probability measures ${m_x}$ and ${m_y}$, i.e. the optimal transportation distance between measures:

\begin{equation}
W_1 \left( m_x, m_y \right) = \inf_{\epsilon \in \Pi \left( m_x, m_y \right)} \left[ \int_{\left( x, y \right) \in X \times Y} d \left( x, y \right) d \epsilon \left( x, y \right) \right],
\end{equation}
for ${\Pi \left( m_x, m_y \right)}$ being the set of couplings between random walks projecting onto measures ${m_x}$ and ${m_y}$ (i.e. the set of measures on the product space ${X \times Y}$ projecting onto measures ${m_x}$ and ${m_y}$). Hence, we are, in effect, generalizing the Riemannian volume measure to an arbitrary probability measure, such that the average distance between points on the surfaces of two balls generalizes to the \textit{1-Wasserstein} distance between the corresponding measures, and therefore the \textit{Ollivier-Ricci} scalar curvature ${\kappa \left( x, y \right)}$ quantifies the discrepancy between the \textit{1-Wasserstein} distance between the measures ${m_x}$ and ${m_y}$ and the ordinary metric distance between $x$ and $y$. Thus, in the particular case where ${\left( X, d \right)}$ is a (pseudo-)Riemannian manifold ${\left( \mathcal{M}, g \right)}$, the \textit{Ollivier-Ricci} scalar curvature ${\kappa \left( x, y \right)}$ and the ordinary Ricci scalar curvature ${R \left( x, y \right)}$ are equivalent up to a multiplicative constant.

In the case where the metric space ${\left( X, d \right)}$ is \textit{discrete}, the integral appearing in the definition of the \textit{1-Wasserstein} distance correspondingly restricts to a discrete sum, and one thus obtains the discrete (or \textit{multi-marginal}) optimal transportation distance between discrete probability measures ${m_x}$ and ${m_y}$:

\begin{equation}
W_1 \left( m_x, m_y \right) = \inf_{\mu_{x, y} \in \Pi \left( m_x, m_y \right)} \left[ \sum_{\left( x^{\prime}, y^{\prime} \right) \in X \times X} d \left( x^{\prime}, y^{\prime} \right) \mu_{x, y} \left( x^{\prime}, y^{\prime} \right) \right],
\end{equation}
where the discrete probability measures ${\mu_{x, y}}$ in the set ${\Pi \left( m_x, m_y \right)}$ satisfy the coupling conditions:

\begin{equation}
\sum_{y^{\prime} \in X} \mu_{x, y} \left( x^{\prime}, y^{\prime} \right) = \mu_x \left( x^{\prime} \right), \qquad \text{ and } \qquad \sum_{x^{\prime} \in X} \mu_{x, y} \left( x^{\prime}, y^{\prime} \right) = \mu_y \left( y^{\prime} \right).
\end{equation}
From here, directed graphs (of the kind represented by Hasse diagrams for causal sets) may be treated as special cases of directed \textit{hypergraphs} ${H = \left( V, E \right)}$, where each hyperedge ${e \in E}$ is generally assumed to represent a directional relation between vertex sets $A$ and $B$ (the \textit{tail set} and the \textit{head set}, respectively), except in this case the sets $A$ and $B$ are each taken to have cardinality exactly 1. In the general case, the discrete/multi-marginal \textit{1-Wasserstein} distance ${W_1 \left( \mu_{A^{in}}, \mu_{B^{out}} \right)}$ between discrete probability measures ${\mu_{A^{in}}}$ and ${\mu_{B^{out}}}$ defined over the directed hypergraph ${H = \left( V, E \right)}$ becomes:

\begin{equation}
W_1 \left( \mu_{A^{in}}, \mu_{B^{out}} \right) = \min_{u \in A^{in} \left( u \to A \right), v \in B^{out} \left( B \to v \right)} \left[ \sum_{u \to A} \sum_{B \to v} d \left( u, v \right) \varepsilon \left( u, v \right) \right].
\end{equation}
Here, ${\varepsilon \left( u, v \right)}$ designates the coupling between vertices $u$ and $v$, such that one effectively minimizes over all couplings ${\varepsilon}$ between the discrete measures ${\mu_{A^{in}}}$ and ${\mu_{B^{out}}}$ satisfying the coupling conditions:

\begin{equation}
\sum_{u \to A} \varepsilon \left( u, v \right) = \sum_{j = 1}^{m} \mu_{y_j} \left( v \right), \qquad \text{ and } \qquad \sum_{B \to v} \varepsilon \left( u, v \right) = \sum_{i = 1}^{n} \mu_{x_i} \left( u \right),
\end{equation}
assuming that the hyperedge $e$ is of the general form:

\begin{equation}
A = \left\lbrace x_1, \dots, x_n \right\rbrace \to^{e} B = \left\lbrace y_1, \dots, y_m \right\rbrace, \qquad \text{ with } \qquad n, m \leq \left\lvert V \right\rvert,
\end{equation}
and the discrete metric ${d \left( u, v \right)}$ corresponds specifically to the number of directed hyperedges that one traverses when traveling between vertices ${u \in A^{in} \left( u \to A \right)}$ and ${v \in B^{out} \left( B \to v \right)}$. If we make the simplest choice of distance metric ${d \left( u, v \right)}$ and coupling constants ${\varepsilon \left( u, v \right)}$, such that all distances/couplings are symmetric (i.e. ${d \left( u, v \right) = d \left( v, u \right)}$ and ${\varepsilon \left( u, v \right) = \varepsilon \left( v, u \right)}$) and each directed hyperedge corresponds to a distance of 1 (which is equivalent to enforcing a torsion-free/metric/Levi-Civita connection on the continuum manifold), then we can write the form of the \textit{Ollivier-Ricci} scalar curvature ${\kappa \left( e \right)}$ function on hyperedges ${e \in E}$ in the following explicit fashion:

\begin{equation}
\kappa \left( e \right) = 1 - W_1 \left( \mu_{A^{in}}, \mu_{B^{out}} \right),
\end{equation}
where the discrete probability measures ${\mu_{A^{in}}}$ and ${\mu_{B^{out}}}$ satisfy the coupling conditions:

\begin{equation}
\mu_{A^{in}} = \sum_{i = 1}^{n} \mu_{x_i}, \qquad \text{ and } \qquad \mu_{B^{out}} = \sum_{j = 1}^{m} \mu_{y_j},
\end{equation}
or, more concretely, one has:

\begin{equation}
\forall 1 \leq i \leq n, z \in V, \qquad \mu_{x_i} \left( z \right) = \begin{cases}
0, \qquad & \text{ if } z = x_i \text{ and } d_{x_{i}^{in}} \neq 0,\\
\frac{1}{n}, \qquad & \text{ if } z = x_i \text{ and } d_{x_{i}^{in}} = 0,\\
\sum\limits_{e^{\prime}: z \to x_i} \frac{1}{n \times d_{x_{i}^{in}} \times \left\lvert \mathrm{Tail} \left( e^{\prime} \right) \right\rvert}, \qquad & \text{ if } z \neq x_i \text{ and } \exists e^{\prime} : z \to x_i,\\
0, \qquad & \text{ if } z \neq x_i \text{ and } \nexists e^{\prime} : z \to x_i,
\end{cases}
\end{equation}
and:

\begin{equation}
\forall 1 \leq j \leq m, z \in V, \qquad \mu_{y_j} \left( z \right) = \begin{cases}
0, \qquad & \text{ if } z = y_j \text{ and } d_{y_{j}^{out}} \neq 0,\\
\frac{1}{m}, \qquad & \text{ if } z = y_j \text{ and } d_{y_{j}^{out}} = 0,\\
\sum\limits_{e^{\prime} : y_j \to z} \frac{1}{m \times d_{y_{j}^{out}} \times \left\lvert \mathrm{Head} \left( e^{\prime} \right) \right\rvert}, \qquad & \text{ if } z \neq y_j \text{ and } \exists e^{\prime} : y_j \to z,\\
0, \qquad & \text{ if } z \neq y_j \text{ and } \nexists e^{\prime} : y_j \to z,
\end{cases}
\end{equation}
respectively. In the above, ${d_{x_{i}^{in}}}$ represents the total number of hyperedges that include a given vertex ${x_i \in A}$ (taken from the \textit{tail set} $A$) as an element of their respective \textit{head set}, and likewise ${d_{y_{j}^{out}}}$ represents the total number of hyperedges that include a given vertex ${y_j \in B}$ (taken from the \textit{head set} $B$) as an element of their respective \textit{tail set}.

\begin{figure}[ht]
\centering
\includegraphics[width=0.395\textwidth]{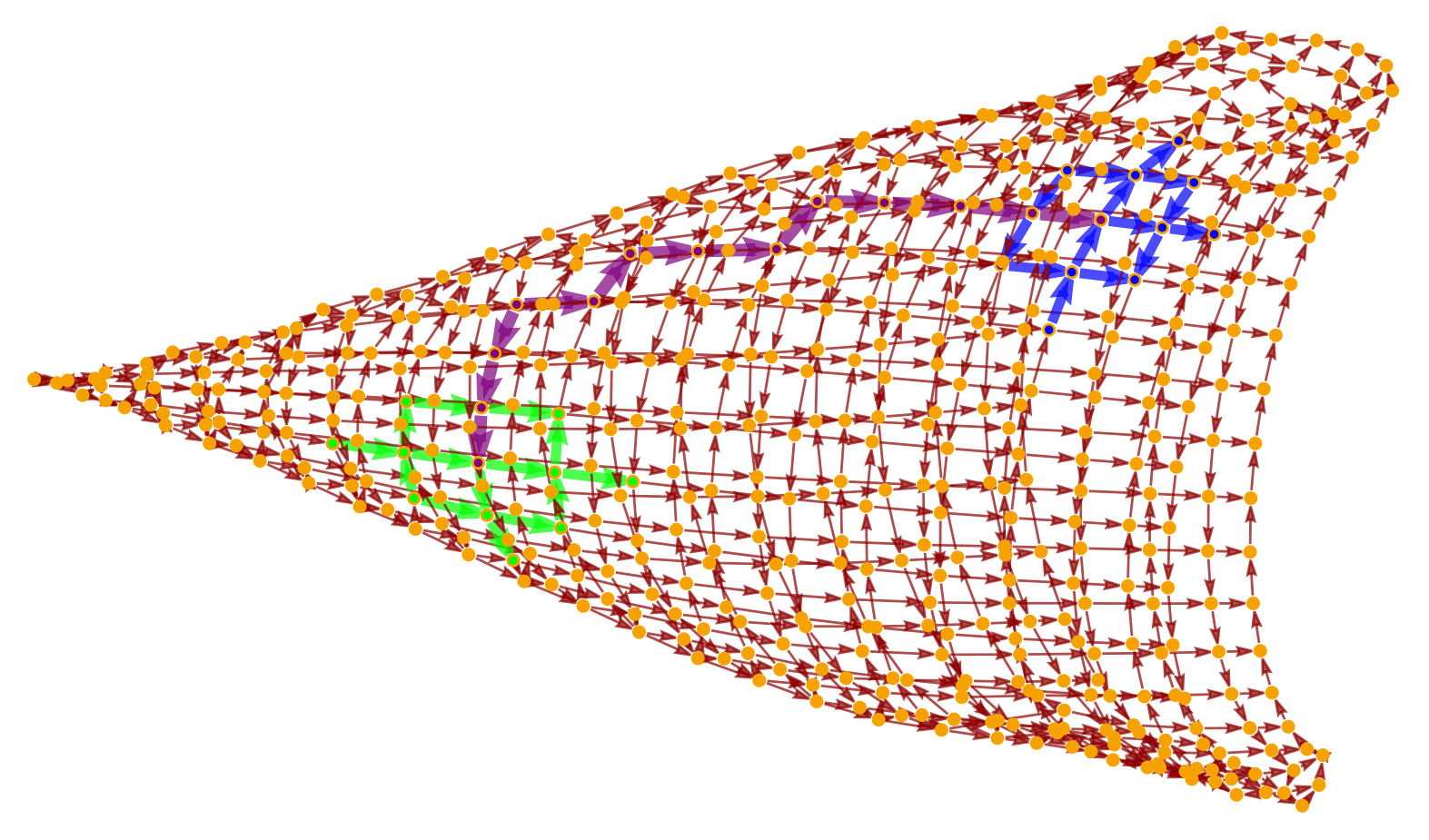}\hspace{0.1\textwidth}
\includegraphics[width=0.395\textwidth]{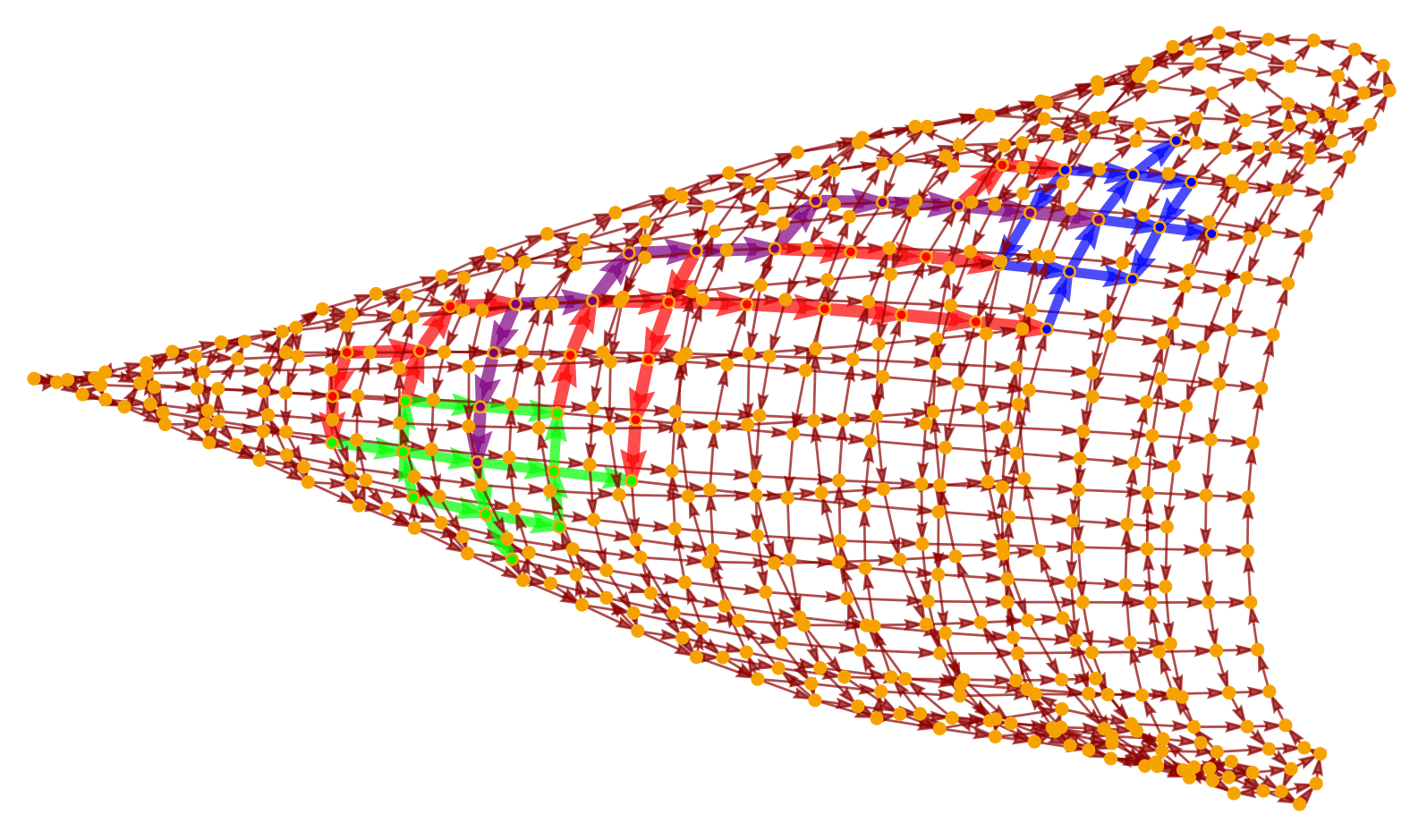}
\caption{On the left, a pair of finite geodesic cones in an asymptotically-flat (directed) causal graph with a ${1 + 1}$-dimensional Lorentzian manifold-like limiting structure, as generated by the hypergraph rewriting rule ${\left\lbrace \left\lbrace x, y, y \right\rbrace, \left\lbrace x, z, u \right\rbrace \right\rbrace \to \left\lbrace \left\lbrace u, v, v \right\rbrace, \left\lbrace v, z, y \right\rbrace, \left\lbrace x, y, v \right\rbrace \right\rbrace}$, with a purple path representing the distance between the centers of the cones. On the right, the family of all possible paths (shown in red) between corresponding points on the surfaces on the surfaces of the two cones after parallel transport. The lack of any net divergence or convergence of these paths implies that the \textit{Ollivier-Ricci} scalar curvature ${\kappa}$ is equal to zero along the purple path.}
\label{fig:Figure6}
\end{figure}

\begin{figure}[ht]
\centering
\includegraphics[width=0.395\textwidth]{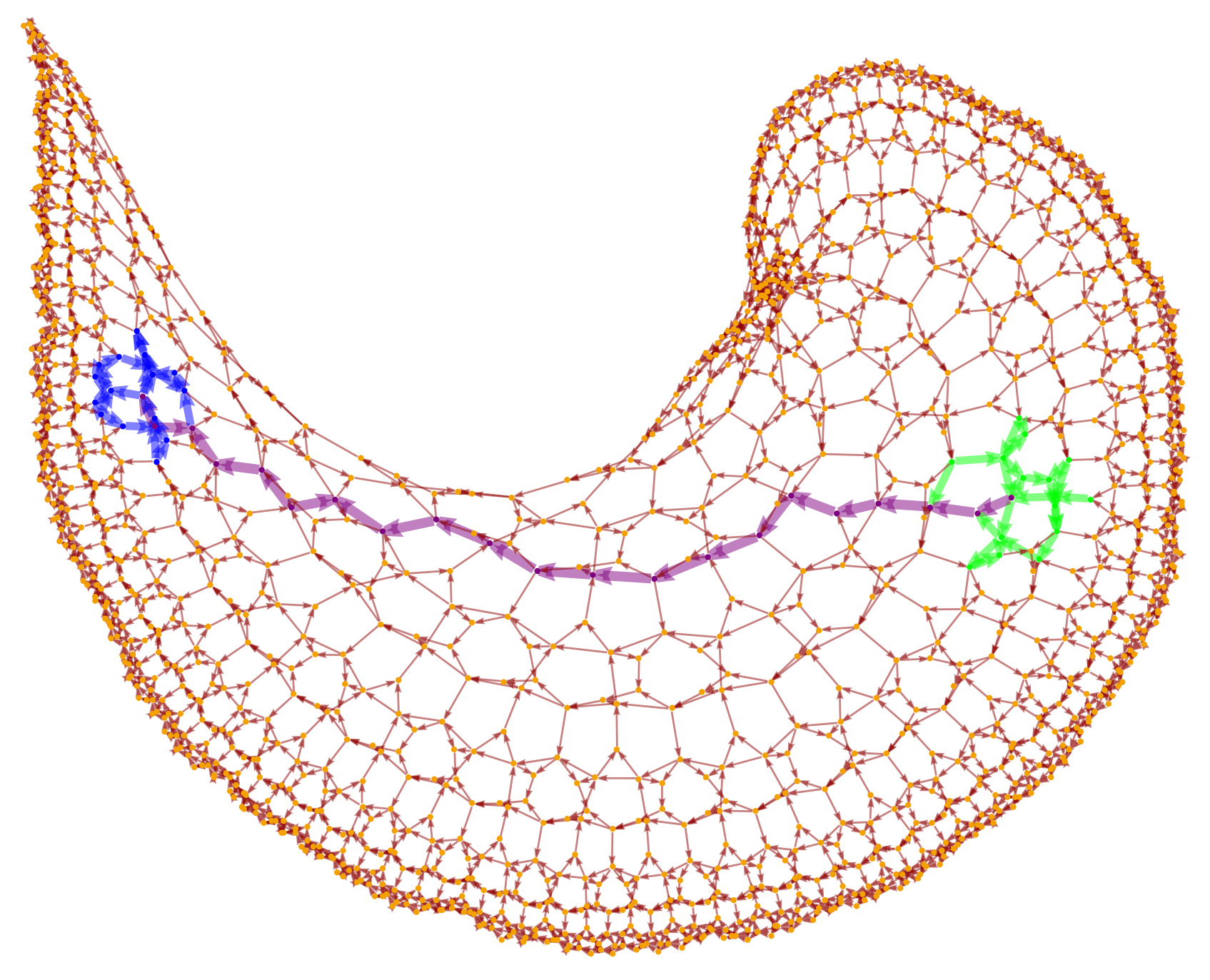}\hspace{0.1\textwidth}
\includegraphics[width=0.395\textwidth]{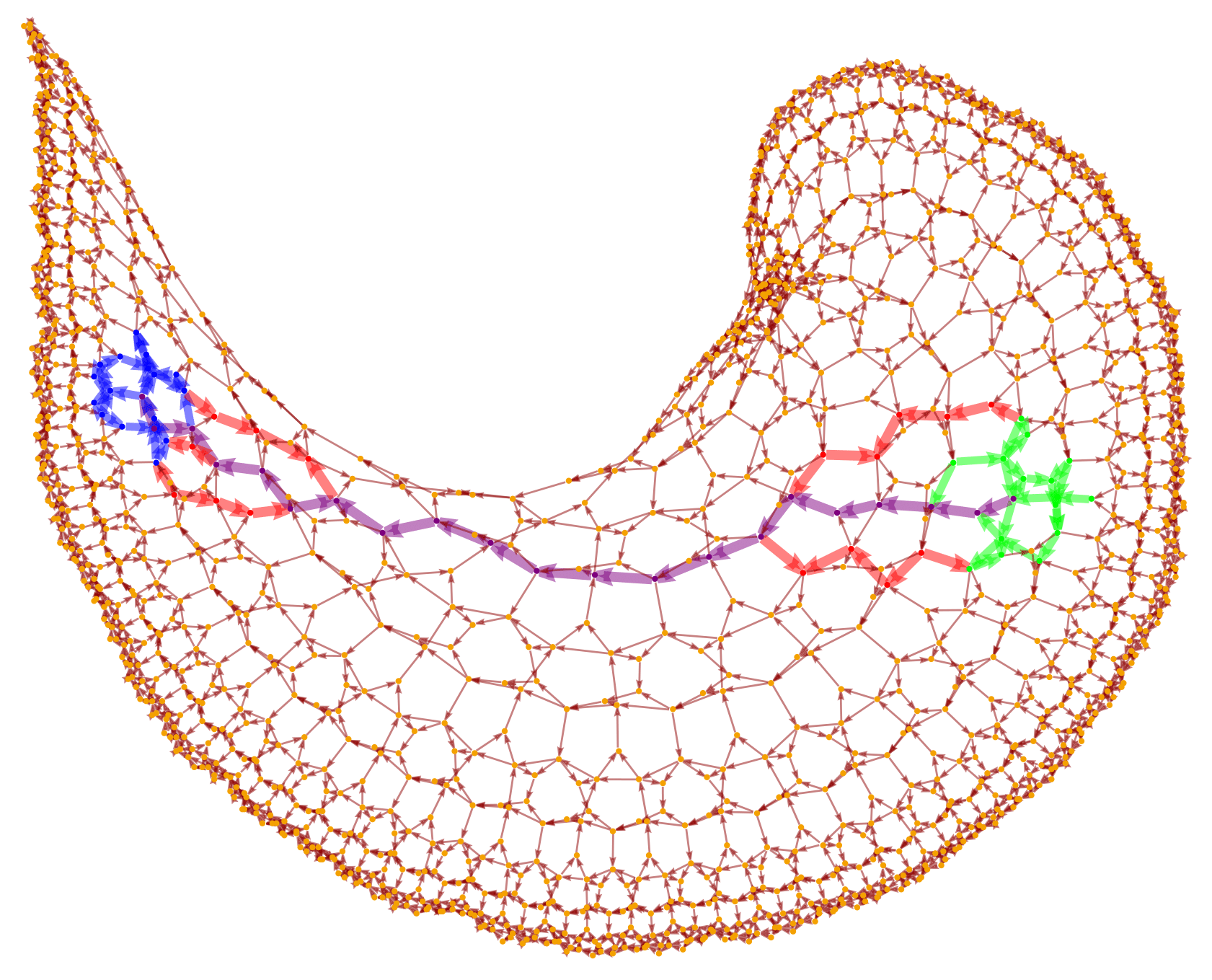}
\caption{On the left, a pair of finite geodesic cones in a (directed) causal graph with a ${1 + 1}$-dimensional Lorentzian manifold-like limiting structure with negative global curvature, as generated by the hypergraph rewriting rule ${\left\lbrace \left\lbrace x, y, x \right\rbrace, \left\lbrace x, z, u \right\rbrace \right\rbrace \to \left\lbrace \left\lbrace u, v, u \right\rbrace, \left\lbrace v, u, z \right\rbrace, \left\lbrace x, y, v \right\rbrace \right\rbrace}$, with a purple path representing the distance between the centers of the cones. On the right, the family of all possible paths (shown in red) between corresponding points on the surfaces of the two cones after parallel transport. The net convergence of these paths implies that the \textit{Ollivier-Ricci} scalar curvature ${\kappa}$ is negative along the purple path.}
\label{fig:Figure7}
\end{figure}

Clearly, in the Lorentzian (directed graph) case, as opposed to the usual Riemannian (undirected graph) case, one must consider discrepancies in distances between directed geodesic \textit{cones}, as opposed merely to geodesic balls, but in all other respects the geometrical intuitions are identical. Examples of how the \textit{Ollivier-Ricci} scalar curvature ${\kappa}$ may be computed for (directed) causal graphs which limit to be asymptotically-flat and asymptotically-positively-curved ${1 + 1}$-dimensional Lorentzian manifold-like structures are shown in Figures \ref{fig:Figure6} and \ref{fig:Figure7}, respectively; the precise details of how such causal graphs are algorithmically generated via hypergraph substitution rules will be outlined shortly. However, these curvature estimation algorithms (which, as discussed, are necessary in order to be able to construct reliable discrete Green's functions for causal sets representing general curved spacetimes) generically require one to know the limiting dimension of the causal set a priori. Clearly, this is not a problem for the case of causal sets generated by sprinkling into a known Lorentzian manifold, but for causal sets generated by an algorithmic procedure such as hypergraph rewriting, the limiting dimension is generically unknown (and quite often very hard to predict), in which case one also requires a reliable algorithm for estimating limiting dimension, as we shall discuss in the following section.

\section{Extending Scalar Field Green's Functions to Non-Integer-Dimensional Spacetimes}
\label{sec:Section2}

\begin{figure}[ht]
\centering
\includegraphics[width=0.395\textwidth]{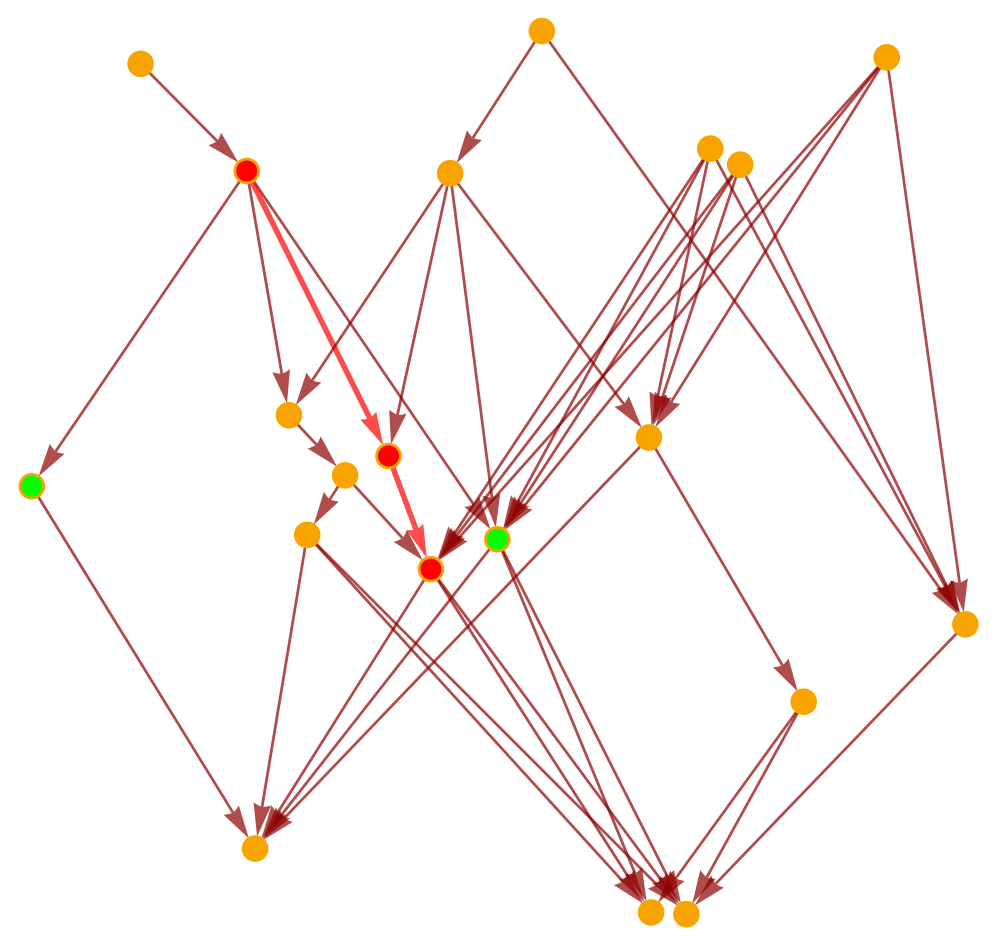}
\caption{The transitive reduction (i.e. the Hasse diagram) of the directed graph generated by connecting all pairs of 20 (uniformly-sprinkled) points that are related by the causal partial order relation ${\prec_{\mathcal{M}}}$ on a rectangular region of a ${1 + 1}$-dimensional flat (Minkowski) spacetime, with a pair of randomly-selected points (shown in red) that are related by the partial order relation, and another pair of randomly-selected points (shown in green) that are not related by the partial order relation.}
\label{fig:Figure8}
\end{figure}

\begin{figure}[ht]
\centering
\includegraphics[width=0.395\textwidth]{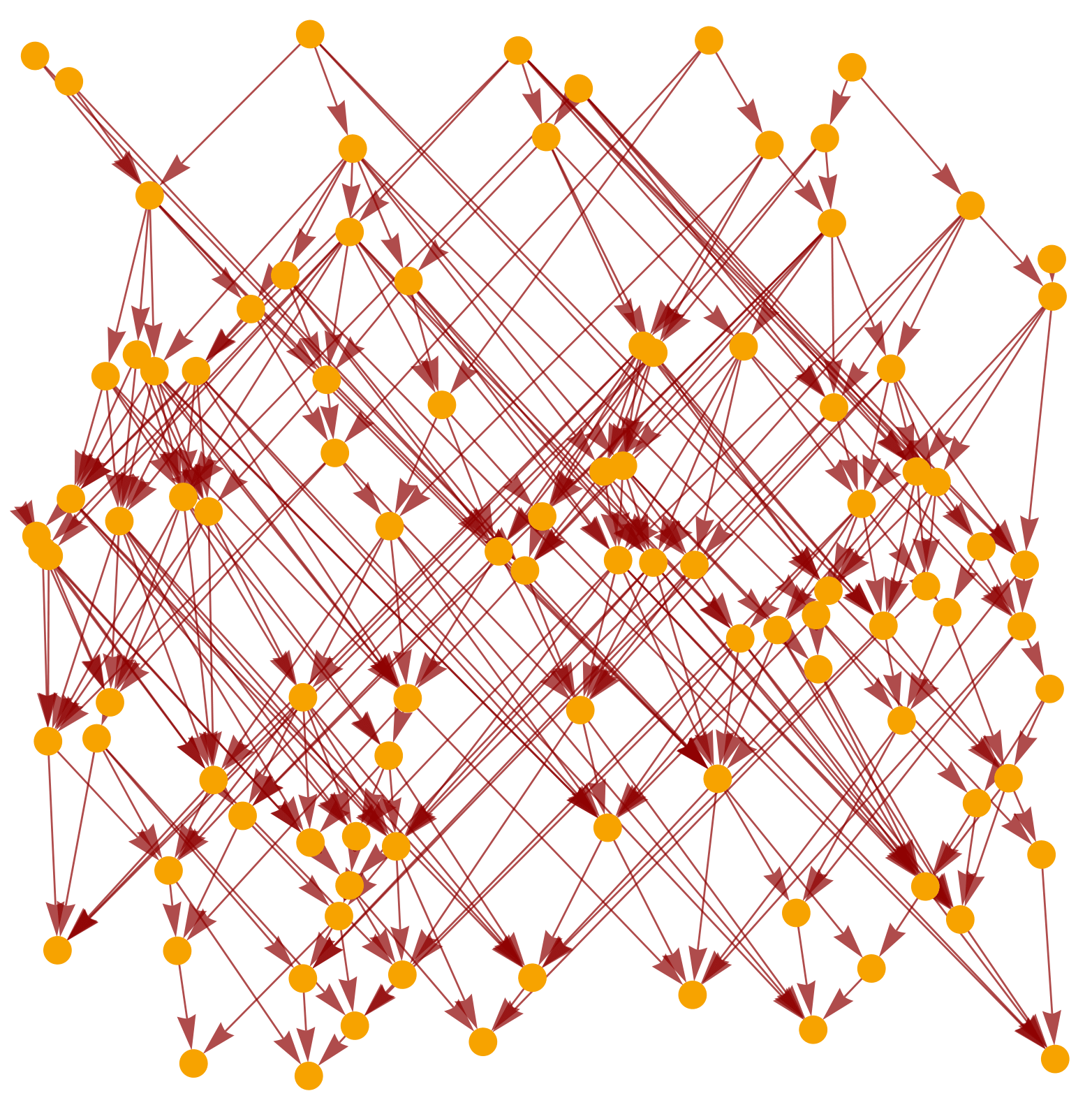}\hspace{0.1\textwidth}
\includegraphics[width=0.395\textwidth]{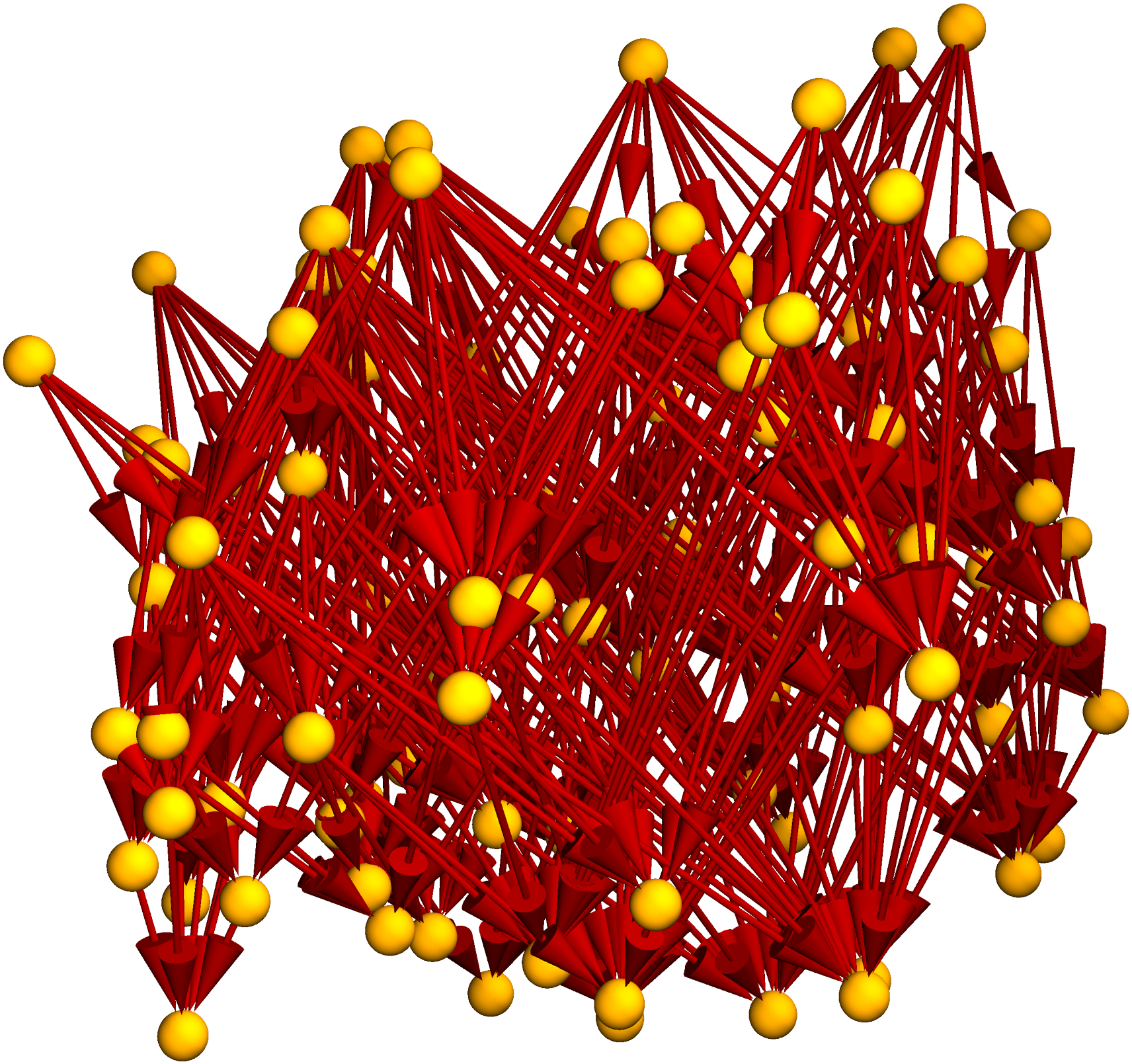}
\caption{The transitive reductions (i.e. the Hasse diagrams) of the directed graphs generated by connecting all pairs of 100 (uniformly-sprinkled) points that are related by the causal partial order relation ${\prec_{\mathcal{M}}}$ on rectangular regions of ${1 + 1}$-dimensional and ${2 + 1}$-dimensional flat (Minkowski) spacetimes, respectively, with estimated dimensions of 2.07 and 2.75, respectively, obtained via the Myrheim-Meyer estimation procedure.}
\label{fig:Figure9}
\end{figure}

Since causal sets generated via algorithmic procedures (such as hypergraph rewriting), rather than via sprinklings into pre-existing Lorentzian manifolds, are not necessarily guaranteed to have integer values for their limiting dimension, it will eventually become necessary for us to ``analytically continue'' the massless discrete Green's functions constructed in the previous section for the case of arbitrary \textit{integer}-dimensional causal sets to the more general non-integer-dimensional case. More formally, it is not the Green's functions themselves that are being analytically continued; rather, we wish to express the contributions to those Green's functions as meromorphic functions of a complex parameter ${d \in \mathbb{C}}$, where this parameter may itself be interpreted as an analytic continuation of the number of spacetime dimensions, through a procedure that is analogous to the dimensional regularization of Feynman integrals in quantum field theory via the methods of 't Hooft and Veltman\cite{hooft}. In order for us to perform this ``analytic continuation'' in an effective manner, we must first introduce a reliable procedure for estimating the limiting dimension of an arbitrary causal set ${\mathcal{C}}$. The approach advocated by Myrheim\cite{myrheim} and Meyer\cite{meyer} begins by computing the expectation value ${\left\langle \hat{R} \right\rangle}$ of the number $R$ of pairs of elements in ${\mathcal{C}}$ that are related by the causal partial order relation ${\prec}$:

\begin{equation}
R = \left\lvert \left\lbrace \left( e_i, e_j \right) : e_i, e_j \in \mathcal{C} \text{ such that } e_i \prec e_j \right\rbrace \right\rvert.
\end{equation}
If we now select a continuum Alexandrov interval ${\mathbf{A} \left[ p, q \right] \subset \mathbb{M}^d}$ between two events $p$ and $q$ (separated by a proper time distance of ${\tau}$) in a $d$-dimensional flat (Minkowski) spacetime, then we know that its volume ${\mathrm{Vol} \left( \mathbf{A} \left[ p, q \right] \right)}$ is given by:

\begin{equation}
\mathrm{Vol} \left( \mathbf{A} \left[ p, q \right] \right) = \frac{\mathcal{V}_{d - 2}}{2^{d - 1} d \left( d - 1 \right)} \tau^d,
\end{equation}
where ${\mathcal{V}_{d - 2}}$ denotes the volume of the unit sphere in ${d - 2}$ dimensions, and where we have exploited Lorentz symmetry to choose the following coordinates for $p$ and $q$ without loss of generality:

\begin{equation}
p = \left( - \frac{\tau}{2}, 0, \dots, 0 \right), \qquad \text{ and } \qquad q = \left( \frac{\tau}{2}, 0, \dots, 0 \right).
\end{equation}
Therefore, the expectation value ${\left\langle \hat{R} \right\rangle}$ for the number of causally-related pairs of elements, when considered over the ensemble ${\Omega}$ of all possible causal sets sprinkled into the interval ${\mathbf{A} \left[ p, q \right]}$, is given (as discussed previously) by the integral:

\begin{equation}
\left\langle \hat{R} \right\rangle = \rho_{c}^{2} \int_{\mathbf{A} \left[ p, q \right]} \int_{J^{+} \left( x_1 \right) \cap J^{-} \left( q \right)} d x_2 d x_1 = \rho_{c}^{2} \frac{\mathcal{V}_{d - 2}}{2^{d - 1} d \left( d - 1 \right)} \int_{\mathbf{A} \left[ p, q \right]} t_{1}^{d} d x_1,
\end{equation}
where we have introduced the parameter ${\tau_1}$ to denote the proper time distance between events ${x_1}$ and $q$, which now evaluates directly to yield:

\begin{equation}
\left\langle \hat{R} \right\rangle = \rho_{c}^{2} \mathrm{Vol} \left( \mathbf{A} \left[ p, q \right] \right) \frac{\Gamma \left( d + 1 \right) \Gamma \left( \frac{d}{2} \right)}{4 \Gamma \left( \frac{3 d}{2} \right)},
\end{equation}
thus allowing us to write the ratio of the expectation value ${\left\langle \hat{R} \right\rangle}$ for the number of pairs of causally-related elements to the square of the expectation value ${\left\langle \hat{n} \right\rangle}$ for the total number of sprinkled elements within the spacetime region ${\mathbf{A} \left[ p, q \right]}$, namely (by the definition of the Poisson point process):

\begin{equation}
\left\langle \hat{n} \right\rangle = \rho_c \mathrm{Vol} \left( \mathbf{A} \left[ p, q \right] \right),
\end{equation}
as:

\begin{equation}
\frac{\left\langle \hat{R} \right\rangle}{\left\langle n \right\rangle^2} = \frac{\Gamma \left( d + 1 \right) \Gamma \left( \frac{d}{2} \right)}{4 \Gamma \left( \frac{3 d}{2} \right)}.
\end{equation}
In consequence, by estimating the proportion of pairs of elements within a given causal set that has been sprinkled into a continuum spacetime region of a specified volume, we are able to invert this equation in order to deduce a corresponding estimate of the continuum spacetime dimension $d$, as shown for a simplified (20 element) example in Figure \ref{fig:Figure8}, and for more realistic (100 element) examples in ${1 + 1}$-dimensional and ${2 + 1}$-dimensional flat (Minkowski) spacetimes in Figure \ref{fig:Figure9}.

Recall, from our previous analysis of the \textit{summing over chains} approach to the computation of discrete path integrals over causal sets, that the abundance ${C_n}$ of chains of length $n$ in a causal set ${\mathcal{C}}$:

\begin{equation}
C_n = \left\lvert \left\lbrace \left( p, r_1, r_2, \dots, r_{n - 1}, q \right) : p, r_1, r_2 \dots, r_{n - 1}, q \in \mathcal{C} \text{ such that } p \prec r_1 \prec r_2 \prec \cdots \prec r_{n - 1} \prec q \right\rbrace \right\rvert,
\end{equation}
has an expectation value ${\left\langle \hat{C_n} \right\rangle}$ (over the ensemble ${\Omega}$ of all possible causal sets sprinkled into the given spacetime) that can be written in the following convenient closed form:

\begin{equation}
\left\langle \hat{C_n} \right\rangle = \frac{\left( \rho_c \mathrm{Vol} \left( \mathbf{A} \left[ q, p \right] \right) \right)^{n - 1}}{n - 1} \left( \frac{\Gamma \left( d + 1 \right)}{2} \right)^{n - 2} \frac{\Gamma \left( \frac{d}{2} \right) \Gamma \left( d \right)}{\Gamma \left( \left( n - 1 \right) \frac{d}{2} \right) \Gamma \left( \frac{n d}{2} \right)}.
\end{equation}
Therefore, the Myheim-Meyer dimension estimation procedure outlined above may be generalized by taking the ratio of the chain abundance expectation values ${\left\langle \hat{C_n} \right\rangle^{\frac{1}{n}}}$ and ${\left\langle \hat{C_{n^{\prime}}} \right\rangle^{\frac{1}{n^{\prime}}}}$, for any ${n, n^{\prime} \in \mathbb{N}}$ such that ${n \neq n^{\prime}}$. For an arbitrary Lorentzian manifold ${\left( \mathcal{M}, g \right)}$ exhibiting non-zero spacetime curvature, and which is therefore not isometric to ${\mathbb{M}^d = \mathbb{R}^{1, d - 1}}$, one can follow the approach of Roy, Sinha and Surya\cite{roy}, namely using the expansion of the metric determinant ${\det \left( g \right)}$ in curved spacetime (in Riemann normal coordinates) to obtain the following series representation for the chain abundance expectation value ${\left\langle \hat{C_n} \right\rangle}$ in powers of the proper time interval ${\tau}$ between events $p$ and $q$:

\begin{equation}
\left\langle \hat{C_n} \right\rangle = \left\langle \hat{C_n} \right\rangle_{\eta} \left[ 1 - \frac{d \left( n - 1 \right)}{12 \left( \left( n - 1 \right) d + 2 \right) \left( n d + 2 \right)} R \left( 0 \right) \tau^2 + \frac{d \left(n - 1 \right)}{12 \left( n d + 2 \right)} R_{0 0} \left( 0 \right) \tau^2 \right] + O \left( \tau^{\left( n - 1 \right) d + 3} \right),
\end{equation}
under the hypothesis that the causal diamond ${\mathbf{A} \left[ p, q \right]}$ is sufficiently small that the following conditions hold:

\begin{equation}
R \left( 0 \right) \tau^2 \ll 1, \qquad \text{ and } \qquad R_{0 0} \left( 0 \right) \tau^2 \ll 1,
\end{equation}
where $R \left( 0 \right)$ designates the Ricci scalar curvature and ${R_{0 0} \left( 0 \right)}$ designates the time-time component of the Ricci curvature tensor (evaluated at the center of the diamond in the Riemann normal coordinate system)\cite{khetrapal}, and where we use the notation ${\left\langle \hat{C_n} \right\rangle_{\eta}}$ to represent the chain abundance expectation value in the flat/Minkowski spacetime case (as presented above). This allows us to derive a generalized dimension estimator for small causal diamonds in arbitrary spacetimes, of the form:

\begin{multline}
\frac{\Gamma \left( d + 1 \right) \Gamma \left( \frac{d}{2} \right)}{4 \Gamma \left( \frac{3 d}{2} \right)} \left( - \frac{1}{3} \frac{\left( d + 2 \right)}{\left( 3 d + 2 \right)} - \frac{\left( 4 d + 2 \right)}{\left( 2 d + 2 \right)} \left( \frac{\left\langle \hat{C_n} \right\rangle}{\frac{1}{3} \left( \frac{\Gamma \left( d + 1 \right)}{2} \right)^2 \frac{\Gamma \left( \frac{d}{2} \right) \Gamma \left( d \right)}{\Gamma \left( \frac{3 d}{2} \right) \Gamma \left( 2 d \right)}} \right)^{\frac{4}{3}} \frac{1}{\left\langle \hat{C_n} \right\rangle^4} \right.\\
\left. + \frac{1}{3} \frac{\left( 4 d + 2 \right) \left( 5 d + 2 \right)}{\left( 2 d + 2 \right) \left( 3 d + 2 \right)} \frac{\left\langle \hat{C_5} \right\rangle}{\frac{1}{4} \left( \frac{\Gamma \left( d + 1 \right)}{2} \right)^{3} \frac{\Gamma \left( \frac{d}{2} \right) \Gamma \left( d \right)}{\Gamma \left( 2 d \right) \Gamma \left( \frac{5 d}{2} \right)}} \frac{1}{\left\langle \hat{C_2} \right\rangle^4} \right) = - \frac{\left\langle \hat{C_3} \right\rangle^2}{\left\langle \hat{C_2} \right\rangle^4}.
\end{multline}

The (generalized) Myrheim-Meyer dimension estimation procedure, however, exhibits a number of characteristics that are somewhat undesirable for our present purposes. Most apparently, its computation necessitates a priori knowledge of the volume ${\mathrm{Vol} \left( \mathbf{A} \left[ p, q \right] \right)}$ of the small causal diamond into which the specified causal set has been sprinkled, which one simply does not possess in the case of causal sets that have been algorithmically grown. More subtly, it suffers from similar issues of tautological definition to those previously seen in the case of the Benincasa-Dowker approach to computing the discrete Ricci scalar ${R \left( x \right)}$, namely that the proper time distance ${\tau}$ between events $p$ and $q$ is conventionally \textit{defined} in terms of the dimension $d$ of the continuum spacetime, via:

\begin{equation}
\tau^{3 d} = \frac{2}{^2 \rho_{c}^{3}} \left( J_1 - 2 J_2 + J_3 \right),
\end{equation}
for the inductively-defined parameters ${J_n}$:

\begin{equation}
J_n = \left( \left( n - 1 \right) d \right) \left( n d + 2 \right) \left( \frac{2^{d - 1} d \left( d - 1 \right)}{\mathcal{V}_{d - 2}} \right)^3 \left( \frac{\left\langle \hat{C_{n - 1}} \right\rangle}{\frac{1}{n} \left( \frac{\Gamma \left( d + 1 \right)}{2} \right)^{n - 1} \frac{\Gamma \left( \frac{d}{2} \right) \Gamma \left( d \right)}{\Gamma \left( \frac{n d}{2} \right) \Gamma \left( \frac{\left( n + 1 \right) d}{2} \right)}} \right)^{\frac{3}{n}}.
\end{equation}
Hence, the estimator is useful if one knows either the continuum dimension $d$ or the proper time distance ${\tau}$ between events and wishes to deduce the other quantity, but not if both quantities are unknown. However, these limitations may nevertheless be surmounted if we instead employ a more direct approximation to the limiting Hausdorff dimensionality of the causal set, by applying a logarithmic distance estimate to the growth rates of volumes of geodesic cones. More specifically, if we select a point ${p \in \mathcal{M}}$ in an arbitrary Riemannian manifold ${\left( \mathcal{M}, g \right)}$ of dimension $d$, then we can expand the infinitesimal volume element (given, as usual, by the square root of the determinant of the metric tensor $g$) around a nearby point ${p + \delta x}$ as a formal series in powers of ${\delta x}$\cite{jost}:

\begin{equation}
\sqrt{\det \left( g \left( p + \delta x \right) \right)} = \sqrt{\det \left( g \left( p \right) \right)} \left( 1 - \frac{1}{6} \sum_{i = 1}^{d} R_{i j} \left( p \right) \delta x^i \delta x^j + O \left( \delta x^3 \right) + \dots \right),
\end{equation}
subject to the assumption that the manifold ${\mathcal{M}}$ is analytic, where ${R_{i j}}$ is the Ricci curvature tensor and ${\delta x^i}$ and ${\delta x^j}$ correspond to the orthogonal components of the vector ${\delta x}$, expressed in contravariant index notation. The volume of a small geodesic ball ${B_{\varepsilon} \left( p \right)}$, centered at point $p$ and of infinitesimal radius ${\varepsilon}$, may therefore be recovered by integrating this infinitesimal volume element:

\begin{equation}
\mathrm{Vol} \left( B_{\varepsilon} \left( p \right) \right) = \int_{B_{\epsilon} \left( p \right)} \sqrt{\det \left( g \left( p + \delta x \right) \right)} d^d \left( \delta x \right);
\end{equation}
note that this integral can be expanded in powers of the radius ${\varepsilon}$ as:

\begin{equation}
\int_{B_{\varepsilon} \left( p \right)} \sqrt{\det \left( g \left( p + \delta x \right) \right)} d^d \left( \delta x \right) = \frac{\pi^{\frac{d}{2}}}{\Gamma \left( \frac{d}{2} + 1 \right)} \varepsilon^d \left( 1 - \frac{\varepsilon^2}{6 \left( d + 2 \right)} \sum_{i = 1}^{d} R_{i}^{i} + O \left( \varepsilon^4 \right) \right),
\end{equation}
where the trace of the curvature tensor ${R_{i}^{i}}$ may be rewritten simply as the Ricci scalar curvature $R$:

\begin{equation}
\sum_{i = 1}^{d} R_{i}^{i} = R.
\end{equation}
If one integrates instead over a small tube ${T_{\varepsilon, \delta x} \left( p \right)}$ emanating from point $p$ along a geodesic oriented in the direction ${\delta x}$, with length ${\delta}$ and of infinitesimal radius ${\varepsilon}$:

\begin{equation}
\mathrm{Vol} \left( T_{\varepsilon, \delta x} \left( p \right) \right) = \int_{T_{\varepsilon, \delta x} \left( p \right)} \sqrt{\det \left( g \left( p + \delta x \right) \right)} d^d \left( \delta x \right),
\end{equation}
one correspondingly obtains an expression for the volume of such a tube as a power series in the radius ${\varepsilon}$\cite{gray}:

\begin{multline}
\int_{T_{\varepsilon, \delta x} \left( p \right)} \sqrt{\det \left( g \left( p + \delta x \right) \right)} d^d \left( \delta x \right) = \frac{\pi^{\frac{d - 1}{2}}}{\Gamma \left( \frac{d + 1}{2} \right)} \varepsilon^{d - 1} \delta x \left( 1 - \left( \frac{d - 1}{d + 1} \right) \left( R - \sum_{i, j = 1}^{d} R_{i j} \hat{\delta x}^i \hat{\delta x}^j \right) \varepsilon^2 \right.\\
\left. + O \left( \varepsilon^3 + \varepsilon^2 \delta x \right) + \dots \right),
\end{multline}
assuming unit vectors along the geodesic with orthogonal components ${\hat{\delta x}^i}$ and ${\hat{\delta x}^j}$ (in contravariant form).

\begin{figure}[ht]
\centering
\includegraphics[width=0.395\textwidth]{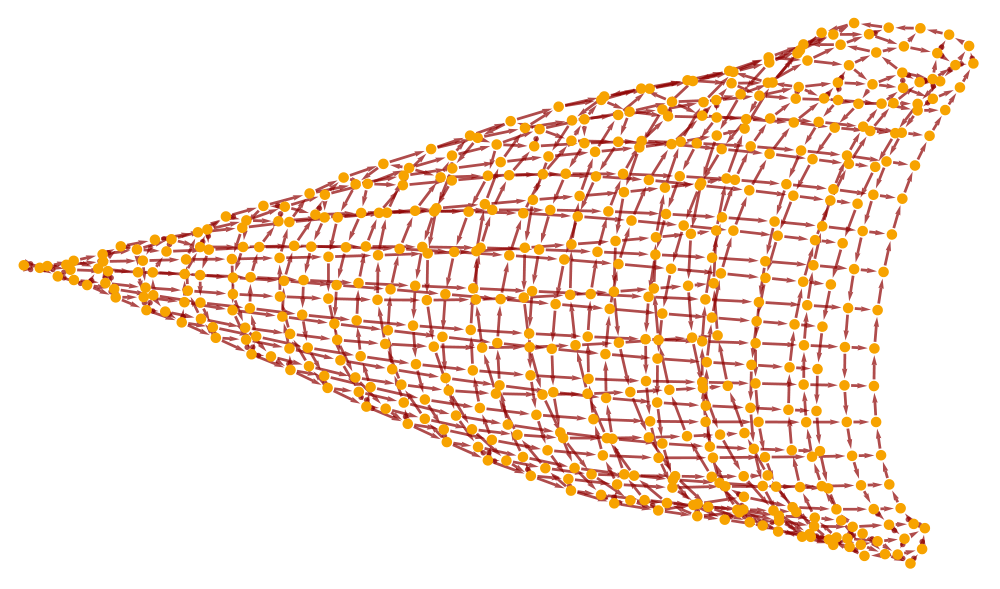}\hspace{0.1\textwidth}
\includegraphics[width=0.395\textwidth]{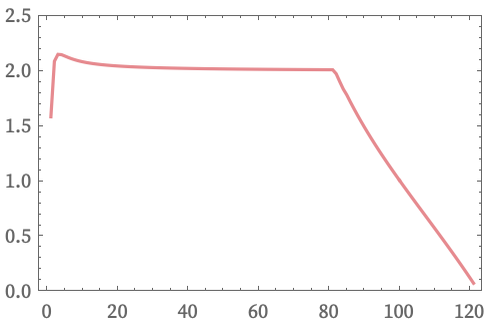}
\caption{Logarithmic difference estimates for the dimension of an asymptotically-flat (directed) causal graph with a ${1 + 1}$-dimensional Lorentzian manifold-like limiting structure, as generated by the hypergraph rewriting rule ${\left\lbrace \left\lbrace x, y, y \right\rbrace, \left\lbrace x, z, u \right\rbrace \right\rbrace \to \left\lbrace \left\lbrace u, v, v \right\rbrace, \left\lbrace v, z, y \right\rbrace, \left\lbrace x, y, v \right\rbrace \right\rbrace}$, showing results that are consistent with a limiting dimension of two.}
\label{fig:Figure10}
\end{figure}

\begin{figure}[ht]
\centering
\includegraphics[width=0.895\textwidth]{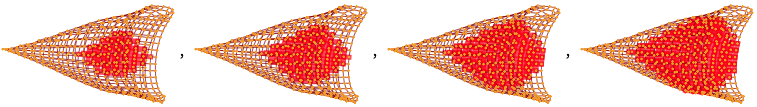}
\caption{The growth of a causal cone (shown in red) in an asymptotically-flat (directed) causal graph with a ${1 + 1}$-dimensonal Lorentzian manifold-like limiting structure, as generated by the hypergraph rewriting rule ${\left\lbrace \left\lbrace x, y, y \right\rbrace, \left\lbrace x, z, u \right\rbrace \right\rbrace \to \left\lbrace \left\lbrace u, v, v \right\rbrace, \left\lbrace v, z, y \right\rbrace, \left\lbrace x, y, v \right\rbrace \right\rbrace}$.}
\label{fig:Figure11}
\end{figure}

\begin{figure}[ht]
\centering
\includegraphics[width=0.395\textwidth]{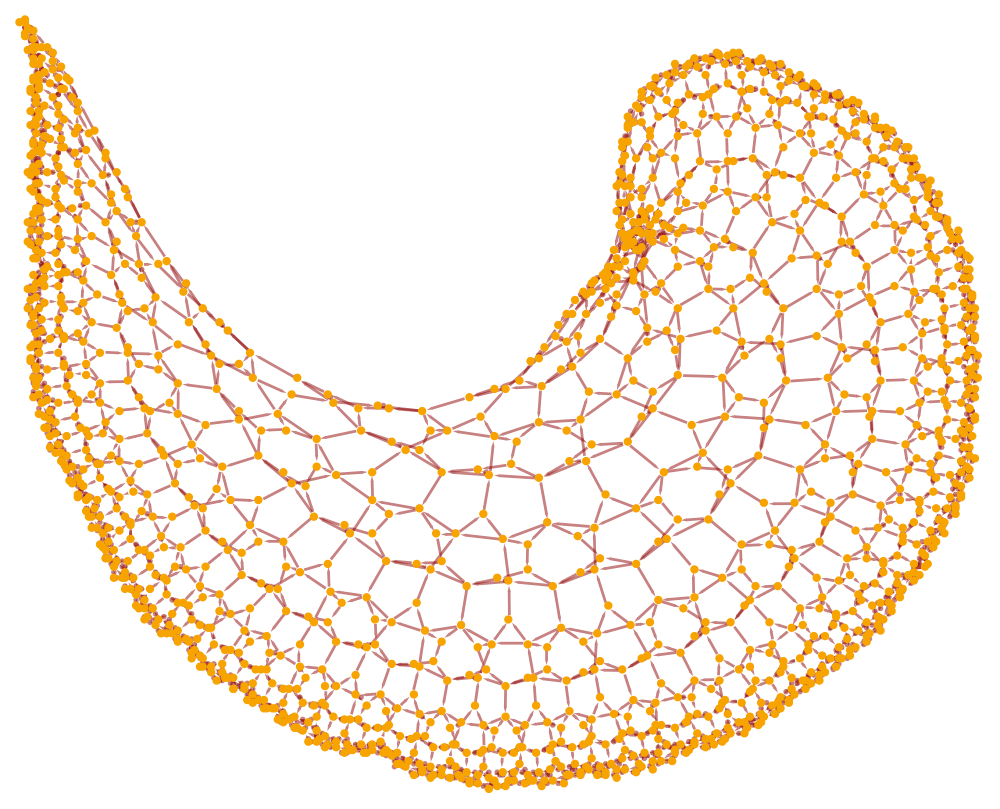}\hspace{0.1\textwidth}
\includegraphics[width=0.395\textwidth]{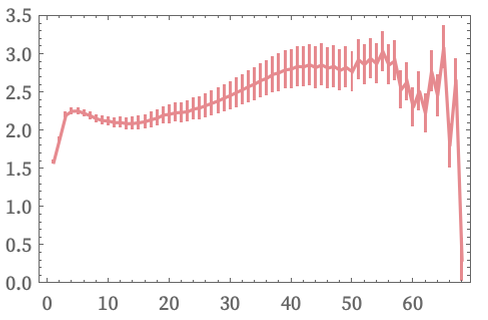}
\caption{Logarithmic difference estimates for the dimension of a (directed) causal graph with a ${1 + 1}$-dimensional Lorentzian manifold-like limiting structure, exhibiting the effects of negative global curvature, as generated by the hypergraph rewriting rule ${\left\lbrace \left\lbrace x, y, x \right\rbrace, \left\lbrace x, z, u \right\rbrace \right\rbrace \to \left\lbrace \left\lbrace u, v, u \right\rbrace, \left\lbrace v, u, z \right\rbrace, \left\lbrace x, y, v \right\rbrace \right\rbrace}$, showing results that are consistent with a limiting dimension of two.}
\label{fig:Figure12}
\end{figure}

\begin{figure}[ht]
\centering
\includegraphics[width=0.895\textwidth]{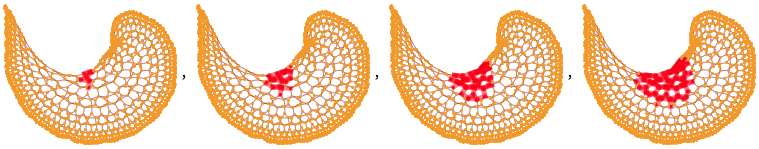}
\caption{The growth of a causal cone (shown in red) in a (directed) causal graph with a ${1 + 1}$-dimensional Lorentzian manifold-like limiting structure, exhibiting the effects of negative global curvature, as generated by the hypergraph rewriting rule ${\left\lbrace \left\lbrace x, y, x \right\rbrace, \left\lbrace x, z, u \right\rbrace \right\rbrace \to \left\lbrace \left\lbrace u, v, u \right\rbrace, \left\lbrace v, u, z \right\rbrace, \left\lbrace x, y, v \right\rbrace \right\rbrace}$.}
\label{fig:Figure13}
\end{figure}

\begin{figure}[ht]
\centering
\includegraphics[width=0.395\textwidth]{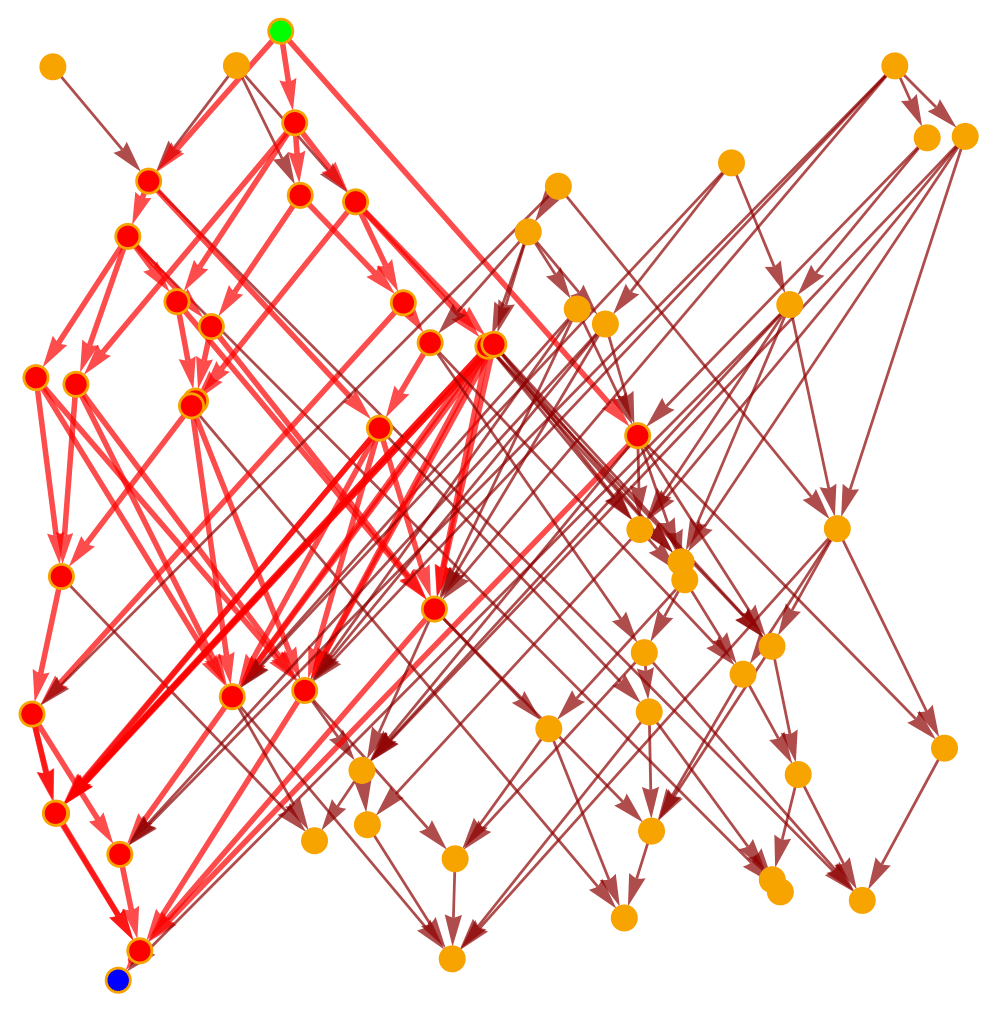}\hspace{0.1\textwidth}
\includegraphics[width=0.395\textwidth]{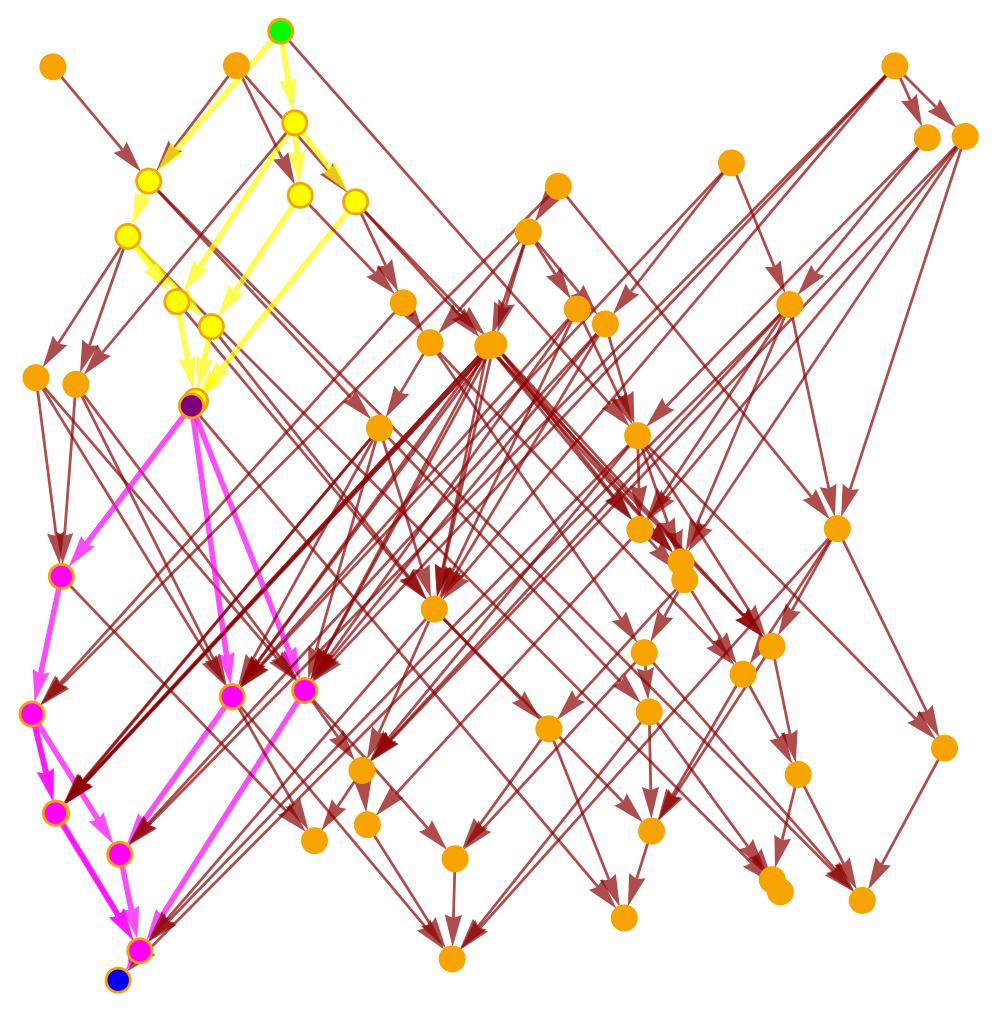}
\caption{On the left, the transitive reduction (i.e. the Hasse diagram) of the directed graph generated by connecting all pairs of 60 (uniformly sprinkled) points that are related by the causal partial order relation ${\prec_{\mathcal{M}}}$ on a rectangular region of a ${1 + 1}$-dimensional flat (Minkowski) spacetime, with a pair of timelike-separated events (shown in green and blue) representing the endpoints of a discrete spacetime order interval (shown in red). On the right, the midpoint of this discrete interval (shown in purple) subdivides it into two discrete subintervals (shown in yellow and magenta), both with cardinality 10, as compared to the overall discrete interval with a cardinality of 28, implying an estimated dimensionality of ${\log_2 \left( \frac{28}{10} \right) \approx 1.49}$.}
\label{fig:Figure14}
\end{figure}

In general for Riemannian manifolds ${\left( \mathcal{M}, g \right)}$, the leading-order term in the expansion for the infinitesimal volume element will be proportional to ${\varepsilon^d}$; indeed, in flat (Euclidean) space ${\mathbb{R}^d}$, it is given precisely by:

\begin{equation}
\mathrm{Vol} \left( B_{\varepsilon} \left( p \right) \right) = \frac{\pi^{\frac{d}{2}}}{\Gamma \left( \frac{d}{2} \right)} \varepsilon^d.
\end{equation}
Therefore, for an undirected graph (considered as the discrete analog of a Riemannian manifold), in which volumes ${\mathrm{Vol} \left( \dots \right)}$ are given instead by cardinalities of vertex sets of subgraphs, as previously discussed, a crude first-order estimator ${\Delta_{\varepsilon} \left( p \right)}$ of the limiting manifold dimension $d$ (given an integer radius ${\varepsilon}$ and a vertex $p$) can be extracted using the following naive logarithmic difference approximation:

\begin{equation}
\Delta_{\varepsilon} \left( p \right) = \frac{\log \left( \mathrm{Vol} \left( B_{\varepsilon + 1} \left( p \right) \right) \right) - \log \left( \mathrm{Vol} \left( B_{\varepsilon} \left( p \right) \right) \right)}{\log \left( \varepsilon + 1 \right) - \log \left( \varepsilon \right)}.
\end{equation}
On the other hand, for a directed graph (considered as a discrete analog of a Lorentzian manifold), this analysis remains unchanged, save for the fact that one must now replace the power series expansion for the volume of an infinitesimal ball ${B_{\varepsilon} \left( p \right)}$ of radius ${\varepsilon}$:

\begin{equation}
\mathrm{Vol} \left( B_{\varepsilon} \left( p \right) \right) = a \varepsilon^d \left[ 1 - \frac{R}{6 \left( d + 2 \right)} \varepsilon^2 + O \left( \varepsilon^4 \right) \right],
\end{equation}
for some constant $a$, with a power series expansion for the volume of an infinitesimal causal cone ${C_t \left( p \right)}$ of proper time length $t$:

\begin{equation}
\mathrm{Vol} \left( C_t \left( p \right) \right) = a t^n \left[ 1 - \frac{1}{6} \sum_{i, j = 1}^{d} R_{i j} t^i t^j + O \left( \left\lVert \mathbf{t} \right\rVert^3 \right) \right],
\end{equation}
for some constant $a$. In the above, the vector ${\mathbf{t}}$ designates the timelike direction in which the cone is oriented, and hence the direction into which the Ricci curvature tensor ${R_{i j}}$ is projected, which (in cases where the metric permits a canonical decomposition) can be written in terms of the ADM gauge variables ${\alpha}$ (lapse) and ${\beta^i}$ (shift) as:

\begin{equation}
t^i = \alpha n^i + \beta^i,
\end{equation}
where ${\mathbf{n}}$ is the timelike unit normal vector to a specified spacelike hypersurface. The computations of the first-order logarithmic difference estimator ${\Delta_t \left( p \right)}$, as a function of the proper time length of the cone $t$, for (directed) causal graphs which limit to be asymptotically-flat and asymptotically-negatively-curved ${1 + 1}$-dimensional Lorentzian manifold-like structures are shown in Figures \ref{fig:Figure10} and \ref{fig:Figure12}, respectively (with the limiting dimension of two identified correctly in both cases), along with a visualization of the growth of the corresponding causal cones shown in Figures \ref{fig:Figure11} and \ref{fig:Figure13}, respectively. As previously noted, this logarithmic difference estimation procedure for computing the Hausdroff dimension of a directed graph is formally equivalent to the \textit{midpoint scaling} dimension estimation procedure devised by Bombelli\cite{bombelli4}\cite{reid}, in which a discrete causal interval ${\mathbf{I} \left[ p, q \right] \subset \mathcal{C}}$ is divided into subintervals ${\mathbf{I} \left[ p, r \right]}$ and ${\mathbf{I} \left[ r, q \right]}$ (for some intermediate element ${r \in \mathcal{C}}$), where the cardinality of the smaller subinterval ${N_{small}}$ is made as large as possible (such that the element $r$ lies as close to the midpoint of the overall discrete interval ${\mathbf{I} \left[ p, q \right]}$ as possible). Therefore, since the volume of a continuum Alexandrov interval ${\mathbf{A} \left[ p, q \right]}$ in $d$-dimensional flat (Minowski) spacetime is given (in terms of the proper time distance ${\tau}$ between events $p$ and $q$) by:

\begin{equation}
\mathrm{Vol} \left( \mathbf{A} \left[ p, q \right] \right) = \frac{\pi^{\frac{d - 1}{2}}}{2^{d - 2} d \left( d - 1 \right) \Gamma \left( \frac{d - 1}{2} \right)} \tau^d,
\end{equation}
it follows that the above procedure, which effectively rescales proper time distances ${\tau_{small}}$ in the smallest spacetime subinterval by a factor of ${\frac{1}{2}}$ (as compared to distances in the overall spacetime interval), will also rescale the volume of the smallest spacetime subinterval by a factor of ${2^{-d}}$:

\begin{equation}
\frac{\tau}{\tau_{small}} = 2, \qquad \implies \qquad \frac{\mathrm{Vol} \left( \mathbf{A} \left[ p, q \right] \right)}{\mathrm{Vol} \left( \mathbf{A} \left[ p, r \right] \right)} = 2^d,
\end{equation}
and therefore, in the causal set case (if $N$ denotes the cardinality of the overall discrete interval ${\mathbf{I} \left[ p, q \right]}$ and ${N_{small}}$ denotes the cardinality of the smallest discrete subinterval), we can construct a first-order logarithmic estimator ${\Delta}$ of the spacetime dimension $d$:

\begin{equation}
\frac{N}{N_{small}} \approx 2^d, \qquad \implies \qquad \Delta = \log_2 \left( \frac{N}{N_{small}} \right).
\end{equation}
The computation of the first-order logarithmic difference estimator ${\Delta}$, for a directed (causal) graph sprinkled into a ${1 + 1}$-dimensional flat (Minkowski) spacetime, is shown in Figure \ref{fig:Figure14}. The only fundamental distinction between the two logarithmic difference estimation procedures described above lies in the fact that, in the former case, one is computing volumes of one-sided (asymmetrical) spacetimes cones ${C_{t} \left( p \right)}$, whereas, in the latter case, one is computing volumes of two-sided (symmetrical) causal diamonds ${\mathbf{I} \left[ p, q \right]}$.

\begin{figure}[ht]
\centering
\includegraphics[width=0.325\textwidth]{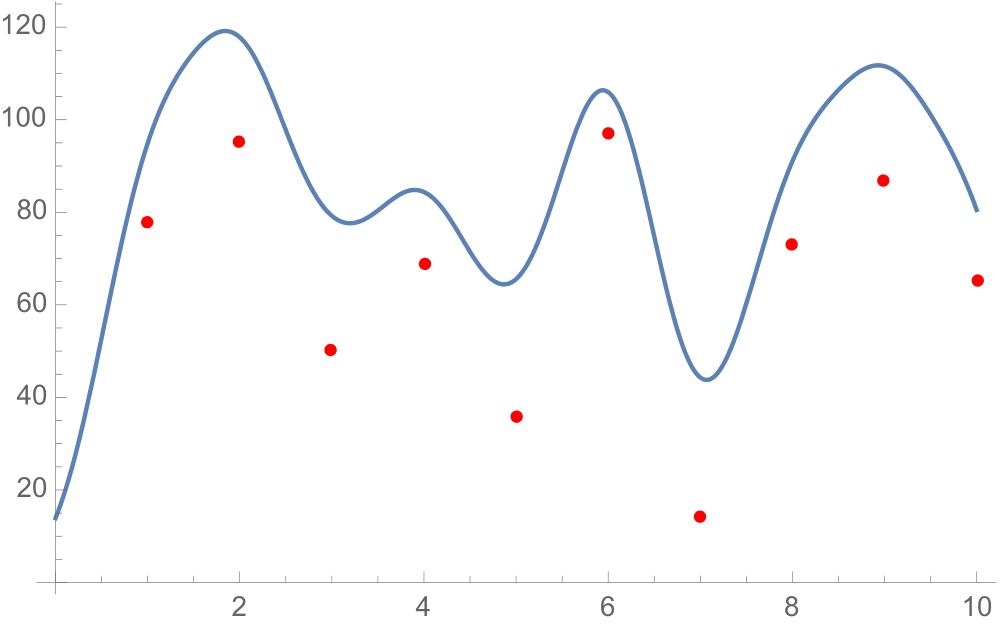}
\includegraphics[width=0.325\textwidth]{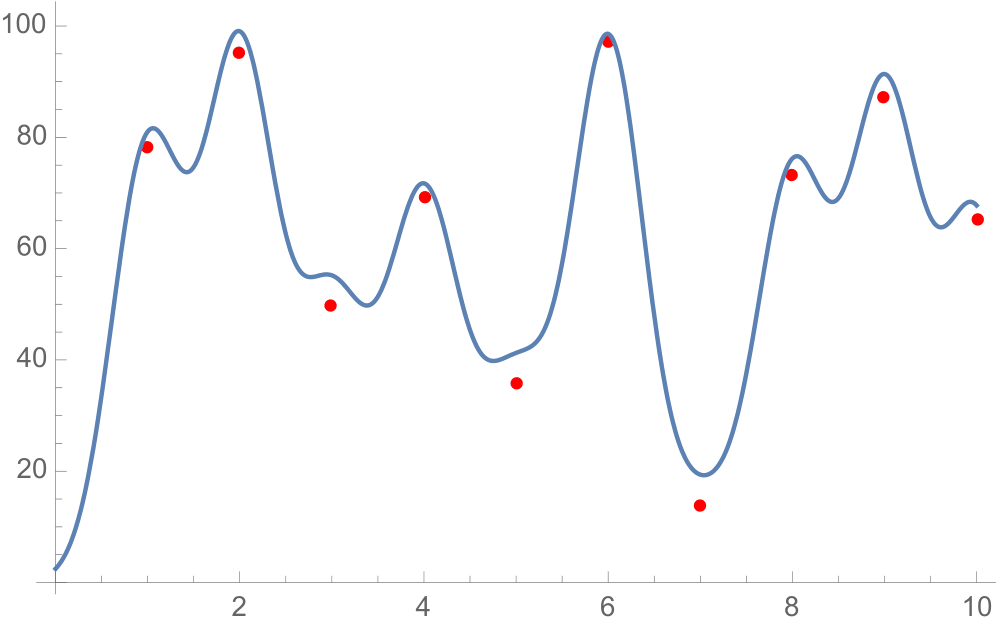}
\includegraphics[width=0.325\textwidth]{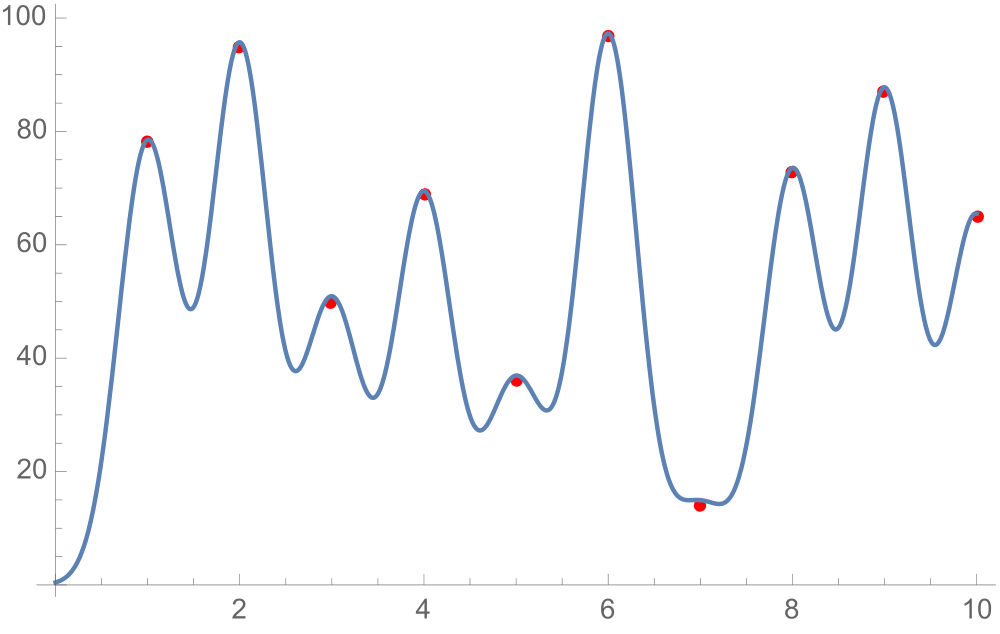}
\caption{The ``analytic continuation'' to the real numbers ${\mathbb{R}}$ (shown in blue) of a discrete function over the natural numbers ${\mathbb{N}}$ (shown in red), assuming exponents ${m_k}$ equal to 10, 20 and 30, respectively (for all ${k \in \mathbb{N}}$).}
\label{fig:Figure15}
\end{figure}

If we consider ${\mathbf{x}}$ to be a fixed basis vector of arbitrary dimensionality, then we may consider the massless retarded Green's function ${\left( G_R \right)_{0}^{\left( d \right)} \left( \mathbf{x} \right)}$ to be a function defined over the natural numbers, since ${d \in \mathbb{N}}$, i.e. one has:

\begin{equation}
\left( G_R \right)_{0}^{\left( d \right)} \left( \mathbf{x} \right) : \mathbb{N} \to \mathbb{R};
\end{equation}
our objective in the present section may therefore be rephrased as a desire to ``analytically continue'' ${\left( G_R \right)_{0}^{\left( d \right)} \left( \mathbf{x} \right)}$ so as to be an analytic function ${\left( \tilde{G_R} \right)_{0}^{\left( d \right)} \left( \mathbf{x} \right)}$ over the real numbers ${\mathbb{R}}$:

\begin{equation}
\left( \tilde{G_R} \right)_{0}^{\left( d \right)} \left( \mathbf{x} \right) : \mathbb{R} \to \mathbb{R}.
\end{equation}
One way to achieve this (although of course this approach is by no means unique) would be to define the continued function ${\left( \tilde{G_R} \right)_{0}^{\left( d \right)} \left( \mathbf{x} \right)}$ as the following sum of ${\mathrm{sinc}}$ functions, evaluated over all natural numbers ${\mathbb{N}}$:

\begin{equation}
\left( \tilde{G_R} \right)_{0}^{\left( d \right)} \left( \mathbf{x} \right) = \sum_{k \in \mathbb{N}} \left( \mathrm{sinc} \left( d - k \right) \right)^{m_k} \left( G_R \right)_{0}^{k} \left( \mathbf{x} \right), \qquad \text{ where } \qquad \mathrm{sinc} \left( d \right) = \frac{\sin \left( \pi d \right)}{\pi d},
\end{equation}
and where the exponents ${m_k}$ are all chosen to be sufficiently large as to guarantee that:

\begin{equation}
\forall \left\lvert d \right\rvert > 1, \qquad \left( \mathrm{sinc} \left( d \right) \right)^{m_k} \left( G_R \right)_{0}^{\left( k \right)} \left( \mathbf{x} \right) < 2^{- \left\lvert k \right\rvert}.
\end{equation}
An example of how this ``analytic continuation'' can be performed, in order to extend a (randomly-generated) discrete function over the natural numbers ${\mathbb{N}}$ to an analytic function over the real numbers ${\mathbb{R}}$, assuming exponents of ${m_k = 10}$, ${m_k = 20}$ and ${m_k = 30}$, respectively (for all ${k \in \mathbb{N}}$), illustrating convergence to a valid continuation for larger exponents, is shown in Figure \ref{fig:Figure15}.

If we wish to guarantee uniqueness of the continuation (at least up to multiples of ${\sin \left( \pi d \right)}$), then it suffices instead to ``analytically continue'' the massless retarded Green's function ${\left( G_R \right)_{0}^{\left( d \right)} \left( \mathbf{x} \right)}$ (considered as a function of dimension $d$ for fixed ${\mathbf{x}}$) to be a meromorphic function ${\left( \tilde{G_R} \right)_{0}^{\left( d \right)} \left( \mathbf{x} \right)}$ over the entire complex plane ${\mathbb{C}}$:

\begin{equation}
\left( \tilde{G_R} \right)_{0}^{\left( d \right)} \left( \mathbf{x} \right) : \mathbb{C} \to \mathbb{R},
\end{equation}
by introducing a sequence of polynomials ${\tilde{G}_n \left( d \right)}$ in $d$ (with ${n \in \mathbb{N}}$) of the general form:

\begin{equation}
\tilde{G}_n \left( d \right) = \left( b + c d \right) d^m \prod_{k = - \left( n - 1 \right)}^{n - 1} \left( d - k \right), \qquad \text{ with } \qquad m, b, c \in \mathbb{C} \text { constant},
\end{equation}
such that the ``analytically-continued'' massless retarded Green's function ${\left( \tilde{G_R} \right)_{0}^{\left( d \right)} \left( \mathbf{x} \right)}$ may be written directly as a sum over all such polynomials:

\begin{equation}
\left( \tilde{G_R} \right)_{0}^{\left( d \right)} \left( \mathbf{x} \right) = \sum_{n \in \mathbb{N}} \tilde{G}_n \left( d \right).
\end{equation}
Clearly, the polynomials ${\tilde{G}_n \left( d \right)}$ must take the appropriate values at ${d = -n}$ and ${d = n}$ to ensure that the correct contributions are yielded within the sum; moreover, the contribution from each polynomial should vanish on all natural-numbered values of $d$ within the closed interval ${\left[ - \left( n - 1 \right), n - 1 \right]}$:

\begin{equation}
\forall d \in \left[ - \left( n - 1 \right), n - 1 \right] \cap \mathbb{N}, \qquad \tilde{G}_n \left( d \right) = 0,
\end{equation}
and the constant ${m \in \mathbb{C}}$ must be chosen to be sufficiently large as to ensure that the value of the polynomial ${\tilde{G}_n \left( d \right)}$ satisfies a suitable exponential decay condition on the disk ${d \in B_{n - 1} \left( 0 \right)}$ (i.e. such that the values of ${\tilde{G}_n \left( d \right)}$ when the modulus of $d$ is no larger than ${n - 1}$ remain sufficiently small, as compared to the values at ${d = -n}$ and ${d = n}$):

\begin{equation}
\forall \left\lvert d \right\rvert \leq n - 1, \qquad \left\lvert \tilde{G}_n \left( d \right) \right\rvert < 2^{-n}.
\end{equation}
Arbitrary factors of ${\sin \left( \pi d \right)}$ may be introduced into the definition of ${\left( \tilde{G_R} \right)_{0}^{\left( d \right)} \left( \mathbf{x} \right)}$ without sacrificing meromorphicity; however, modulo such factors, this continuation from ${\mathbb{N}}$ to ${\mathbb{C}}$ is guaranteed to be unique (subject to certain technical assumptions, and modulo all singularities) by virtue of Carlson's theorem in complex analysis\cite{carlson}.

Recall that, for any function $f$ that is holomorphic over the entire complex plane ${\mathbb{C}}$, Carlson's theorem states that if $f$ is of \textit{exponential type}, meaning that its modulus is bounded exponentially in the following way:

\begin{equation}
\forall z \in \mathbb{C}, \qquad \left\lvert f \left( z \right) \right\rvert \leq C e^{\tau \left\lvert z \right\rvert}, \qquad \text{ with } \qquad C, \tau \in \mathbb{R} \text{ constant},
\end{equation}
if the growth of $f$ is suitably bounded at (imaginary) infinity, meaning that there exists some constant ${c \in \mathbb{R}}$ such that the following statement holds:

\begin{equation}
\forall y \in \mathbb{R}, \qquad \left\lvert f \left( i y \right) \right\rvert \leq C e^{c \left\lvert y \right\rvert}, \qquad \text{ with } \qquad c < \pi \text{ constant},
\end{equation}
and if ${f \left( n \right) = 0}$ for every non-negative integer ${n \in \mathbb{N}}$, then ${f \left( z \right)}$ is identically equal to zero for all ${z \in \mathbb{C}}$. Given the massless retarded Green's function ${\left( G_R \right)_{0}^{\left( d \right)} \left( \mathbf{x} \right)}$ (as a function of ${d \in \mathbb{N}}$ for fixed ${\mathbf{x}}$) and its ``analytic continuation'' as a meromorphic function ${\left( \tilde{G_R} \right)_{0}^{\left( d \right)} \left( \mathbf{x} \right)}$ over ${\mathbb{C}}$, we know (by definition of the ``analytic continuation'') that the two functions must agree for all natural numbers:

\begin{equation}
\forall d \in \mathbb{N}, \qquad \left( \tilde{G_R} \right)_{0}^{\left( d \right)} \left( \mathbf{x} \right) = \left( G_R \right)_{0}^{\left( d \right)} \left( \mathbf{x} \right),
\end{equation}
at least upon removal of all singularities, and, moreover, the polynomials ${\tilde{G}_n \left( d \right)}$ and constant ${m \in \mathbb{C}}$ have been specifically chosen to ensure that the growth conditions:

\begin{equation}
\forall d \in \mathbb{C}, \qquad \left\lvert \left( \tilde{G_R} \right)_{0}^{\left( d \right)} \left( \mathbf{x} \right) \right\rvert \leq C e^{\tau \left\lvert d \right\rvert}, \qquad \text{ with } \qquad C, \tau \in \mathbb{R} \text{ constant},
\end{equation}
and:

\begin{equation}
\forall d \in \mathbb{R}, \qquad \left\lvert \left( \tilde{G_R} \right)_{0}^{\left( i d \right)} \left( \mathbf{x} \right) \right\rvert \leq C e^{c \left\lvert d \right\rvert}, \qquad \text{ with } \qquad c \in \mathbb{R} \text{ such that } c < \pi,
\end{equation}
must be satisfied. Therefore, the ``analytic continuation'' ${\left( \tilde{G_R} \right)_{0}^{\left( d \right)} \left( \mathbf{x} \right)}$ (modulo singularities) must be unique, up to factors of ${\sin \left( \pi d \right)}$, for suppose, on the contrary, that there exist two distinct functions, ${\left( \tilde{G_R} \right)_{0}^{\left( d \right)} \left( \mathbf{x} \right)}$ and ${\left( \tilde{H_R} \right)_{0}^{\left( d \right)} \left( \mathbf{x} \right)}$, which both obey the requisite growth conditions, and which both agree with the original massless retarded Green's function ${\left( G_R \right)_{0}^{\left( d \right)} \left( \mathbf{x} \right)}$ for all ${d \in \mathbb{N}}$. Then, if we define a third function ${\left( \tilde{I_R} \right)_{0}^{\left( d \right)} \left( \mathbf{x} \right)}$ to be the difference between these two:

\begin{equation}
\left( \tilde{I_R} \right)_{0}^{\left( d \right)} \left( \mathbf{x} \right) = \left( \tilde{G_R} \right)_{0}^{\left( d \right)} \left( \mathbf{x} \right) - \left( \tilde{H_R} \right)_{0}^{\left( d \right)} \left( \mathbf{x} \right),
\end{equation}
then, by Carlson's theorem, ${\left( \tilde{I_R} \right)_{0}^{\left( d \right)} \left( \mathbf{x} \right)}$ must be identically zero for all ${d \in \mathbb{C}}$, and therefore the two functions ${\left( \tilde{G_R} \right)_{0}^{\left( d \right)} \left( \mathbf{x} \right)}$ and ${\left( \tilde{H_R} \right)_{0}^{\left( d \right)} \left( \mathbf{x} \right)}$ must be identical for all ${d \in \mathbb{C}}$. This completes the proof.

Note that, as mentioned as the start of this section, our reason for putting the term ``analytic continuation'' in quotation marks is to make clear that it is not actually the massless Green's functions themselves that are being analytically continued (although it is a useful abuse of terminology to describe it as such), but rather the parameter $d$ that is being continued from natural to complex values. Moreover, note that when one performs this continuation for the massless Green's functions directly, it effectively corresponds to taking the (modified) Bessel functions ${I_{\alpha} \left( x \right)}$, ${Y_{\alpha} \left( x \right)}$, ${J_{\alpha} \left( x \right)}$ and ${K_{\alpha} \left( x \right)}$, as well as the corresponding Hankel functions ${H_{\alpha}^{\left( 1 \right)} \left( x \right)}$ and ${H_{\alpha}^{\left( 2 \right)} \left( x \right)}$, that appear in the definitions of the massless Green's functions presented in the preceding section, and continuing them to the case of general complex (as opposed to integer) values of the order parameter ${\alpha}$. The technical assumptions placed on the ``analytic continuation'' presented above (i.e. those necessary for Carlson's theorem to hold and guarantee uniqueness) effectively amount to choosing a branch cut on the corresponding special functions, since these functions become multi-valued (and hence define a Riemann surface) for non-integer values of the order parameter.

\section{Wightman Functions, the Sorkin-Johnston Vacuum and Spacetime Entanglement Entropies}
\label{sec:Section3}

Throughout this section, we broadly follow the notational and axiomatic conventions for algebraic quantum field theory in the Heisenberg picture, as defined by Haag and Kastler\cite{haag}, and later refined by Wightman and Gardin\cite{wightman}, and as employed in the causal set theory context by Johnston\cite{johnston}\cite{johnston2}\cite{johnston3}. We begin by briefly reviewing this formalism. Recall that, if ${F_{+} \left( \mathcal{H} \right)}$ denotes the bosonic (symmetric) Fock space for the Hilbert space ${\mathcal{H}}$, i.e. the Hilbert space given by the metric space completion of the direct sum of symmetrized tensor powers of ${\mathcal{H}}$:

\begin{equation}
F_{+} \left( \mathcal{H} \right) = \overline{\bigoplus_{n = 0}^{\infty} S_{+} \mathcal{H}^{\otimes n}},
\end{equation}
where ${S_{+}}$ denotes the tensor symmetrization operator and the overline designates the completion of the resulting Hilbert space, then we can define a family of free, bosonic scalar field operators ${\hat{\Phi} \left( \mathbf{x} \right)}$ in $d$ spacetime dimensions (with mass $m$) acting on the Fock space ${F_{+} \left( \mathcal{H} \right)}$, which must satisfy the axioms of self-adjointness/Hermiticity:

\begin{equation}
\forall \mathbf{x} \in F_{+} \left( \mathcal{H} \right), \qquad \hat{\Phi} \left( \mathbf{x} \right) = \left( \hat{\Phi} \left( \mathbf{x} \right) \right)^{\dagger},
\end{equation}
commutation with respect to the \textit{Pauli-Jordan operator} ${\hat{\Delta} \left( \mathbf{x}, \mathbf{y} \right)}$:

\begin{equation}
\forall \mathbf{x}, \mathbf{y} \in F_{+} \left( \mathcal{H} \right), \qquad \left[ \hat{\Phi} \left( \mathbf{x} \right), \hat{\Phi} \left( \mathbf{y} \right) \right] = \hat{\Phi} \left( \mathbf{x} \right) \hat{\Phi} \left( \mathbf{y} \right) - \hat{\Phi} \left( \mathbf{y} \right) \hat{\Phi} \left( \mathbf{x} \right) = i \hat{\Delta} \left( \mathbf{x}, \mathbf{y} \right),
\end{equation}
where the Pauli-Jordan operator ${\hat{\Delta} \left( \mathbf{x}, \mathbf{y} \right)}$ here simply corresponds to the difference between the retarded and advanced Green's functions ${\left( G_R \right)_{m}^{\left( d \right)}}$ and ${\left( G_A \right)_{m}^{\left( d \right)}}$ (otherwise known as the covariantly-defined Peierls bracket):

\begin{equation}
\forall \mathbf{x}, \mathbf{y} \in F_{+} \left( \mathcal{H} \right), \qquad \hat{\Delta} \left( \mathbf{x}, \mathbf{y} \right) = \left( G_R \right)_{m}^{\left( d \right)} \left( \mathbf{x}, \mathbf{y} \right) - \left( G_A \right)_{m}^{\left( d \right)} \left( \mathbf{x}, \mathbf{y} \right),
\end{equation}
as well as satisfaction of the (massive) Klein-Gordon equation:

\begin{equation}
\forall \mathbf{x} \in F_{+} \left( \mathcal{H} \right), \qquad \left( \Box + m^2 \right) \hat{\Phi} \left( \mathbf{x} \right) = 0.
\end{equation}
Assuming that the Fock space ${F_{+} \left( \mathcal{H} \right)}$ admits a vacuum state ${\ket{0} \in F_{+} \left( \mathcal{H} \right)}$ that is invariant under the action of the Poincar\'e group ${\mathbb{R}^{1, d - 1} \rtimes O \left( 1, d - 1 \right)}$, we can define a \textit{Wightman function} (i.e. a two-point correlation function) ${\hat{W}_{m}^{\left( d \right)} \left( \mathbf{x}, \mathbf{y} \right)}$ as the vacuum expectation value of the (non-time-ordered) product of field operators ${\hat{\Phi} \left( \mathbf{x} \right)}$ and ${\hat{\Phi} \left( \mathbf{y} \right)}$:

\begin{equation}
\forall \mathbf{x}, \mathbf{y} \in F_{+} \left( \mathcal{H} \right), \qquad \hat{W}_{m}^{\left( d \right)} \left( \mathbf{x}, \mathbf{y} \right) = \bra{0} \hat{\Phi} \left( \mathbf{x} \right) \hat{\Phi} \left( \mathbf{y} \right) \ket{0}.
\end{equation}

This Wightman function ${\hat{W}_{m}^{\left( d \right)} \left( \mathbf{x}, \mathbf{y} \right)}$ may be considered to be (up to a factor of $i$) a non-time-ordered version of the massive $d$-dimensional Feynman propagator ${\left( G_F \right)_{m}^{\left( d \right)} \left( \mathbf{x} \right)}$ for the Klein-Gordon equation (assuming flat/Minkowsi spacetime):

\begin{equation}
\left( \Box + m^2 \right) \left( G_F \right)_{m}^{\left( d \right)} \left( \mathbf{x} \right) = \delta^d \left( \mathbf{x} \right),
\end{equation}
from which, by applying the same modified Fourier transform operator ${F^{*}}$ as previously used in the derivation of the massive retarded Green's function ${\left( G_R \right)_{m}^{\left( d \right)} \left( \mathbf{x} \right)}$, we are able to derive the following trivial solution in momentum space:

\begin{equation}
\left( \tilde{G_F} \right)_{m}^{\left( d \right)} \left( \mathbf{p} \right) = - \frac{1}{\mathbf{p}^2 - m^2},
\end{equation}
yielding a corresponding solution in position space:

\begin{equation}
\left( G_F \right)_{m}^{\left( d \right)} \left( \mathbf{x} \right) = - \frac{1}{\left( 2 \pi \right)^d} \int_{- \infty}^{\infty} \frac{e^{- i \mathbf{p} \cdot \mathbf{x}}}{\mathbf{p}^2 - m^2} d^d \mathbf{p},
\end{equation}
whose poles in the integrand, namely at ${\mathbf{p}^2 = m^2}$, may be avoided by choosing an integration contour that goes \textit{under} the left pole and \textit{over} the right pole. This yields a definition for the massive Feynman propagator ${\left( G_F \right)_{m}^{\left( d \right)} \left( \mathbf{x} \right)}$ in terms of the following (distributional) limit:

\begin{equation}
\left( G_F \right)_{m}^{\left( d \right)} \left( \mathbf{x} \right) = \lim_{\epsilon \to 0^{+}} \left[ - \frac{1}{\left( 2 \pi \right)^d} \int_{- \infty}^{\infty} \frac{e^{- i \mathbf{p} \cdot \mathbf{x}}}{\mathbf{p}^2 - m^2 + i \epsilon} d^d \mathbf{p} \right].
\end{equation}
Note that, in general, the Feynman propagator ${\left( G_F \right)_{m}^{\left( d \right)} \left( \mathbf{x} \right)}$ is given by the real part (or, more precisely, the part that does not involve modified Bessel functions of the second kind ${K_{\alpha} \left( x \right)}$) of the causal Green's function ${\left( G_C \right)_{m}^{\left( d \right)} \left( \mathbf{x} \right)}$; for instance, in the ${d = 4}$ example analyzed previously, we found the following explicit form for the causal Green's function, namely:

\begin{equation}
\left( G_C \right)_{m}^{\left( d \right)} \left( \mathbf{x} \right) = \frac{\delta \left( \tau^2 \right)}{4 \pi} - \frac{m}{8 \pi} \theta \left( \tau^2 \right) \frac{H_{1}^{\left( 2 \right)} \left( m \sqrt{\tau^2} \right)}{\sqrt{\tau^2}} + \frac{i m}{4 \pi^2} \theta \left( - \tau^2 \right) \frac{K_1 \left( m \sqrt{- \tau^2} \right)}{\sqrt{- \tau^2}},
\end{equation}
with ${\tau^2 = x_{0}^{2} - x_{1}^{2} - x_{2}^{2} - x_{3}^{2}}$, implying a Feynman propagator ${\left( G_F \right)_{m}^{\left( 4 \right)} \left( \mathbf{x} \right)}$ of the form:

\begin{equation}
\left( G_F \right)_{m}^{\left( 4 \right)} \left( \mathbf{x} \right) = \frac{\delta \left( s^2 \right)}{4 \pi} - \frac{m}{8 \pi} \frac{H_{1}^{\left( 2 \right)} \left( m s \right)}{s},
\end{equation}
where we have introduced the parameter $s$ in order to eliminate explicit dependence on the Heaviside step function ${\theta \left( \alpha \right)}$:

\begin{equation}
s = \lim_{\epsilon \to 0^{+}} \left[ \sqrt{\tau^2 - i \epsilon} \right] = \begin{cases}
\sqrt{x_{0}^{2} - x_{1}^{2} - x_{2}^{2} - x_{3}^{2}}, \qquad & \text{ if } x_{0}^{2} \geq x_{1}^{2} + x_{2}^{2} + x_{3}^{2},\\
- i \sqrt{x_{1}^{2} + x_{2}^{2} + x_{3}^{2} - x_{0}^{2}}, \qquad & \text{ if } x_{0}^{2} < x_{1}^{2} + x_{2}^{2} + x_{3}^{2}.
\end{cases}
\end{equation}
Thus, as noted by Birrell and Davies\cite{birrell}, the integral inside the distributional limit may be evaluated via Fourier-analytic methods, in order to yield the following general form of the massive Feynman propagator ${\left( G_F \right)_{m}^{\left( d \right)} \left( \mathbf{x} \right)}$ in $d$ spacetime dimensions:

\begin{equation}
\left( G_F \right)_{m}^{\left( d \right)} \left( \mathbf{x} \right) = \lim_{\epsilon \to 0^{+}} \left[ \frac{\pi}{2} \frac{\left( -1 \right)^d}{\left( 2 \pi \right)^{\frac{d}{2}}} \left( \frac{m}{\sqrt{\tau^2 - i \epsilon}} \right)^{\frac{d}{2} - 1} H_{\frac{d}{2} - 1}^{\left( 2 \right)} \left( m \sqrt{\tau^2 - i \epsilon} \right) \right],
\end{equation}
such that the real part of the Feynman propagator is always equal to the average of the advanced and retarded massive Green's functions ${\left( G_A \right)_{m}^{\left( d \right)} \left( \mathbf{x} \right)}$ and ${\left( G_R \right)_{m}^{\left( d \right)} \left( \mathbf{x} \right)}$:

\begin{equation}
\mathrm{Re} \left( \left( G_F \right)_{m}^{\left(d \right)} \left( \mathbf{x} \right) \right) = \frac{\left( G_A \right)_{m}^{\left( d \right)} \left( \mathbf{x} \right) + \left( G_R \right)_{m}^{\left( d \right)} \left( \mathbf{x} \right)}{2}.
\end{equation}
Therefore, we have that the massive $d$-dimensional Feynman propagator ${\left( G_F \right)_{m}^{\left( d \right)} \left( \mathbf{x}, \mathbf{y} \right)}$ is simply the vacuum expectation value of the product of field operators ${\hat{\Phi} \left( \mathbf{x} \right)}$ and ${\hat{\Phi} \left( \mathbf{y} \right)}$ (just like the Wightman function ${\hat{W}_{m}^{\left( d \right)} \left( \mathbf{x}, \mathbf{y} \right)}$), albeit now with time-ordering going from right to left:

\begin{equation}
\left( G_F \right)_{m}^{\left( d \right)} \left( \mathbf{x}, \mathbf{y} \right) = i \bra{0} \mathcal{T} \hat{\Phi} \left( \mathbf{x} \right) \hat{\Phi} \left( \mathbf{y} \right) \ket{0},
\end{equation}
for time-ordering ``operator'' ${\mathcal{T}}$, as required.

Now, if we consider a finite causal set ${\mathcal{C}}$ of cardinality ${\left\lvert \mathcal{C} \right\vert = p}$ (as before), then we can define a discrete analog of the Pauli-Jordan operator ${\hat{\Delta}_{\mathcal{C}} \left( x, y \right)}$ on causal set elements $x$ and $y$ by means of the following matrix:

\begin{equation}
\forall x, y \in \mathcal{C}, \qquad \hat{\Delta}_{\mathcal{C}} \left( x, y \right) = \left( K_R \right)_{m}^{\left( d \right)} \left( x, y \right) - \left( K_A \right)_{m}^{\left( d \right)} \left( x, y \right),
\end{equation}
where ${\left( K_R \right)_{m}^{\left( d \right)} \left( x, y \right)}$ is the discrete retarded propagator matrix for a field of mass $m$; for instance in the massless case ${m = 0}$, we have shown previously that this is given for the ${d = 2}$ and ${d = 4}$ cases by:

\begin{equation}
\left( K_R \right)_{0}^{\left( d \right)} \left( x, y \right) = \frac{1}{2} C_0 \left( x, y \right), \qquad \text{ and } \qquad \left( K_R \right)_{0}^{\left( d \right)} \left( x, y \right) = \frac{1}{2 \pi} \sqrt{\frac{1}{6}} L_0 \left( x, y \right),
\end{equation}
respectively, for causal matrix ${C_0 \left( x, y \right)}$ and link matrix ${L_0 \left( x, y \right)}$. Moreover, since these propagators are all finite-dimensional matrices, the discrete advanced propagator ${\left( K_A \right)_{m}^{\left( d \right)} \left( x, y \right)}$ matrix can, in general, be obtained simply by taking the transpose of the retarded one:

\begin{equation}
\left( K_A \right)_{m}^{\left( d \right)} \left( x, y \right) = \left( \left( K_R \right)_{m}^{\left( d \right)} \left( x, y \right) \right)^{\intercal} = \left( K_R \right)_{m}^{\left( d \right)} \left( y, x \right).
\end{equation}
Both the discrete Pauli-Jordan operator ${\hat{\Delta}_{\mathcal{C}} \left( x, y \right)}$ and its imaginary variant ${i \hat{\Delta}_{\mathcal{C}} \left( x, y \right)}$ are skew-symmetric matrices:

\begin{equation}
\left( \hat{\Delta}_{\mathcal{C}} \left( x, y \right) \right)^{\intercal} = \hat{\Delta}_{\mathcal{C}} \left( y, x \right) = - \hat{\Delta}_{\mathcal{C}} \left( x, y \right), \qquad \text{ and } \qquad \left( i \hat{\Delta}_{\mathcal{C}} \left( x, y \right) \right)^{\intercal} = i \hat{\Delta}_{\mathcal{C}} \left( y, x \right) = - i \hat{\Delta}_{\mathcal{C}} \left( x, y \right),
\end{equation}
and, moreover, the imaginary variant ${i \hat{\Delta}_{\mathcal{C}} \left( x, y \right)}$ is guaranteed to be Hermitian:

\begin{equation}
\left( i \hat{\Delta}_{\mathcal{C}} \left( x, y \right) \right)^{\dagger} = \left( i \hat{\Delta}_{\mathcal{C}} \left( y, x \right) \right)^{*} = i \hat{\Delta}_{\mathcal{C}} \left( x, y \right),
\end{equation}
where the asterisk designates complex conjugation, implying that the rank $s$ of the discrete Pauli-Jordan operator ${i \hat{\Delta} \left( x, y \right)}$ is always even, and that its eigenspectrum ${\left\lbrace \lambda_i \right\rbrace}$:

\begin{equation}
i \hat{\Delta}_{\mathcal{C}} \circ \mathbf{v}_{i} \left( x \right) = \sum_{y \in \mathcal{C}} i \hat{\Delta}_{\mathcal{C}} \left( x, y \right) \mathbf{v}_{i} \left( y \right) = \lambda_i \mathbf{v}_i \left( x \right),
\end{equation}
decomposes canonically into positive and negative parts ${\lambda_i}$ and ${- \lambda_i}$ (for ${\lambda_i > 0}$), with corresponding eigenfunctions ${\mathbf{v}_{i}^{+} \left( x \right)}$ and ${\mathbf{v}_{i}^{-} \left( x \right)}$, respectively, i.e. one has:

\begin{equation}
i \hat{\Delta}_{\mathcal{C}} \circ \mathbf{v}_{i}^{\pm} \left( x \right) = \sum_{y \in \mathcal{C}} i \hat{\Delta}_{\mathcal{C}} \left( x, y \right) \mathbf{v}_{i}^{\pm} \left( y \right) = \pm \lambda_i \mathbf{v}_{i}^{\pm} \left( x \right),
\end{equation}
along with a class of eigenfunctions ${\mathbf{v}_{j}^{0} \left( x \right)}$ associated with the zero eigenvalues:

\begin{equation}
i \hat{\Delta}_{\mathcal{C}} \circ \mathbf{v}_{j}^{0} \left( x \right) = \sum_{y \in \mathcal{C}} i \hat{\Delta}_{\mathcal{C}} \left( x, y \right) \mathbf{v}_{j}^{0} \left( y \right) = \mathbf{0},
\end{equation}
for ${i = 1, 2, \dots, \frac{s}{2}}$, ${j = 1, 2, \dots, p - s}$. If we now select the eigenfunctions ${\mathbf{v}_{i}^{+} \left( x \right)}$, ${\mathbf{v}_{i}^{-} \left( x \right)}$ and ${\mathbf{v}_{i}^{0} \left( x \right)}$ so as to guarantee that the conjugation conditions:

\begin{equation}
\mathbf{v}_{i}^{+} \left( x \right) = \left( \mathbf{v}_{i}^{-} \left( x \right) \right)^{*}, \qquad \text{ and } \qquad \mathbf{v}_{k}^{0} \left( x \right) = \left( \mathbf{v}_{k}^{0} \left( x \right) \right)^{*},
\end{equation}
as well as the inner product conditions:

\begin{equation}
\sum_{x \in \mathcal{C}} \mathbf{v}_{i}^{+} \left( x \right) \mathbf{v}_{j}^{+} \left( x \right) = \sum_{x \in \mathcal{C}} \mathbf{v}_{i}^{-} \left( x \right) \mathbf{v}_{j}^{-} \left( x \right) = \delta_{i j}, \qquad \sum_{x \in \mathcal{C}} \mathbf{v}_{k}^{0} \left( x \right) \mathbf{v}_{l}^{0} \left( x \right) = \delta_{k l},
\end{equation}
and:

\begin{equation}
\sum_{x \in \mathcal{C}} \mathbf{v}_{i}^{+} \left( x \right) \mathbf{v}_{j}^{-} \left( x \right) = \sum_{x \in \mathcal{C}} \mathbf{v}_{k}^{0} \left( x \right) \mathbf{v}_{i}^{+} \left( x \right) = \sum_{x \in \mathcal{C}} \mathbf{v}_{k}^{0} \left( x \right) \mathbf{v}_{i}^{-} \left( x \right) = 0,
\end{equation}
for ${i, j = 1, 2, \dots, \frac{s}{2}}$, ${k = l = 1, 2, p - s}$, are satisfied, then these eigenfunctions consequently form an orthonormal basis for the finite-dimensional Hilbert space ${\mathbb{C}^{p}}$ of the causal set ${\mathcal{C}}$, thus allowing us to write the following explicit spectral decomposition of the (imaginary variant of the) discrete Pauli-Jordan operator ${i \hat{\Delta}_{\mathcal{C}} \left( x, y \right)}$ as the difference between the positive and negative parts of the eigenspectrum:

\begin{equation}
i \hat{\Delta}_{\mathcal{C}} \left( x, y \right) = \left( \sum_{i = 1}^{\frac{s}{2}} \lambda_i \mathbf{v}_{i}^{+} \left( x \right) \left( \mathbf{v}_{i}^{+} \left( y \right) \right)^{*} \right) - \left( \sum_{i = 1}^{\frac{s}{2}} \lambda_i \mathbf{v}_{i}^{-} \left( x \right) \left( \mathbf{v}_{i}^{-} \left( y \right) \right)^{*} \right).
\end{equation}

This canonical decomposition of the eigenspectrum of the imaginary discrete Pauli-Jordan operator ${i \hat{\Delta}_{\mathcal{C}} \left( x, y \right)}$ into positive and negative parts may be interpreted as being the discrete analog of the decomposition of the space of solutions for the continuum Klein-Gordon operator, namely ${\ker \left( \Box - m^2 \right)}$, into positive and negative frequency classes of modes in flat (Minkowski) spacetime ${\mathbb{M}^d = \mathbb{R}^{1, d - 1}}$. Hence, just as in the continuum case, we can use this decomposition of the eigenspectrum to define a unique vacuum state for our causal set quantum field theory, namely the \textit{Sorkin-Johnston} (or SJ) vacuum\cite{sorkin6}. By restricting the imaginary discrete Pauli-Jordan operator ${i \hat{\Delta}_{\mathcal{C}} \left( x, y \right)}$ purely to the positive eigenspace, i.e. the space spanned by the eigenfunctions ${\mathbf{v}_{i}^{+} \left( x \right)}$ with positive associated eigenvalues, we are able to construct the discrete Sorkin-Johnston Wightman function (or two-point correlation function) ${\hat{W}_{SJ} \left( x, y \right)}$:

\begin{equation}
\hat{W}_{SJ} \left( x, y \right) = \sum_{i = 1}^{\frac{s}{2}} \lambda_i \mathbf{v}_{i}^{+} \left( x \right) \left( \mathbf{v}_{i}^{+} \left( y \right) \right)^{*},
\end{equation}
such that the Pauli-Jordan operator itself may be reconstructed as:

\begin{multline}
i \hat{\Delta}_{\mathcal{C}} \left( x, y \right) = \left( \sum_{i = 1}^{\frac{s}{2}} \lambda_i \mathbf{v}_{i}^{+} \left( x \right) \left( \mathbf{v}_{i}^{+} \left( y \right) \right)^{*} \right) - \left( \sum_{i = 1}^{\frac{s}{2}} \lambda_i \mathbf{v}_{i}^{+} \left( x \right) \left( \mathbf{v}_{i}^{+} \left( y \right) \right)^{*} \right)^{*}\\
= \hat{W}_{SJ} \left( x, y \right) - \left( \hat{W}_{SJ} \left( x, y \right) \right)^{*} = \hat{W}_{SJ} \left( x, y \right) - \left( \hat{W}_{SJ} \left( x, y \right) \right)^{\intercal}.
\end{multline}
If we now define a family of free (bosonic) discrete scalar field operators ${\hat{\phi} \left( x \right)}$ over our causal set ${\mathcal{C}}$, effectively associating every element ${x \in \mathcal{C}}$ with an operator ${\hat{\phi} \left( x \right)}$ acting on some arbitrary Hilbert space ${\mathcal{H}}$, then the standard axioms of self-adjointness/Hermiticity:

\begin{equation}
\forall x \in \mathcal{C}, \qquad \hat{\phi} \left( x \right) = \left( \hat{\phi} \left( x \right) \right)^{\dagger},
\end{equation}
and commutation with respect to the discrete Pauli-Jordan operator ${\hat{\Delta}_{\mathcal{C}} \left( x, y \right)}$:

\begin{equation}
\forall x, y \in \mathcal{C}, \qquad \left[ \hat{\phi} \left( x \right), \hat{\phi} \left( y \right) \right] = \hat{\phi} \left( x \right) \hat{\phi} \left( y \right) - \hat{\phi} \left( y \right) \hat{\phi} \left( x \right) = i \hat{\Delta}_{\mathcal{C}} \left( x, y \right),
\end{equation}
are essentially trivial to extend to the discrete case, while the causal set analog of satisfaction of the (massive) Klein-Gordon equation in the continuum theory:

\begin{equation}
\forall \mathbf{x} \in \mathcal{F}_{+} \left( \mathcal{H} \right), \qquad \left( \Box + m^2 \right) \hat{\Phi} \left( \mathbf{x} \right) = 0,
\end{equation}
is a little more subtle. In the continuum case, if we apply the massive Klein-Gordon operator ${\left( \Box + m^2 \right)}$ to the commutator appearing in the Pauli-Jordan operator axiom discussed previously, then we can easily deduce (as emphasized by Noldus\cite{johnston3}) that the satisfaction of the massive Klein-Gordon equation implies that the resulting operator ${\left( \Box + m^2 \right) \hat{\Phi} \left( \mathbf{x} \right)}$ must commute with all the ${\hat{\Phi} \left( y \right)}$ field operators:

\begin{equation}
\forall \mathbf{x}, \mathbf{y} \in \mathcal{F}_{+} \left( \mathcal{H} \right), \qquad \left[ \left( \Box+ m^2 \right) \hat{\Phi} \left( \mathbf{x} \right), \hat{\Phi} \left( \mathbf{y} \right) \right] = \left( \Box + m^2 \right) i \hat{\Delta} \left( \mathbf{x}, \mathbf{y} \right) = 0.
\end{equation}
In the causal set case, we can now introduce a $p$-dimensional vector ${\mathbf{w}}$, such that:

\begin{equation}
i \hat{\Delta}_{\mathcal{C}} \circ \mathbf{w} \left( x \right) = \sum_{y \in \mathcal{C}} i \hat{\Delta}_{\mathcal{C}} \left( x, y \right) \mathbf{w} \left( y \right) = \mathbf{0},
\end{equation}
which may, in turn, be rewritten in terms of the causal set field operators ${\hat{\phi} \left( x \right)}$ and ${\hat{\phi} \left( y \right)}$ using the commutator for the discrete Pauli-Jordan operator axiom above:

\begin{equation}
i \hat{\Delta}_{\mathcal{C}} \circ \mathbf{w} \left( x \right) = \mathbf{0}, \qquad \implies \qquad \left[ \left( \sum_{x^{\prime} \in \mathcal{C}} \mathbf{w} \left( x^{\prime} \right) \hat{\phi} \left( x^{\prime} \right) \right), \hat{\phi} \left( y \right) \right] = \left( \left( \mathbf{w} \left( x \right) \right)^{\intercal} \circ i \hat{\Delta}_{\mathcal{C}} \right) \left( y \right) = \mathbf{0},
\end{equation}
from which we can deduce the causal set analog of the (massive) Klein-Gordon equation axiom, i.e. the axiom that guarantees that any linear combination of the ${\hat{\phi} \left( x \right)}$ discrete field operators that commutes with all of the ${\hat{\phi} \left( y \right)}$ discrete field operators must be identically zero, namely:

\begin{equation}
\forall x \in \mathcal{C}, \qquad \sum_{y \in \mathcal{C}} i \hat{\Delta}_{\mathcal{C}} \left( x, y \right) \mathbf{w} \left( y \right) = 0, \qquad \implies \qquad \sum_{x^{\prime} \in \mathcal{C}} \mathbf{w} \left( x^{\prime} \right) \hat{\phi} \left( x^{\prime} \right) = 0.
\end{equation}

The causal set field operators ${\hat{\phi} \left( x \right)}$ thus allow us to construct discrete analogs of various familiar operator families from quantum field theory, as first shown by Johnston\cite{johnston}\cite{johnston2}\cite{johnston3}, including the creation and annihilation operators ${\left( \hat{a}_i \right)^{\dagger}}$ and ${\hat{a}_i}$, respectively, which we may define in terms of the positive and negative parts of the eigenspectrum, respectively, as follows:

\begin{equation}
\left( \hat{a}_i \right)^{\dagger} = \sum_{x \in \mathcal{C}} \frac{1}{\lambda_i} \mathbf{v}_{i}^{+} \left( x \right) \hat{\phi} \left( x \right), \qquad \text{ and} \qquad \hat{a}_i = \sum_{x \in \mathcal{C}} \frac{1}{\lambda_i} \mathbf{v}_{i}^{-} \left( x \right) \hat{\phi} \left( x \right),
\end{equation}
for ${i = 1, 2, \dots, \frac{s}{2}}$, and with $s$ being (as above) the rank of the imaginary variant of the discrete Pauli-Jordan operator ${i \hat{\Delta}_{\mathcal{C}} \left( x, y \right)}$. From here, it is straightforward to verify that the standard commutation relations that define creation and annihilation operators in continuum quantum field theories:

\begin{equation}
\left[ \left( \hat{a}_i \right)^{\dagger}, \left( \hat{a}_j \right)^{\dagger} \right] = 0, \qquad \left[ \hat{a}_i, \hat{a}_j \right] = 0, \qquad \left[ \hat{a}_i, \left( \hat{a}_j \right)^{\dagger} \right] = \delta_{i j},
\end{equation}
hold, as a direct consequence of the assumed orthonormality of the eigenfunctions ${\mathbf{v}_{i}^{+} \left( x \right)}$, ${\mathbf{v}_{i}^{-} \left( x \right)}$ and ${\mathbf{v}_{i}^{0} \left( x \right)}$ as a basis for the causal set Hilbert space ${\mathbb{C}^p}$:

\begin{multline}
\left[ \left( \hat{a}_i \right)^{\dagger}, \left( \hat{a}_j \right)^{\dagger} \right] = \sum_{x \in \mathcal{C}} \frac{\mathbf{v}_{i}^{+} \left( x \right) i \hat{\Delta}_{\mathcal{C}} \circ \mathbf{v}_{j}^{+} \left( x \right)}{\sqrt{\lambda_i \lambda_j}} = \sum_{x \in \mathcal{C}} \sum_{y \in \mathcal{C}} \frac{\mathbf{v}_{i}^{+} \left( x \right) i \hat{\Delta}_{\mathcal{C}} \left( x, y \right) \mathbf{v}_{j}^{+} \left( y \right)}{\sqrt{\lambda_i \lambda_j}}\\
= \sum_{x \in \mathcal{C}} \frac{\lambda_j \mathbf{v}_{i}^{-} \left( x \right) \mathbf{v}_{j}^{+} \left( x \right)}{\sqrt{\lambda_i \lambda_j}} = 0,
\end{multline}

\begin{multline}
\left[ \hat{a}_i, \hat{a}_j \right] = \sum_{x \in \mathcal{C}} \frac{\mathbf{v}_{i}^{-} \left( x \right) i \hat{\Delta}_{\mathcal{C}} \circ \mathbf{v}_{j}^{-} \left( x \right)}{\sqrt{\lambda_i \lambda_j}} = \sum_{x \in \mathcal{C}} \sum_{y \in \mathcal{C}} \frac{\mathbf{v}_{i}^{-} \left( x \right) i \hat{\Delta}_{\mathcal{C}} \left( x, y \right) \mathbf{v}_{j}^{-} \left( y \right)}{\sqrt{\lambda_i \lambda_j}}\\
= \sum_{x \in \mathcal{C}} \frac{- \lambda_j \mathbf{v}_{i}^{+} \left( x \right) \mathbf{v}_{j}^{-} \left( x \right)}{\sqrt{\lambda_i \lambda_j}} = 0,
\end{multline}
and:

\begin{multline}
\left[ \hat{a}_i, \left( \hat{a}_j \right)^{\dagger} \right] = \sum_{x \in \mathcal{C}} \frac{\mathbf{v}_{i}^{-} \left( x \right) i \hat{\Delta}_{\mathcal{C}} \circ \mathbf{v}_{j}^{+} \left( x \right)}{\sqrt{\lambda_i \lambda_j}} = \sum_{x \in \mathcal{C}} \sum_{y \in \mathcal{C}} \frac{\mathbf{v}_{i}^{-} \left( x \right) i \hat{\Delta}_{\mathcal{C}} \left( x, y \right) \mathbf{v}_{j}^{+} \left( y \right)}{\sqrt{\lambda_i \lambda_j}}\\
= \sum_{x \in \mathcal{C}} \frac{\lambda_j \mathbf{v}_{i}^{+} \left( x \right) \mathbf{v}_{j}^{+} \left( x \right)}{\sqrt{\lambda_i \lambda_j}} = \delta_{i j},
\end{multline}
for ${i, j = 1, 2, \dots, \frac{s}{2}}$, as required. Moreover, since these eigenfunctions form an orthonormal basis for ${\mathbb{C}^p}$, it is possible to write an explicit spectral decomposition of the field operators ${\hat{\phi} \left( x \right)}$ in terms of the creation and annihilation operators ${\left( \hat{a}_i \right)^{\dagger}}$ and ${\hat{a}_i}$ as:

\begin{equation}
\hat{\phi} \left( x \right) = \left( \sum_{i = 1}^{\frac{s}{2}} \sqrt{\lambda_i} \mathbf{v}_{i}^{+} \left( x \right) \hat{a}_i \right) + \left( \sum_{i = 1}^{\frac{s}{2}} \sqrt{\lambda_i} \mathbf{v}_{i}^{-} \left( x \right) \left( \hat{a}_i \right)^{\dagger} \right),
\end{equation}
using the existing spectral decomposition of the discrete Pauli-Jordan operator ${i \hat{\Delta}_{\mathcal{C}} \left( x, y \right)}$, as well as the relationship described above between the commutation relations for the causal set field operators ${\hat{\phi} \left( x \right)}$ and ${\hat{\phi} \left( y \right)}$, and those for the discrete Pauli-Jordan operator itself. The Sorkin-Johnston vacuum state ${\ket{0}_{SJ} \in \mathcal{H}}$ may therefore be defined formally by how the annihilation operators act upon it:

\begin{equation}
\forall i \in \left\lbrace 1, 2, \dots, \frac{s}{2} \right\rbrace, \qquad \hat{a}_i \ket{0}_{SJ} = \mathbf{0}, \text{ such that } \bra{0}_{SJ} \ket{0}_{SJ} = 1,
\end{equation}
thus allowing us to express ${\mathcal{H}}$ as a Fock space with basis vectors given by the products of creation operators acting on the Sorkin-Johnston vacuum state, of the general form:

\begin{equation}
\left( \left( \hat{a}_1 \right)^{\dagger} \right)^{n_1} \left( \left( \hat{a}_2 \right)^{\dagger} \right)^{n_2} \cdots \left( \left( \hat{a}_s \right)^{\dagger} \right)^{n_s} \ket{0}_{SJ}, \qquad \text{ for } \qquad i \in \left\lbrace 1, 2, \dots \frac{s}{2} \right\rbrace \text{ such that } n_i \geq 0.
\end{equation}

In order to see how the entanglement entropy of a given subset of a causal set (considered to be a reduced subsystem of the overall causal set ${\mathcal{C}}$, with ${\mathcal{C}}$ conceptualized here as a purely quantum mechanical system), we revisit how ordinary quantum mechanical entanglement entropies may be calculated in the case of continuous spacetimes. Following Sorkin\cite{sorkin}\cite{sorkin4}, consider first the minimal case of an entanglement entropy for a spacetime consisting of exactly two subsystems, each with a single degree of freedom, as described by the Wightman function:

\begin{equation}
\hat{W}_{m}^{\left( d \right)} \left( \mathbf{x}, \mathbf{y} \right) = \bra{0} \hat{\Phi} \left( \mathbf{x} \right) \hat{\Phi} \left( \mathbf{y} \right) \ket{0} = \begin{bmatrix}
\left\langle \hat{q} \hat{q} \right\rangle & \left\langle \hat{q} \hat{p} \right\rangle\\
\left\langle \hat{p} \hat{q} \right\rangle & \left\langle \hat{p} \hat{p} \right\rangle
\end{bmatrix},
\end{equation}
and the Pauli-Jordan operator:

\begin{equation}
\hat{\Delta} \left( \mathbf{x}, \mathbf{y} \right) = 2 \mathrm{Im} \left( \hat{W}_{m}^{\left( d \right)} \left( \mathbf{x}, \mathbf{y} \right) \right) = 2 \mathrm{Im} \left( \begin{bmatrix}
\left\langle \hat{q} \hat{q} \right\rangle & \left\langle \hat{q} \hat{p} \right\rangle\\
\left\langle \hat{p} \hat{q} \right\rangle & \left\langle \hat{p} \hat{p} \right\rangle
\end{bmatrix} \right) = \begin{bmatrix}
0 & 1\\
-1 & 0
\end{bmatrix},
\end{equation}
in explicit matrix form, where our single degree of freedom has been defined here in terms of a canonically-conjugate pair of variables ${\hat{q}}$ and ${\hat{p}}$, i.e. a pair of variables satisfying the canonical commutation relation:

\begin{equation}
\left[ \hat{q}, \hat{p} \right] = i,
\end{equation}
with a corresponding Gaussian density matrix ${\rho}$, expressed in the ${\hat{q}}$ basis as:

\begin{equation}
\bra{\hat{q}} \rho \ket{\hat{q}^{\prime}} = \exp \left( - \frac{A}{2} \left( \hat{q}^2 + \left( \hat{q}^{\prime} \right)^2 \right) + \frac{i B}{2} \left( \hat{q}^2 - \left( \hat{q}^{\prime} \right)^2 \right) - \frac{C}{2} \left( \hat{q} - \left( \hat{q}^{\prime} \right)^2 \right) \right),
\end{equation}
for some (as yet undetermined) parameters $A$, $B$ and $C$, and with ${\left\langle \hat{q} \hat{q} \right\rangle}$, ${\left\langle \hat{q} \hat{p} \right\rangle}$, ${\left\langle \hat{p} \hat{q} \right\rangle}$ and ${\left\langle \hat{p} \hat{p} \right\rangle}$ being the associated correlation functions. If ${\rho_{red}}$ designates the reduced density matrix obtained by partially tracing out one subsystem from the full density matrix ${\rho}$ (for the case of spacetime entanglement entropies, at least for globally hyperbolic spacetimes, this necessitates tracing out a subregion of a given spacelike hypersurface/Cauchy surface ${\Sigma}$), then the standard von Neumann entropy ${S \left( \rho_{red} \right)}$ may be computed as:

\begin{equation}
S \left( \rho_{red} \right) = - \mathrm{Tr} \left( \rho_{red} \log \left( \rho_{red} \right) \right).
\end{equation}

In order to evaluate the entanglement entropy in this minimal case, we follow the techniques developed by Bombelli, Koul, Lee and Sorkin\cite{bombelli}, by imagining the overall system as consisting of a pair of oscillators, each possessing a single degree of freedom, with corresponding annihilation operators ${\hat{a}}$ and ${\hat{b}}$, such that its overall state vector ${\ket{\psi}}$ is given by:

\begin{equation}
\ket{\psi} = C e^{\sqrt{\mu} \left( \hat{a} \right)^{\dagger} \left( \hat{b} \right)^{\dagger}} \ket{0}_{\hat{a}} \otimes \ket{0}_{\hat{b}} = C \sum_{n = 0}^{\infty} \mu^{\frac{n}{2}} \ket{n}_{\hat{a}} \otimes \ket{n}_{\hat{b}},
\end{equation}
where ${\mu}$ is an arbitrary parameter (to be determined), and $C$ is a normalization constant of the form:

\begin{equation}
C = \sqrt{1 - \mu}.
\end{equation}
The density matrix ${\rho}$ and reduced density matrix ${\rho_{red}}$ (obtained by partially tracing out the oscillator whose annihilation operator is ${\hat{b}}$) are therefore given by:

\begin{equation}
\rho = \ket{\psi} \bra{\psi}, \qquad \text{ and } \qquad \rho_{red} = \sum_{m = 0}^{\infty} {}_{\hat{b}} \braket{m}{\psi} \braket{\psi}{m}_{\hat{b}} = \sum_{m = 0}^{\infty} C^2 \mu^m \ket{m}_{\hat{a}} \bra{m},
\end{equation}
respectively, with the von Neumann entropy ${S \left( \rho_{red} \right)}$ therefore being of the form:

\begin{equation}
S \left( \rho_{red} \right) = - \mathrm{Tr} \left( \rho_{red} \log \left( \rho_{red} \right) \right) = - \log \left( 1 - \mu \right) - \frac{\mu}{1 - \mu} \log \left( \mu \right).
\end{equation}
Thinking about this problem in the Schr\"{o}dinger picture, we can write the wave function ${\psi \left( x, y \right)}$ in explicit coordinates $x$ and $y$, at least up to some multiplicative constant $K$ (to be selected so as to guarantee the eventual normalization of the state vector ${\ket{\psi}}$), as:

\begin{equation}
\psi \left( x, y \right) = K \exp \left[ - \frac{1 + \mu}{1 - \mu} \frac{x^2 + y^2}{2} - \frac{2 \sqrt{\mu}}{1 - \mu} x y \right],
\end{equation}
such that the overall density matrix ${\rho \left[ \left( x, y \right), \left( x^{\prime}, y^{\prime} \right) \right]}$, in coordinates $x$, $y$, ${x^{\prime}}$ and ${y^{\prime}}$ is given simply by:

\begin{equation}
\rho \left[ \left( x, y \right), \left( x^{\prime}, y^{\prime} \right) \right] = \psi \left( x, y \right) \left( \psi \left( x^{\prime}, y^{\prime} \right) \right)^{\dagger},
\end{equation}
and therefore, by again partially tracing out the oscillator with annihilation operator ${\hat{b}}$ (and thus with coordinates $y$, ${y^{\prime}}$), we consequently obtain the associated reduced density matrix ${\rho_{red} \left( x, x^{\prime} \right)}$:

\begin{multline}
\rho_{red} \left( x, x^{\prime} \right) = \int K^2 \exp \left[ - \frac{1 + \mu}{1 - \mu} \frac{x^2 + y^2}{2} - \frac{2 \sqrt{\mu}}{1 - \mu} x y - \frac{1 + \mu}{1 - \mu} \frac{\left( x^{\prime} \right)^2 + y^2}{2} - \frac{2 \sqrt{\mu}}{1 - \mu} x^{\prime} y \right] d y\\
= K^2 \exp \left[ - \frac{1 + \mu}{1 - \mu} \frac{x^2 + \left( x^{\prime} \right)^2}{2} \right] \sqrt{\frac{\pi \left( 1 - \mu \right)}{1 + \mu}} \exp \left[ \frac{\mu \left( x + x^{\prime} \right)^2}{\left( 1 - \mu \right) \left( 1 + \mu \right)} \right],
\end{multline}
or, with $K$ now selected so as to guarantee normalization of the wave function ${\psi \left( x, y \right)}$:

\begin{equation}
\rho_{red} \left( x, x^{\prime} \right) = \sqrt{\frac{1 - \mu}{\pi \left( 1 + \mu \right)}} \exp \left[ - \frac{1 + \mu}{1 - \mu} \frac{x^2 + \left( x^{\prime} \right)^2}{2} + \frac{\mu}{1 - \mu^2} \left( x + x^{\prime} \right)^2 \right].
\end{equation}

Noting now that, for our original conjugate variables ${\hat{q}}$ and ${\hat{p}}$, one has:

\begin{equation}
\left\langle \hat{q} \hat{q} \right\rangle \left\langle \hat{p} \hat{p} \right\rangle - \mathrm{Re} \left( \left\langle \hat{q} \hat{p} \right\rangle \right)^2 = \frac{C}{2 A} + \frac{1}{4},
\end{equation}
indicating, as noted by Sorkin\cite{sorkin4}, that the entanglement entropy should depend purely upon the ratio of parameters $C$ and $A$, with all dependence upon the $B$ parameter being eliminated (and therefore we can, without loss of generality, assume ${B = 0}$ henceforth). Consequently, by setting parameters $A$ and $C$ to be:

\begin{equation}
A = \frac{C \left( \mu - 1 \right)^2}{2 \mu}, \qquad \text{ and } \qquad C = \frac{2 A \mu}{\left( \mu - 1 \right)^2},
\end{equation}
respectively, such that the particular ratio of relevance for the entropy calculation is simply:

\begin{equation}
\frac{C}{2 A} = \frac{\mu}{\left( \mu - 1 \right)^2},
\end{equation}
we obtain the density matrix ${\rho_{red} \left( x, x^{\prime} \right)}$ in such a form that its von Neumann entropy ${S \left( \rho_{red} \left( x, x^{\prime} \right) \right)}$ may be written quite straightforwardly as:

\begin{equation}
S \left( \rho_{red} \left( x, x^{\prime} \right) \right) = - \mathrm{Tr} \left( \rho_{red} \left( x, x^{\prime} \right) \log \left( \rho_{red} \left( x, x^{\prime} \right) \right) \right) = - \frac{\mu \log \left( \mu \right) + \left( 1 - \mu \right) \log \left( 1 - \mu \right)}{1 - \mu}.
\end{equation}
The definitions given for the parameters $A$ and $C$ above can now be inverted to yield the following explicit form for ${\mu}$ in terms of the parameters appearing in the original definition of the Gaussian density matrix in the ${\hat{q}}$ basis:

\begin{equation}
\mu = \frac{\sqrt{1 + \frac{2 C}{A}} - 1}{\sqrt{1 + \frac{2 C}{A}} + 1}.
\end{equation}
Considering now the spectral decomposition for the product of the inverse of the Pauli-Jordan operator ${\hat{\Delta} \left( \mathbf{x}, \mathbf{y} \right)}$ and the real part of the Wightman function ${\hat{W}_{m}^{\left( d \right)} \left( \mathbf{x}, \mathbf{y} \right)}$:

\begin{multline}
\left( \hat{\Delta} \left( \mathbf{x}, \mathbf{y}  \right) \right)^{-1} \mathrm{Re} \left( \hat{W}_{m}^{\left( d \right)} \left( \mathbf{x}, \mathbf{y} \right) \right) = \left( \begin{bmatrix}
0 & 1\\
-1 & 0
\end{bmatrix} \right)^{-1} \mathrm{Re} \left( \begin{bmatrix}
\left\langle \hat{q} \hat{q} \right\rangle & \left\langle \hat{q} \hat{p} \right\rangle\\
\left\langle \hat{p} \hat{q} \right\rangle & \left\langle \hat{p} \hat{p} \right\rangle
\end{bmatrix} \right)\\
= \begin{bmatrix}
0 & -1\\
1 & 0
\end{bmatrix} \begin{bmatrix}
\left\langle \hat{q} \hat{q} \right\rangle & \mathrm{Re} \left( \left\langle \hat{q} \hat{p} \right\rangle \right)\\
\mathrm{Re} \left( \left\langle \hat{q} \hat{p} \right\rangle \right) & \left\langle \hat{p} \hat{p} \right\rangle
\end{bmatrix} = \begin{bmatrix}
- \mathrm{Re} \left( \left\langle \hat{q} \hat{p} \right\rangle \right) & - \left\langle \hat{p} \hat{p} \right\rangle\\
\left\langle \hat{q} \hat{q} \right\rangle & \mathrm{Re} \left( \left\langle \hat{q} \hat{p} \right\rangle \right)
\end{bmatrix},
\end{multline}
we can see immediately that its eigenvalues are purely imaginary (since the resulting matrix is skew-Hermitian), and therefore may be written in the form ${\lambda = \pm i \sigma}$ for some ${\sigma \in \mathbb{R}}$, allowing us to write the von Neumann entropy for the reduced density matrix ${\rho_{red} \left( x, x^{\prime} \right)}$ as:

\begin{equation}
S \left( \rho_{red} \left( x, x^{\prime} \right) \right) = \left( \sigma + \frac{1}{2} \right) \log \left( \sigma + \frac{1}{2} \right) - \left( \sigma - \frac{1}{2} \right) \log \left( \sigma - \frac{1}{2} \right).
\end{equation}
We can eliminate the ${\frac{1}{2}}$ terms by noting that the Wightman function ${\hat{W}_{m}^{\left( d \right)} \left( \mathbf{x}, \mathbf{y} \right)}$ itself may trivially be obtained from its purely real part via the transformation:

\begin{equation}
\hat{W}_{m}^{\left( d \right)} \left( \mathbf{x}, \mathbf{y} \right) = \mathrm{Re} \left( \hat{W}_{m}^{\left( d \right)} \left( \mathbf{x}, \mathbf{y} \right) \right) + \frac{i}{2} \hat{\Delta} \left( \mathbf{x}, \mathbf{y} \right),
\end{equation}
enabling us to consider instead the spectrum of the product of the inverse of the Pauli-Jordan operator ${\hat{\Delta} \left( \mathbf{x}, \mathbf{y} \right)}$ and the full Wightman function ${\hat{W}_{m}^{\left( d \right)} \left( \mathbf{x}, \mathbf{y} \right)}$:

\begin{equation}
\left( \hat{\Delta} \left( \mathbf{x}, \mathbf{y} \right) \right)^{-1} \hat{W}_{m}^{\left( d \right)} \left( \mathbf{x}, \mathbf{y} \right) = \begin{bmatrix}
0 & -1\\
1 & 0
\end{bmatrix} \begin{bmatrix}
\left\langle \hat{q} \hat{q} \right\rangle & \left\langle \hat{q} \hat{p} \right\rangle\\
\left\langle \hat{p} \hat{q} \right\rangle & \left\langle \hat{p} \hat{p} \right\rangle
\end{bmatrix} = \begin{bmatrix}
- \left\langle \hat{p} \hat{q} \right\rangle & - \left\langle \hat{p} \hat{p} \right\rangle\\
\left\langle \hat{q} \hat{q} \right\rangle & \left\langle \hat{q} \hat{p} \right\rangle
\end{bmatrix},
\end{equation}
which is also clearly skew-Hermitian, and therefore its eigenvalues may be written in the form ${\lambda = i \omega_{\pm}}$ for some ${\omega_{+}, \omega_{-} \in \mathbb{R}}$ such that ${i \omega_{\pm} = i \left( \frac{1}{2} \pm \sigma \right)}$, and such that the expression for the von Neumann entropy ${S \left( \rho_{red} \left( x, x^{\prime} \right) \right)}$ simply reduces to:

\begin{equation}
S \left( \rho_{red} \left( x, x^{\prime} \right) \right) = \omega_{+} \log \left( \omega_{+} \right) - \omega_{-} \log \left( \omega_{-} \right).
\end{equation}

\begin{figure}[ht]
\centering
\includegraphics[width=0.395\textwidth]{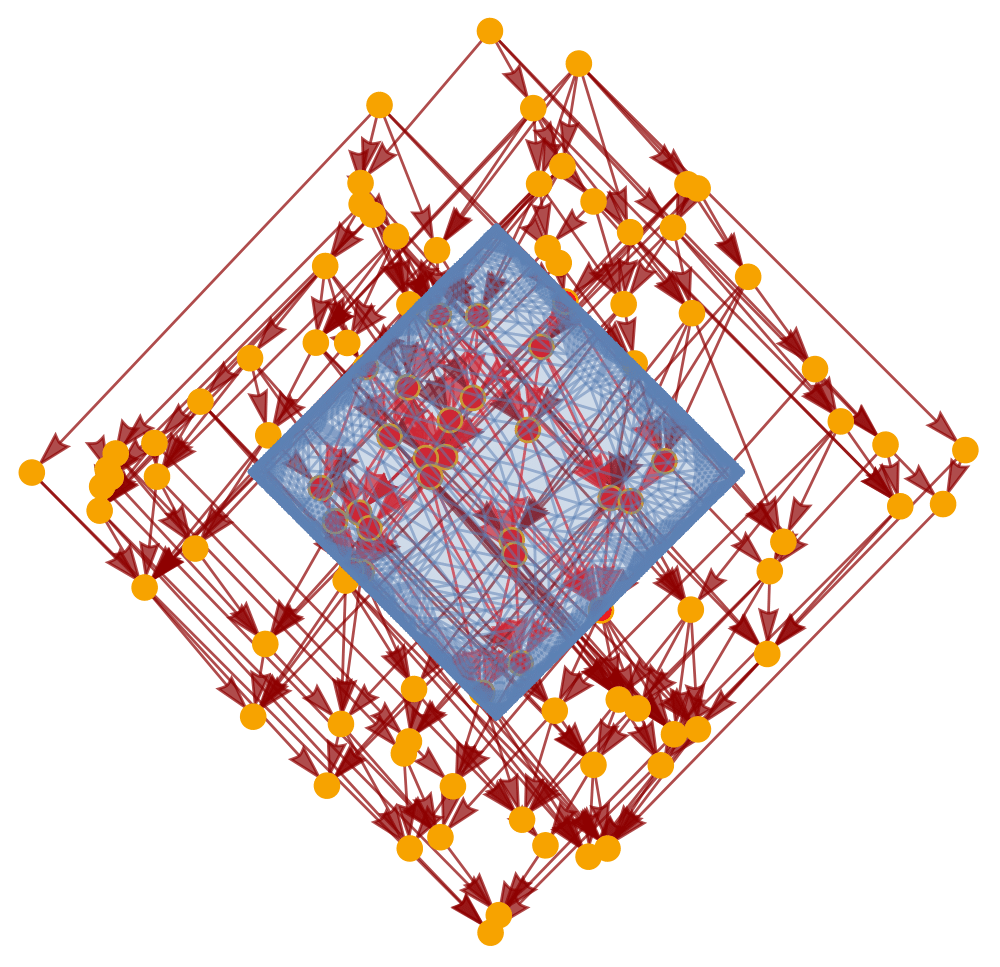}\hspace{0.1\textwidth}
\includegraphics[width=0.395\textwidth]{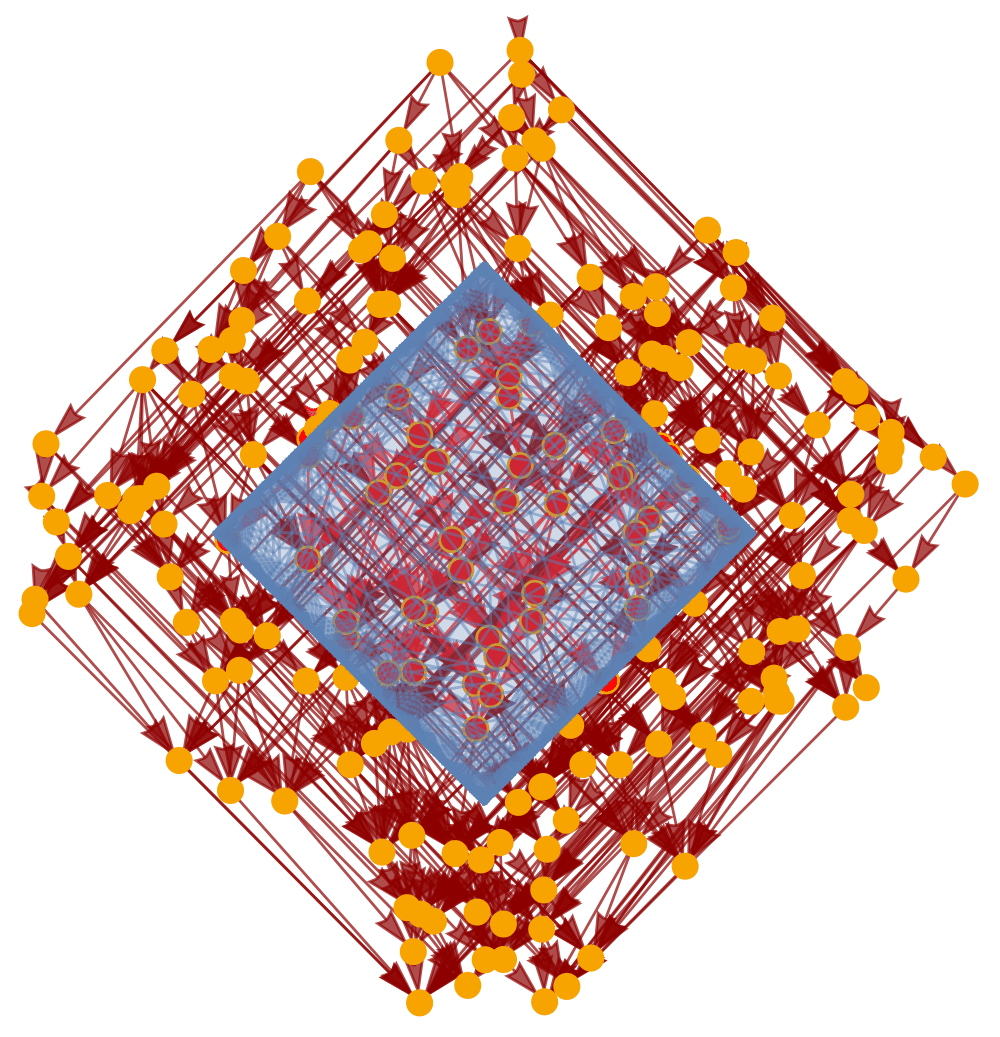}
\caption{On the left, the transitive reduction (i.e. the Hasse diagram) of the directed graph generated by connecting all pairs of 100 (uniformly-sprinkled) points that are related by the causal partial order relation ${\prec_{\mathcal{M}}}$ on a diamond-shaped region of a ${1 + 1}$-dimensional flat (Minkowski) spacetime, with a smaller diamond-shaped subregion selected within it (shown in red and highlighted within the blue diamond, with side length equal to ${\frac{1}{2}}$ the side length of the overall diamond), for which the discrete Pauli-Jordan operator ${\hat{\Delta}_{\mathcal{C}} \left( x, y \right)}$ is not invertible, and therefore for which the entanglement entropy $S$ is undefined in the simple/naive case, and is equal to 0.399 in the generalized/robust case. On the right, a sprinkling into the same region of ${1 + 1}$-dimensional flat (Minkowski) spacetime, but with double the sprinkling density (200 points), for which the discrete Pauli-Jordan operator ${\hat{\Delta}_{\mathcal{C}} \left( x, y \right)}$ \textit{is} invertible, yielding an entanglement entropy estimate $S$ of 0.622 in the simple/naive case, and 0.590 in the generalized/robust case.}
\label{fig:Figure16}
\end{figure}

Treating the case analyzed above (namely of a spacetime consisting of two identical subsystems, each with a single degree of freedom) as the base case of a more general inductive construction, we consider now taking a general Wightman function ${\hat{W}_{m}^{\left( d \right)} \left( \mathbf{x}, \mathbf{y} \right)}$ and performing a block-diagonalization of its matrix representation, such that each block is a ${2 \times 2}$ matrix of the form already discussed. Thus, following Sorkin\cite{sorkin4}, we introduce the operator ${\hat{L} \left( \mathbf{x}, \mathbf{y} \right)}$ which is equal (up to a factor of $i$) to the product of the inverse of the Pauli-Jordan operator ${\hat{\Delta} \left( \mathbf{x}, \mathbf{y} \right)}$ and the overall Wightman function ${\hat{W}_{m}^{\left( d \right)} \left( \mathbf{x}, \mathbf{y} \right)}$:

\begin{equation}
i \hat{L} \left( \mathbf{x}, \mathbf{y} \right) = \left( \hat{\Delta} \left( \mathbf{x}, \mathbf{y} \right) \right)^{-1} \hat{W}_{m}^{\left( d \right)} \left( \mathbf{x}, \mathbf{y} \right).
\end{equation}
Due to the presence of the factor of $i$, all $n$ eigenvalues ${\lambda_1, \lambda_2, \dots, \lambda_n}$ of ${i \hat{L} \left( \mathbf{x}, \mathbf{y} \right)}$ are now guaranteed to be real, allowing us to write the von Neumann entropy of the reduced density matrix ${\rho_{red}}$ in the general case as a sum over the entropies for each block in the block-diagonalization, i.e:

\begin{equation}
S \left( \rho_{red} \right) = \sum_{i = 1}^{n} \lambda_i \log \left( \left\lvert \lambda_i \right\rvert \right),
\end{equation}
where every negative eigenvalue ${\lambda_i}$ is paired with exactly one positive eigenvalue ${1 - \lambda_i}$. Evidently, this computation has a natural discrete analog for a (finite) causal set ${\mathcal{C}}$, in which the discrete form of the ${\hat{L}}$ operator, i.e. ${\hat{L}_{\mathcal{C}} \left( x, y \right)}$, is defined in terms of the discrete Pauli-Jordan operator ${\hat{\Delta}_{\mathcal{C}} \left( x, y \right)}$ and the Sorkin-Johnston Wightman function ${\hat{W}_{SJ} \left( x,  y \right)}$ as:

\begin{equation}
\forall x, y \in \mathcal{C}, \qquad i \hat{L}_{\mathcal{C}} \left( x, y \right) = \left( \hat{\Delta}_{\mathcal{C}} \left( x, y \right) \right)^{-1} \hat{W}_{SJ} \left( x, y \right),
\end{equation}
with the spacetime entanglement entropy being computed using the eigenvalues ${\lambda_i}$ of the discrete ${i \hat{L}_{\mathcal{C}} \left( x, y \right)}$ operator precisely as before, i.e. by means of the causal set eigenvalue equation:

\begin{equation}
\hat{L}_{\mathcal{C}} \circ \mathbf{v}_i \left( x \right) = \sum_{y \in \mathcal{C}} \hat{L}_{\mathcal{C}} \left( x, y \right) \mathbf{v}_i \left( y \right) = \lambda_i \mathbf{v}_i \left( x \right).
\end{equation}
However, this inductive argument still neglects an important subtlety, namely that the Pauli-Jordan operator ${\hat{\Delta} \left( \mathbf{x}, \mathbf{y} \right)}$ (and, by extension, its discrete counterpart ${\hat{\Delta}_{\mathcal{C}} \left( x, y \right)}$) will not, in general, be invertible, due to the presence of blocks within the block-diagonalization that may consist entirely of zeroes. This problem can be circumvented by modifying the eigenvalue equation so as to ignore such blocks of zeroes, yielding the following generalized eigenvalue problem in the continuum case:

\begin{equation}
\hat{W}_{m}^{\left( d \right)} \left( \mathbf{x}, \mathbf{y} \right) \mathbf{v}_i = i \lambda_i \hat{\Delta} \left( \mathbf{x}, \mathbf{y} \right) \mathbf{v}_i,
\end{equation}
or, in the discrete/causal set case:

\begin{equation}
\hat{W}_{SJ} \circ \mathbf{v}_i \left( x \right) = \sum_{y \in \mathcal{C}} \hat{W}_{SJ} \left( x, y \right) \mathbf{v}_i \left( y \right) = i \lambda_i \hat{\Delta}_{\mathcal{C}} \circ \mathbf{v}_i \left( x \right) = \sum_{y \in \mathcal{C}} i \lambda_i \hat{\Delta}_{\mathcal{C}} \left( \mathbf{x}, \mathbf{y} \right) \mathbf{v}_i \left( y \right).
\end{equation}
Thus, the naive eigenvalue approach is generally less computationally demanding (since the only real bottleneck lies in the inversion of the discrete Pauli-Jordan operator ${\hat{\Delta}_{\mathcal{C}} \left( x, y \right)}$, which can be done reasonably efficiently using symbolic linear algebra software), but considerably more fragile, whilst the generalized eigenvalue approach is much more expensive (since it effectively involves solving a large linear optimization problem), but far more robust. A comparison between these two approaches to computing the causal set spacetime entanglement entropy, for examples of sprinkled causal sets in which the discrete Pauli-Jordan operator ${\hat{\Delta}_{\mathcal{C}} \left( x, y \right)}$ both is and is not invertible (100 and 200 element sprinklings, respectively, both performed into a diamond-shaped region of a ${1 + 1}$-dimensional flat/Minkowski spacetime, with a smaller diamond-shaped subregion of half the side length selected within it), is shown in Figure \ref{fig:Figure16}.

\begin{figure}[ht]
\centering
\includegraphics[width=0.495\textwidth]{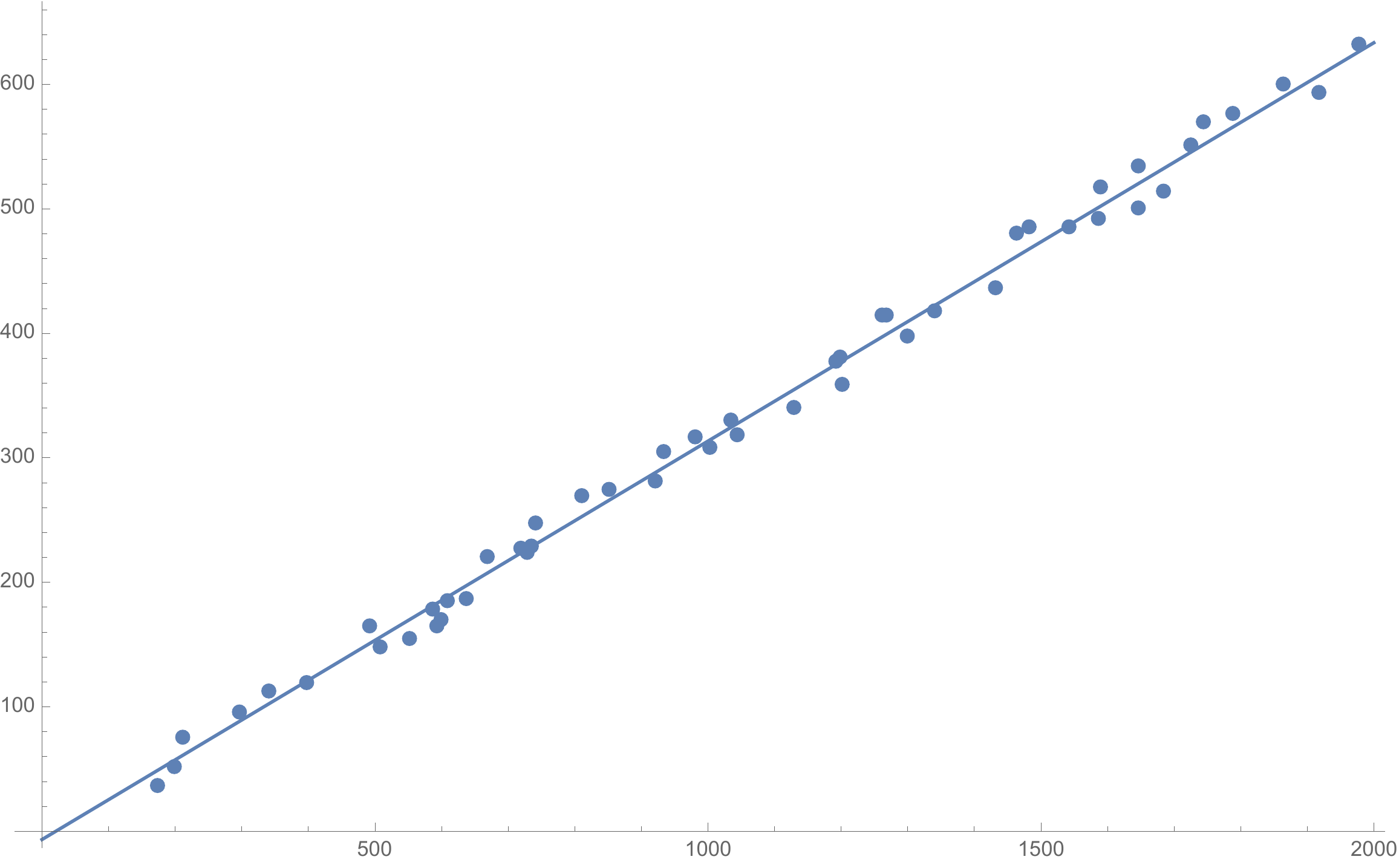}
\includegraphics[width=0.495\textwidth]{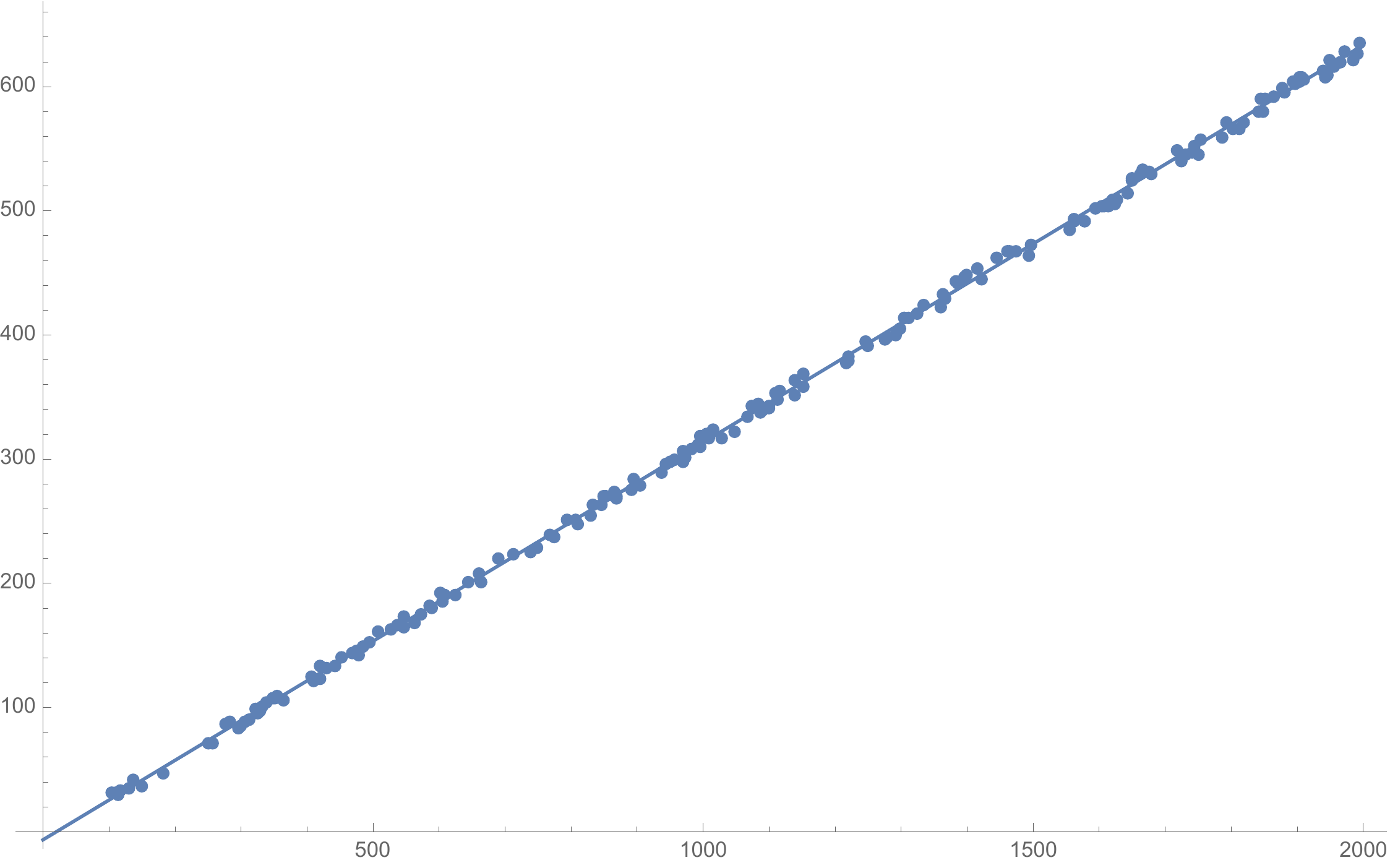}
\caption{Computations of the spacetime entanglement entropy $S$ for 2000 randomly-selected causal sets, sprinkled into a diamond-shaped region of a ${1 + 1}$-dimensional flat (Minkowski) spacetime, with a smaller diamond-shaped subregion (with side length equal to ${\frac{1}{2}}$ of the side length of the overall diamond, and with the cardinalities $N$ of the interior diamond region ranging between 100 and 2000) selected inside, demonstrating the expected linear scaling relation ${S = 0.32 N - 6.64}$. The entanglement entropy estimates are produced using the simple/naive approach (left) and the generalized/robust approach (right), with all points for which the entanglement entropy is undefined being omitted, showing strong quantitative agreement between the two approaches.}
\label{fig:Figure17}
\end{figure}

When these two entanglement entropy estimation procedures are applied to 2000 randomly-selected causal sets that have been sprinkled into a diamond-shaped region of a ${1 + 1}$-dimensional flat/Minkowski spacetime, with a smaller diamond-shaped subregion selected within it whose side length is exactly half that of the larger diamond, we obtain the same linear relationship discovered by Sorkin and Yazdi\cite{sorkin7} between the cardinality of the interior diamond region $N$ and the estimated entanglement entropy $S$, namely ${S = 0.32 N - 6.64}$, as shown in Figure \ref{fig:Figure17}. Following their approach, we seek to recover the desired logarithmic relationship that is indicative of a (spatial) area law rather than a volume law by truncating the spectrum of the discrete Pauli-Jordan operator ${\hat{\Delta}_{\mathcal{C}} \left( x, y \right)}$ (and, by extension, the spectrum of the Sorkin-Johnston Wightman function ${\hat{W}_{SJ} \left( x, y \right)}$), excluding those eigenvalue pairs ${\lambda_i}$ and ${1 - \lambda_i}$ whose corresponding eigenmodes have wavelengths falling below a specified ultraviolet cutoff. The appropriate choice of cutoff may be determined by performing a full eigendecomposition of the imaginary variant of the continuum Pauli-Jordan operator ${i \hat{\Delta} \left( \mathbf{x}, \mathbf{y} \right)}$ in order to see which eigenvalue pairs are most naturally excluded, following the methods of Johnston\cite{johnston3}, which we shall briefly review here. Recall that, for any given spacetime volume ${\mathcal{V}}$, the imaginary variant of the Pauli-Jordan operator ${i \hat{\Delta}}$ may be interpreted as defining an integral operator on the Hilbert space ${L^2 \left( \mathcal{V} \right)}$ of square-integrable functions ${\psi \in L^2 \left( \mathcal{V} \right)}$ on ${\mathcal{V}}$, namely:

\begin{equation}
\left( i \hat{\Delta} \psi \right) \left( \mathbf{x} \right) = \int_{\mathcal{V}} i \Delta \left( \mathbf{x}, \mathbf{y} \right) \psi \left( \mathbf{y} \right) d \mathbf{y}.
\end{equation}
This integral is guaranteed to be well-defined for all functions ${\psi \in L^2 \left( \mathcal{V} \right)}$, at least in the case where the operator ${i \hat{\Delta}}$ and the region ${\mathcal{V}}$ have been specially selected so as to ensure that ${i \hat{\Delta}}$ is a \textit{Hilbert-Schmidt integral kernel}, i.e. to ensure that the integral of the modulus squared of ${i \hat{\Delta}}$ is finite:

\begin{equation}
\int_{\mathcal{V}} \int_{\mathcal{V}} \left\lvert i \hat{\Delta} \left( \mathbf{x}, \mathbf{y} \right) \right\rvert^2 d \mathbf{y} d \mathbf{x} < \infty.
\end{equation}
The resulting operator ${i \hat{\Delta}}$ on the Hilbert space ${L^2 \left( \mathcal{V} \right)}$ is known as a \textit{Hilbert-Schmidt operator}, since by the above statement it is now guaranteed to be bounded, and since:

\begin{equation}
i \hat{\Delta} \left( \mathbf{x}, \mathbf{y} \right) = \left( i \hat{\Delta} \left( \mathbf{x}, \mathbf{y} \right) \right)^{\dagger} = \left( i \hat{\Delta} \left( \mathbf{y}, \mathbf{x} \right) \right)^{*},
\end{equation}
by definition, ${i \hat{\Delta}}$ is also guaranteed to be self-adjoint/Hermitian\cite{stone}. This property has the welcome consequence of ensuring that the operator ${i \hat{\Delta}}$ has a finite, or at most countably infinite, set of eigenvalues ${\lambda_n}$, satisfying:

\begin{equation}
\sum_{n} \lambda_{n}^{2} = \int_{\mathcal{V}} \int_{\mathcal{V}} \left\lvert i \hat{\Delta} \left( \mathbf{x}, \mathbf{y} \right) \right\rvert^2 d \mathbf{y} d \mathbf{x},
\end{equation}
with the corresponding eigenfunctions ${\psi_n \in L^2 \left( \mathcal{V} \right)}$ given by:

\begin{equation}
\int_{\mathcal{V}} i \hat{\Delta} \left( \mathbf{x}, \mathbf{y} \right) \psi_n \left( \mathbf{y} \right) d \mathbf{y} = \lambda_n \psi_n \left( \mathbf{x} \right),
\end{equation}
for some $n$. If we now apply the (massive) Klein-Gordon operator ${\left( \Box + m^2 \right)}$ to both sides of this eigenfunction equation:

\begin{equation}
\left( \Box + m^2 \right) \int_{\mathcal{V}} i \hat{\Delta} \left( \mathbf{x}, \mathbf{y} \right) \psi_n \left( \mathbf{y} \right) d \mathbf{y}  = \int_{\mathcal{V}} \left( \Box + m^2 \right) i \hat{\Delta} \left( \mathbf{x}, \mathbf{y} \right) \psi_n \left( \mathbf{y} \right) d \mathbf{y} = \lambda_n \left( \Box + m^2 \right) \psi_n \left( \mathbf{x} \right),
\end{equation}
and exploit the fact that the imaginary variant of the Pauli-Jordan operator ${i \hat{\Delta}}$ satisfies the massive Klein-Gordon equation by definition (since the Pauli-Jordan operator ${\hat{\Delta}}$ itself is a difference of retarded and advanced Green's functions):

\begin{equation}
\left( \Box + m^2 \right) i \hat{\Delta} \left( \mathbf{x}, \mathbf{y} \right) = 0,
\end{equation}
we see immediately that the eigenfunctions ${\psi_n \in L^2 \left( \mathcal{V} \right)}$ whose corresponding eigenvalues ${\lambda_n}$ are non-zero must also satisfy the massive Klein-Gordon equation themselves:

\begin{equation}
\left( \Box + m^2 \right) \psi_n \left( \mathbf{x} \right) = 0,
\end{equation}
from which the eigenfunctions can be determined up to a normalization constant and a phase factor. The (massive) $d$-dimensional continuum Wightman function ${\hat{W}_{m}^{\left( d \right)} \left( \mathbf{x}, \mathbf{y} \right)}$ is then given by the following sum of non-zero eigenvalue-eigenfunction pairs:

\begin{equation}
\hat{W}_{M}^{\left( d \right)} \left( \mathbf{x}, \mathbf{y} \right) = \sum_{n \in \mathbb{N} : \lambda_n > 0} \lambda_n \psi_n \left( \mathbf{x} \right) \left( \psi_n \left( \mathbf{y} \right) \right)^{*},
\end{equation}
under the additional assumption that the eigenfunctions ${\psi_n \in L^2 \left( \mathcal{V} \right)}$ obey the normalization convention:

\begin{equation}
\forall n \in \mathbb{N} : \lambda_n > 0, \qquad \left\lVert \psi_n \right\rVert = 1, \qquad \text{ where } \qquad \left\lVert \psi_n \right\rVert^2 = \int_{\mathcal{V}} \left\lvert \psi_n \left( \mathbf{x} \right) \right\rvert^2 d \mathbf{x}.
\end{equation}

As an illustrative example of how the eigendecomposition of the imaginary variant of the continuum Pauli-Jordan operator ${i \hat{\Delta} \left( \mathbf{x}, \mathbf{y} \right)}$ may be performed, we consider the case of a ${1 + 1}$-dimensional flat (Minkowski) spacetime ${\mathbb{M}^2 = \mathbb{R}^{1, 1}}$ equipped with a massless scalar field ${\phi}$, henceforth adopting the light cone coordinate system ${\left( u, v \right)}$:

\begin{equation}
u = \frac{t + x}{\sqrt{2}}, \qquad \text{ and } \qquad v = \frac{t - x}{\sqrt{2}},
\end{equation}
such that the ${1 + 1}$-dimensional d'Alembertian operator ${\Box}$ reduces to the following elegant form:

\begin{equation}
\Box = 2 \frac{\partial^2}{\partial u \partial v}.
\end{equation}
Recalling that the massless retarded Green's function ${\left( G_R \right)_{0}^{\left( 2 \right)} \left( \mathbf{x} \right)}$ in a ${1 + 1}$-dimensional flat (Minkowksi) spacetime is given by:

\begin{equation}
\left( G_R \right)_{0}^{\left( 2 \right)} \left( \mathbf{x} \right) = \frac{1}{2} \theta \left( t \right) \theta \left( \tau^2 \right),
\end{equation}
for Heaviside step function ${\theta \left( \alpha \right)}$, and where ${\tau = \sqrt{t^2 - x^2}}$ in spacetime coordinates ${\left( t, x \right)}$, we obtain the following explicit form of the Pauli-Jordan operator ${\hat{\Delta} \left( u, v \right)}$:

\begin{equation}
\hat{\Delta} \left( u, v \right) = \left( G_R \right)_{0}^{\left( 2 \right)} \left( u, v \right) - \left( G_A \right)_{0}^{\left( 2 \right)} \left( u, v \right) = \frac{1}{2} \left[ \theta \left( u \right) \theta \left( v \right) - \theta \left( -u \right) \theta \left( -v \right) \right] = \frac{1}{2} \left[ \theta \left( u \right) + \theta \left( v \right) - 1 \right].
\end{equation}
Within any finite diamond-shaped region of spacetime with side length equal to $L$, namely ${\mathcal{V} = \left[ -L, L \right] \times \left[ -L, L \right]}$ in the light cone coordinates ${\left( u, v \right)}$, it is trivial to see that the imaginary variant of the Pauli-Jordan operator ${i \hat{\Delta} \left( u, v \right)}$ is a Hilbert-Schmidt operator, since the relevant integral now reduces to:

\begin{equation}
\int_{\mathcal{V}} \int_{\mathcal{V}} \left\lvert i \hat{\Delta} \left( \mathbf{x}, \mathbf{y} \right) \right\rvert^2 d \mathbf{y} d \mathbf{x} = \int_{-L}^{L} \int_{-L}^{L} \int_{-L}^{L} \int_{-L}^{L} \left\lvert i \hat{\Delta} \left( u - u^{\prime}, v - v^{\prime} \right) \right\rvert^2 d v^{\prime} d v d u^{\prime} d u = 2 L^4 < \infty.
\end{equation}
Moreover, we also know that any eigenfunctions ${\psi_n \in L^2 \left( \left[ -L, L \right] \times \left[ -L, L \right] \right)}$ of the imaginary variant of the Pauli-Jordan operator ${i \hat{\Delta} \left( u, v \right)}$ with non-zero eigenvalues must satisfy the massless Klein-Gordon equation:

\begin{equation}
\Box \psi_n \left( u, v \right) = 0,
\end{equation}
and therefore they will all be of the general form:

\begin{equation}
\psi_n \left( u, v \right) = \left( \psi_n \right)_1 \left( u \right) + \left( \psi_n \right)_2 \left( v \right), \qquad \text{ where } \qquad \left( \psi_n \right)_1, \left( \psi_n \right)_2 \in L^2 \left( \left[ -L, L \right] \right).
\end{equation}
These eigenfunctions with non-zero eigenvalues may consequently be computed by observing the action of the imaginary variant of the Pauli-Jordan operator ${i \hat{\Delta}}$ on an appropriate set of basis functions; following Johnston\cite{johnston3}, we choose ${\psi_k \left( u, v \right) = e^{i k u}}$, ${\psi_k \left( u, v \right) = e^{i k v}}$ and ${\psi_k \left( u, v \right) = 1}$, yielding:

\begin{equation}
\psi_k \left( u, v \right) = e^{i k u}, \qquad \implies \qquad \left( i \hat{\Delta} \psi_k \right) \left( u, v \right) = \frac{L}{k} e^{i k u} - \frac{L}{k} \cos \left( k L \right) + i \frac{v}{k} \sin \left( k L \right),
\end{equation}
\begin{equation}
\psi_k \left( u, v \right) = e^{i k v}, \qquad \implies \qquad \left( i \hat{\Delta} \psi_k \right) \left( u, v \right)= \frac{L}{k} e^{i k v} - \frac{L}{k} \cos \left( k L \right) + i \frac{u}{k} \sin \left( k L \right),
\end{equation}
and:

\begin{equation}
\psi_k \left( u, v \right) = 1, \qquad \implies \qquad \left( i \hat{\Delta} \psi_k \right) \left( u, v \right) = i L \left( u + v\right),
\end{equation}
respectively.

\begin{figure}[ht]
\centering
\includegraphics[width=0.495\textwidth]{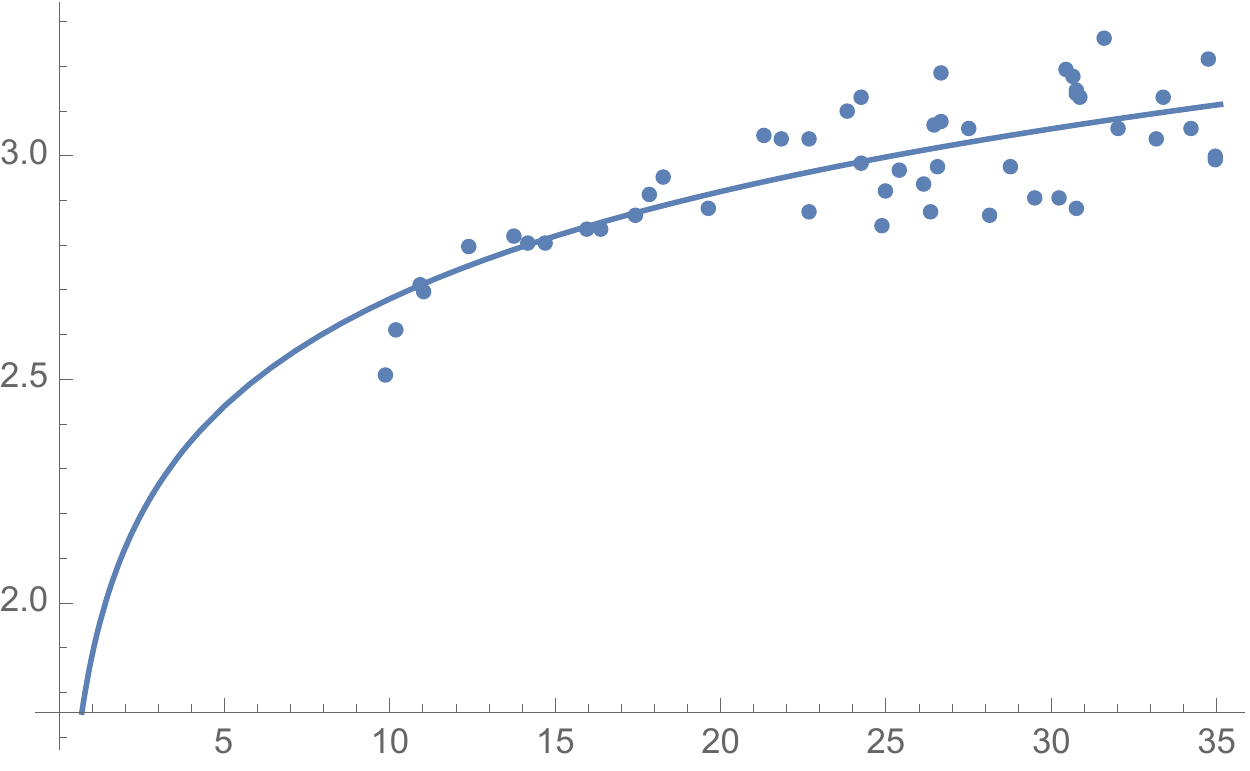}
\includegraphics[width=0.495\textwidth]{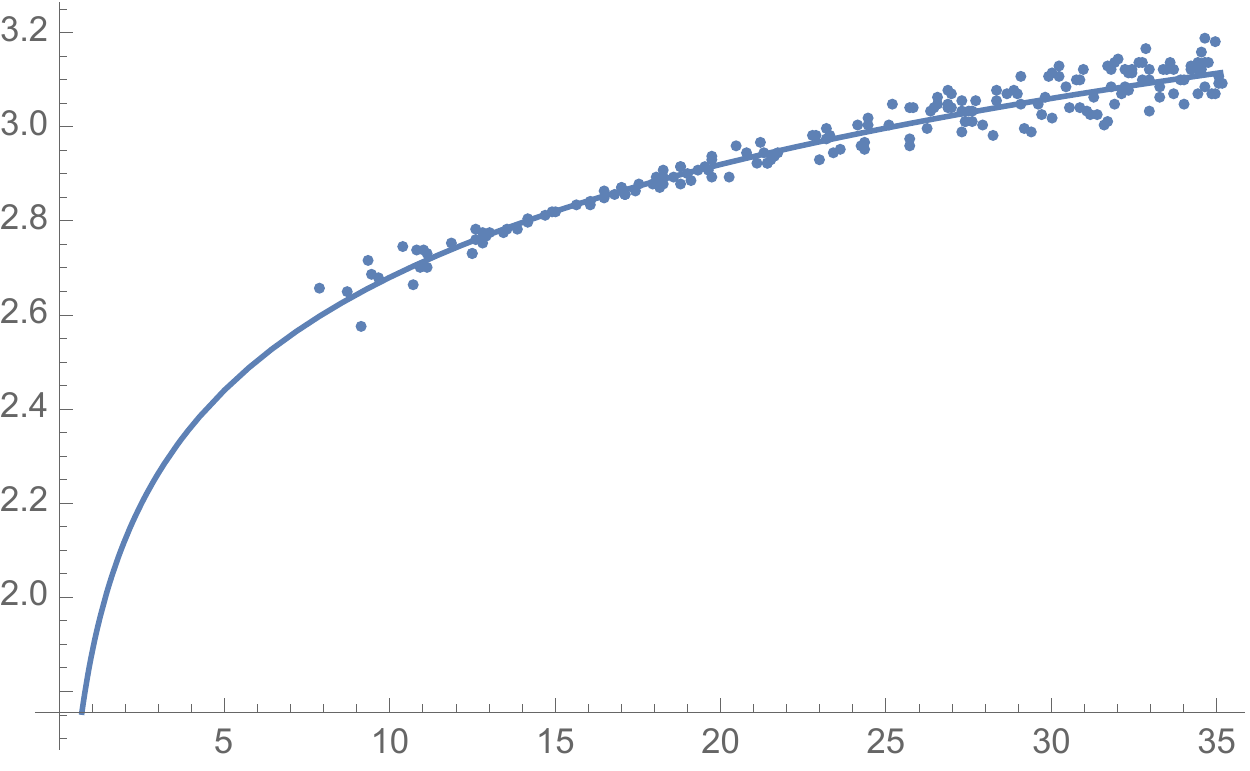}
\caption{Computations of the spacetime entanglement entropy $S$ for 2000 randomly-selected causal sets, sprinkled into a diamond-shaped region of a ${1 + 1}$-dimensional flat (Minkowski) spacetime, with a smaller diamond-shaped subregion (with side length equal to ${\frac{1}{2}}$ of the side length of the overall diamond, and with the cardinalities $N$ of the interior diamond region ranging between 100 and 2000) selected inside, demonstrating the expecting logarithmic scaling relation ${S = 0.346 \log \left( \frac{\sqrt{N}}{4 \pi} \right) + 1.883}$, with the $x$-axis here being labeled by ${\frac{\sqrt{N}}{4 \pi}}$. The entanglement entropy estimates are produced using the simple/naive approach (left) and the generalized/robust approach (right), with all points for which the entanglement entropy is undefined being omitted, showing strong quantitative agreement between the two approaches.}
\label{fig:Figure18}
\end{figure}

Thus, if we now introduce two families of functions, denoted ${\left( \psi_k \right)_1 \left( u, v \right)}$ and ${\left( \psi_k \right)_2 \left( u, v \right)}$, and given by:

\begin{equation}
\left( \psi_k \right)_1 \left( u, v \right) = e^{i k u} - e^{i k v}, \qquad \text{ where } \qquad k = \frac{n \pi}{L}, \qquad n \in \mathbb{N},
\end{equation}
and:

\begin{equation}
\left( \psi_k \right)_2 \left( u, v \right) = e^{i k u} + e^{i k v} - 2 \cos \left( k L \right), \qquad \text{ where } \qquad \tan \left( k L \right) = 2 k L, \qquad k \neq 0,
\end{equation}
respectively, then it is straightforward to see that the imaginary variant of the Pauli-Jordan operator ${i \hat{\Delta}}$ acts upon both families in the required way, namely yielding:

\begin{equation}
\left( i \hat{\Delta} \left( \psi_k \right)_1 \right) \left( u, v \right) = \frac{L}{k} \left( \psi_k \right)_1 \left( u, v \right), \qquad \text{ and } \qquad \left( i \hat{\Delta} \left( \psi_k \right)_2 \right) \left( u, v \right) = \frac{L}{k} \left( \psi_k \right)_2 \left( u, v \right),
\end{equation}
respectively, hence allowing us to conclude that ${\left( \psi_k \right)_1 \left( u, v \right)}$ and ${\left( \psi_k \right)_2 \left( u, v \right)}$ are indeed valid families of eigenfunctions for ${i \hat{\Delta} \left( u, v \right)}$ (with non-zero eigenvalues). These eigenfunctions may now be appropriately normalized by noting that their ${L^2 \left( \left[ -L, L \right] \right)}$ norms are given by:

\begin{equation}
\left\lVert \left( \psi_k \right)_1 \right\rVert^2 = 8 L^2, \qquad \text{ and } \qquad \left\lVert \left( \psi_k \right)_2 \right\rVert^2 = 8 L^2 - 16 L^2 \cos^2 \left( k L \right),
\end{equation}
respectively. Furthermore, we can ascertain that ${\left( \psi_k \right)_1 \left( u, v \right)}$ and ${\left( \psi_k \right)_2 \left( u, v \right)}$ constitute the \textit{only} families of eigenfunctions with non-zero eigenvalues by recalling that the eigenvalues ${\lambda_n}$ and eigenfunctions ${\psi_n \in L^2 \left( \left[ -L, L \right] \times \left[ -L, L \right] \right)}$ must satisfy the following summation condition:

\begin{equation}
\sum_{n} \lambda_{n}^{2} = \int_{\mathcal{V}} \int_{\mathcal{V}} \left\lvert i \hat{\Delta} \left( \mathbf{x}, \mathbf{y} \right) \right\rvert^2 d \mathbf{y} d \mathbf{x}.
\end{equation}
The sum of all the non-zero eigenvalues associated with the families ${\left( \psi_k \right)_1 \left( u, v \right)}$ and ${\left( \psi_k \right)_2 \left( u, v \right)}$ is given by:

\begin{equation}
\sum_{n = - \infty : n \neq 0}^{\infty} \left( \frac{L^2}{\pi n} \right)^2 + \sum_{\tan \left( x \right) = 2 x : x \neq 0} \left( \frac{L^2}{x} \right)^2 = 2 L^4 \left( \sum_{n = 1}^{\infty} \frac{1}{\left( \pi n \right)^2} + \sum_{\tan \left( x \right) = 2 x : x > 0} \frac{1}{x^2} \right),
\end{equation}
In the above, the first sum arises from the ${\left( \psi_k \right)_1 \left( u, v \right)}$ eigenfunction family (defined for all ${k = \frac{n \pi}{L}}$, with ${n \in \mathbb{N}}$), and evaluates, via the solution to the Basel problem, to yield

\begin{equation}
\sum_{n = 1}^{\infty} \frac{1}{\left( \pi n \right)^2} = \frac{1}{6},
\end{equation}
whilst the second sum arises from the ${\left( \psi_k \right)_2 \left( u, v \right)}$ eigenfunction family (defined for all $k$ satisfying the transcendental equation ${\tan \left( k L \right) = 2 k L}$, to which which there exists a countably infinite set of real solutions), and evaluates to yield\cite{spiegel}:

\begin{equation}
\sum_{\tan \left( x \right) = 2 x : x > 0} \frac{1}{x^2} = \frac{5}{6}.
\end{equation}
Therefore, we have shown that, indeed:

\begin{multline}
\sum_{n} \lambda_{n}^{2} = \sum_{n = - \infty : n \neq 0}^{\infty} \left( \frac{L^2}{\pi n} \right)^2 + \sum_{\tan \left( x \right) = 2 x : x \neq 0} \left( \frac{L^2}{x} \right)^2 = 2 L^4\\
= \int_{-L}^{L} \int_{-L}^{L} \int_{-L}^{L} \int_{-L}^{L} \left\lvert i \hat{\Delta} \left( u - u^{\prime}, v - v^{\prime} \right) \right\rvert^2 d v^{\prime} d v d u^{\prime} d u = \int_{\mathcal{V}} \int_{\mathcal{V}} \left\lvert i \hat{\Delta} \left( \mathbf{x}, \mathbf{y} \right) \right\rvert^2 d \mathbf{y} d \mathbf{x},
\end{multline}
as required, and consequently all non-zero eigenvalues (and their corresponding eigenfunctions) have been successfully identified. Since the non-zero eigenvalues ${\lambda_k}$ are therefore given by ${\lambda_k = \frac{L}{k}}$ (along with their corresponding partners ${1 - \lambda_k}$), we note, following Sorkin and Yazdi\cite{sorkin7}, that a spectral truncation which preserves only those eigenvalues ${\lambda}$ (and the corresponding ${1 - \lambda}$ partners) whose magnitudes are at least ${\lambda_{min} \sim \frac{\sqrt{N}}{4 \pi}}$ (where $N$ denotes either the cardinality of the interior or the exterior diamond region, depending upon whether it is the eigenvalues of operators defined on the interior or the exterior diamond that are being truncated) constitutes a natural choice of ultraviolet cutoff, since it is consistent with both the choice of sprinkling density of the underlying causal set and the expected dimensions of an entropic area law. Upon imposing this truncation on the spectrum of the discrete Pauli-Jordan operator ${\hat{\Delta}_{\mathcal{C}} \left( x, y \right)}$ (and, by extension, on the Sorkin-Johnston Wightman function ${\hat{W}_{SJ} \left( x, y \right)}$), we obtain the same logarithmic relationship (indicative of a spatial area law) between the cardinality of the interior diamond region $N$ and the estimated entanglement entropy $S$, namely ${S = 0.346 \log \left( \frac{\sqrt{N}}{4 \pi} \right) + 1.883}$, as shown in Figure \ref{fig:Figure18}.

\section{The Wolfram Model/Hypergraph Rewriting Case}
\label{sec:Section4}

\begin{figure}[ht]
\centering
\includegraphics[width=0.345\textwidth]{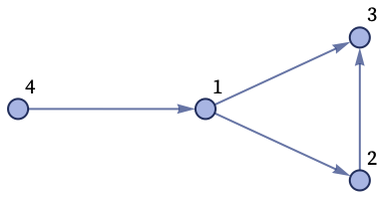}\hspace{0.2\textwidth}
\includegraphics[width=0.345\textwidth]{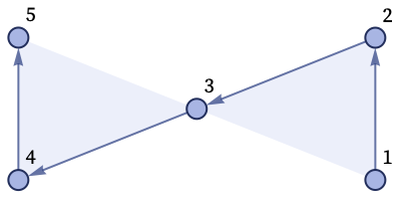}
\caption{Examples of simple directed hypergraphs, corresponding to the finite collections of ordered relations between elements ${\left\lbrace \left\lbrace 1, 2 \right\rbrace, \left\lbrace 1, 3 \right\rbrace, \left\lbrace 2, 3 \right\rbrace, \left\lbrace 4, 1 \right\rbrace \right\rbrace}$ and ${\left\lbrace \left\lbrace 1, 2, 3 \right\rbrace, \left\lbrace 3, 4, 5 \right\rbrace \right\rbrace}$, respectively.}
\label{fig:Figure19}
\end{figure}

\begin{figure}[ht]
\centering
\includegraphics[width=0.495\textwidth]{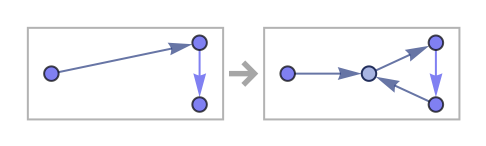}
\caption{An example of a simple hypergraph rewriting rule, corresponding to the set substitution rule ${\left\lbrace \left\lbrace x, y \right\rbrace, \left\lbrace y, z \right\rbrace \right\rbrace \to \left\lbrace \left\lbrace w, y \right\rbrace, \left\lbrace y, z \right\rbrace, \left\lbrace z, w \right\rbrace, \left\lbrace x, w \right\rbrace \right\rbrace}$.}
\label{fig:Figure20}
\end{figure}

\begin{figure}[ht]
\centering
\includegraphics[width=0.695\textwidth]{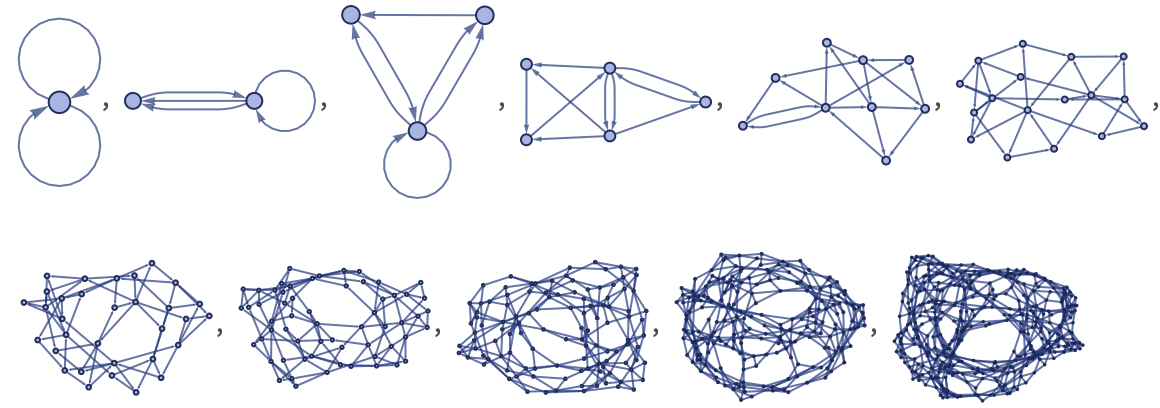}
\caption{An example of a possible 10 step evolution history for the hypergraph rewriting/set substitution rule ${\left\lbrace \left\lbrace x, y \right\rbrace, \left\lbrace y, z \right\rbrace \right\rbrace \to \left\lbrace \left\lbrace w, y \right\rbrace, \left\lbrace y, z \right\rbrace, \left\lbrace z, w \right\rbrace, \left\lbrace x, w \right\rbrace \right\rbrace}$, starting from the double self-loop initial condition ${\left\lbrace \left\lbrace 0, 0 \right\rbrace, \left\lbrace 0, 0 \right\rbrace \right\rbrace}$.}
\label{fig:Figure21}
\end{figure}

Generating \textit{algorithmic} causal sets dynamically via Wolfram model evolution is a relatively straightforward process\cite{gorard3}; one starts with an initial (finite, usually directed) \textit{hypergraph} ${H = \left( V, E \right)}$, i.e. a generalization of an ordinary (directed) graph in which \textit{hyper}edges can connect arbitrary (non-empty) subsets of vertices\cite{wolfram}\cite{wolfram2}\cite{gorard}\cite{gorard2}:

\begin{equation}
E \subseteq \mathcal{P} \left( V \right) \setminus \left\lbrace \varnothing \right\rbrace,
\end{equation}
where ${\mathcal{P}}$ denotes the power set function. Examples of simple (directed) hypergraphs, represented abstractly as finite collections of (ordered) relations between elements, are shown in Figure \ref{fig:Figure19}. The dynamics of the Wolfram model system are then determined by means of an \textit{abstract rewriting rule} $R$ of the form:

\begin{equation}
R : \qquad H_1 = \left( V_1, E_1 \right) \to H_2 = \left( V_2, E_2 \right),
\end{equation}
in which, loosely speaking, a subhypergraph matching the pattern given by ${H_1}$ is replaced with a distinct subhypergraph matching the pattern given by ${H_2}$. A fully rigorous description of the hypergraph rewriting semantics of the Wolfram model can be given in terms of double-pushout rewriting rules over (selective) adhesive categories\cite{gorard5}\cite{gorard6}\cite{gorard7}. An example of a simple (directed) hypergraph rewriting rule, represented abstractly as a set substitution system (i.e. a rewriting rule defined over subsets of the collection of ordered relations between elements), is shown in Figure \ref{fig:Figure20}. The resulting evolution of the Wolfram model system, obtained by iteratively applying this rule to a simple hypergraph initial condition (consisting of a pair of ``self-loops'' of the form ${\left\lbrace \left\lbrace 0, 0 \right\rbrace, \left\lbrace 0, 0 \right\rbrace \right\rbrace}$) is shown in Figure \ref{fig:Figure21} - note that, at each step, the rule is applied to the maximal non-overlapping set of matching subhypergraphs simultaneously.

\begin{figure}[ht]
\centering
\includegraphics[width=0.395\textwidth]{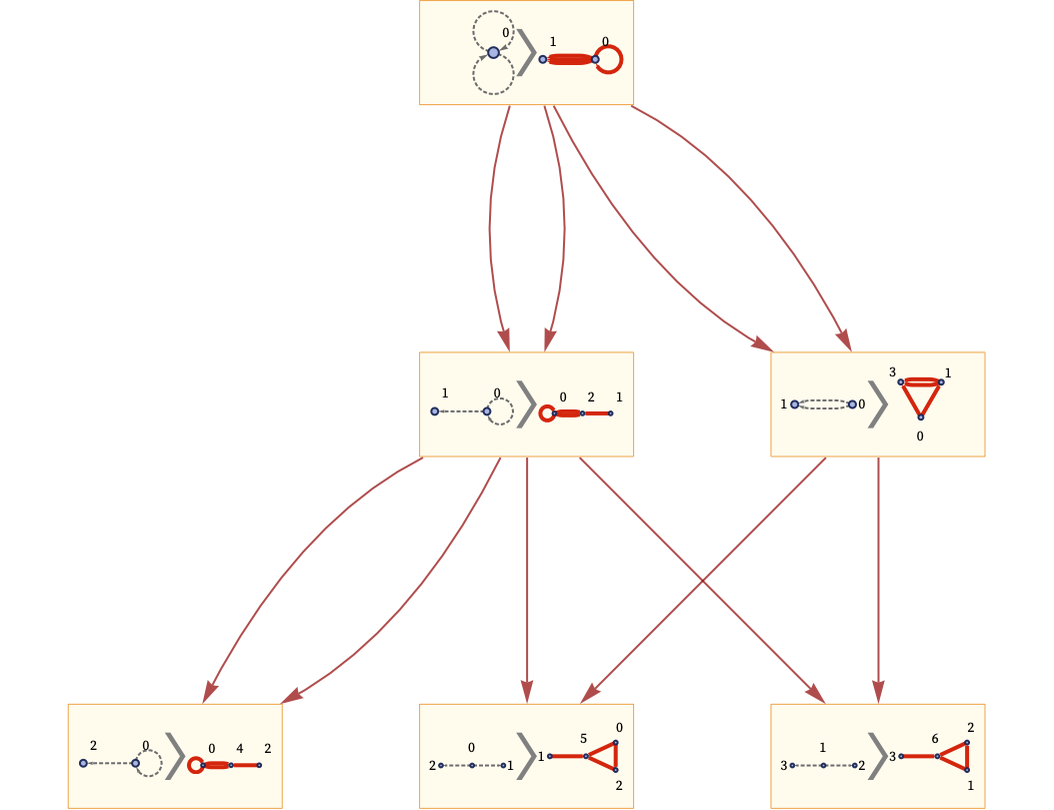}
\caption{The causal graph obtained after 3 steps of the hypergraph rewriting/set substitution rule ${\left\lbrace \left\lbrace x, y \right\rbrace, \left\lbrace y, z \right\rbrace \right\rbrace \to \left\lbrace \left\lbrace x, y \right\rbrace, \left\lbrace x, w \right\rbrace, \left\lbrace y, w \right\rbrace, \left\lbrace z, w \right\rbrace \right\rbrace}$, starting from the double self-loop initial condition ${\left\lbrace \left\lbrace 0, 0 \right\rbrace, \left\lbrace 0, 0 \right\rbrace \right\rbrace}$.}
\label{fig:Figure22}
\end{figure}

\begin{figure}[ht]
\centering
\includegraphics[width=0.695\textwidth]{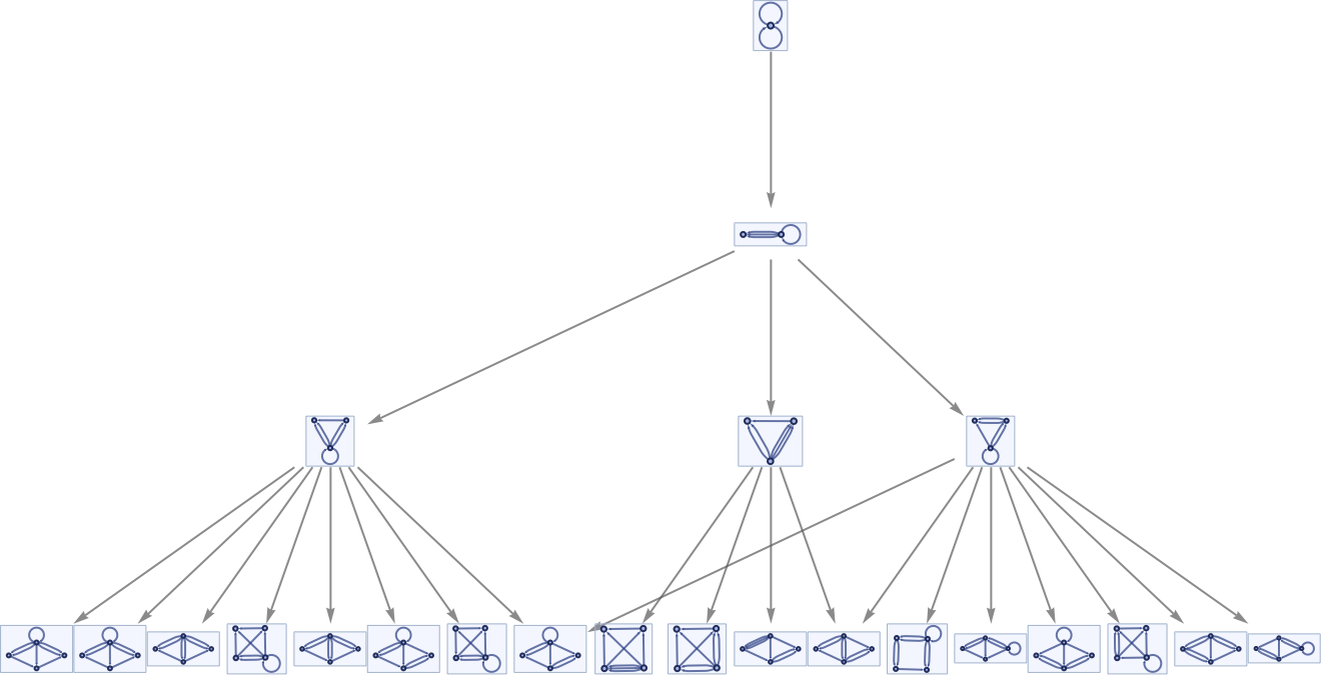}
\caption{A multiway evolution graph representing the non-deterministic evolution of the hypergraph rewriting/set substitution rule ${\left\lbrace \left\lbrace x, y \right\rbrace, \left\lbrace y, z \right\rbrace \right\rbrace \to \left\lbrace \left\lbrace w, y \right\rbrace, \left\lbrace y, z \right\rbrace, \left\lbrace z, w \right\rbrace, \left\lbrace x, w \right\rbrace \right\rbrace}$, starting from the double self-loop initial condition ${\left\lbrace \left\lbrace 0, 0 \right\rbrace, \left\lbrace 0, 0 \right\rbrace \right\rbrace}$.}
\label{fig:Figure23}
\end{figure}

One can now represent the causal interactions between successive applications of the rewriting rule by means of a \textit{causal graph}: a directed, acyclic graph in which each vertex corresponds to an application of the rewriting rule and each edge corresponds to a causal relationship between those rewrites. More concretely, the directed edge ${A \to B}$ exists if and only if:

\begin{equation}
\mathrm{In} \left( B \right) \cap \mathrm{Out} \left( A \right) \neq \varnothing,
\end{equation}
i.e. if and only if the input for rule application $B$ uses hyperedges that were produced by the output of rule application $A$, and therefore if application $B$ could not have occurred unless application $A$ had previously occurred. Once again, a fully rigorous description of the causal semantics of hypergraph rewriting systems may be given in terms of causal 2-categories and multiway evolution causal graphs\cite{gorard7}. Note that the transitive reduction of a Wolfram model causal graph yields (the Hasse diagram for) a corresponding causal set. In general, however, there does not exist a single, canonical rewriting order (and hence a single, unique evolution history) for a given Wolfram model rule, since at any given time step there can exist many possible maximal non-overlapping sets of matching subhypergraphs; physically, this corresponds to the fact that there does not exist a universally-preferred choice of spacetime gauge\cite{gorard}\cite{gorard3}\cite{gorard4}. For this reason, it is helpful to parametrize the set of possible evolution histories by means of a \textit{multiway system}, or, more exactly, a \textit{multiway evolution graph}: a directed, acyclic graph in which each vertex corresponds to a global state of the hypergraph and each edge corresponds to a rewrite. More concretely, the directed edge ${A \to B}$ exists if and only if there exists an application of the rewriting rule that transforms hypergraph $A$ to hypergraph $B$. Examples of causal and multiway evolution graphs, representing the evolution of the Wolfram model system described above, are shown in Figures \ref{fig:Figure22} and \ref{fig:Figure23}, respectively; note that in these figures, as subsequently, pairs of isomorphic hypergraphs appearing in the multiway evolution graph are detected and merged, using a generalized version of the ``uniqueness trees'' algorithm\cite{gorard9} for graph isomorphism and canonicalization. The fully general-relativistic formulation of the Wolfram model depends upon the underlying hypergraph rewriting rule being \textit{causal invariant} (the discrete analog of general covariance for hypergraph rewriting systems), meaning that the causal graph is always isomorphic, irrespective of which path through the multiway system, and therefore which updating order/evolution history, is chosen\cite{gorard}\cite{gorard3}\cite{gorard4}.

\begin{figure}[ht]
\centering
\includegraphics[width=0.745\textwidth]{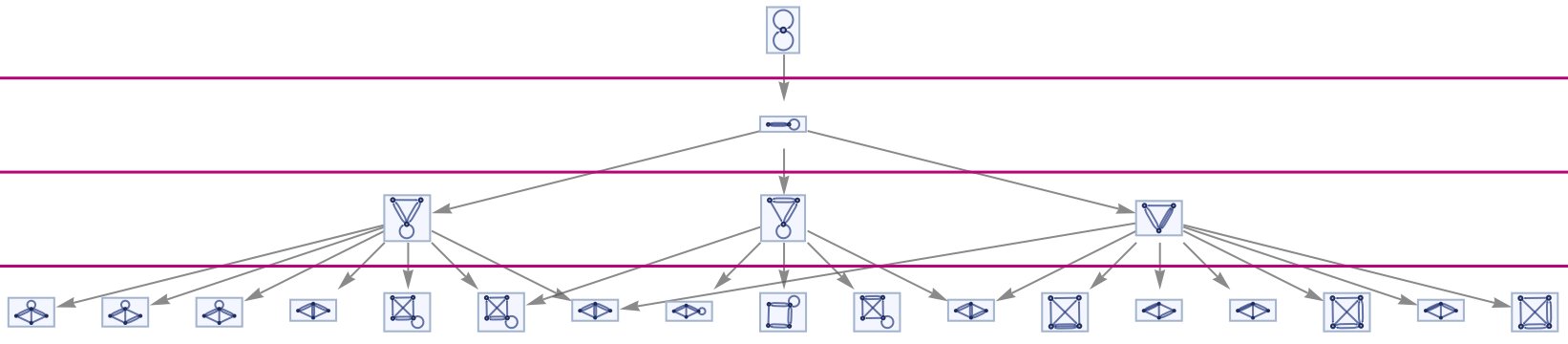}
\caption{A typical ``foliation'' of the multiway evolution graph representing the non-deterministic evolution of the hypergraph rewriting/set substitution rule ${\left\lbrace \left\lbrace x, y \right\rbrace, \left\lbrace y, z \right\rbrace \right\rbrace \to \left\lbrace \left\lbrace w, y \right\rbrace, \left\lbrace y, z \right\rbrace, \left\lbrace z, w \right\rbrace, \left\lbrace x, w \right\rbrace \right\rbrace}$, starting from the double self-loop initial condition ${\left\lbrace \left\lbrace 0, 0 \right\rbrace, \left\lbrace 0, 0 \right\rbrace \right\rbrace}$.}
\label{fig:Figure24}
\end{figure}

\begin{figure}[ht]
\centering
\includegraphics[width=0.345\textwidth]{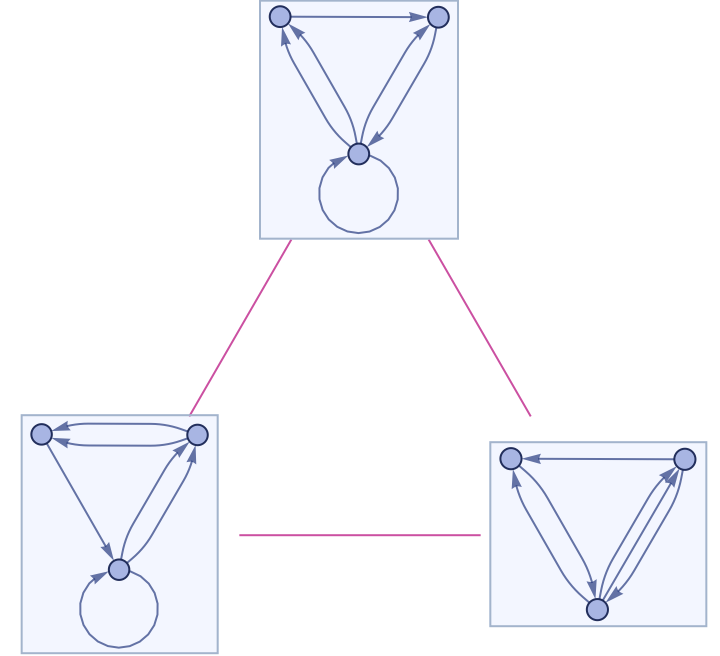}\hspace{0.1\textwidth}
\includegraphics[width=0.395\textwidth]{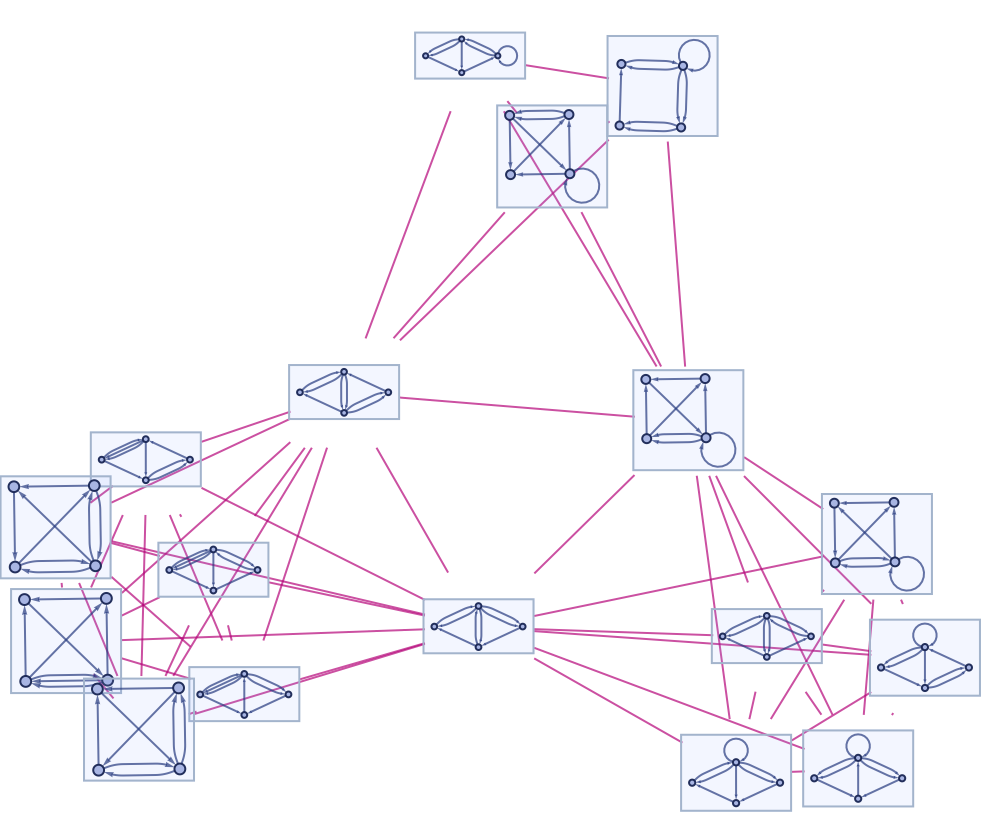}
\caption{A time-ordered sequence of branchial graphs corresponding to a typical ``foliation'' of the multiway evolution graph representing the non-deterministic evolution of the hypergraph rewriting/set substitution rule ${\left\lbrace \left\lbrace x, y \right\rbrace, \left\lbrace y, z \right\rbrace \right\rbrace \to \left\lbrace \left\lbrace w, y \right\rbrace, \left\lbrace y, z \right\rbrace, \left\lbrace z, w \right\rbrace, \left\lbrace x, w \right\rbrace \right\rbrace}$, starting from the double self-loop initial condition ${\left\lbrace \left\lbrace 0, 0 \right\rbrace, \left\lbrace 0, 0 \right\rbrace \right\rbrace}$.}
\label{fig:Figure25}
\end{figure}

\begin{figure}[ht]
\centering
\includegraphics[width=0.495\textwidth]{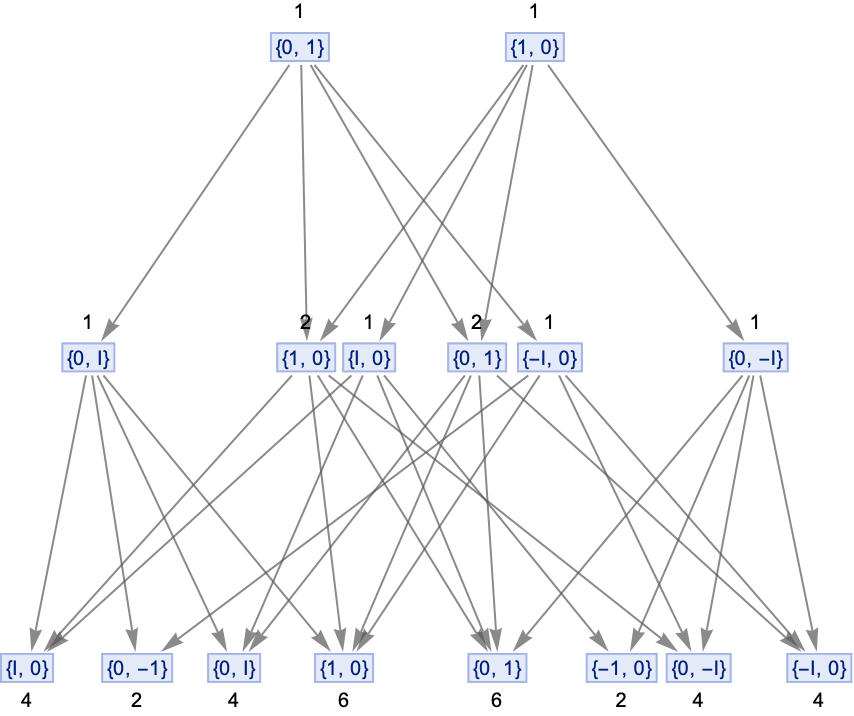}
\caption{A multiway evolution graph representing the quantum evolution of the root-NOT quantum gate (${\sqrt{NOT}}$), applied to an initial superposition ${\frac{1}{\sqrt{2}} \left( \ket{0} + \ket{1} \right)}$ of the pure eigenstates ${\ket{0}}$ and ${\ket{1}}$, with vertex weights corresponding to the number of distinct evolution paths leading to a given state (i.e. magnitudes of the corresponding amplitudes).}
\label{fig:Figure26}
\end{figure}

\begin{figure}[ht]
\centering
\includegraphics[width=0.395\textwidth]{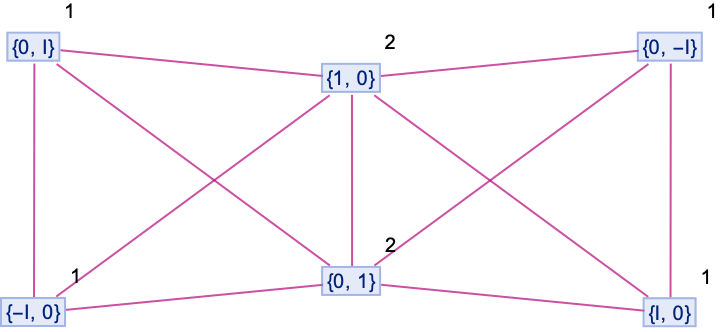}\hspace{0.1\textwidth}
\includegraphics[width=0.395\textwidth]{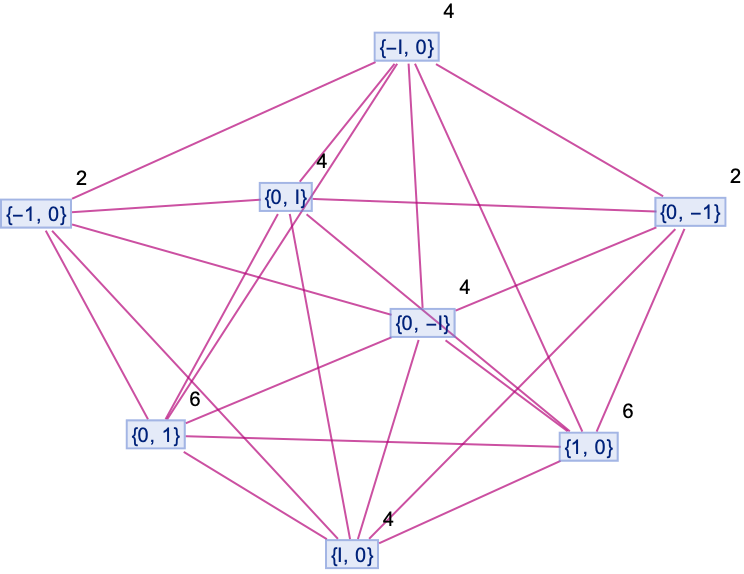}
\caption{A time-ordered sequence of branchial graphs corresponding to a typical ``foliation'' of the multiway evolution graph representing the quantum evolution of the root-NOT quantum gate (${\sqrt{NOT}}$), applied to an initial superposition ${\frac{1}{\sqrt{2}} \left( \ket{0} + \ket{1} \right)}$ of the pure eigenstates ${\ket{0}}$ and ${\ket{1}}$, with vertex weights corresponding to the number of distinct evolution paths leading to a given state (i.e. magnitudes of the corresponding amplitudes).}
\label{fig:Figure27}
\end{figure}

The resulting multiway system possesses the algebraic structure of a dagger-symmetric, compact-closed monoidal category\cite{gorard5}\cite{gorard6}, indicating that it is equipped with a symmetric tensor product operation ${\otimes}$ (generalizing the usual tensor product of finite-dimensional Hilbert spaces), an involutive ``dagger'' operation ${\dag}$ (generalizing the Hermitian adjoint/conjugate transpose operation on linear maps) and a compact/dual structure ${*}$ (generalizing the concept of a dual space in the category of finite-dimensional vector spaces), all obeying the standard associativity, unitality, coherence, etc. axioms that one would expect from such operations. Loosely speaking, the tensor product structure ${\otimes}$ represents the parallel composition of hypergraph states/rewrite applications on neighboring branches of the multiway system, the dagger structure ${\dag}$ represents the inversion of multiway evolution edges and the compact structure ${*}$ represents the swapping of the vertex and hyperedge sets in a given hypergraph (so as to obtain a formal notion of a hypergraph dual). These basic category-theoretic primitives then allow one to define all of the standard algebraic operations of (finite-dimensional) quantum mechanics, following in the spirit of Abramsky and Coecke's formulation of \textit{categorical quantum mechanics}\cite{abramsky}\cite{abramsky2}. The tensor product structure of a given multiway evolution may be neatly visualized using the formalism of \textit{branchial graphs}, in which a multiway evolution graph ${G_{multiway} = \left( V_{multiway}, E_{multiway} \right)}$ is ``foliated'' into a time-ordered sequence of discrete ``branchlike hypersurfaces'' (akin to spacelike hypersurfaces/Cauchy surfaces in foliations of spacetimes) ${\Sigma_t}$, where $t$ is a universal time function ${t : V_{multiway} \to \mathbb{Z}}$ assigning hypergraph states in the multiway system to corresponding discrete time coordinates, such that the branchlike hypersurfaces are exactly the level surfaces of this function, satisfying:

\begin{equation}
\forall t_0 \in \mathbb{Z}, \qquad \Sigma_{t_0} = \left\lbrace p \in V_{multiway} : t \left( p \right) = t_0 \right\rbrace,
\end{equation}
and:

\begin{equation}
\forall t_1, t_2 \in \mathbb{Z}, \qquad \Sigma_{t_1} \cap \Sigma_{t_2} = \varnothing \qquad \Leftrightarrow \qquad t_1 \neq t_2,
\end{equation}
i.e. the hypersurfaces should not intersect. The resulting \textit{branchial graphs} represent these abstract branchlike hypersurfaces combinatorially, by effectively showing the ancestry distance between hypergraph states for a given value of the universal time function $t$; specifically, the undirected edge ${A \leftrightarrow B}$ exists in the branchial graph if and only if the corresponding hypergraph states for vertices $A$ and $B$ share some common ancestor $C$ in the multiway evolution graph. An example of such a multiway ``foliation'' is shown in Figure \ref{fig:Figure24}, for the case of the non-deterministic Wolfram model evolution described above, with the corresponding sequence of branchial graphs/branchlike hypersurfaces shown in Figure \ref{fig:Figure25}. Interpreting each vertex in a multiway evolution graph/branchial graph as a pure quantum eigenstate, each vertex may be assigned a numerical weight based on the number of distinct evolution paths through the multiway evolution graph leading to that state, interpreted as a (magnitude of a) quantum amplitude, in such a way that each branchial graph represents an instantaneous superposition of all of the eigenstates within its vertex set\cite{gorard2}. If the weights of the vertices in the branchial graph represent the amplitudes of eigenstates the superposition, then the combinatorial structure of the branchial graph as a whole represents the tensor product structure of the superposition: more specifically, each undirected branchial edge represents a pair of pure eigenstates that have been ``tensored'' together, and which therefore no longer constitute a separable multipartite quantum state\cite{gorard5}\cite{gorard6}. To make this intuition more explicit, we can consider constructing a multiway evolution graph by iteratively applying a root-NOT quantum gate:

\begin{equation}
\sqrt{NOT} = \frac{1}{2} \begin{bmatrix}
1 + i & 1 - i\\
1 - i & 1 + i
\end{bmatrix},
\end{equation}
to an initial superposition ${\frac{1}{\sqrt{2}} \left( \ket{0} + \ket{1} \right)}$ of pure eigenstates ${\ket{0}}$ and ${\ket{1}}$. In Figure \ref{fig:Figure26}, we see the canonical ``foliation'' of the resulting multiway evolution graph, and in Figure \ref{fig:Figure27} we see the evolution of the amplitudes of the eigenstates in the resulting superpositions, as well as the evolution of their tensor product structure, by means of a time-ordered sequence of branchial graphs.

Due to this correspondence between the combinatorial structure of branchial graphs and the tensor product structure of quantum states, the induced metric/line element on a branchial graph converges, in the limit as the cardinality of its vertex set goes to infinity, to\cite{gorard2}\cite{gorard5}:

\begin{equation}
ds^2 = \frac{\braket{d \psi}{d \psi}}{\braket{\psi}{\psi}} - \frac{\braket{d \psi}{\psi} \braket{\psi}{d \psi}}{\left( \braket{\psi}{\psi} \right)^2},
\end{equation}
where ${\ket{\psi}}$ represents a pure state:

\begin{equation}
\ket{\psi} = \sum_{k = 0}^{n} Z_k \ket{e_k} = \left[ Z_0 : Z_1 : \cdots : Z_n \right],
\end{equation}
${\ket{d \psi}}$ represents its infinitesimal variation, and where ${\left\lbrace \ket{e_k} \right\rbrace}$ is an orthonormal basis set for some $n$-dimensional Hilbert space ${\mathcal{H}}$. We can see that this is none other than a special case of the (line element of the) \textit{Fubini-Study metric} on the complex projective Hilbert space ${\mathbb{CP}^{n}}$:

\begin{equation}
ds^2 = \frac{\left\lvert \mathbf{Z} \right\rvert^2 \left\lvert d \mathbf{Z} \right\rvert^2 - \left( \mathbf{\bar{Z}} \cdot d \mathbf{Z} \right) \left( \mathbf{Z} \cdot d \mathbf{\bar{Z}} \right)}{\left\lvert \mathbf{Z} \right\rvert^4} = \frac{Z_{\alpha} \bar{Z}^{\alpha} d Z_{\beta} z \bar{Z}^{\beta} - \bar{Z}^{\alpha} Z_{\beta} d Z_{\alpha} \bar{Z}^{\beta}}{\left( Z_{\alpha} \bar{Z}^{\alpha} \right)^2},
\end{equation}
with ${\mathbf{Z} = \left[ Z_0, \dots, Z_n \right]}$ playing the role of the homogeneous coordinates for a projective variety. In order to make manifest the connection between this limiting branchial metric and standard measures of quantum entanglement, we follow the general construction of Cocchiarella et al.\cite{cocchiarella} of an entanglement distance for generic, finite-dimensional, hybrid quantum systems. Note that, in what follows, our use of upper and lower tensor indices does not correspond to a distinction between contravariant and covariant transformations of components, but rather to a distinction between the indexing of families of matrices and vectors (upper indices) and the indexing of the \textit{components} of elements of such families (lower indices); for instance, in our notational convention, ${A^{\mu}}$ designates the ${\mu}$-th matrix in a family of matrices ${\left\lbrace A^{\mu} \right\rbrace_{\mu}}$, ${A_{i j}^{\mu}}$ designates the ${\left( i, j \right)}$-th entry in that matrix, and ${\mathbf{A}_{i}^{\mu}}$ designates the row vector found at position $i$ in that matrix, etc. Assuming that the overall Hilbert space ${\mathcal{H}}$ can be decomposed into a tensor product of $M$ finite-dimensional Hilbert spaces ${\mathcal{H}_{d_{\mu}}}$, each of dimension ${d_{\mu}}$ (for ${\mu = 0, 1, \dots M - 1}$):

\begin{equation}
\mathcal{H} = \mathcal{H}_{d_0} \otimes \mathcal{H}_{d_1} \otimes \cdots \otimes \mathcal{H}_{d_{M - 1}},
\end{equation}
we can measure the entanglement ${E \left( \ket{s} \right)}$ of a generic state ${\ket{s}}$ in ${\mathcal{H}}$ via:

\begin{equation}
E \left( \ket{s} \right) = \sum_{\mu = 0}^{M - 1} \left[ \mathrm{tr} \left( A^{\mu} \right) - 2 \left( d_{\mu} - 1 \right) \right],
\end{equation}
where the ${A^{\mu}}$ form a set of $M$ matrices, with explicit components given by:

\begin{equation}
A_{i j}^{\mu} = \bra{s} T^{\mu i} T^{\mu j} \ket{s} - \bra{s} T^{\mu i} \ket{s} \bra{s} T^{\mu j} \ket{s},
\end{equation}
and where the ${T^{\mu \ell}}$ are \textit{generalized Gell-Mann matrices}: a set of ${d_{\mu}^2 - 1}$ matrices, each of dimension ${d_{\mu} \times d_{\mu}}$, forming a fundamental representation for the generators of the Lie algebra ${\mathfrak{su} \left( d_{\mu} \right)}$ associated to the special unitary group ${SU \left( d_{\mu} \right)}$ for ${d_{\mu} \geq 2}$. Concretely, if we define ${E^{j k}}$ (where ${j, k = 1, \dots, d_{\mu}}$) to be the matrix with a 1 as the ${\left( j, k \right)}$-th entry, and a 0 everywhere else, then we can write the generalized Gell-Mann matrices directly as a collection of \textit{symmetric} matrices of the form:

\begin{equation}
T^{\mu \ell} = \left( E^{j k} + E^{k j} \right),
\end{equation}
where:

\begin{equation}
\ell = 2 \left( k - j \right) + \left( j - 1 \right) \left( 2 d_{\mu} - j \right) - 1, \qquad j = 1, \dots, d_{\mu} - 1, \qquad k = j + 1, \dots, d_{\mu},
\end{equation}
a collection of \textit{antisymmetric} matrices of the form:

\begin{equation}
T^{\mu \ell} = - i \left( E^{j k} - E^{k j} \right),
\end{equation}
where:

\begin{equation}
\ell = 2 \left( k - j \right) + \left( j - 1 \right) \left( 2 d_{\mu} - j \right), \qquad j = 1, \dots, d_{\mu} - 1, \qquad k = j + 1, \dots, d_{\mu},
\end{equation}
and a collection of \textit{diagonal} matrices of the form:

\begin{equation}
T^{\mu \ell} = \sqrt{\frac{2}{k \left( k + 1 \right)}} \left[ \left( \sum_{j = 1}^{k} E^{j j} \right) - k E^{k + 1, k + 1} \right],
\end{equation}
where:

\begin{equation}
\ell = d_{\mu} \left( d_{\mu} - 1 \right) + k, \qquad k = 1, \dots, d_{\mu} - 1.
\end{equation}
This generalized definition reproduces the standard Pauli and Gell-Mann matrices in the ${d_{\mu} = 2}$ and ${d_{\mu} = 3}$ cases, respectively.

To prove that ${E \left( \ket{s} \right)}$ does indeed satisfy the requisite axioms of an entanglement monotone\cite{cocchiarella}, it suffices to show that the value of ${E \left( \ket{s} \right)}$ is bounded both above and below, with its minimum value (${E \left( \ket{s} \right) = 0}$) being obtained whenever ${\ket{s}}$ is a fully-separable state, and its maximum value being obtained whenever ${\ket{s}}$ is a maximally-entangled state, and, moreover, to show that the value of ${E \left( \ket{s} \right)}$ is invariant under the application of local unitary operators\cite{vedral}. Firstly, from the following general identity for the generalized Gell-Mann matrices ${T^{\mu k}}$:

\begin{equation}
\sum_{k = 1}^{d_{\mu}^{2} - 1} T^{\mu k} T^{\mu k} = \frac{2 \left( d_{\mu}^{2} - 1 \right)}{d_{\mu}} \mathbb{I},
\end{equation}
where ${\mathbb{I}}$ denotes the identity tensor, we can express the trace of the matrix ${A^{\mu}}$ directly as:

\begin{equation}
\mathrm{tr} \left( A^{\mu} \right) = \frac{2 \left( d_{\mu}^{2} - 1 \right)}{d_{\mu}} - \sum_{k = 1}^{d_{\mu}^{2} - 1} \left( \bra{s} T^{\mu k} \ket{s} \right)^2,
\end{equation}
thus allowing us to express the overall entanglement measure ${E \left( \ket{s} \right)}$ as:

\begin{equation}
E \left( \ket{s} \right) = \sum_{\mu = 0}^{M - 1} \left[ \frac{2 \left( 2_{\mu} - 1 \right)}{d_{\mu}} - \sum_{k = 1}^{d_{\mu}^{2} - 1} \left( \bra{s} T^{\mu k} \ket{s} \right)^2 \right].
\end{equation}
Moreover, we have that, for a normalized state ${\ket{s_{\mu}} \in \mathcal{H}_{d_{\mu}}}$, the following identity on the generalized Gell-Mann matrices ${T^{\mu k}}$ holds:

\begin{equation}
\sum_{k = 1}^{d_{\mu}^{2} - 1} \left( \bra{s_{\mu}} T^{\mu k} \ket{s_{\mu}} \right)^2 = \frac{2 \left( d_{\mu} - 1 \right)}{d_{\mu}},
\end{equation}
i.e. the largest absolute eigenvalue across the entire set of matrices ${\left\lbrace T^{\mu k} \right\rbrace_k}$ is equal to ${\sqrt{\frac{2 \left( d_{\mu} - 1 \right)}{d_{\mu}}}}$, and so one obtains the following bound on the trace of ${A^{\mu}}$:

\begin{equation}
\mathrm{tr} \left( A^{\mu} \right) \geq \frac{2 \left( d_{\mu}^{2} - 1 \right)}{d_{\mu}} - \frac{2 \left( d_{\mu} - 1 \right)}{d_{\mu}} = 2 \left( d_{\mu} - 1 \right),
\end{equation}
and therefore:

\begin{equation}
\mathrm{tr} \left( A^{\mu} \right) - 2 \left( d_{\mu} - 1 \right) \geq 0, \qquad \implies \qquad E \left( \ket{s} \right) \geq 0,
\end{equation}
since ${E \left( \ket{s} \right)}$ reduces to a sum of non-negative real values, as required. This maximum absolute eigenvalue is only obtained when ${\ket{s}}$ is a normalized tensor product of basis vectors, which is only possible when ${\ket{s}}$ is a fully-separable state of the form:

\begin{equation}
\ket{s} = \ket{s_0} \otimes \cdots \otimes \ket{s_{M - 1}},
\end{equation}
in which case ${\mathrm{tr} \left( A^{\mu} \right) = 2 \left( d_{\mu} - 1 \right)}$, and therefore ${E \left( \ket{s} \right) = 0}$. Secondly, since the second term ${\sum_{k = 1}^{d_{\mu}^{2} - 1} \left( \bra{s} T^{\mu k} \ket{s} \right)^2}$ in the definition of the entanglement measure ${E \left( \ket{s} \right)}$ is a sum of non-negative real values, it follows that ${E \left( \ket{s} \right)}$ is bounded above by:

\begin{equation}
E \left( \ket{s} \right) \leq \sum_{\mu = 0}^{M - 1} \frac{2 \left( d_{\mu} - 1 \right)}{d_{\mu}},
\end{equation}
with this bound only being obtained for a maximally-entangled state ${\ket{s}}$, since only then does the following identity for the generalized Gell-Mann matrices ${T^{\mu k}}$ hold:

\begin{equation}
\forall \mu \in \left\lbrace 0, \dots M - 1 \right\rbrace \text{ and } k \in \left\lbrace 1, \dots, d_{\mu}^{2} - 1 \right\rbrace, \qquad \bra{s} T^{\mu k} \ket{s} = 0,
\end{equation}
and therefore:

\begin{equation}
E \left( \ket{s} \right) = \sum_{\mu = 0}^{M - 1} \frac{2 \left( d_{\mu} - 1 \right)}{d_{\mu}},
\end{equation}
as required.

Finally, if ${\ket{s} \in \mathcal{H}}$ is a normalized state and ${U^{\mu}}$ denotes a local unitary operator acting on the ${\mu}$-th subsystem in the tensor product, of the general form:

\begin{equation}
U^{\mu} : \mathcal{H}_{d_{\mu}} \to \mathcal{H}_{d_{\mu}},
\end{equation}
i.e. ${U^{\mu}}$ is an arbitrary element of ${SU \left( d_{\mu} \right)}$ (assuming unit determinant), then we have the following invariance property of the generalized Gell-Mann matrices ${T^{\mu k}}$ under the action of ${U^{\mu}}$:

\begin{equation}
\sum_{k = 1}^{d_{\mu}^{2} - 1} \left( \bra{s} \left( U^{\mu} \right)^{\dag} T^{\mu k} U^{\mu} \ket{s} \right)^2 = \sum_{k = 1}^{d_{\mu}^{2} - 1} \sum_{\alpha = 1}^{d_{\mu}^{2} -1} \left( n_{\alpha}^{k} \right)^2 \left( \bra{s} T^{\mu \alpha} \ket{s} \right)^2 = \sum_{\alpha = 1}^{d_{\mu}^{2} - 1} \left( \bra{s} T^{\mu \alpha} \ket{s} \right)^2 \sum_{k = 1}^{d_{\mu}^2 - 1} \left( n_{\alpha}^{k} \right)^2,
\end{equation}
where the ${n_{\alpha}^{k}}$ designate components of a unit vector (see below), and therefore:

\begin{equation}
\sum_{\alpha = 1}^{d_{\mu}^{2} - 1} \left( \bra{s} T^{\mu \alpha} \ket{s} \right)^2 \sum_{k = 1}^{d_{\mu}^{2} - 1} \left( n_{\alpha}^{k} \right)^2 = \sum_{\alpha = 1}^{d_{\mu}^{2} - 1} \left( \bra{s} T^{\mu \alpha} \ket{s} \right)^2.
\end{equation}
Consequently, our entanglement measure ${E \left( \ket{s} \right)}$ is invariant under such local unitary transformations, and so, given a family of such operators ${U^{\mu}}$ for ${\mu = 0, \dots, M - 1}$ acting on ${\ket{s}}$, we obtain the following equivalence class of states sharing the same degree of entanglement:

\begin{equation}
\ket{U, s} = \prod_{\mu = 0}^{M - 1} U^{\mu} \ket{s},
\end{equation}
and, by extension, the following equivalence class of infinitesimal variations of states:

\begin{equation}
\ket{ d U, s} = \sum_{\mu = 0}^{M - 1} d \tilde{U}^{\mu} \ket{U, s}.
\end{equation}
This equivalence class allows us to deduce the relationship between the entanglement measure ${E \left( \ket{s} \right)}$ and the original limiting Fubini-Study metric on our branchial graphs:

\begin{equation}
ds^2 = \braket{d \psi}{d \psi} - \frac{1}{4} \left( \left\lvert \braket{\psi}{d \psi} - \braket{d \psi}{\psi} \right\rvert \right)^2,
\end{equation}
since the infinitesimal ${SU \left( d_{\mu} \right)}$ transformation on the ${\mu}$-th subsystem ${d \tilde{U}^{\mu}}$ may be written in the form:

\begin{equation}
d \tilde{U}^{\mu} = - i \left( \mathbf{n}^{\mu} \cdot \mathbf{T}^{\mu} \right) d \xi_{\mu},
\end{equation}
where ${\mathbf{n}^{\mu}}$ is a unit vector in ${\mathbb{R}^{d_{\mu}}}$ and ${\mathbf{T}^{\mu}}$ designates the vector formed from the generators ${T^{\mu \alpha}}$, ${\alpha = 1, \dots, d_{\mu}^{2} - 1}$ of the ${\mathfrak{su} \left( d_{\mu} \right)}$ Lie algebra. The components ${g_{\mu \nu} \left( \mathbf{v} \right)}$ of the Fubini-Study metric tensor ${g \left( \mathbf{v} \right)}$ may therefore be written as:

\begin{equation}
\sum_{\mu, \nu} g_{\mu \nu} \left( \mathbf{v} \right) d \xi_{\mu} d \xi_{\mu} = \sum_{\mu, \nu} \left( \bra{s} \left( \mathbf{v}^{\mu} \cdot \mathbf{T}^{\mu} \right) \left( \mathbf{v}^{\nu} \cdot \mathbf{T}^{\nu} \right) \ket{s} - \bra{s} \left( \mathbf{v}^{\mu} \cdot \mathbf{T}^{\mu} \right) \ket{s} \bra{s} \left( \mathbf{v}^{\mu} \cdot \mathbf{T}^{\nu} \right) \ket{s} \right) d \xi_{\mu} d \xi_{\nu},
\end{equation}
where the unit vectors ${\mathbf{v}^{\mu}}$ in ${\mathbb{R}^{d_{\mu}}}$ are simply rotations of ${\mathbf{n}^{\mu}}$:

\begin{equation}
\mathbf{v}^{\nu} \cdot \mathbf{T}^{\nu} = \left( U^{\nu} \right)^{\dag} \mathbf{n}^{\nu} \cdot \mathbf{T}^{\nu} U^{\nu}.
\end{equation}
This yields the following explicit relationship between the components ${g_{\mu \mu} \left( \mathbf{v}^{\mu} \right)}$ of the Fubini-Study metric tensor for a generic state ${\ket{s}}$ and the components ${A_{i j}^{\mu}}$ of the matrices ${A^{\mu}}$ derived from the generalized Gell-Mann matrices ${T^{\mu k}}$:

\begin{equation}
g_{\mu \mu} \left( \mathbf{v}^{\mu} \right) = \sum_{i, j} v_{i}^{\mu} v_{j}^{\mu} A_{i j}^{\mu},
\end{equation}
hence justifying the existence of a monotonic relationship between the trace of the matrices ${A^{\mu}}$ (and therefore the entanglement monotone ${E \left( \ket{s} \right)}$) and the limiting Fubini-Study metric ${g \left( \mathbf{v} \right)}$ on branchial graphs.

\begin{figure}[ht]
\centering
\includegraphics[width=0.295\textwidth]{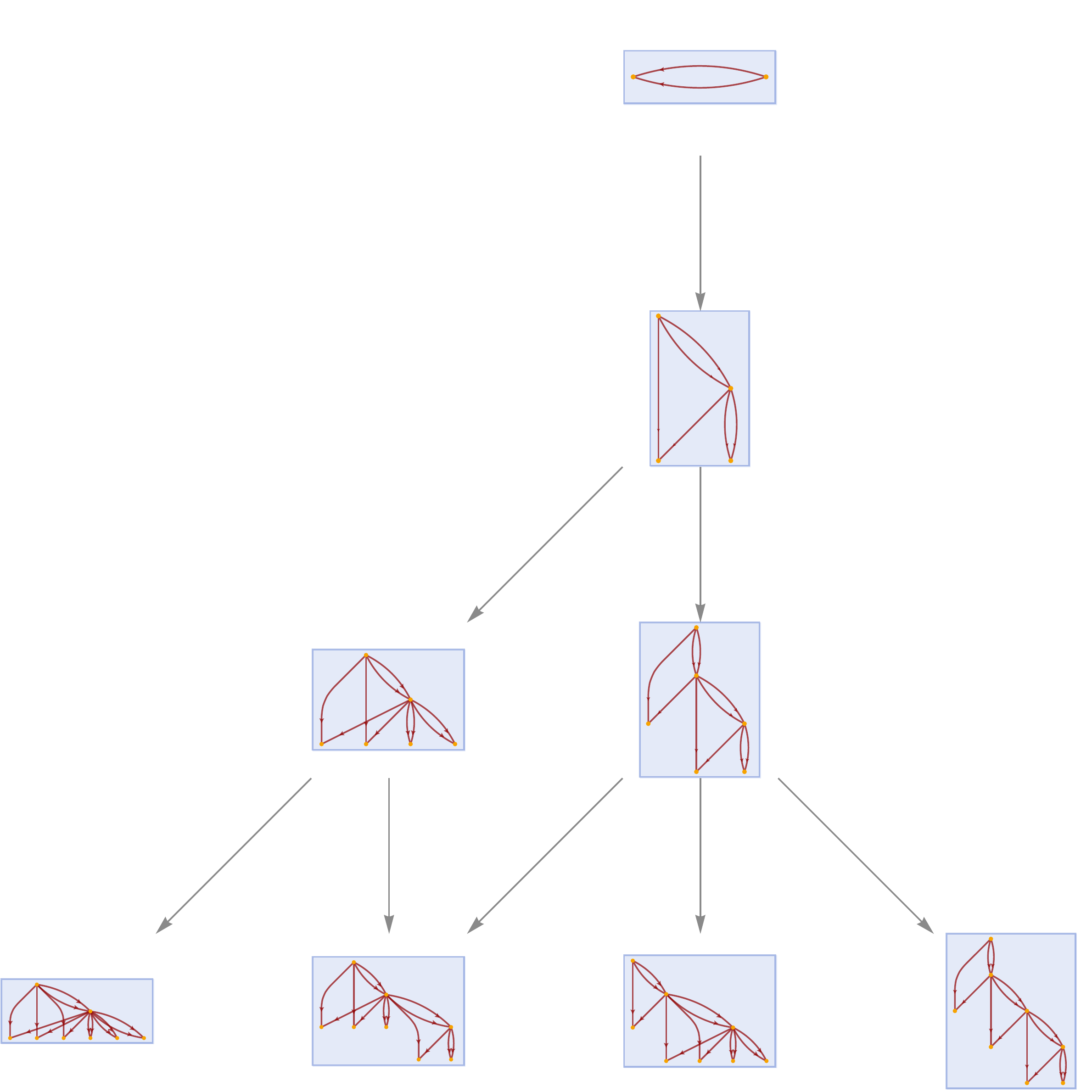}\hspace{0.1\textwidth}
\includegraphics[width=0.445\textwidth]{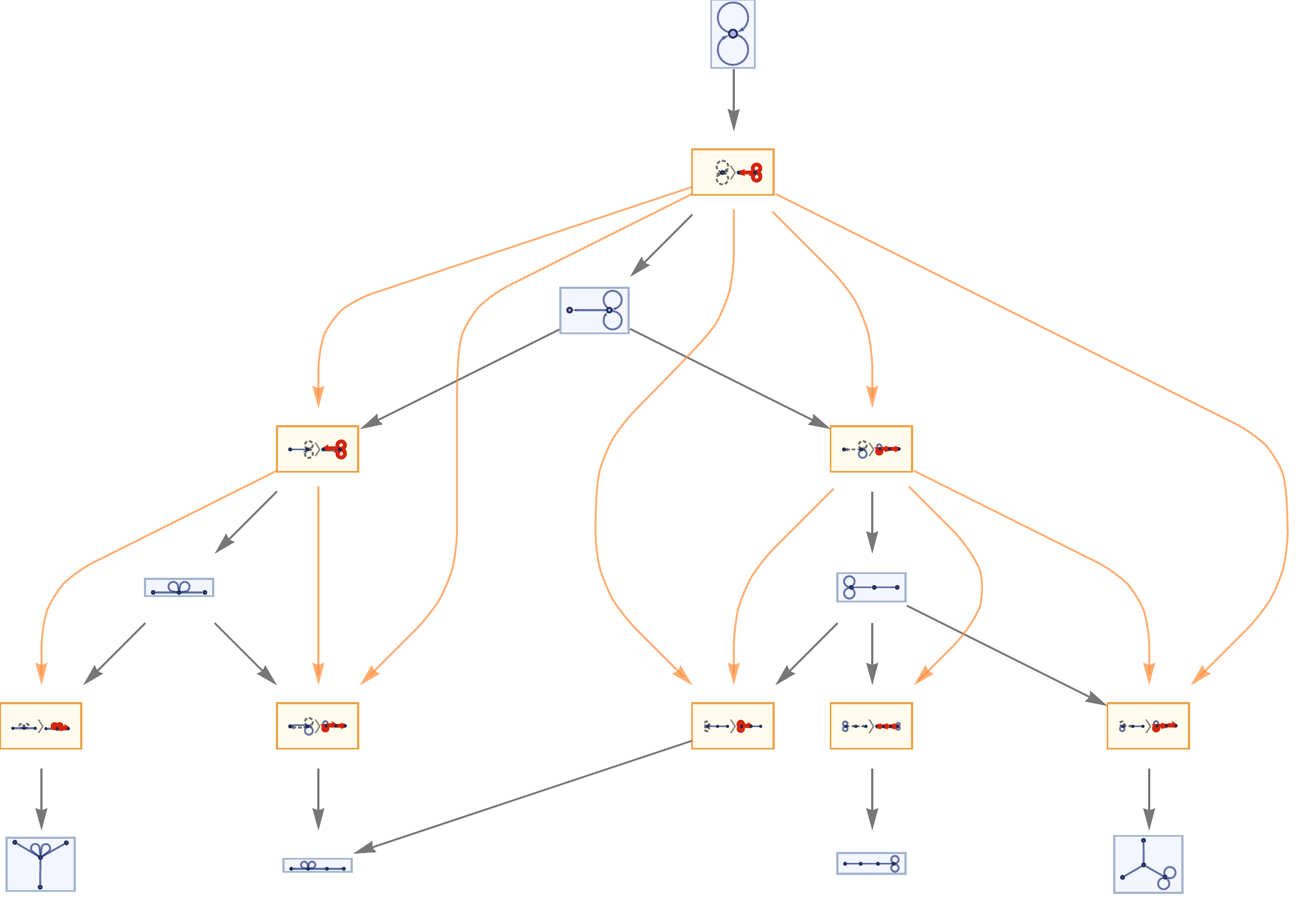}
\caption{On the left, a causal multiway evolution graph representing the non-deterministic evolution of the hypergraph rewriting/set substitution system ${\left\lbrace \left\lbrace x, y \right\rbrace, \left\lbrace y, z \right\rbrace \right\rbrace \to \left\lbrace \left\lbrace x, y \right\rbrace, \left\lbrace y, z \right\rbrace, \left\lbrace z, w \right\rbrace \right\rbrace}$, starting from the double self-loop initial condition ${\left\lbrace \left\lbrace 0, 0 \right\rbrace, \left\lbrace 0, 0 \right\rbrace \right\rbrace}$. On the right, the corresponding multiway evolution causal graph (with evolution edges shown in gray and causal edges shown in orange) for the same system.}
\label{fig:Figure28}
\end{figure}

\begin{figure}[ht]
\centering
\includegraphics[width=0.395\textwidth]{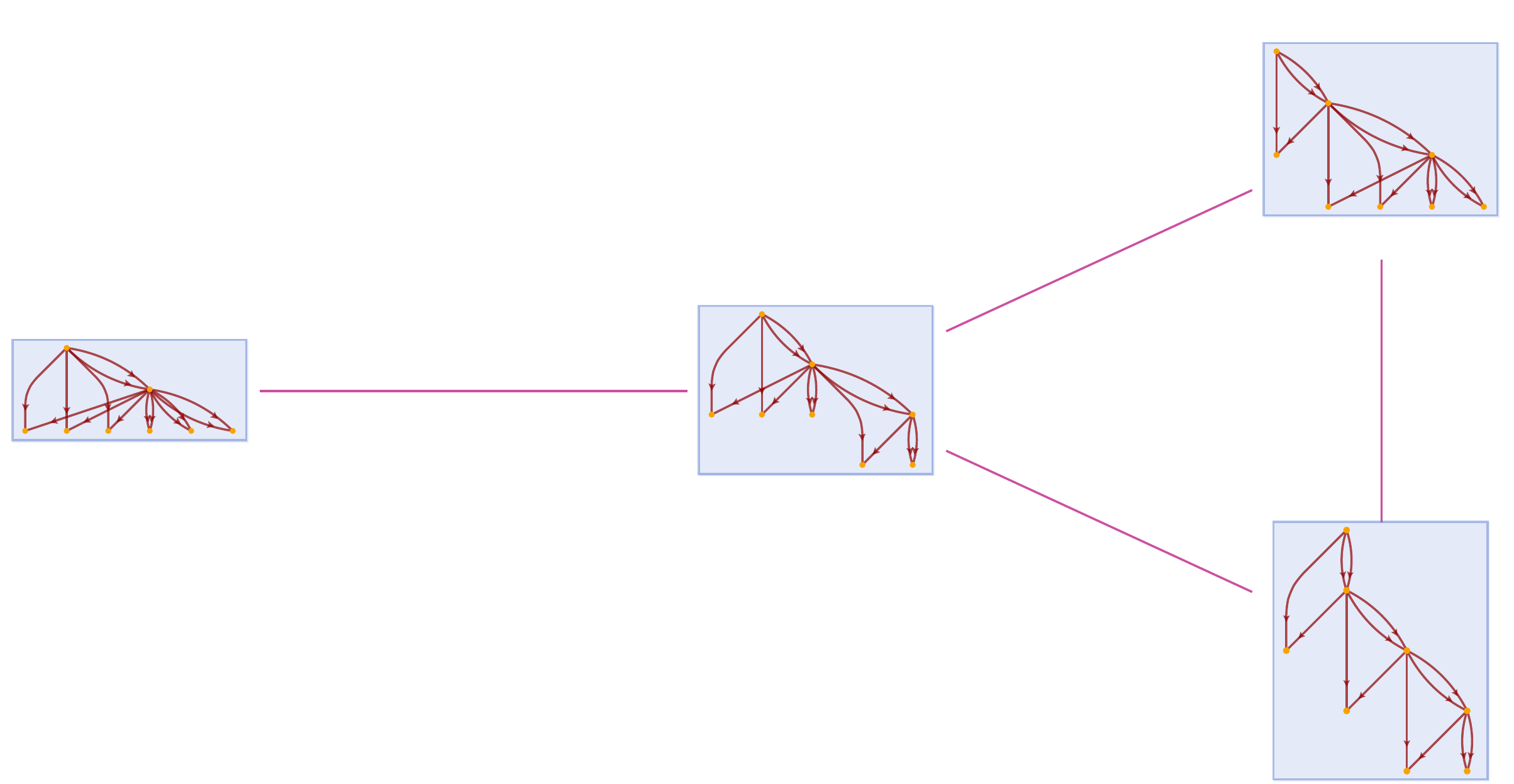}\hspace{0.1\textwidth}
\includegraphics[width=0.495\textwidth]{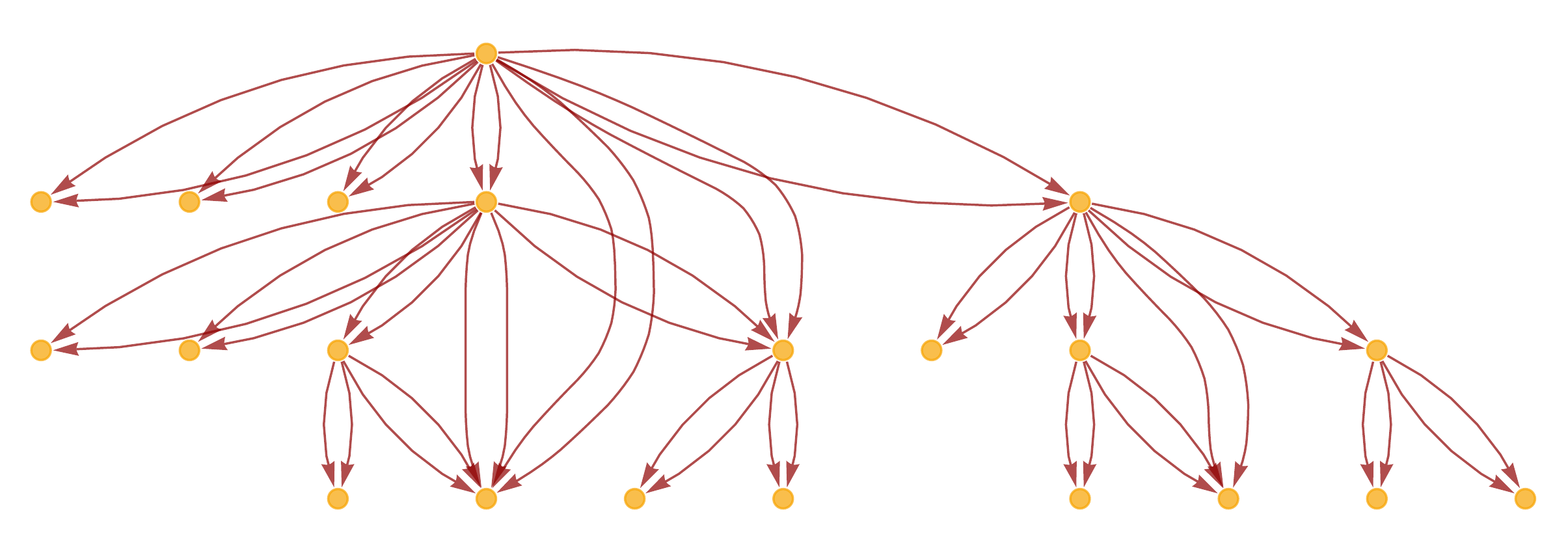}
\caption{On the left, a causal branchial graph corresponding to a typical ``foliation'' of the causal multiway evolution graph representing the non-deterministic evolution of the hypergraph rewriting/set substitution system ${\left\lbrace \left\lbrace x, y \right\rbrace, \left\lbrace y, z \right\rbrace \right\rbrace \to \left\lbrace \left\lbrace x, y \right\rbrace, \left\lbrace y, z \right\rbrace, \left\lbrace z, w \right\rbrace \right\rbrace}$, starting from the double self-loop initial condition ${\left\lbrace \left\lbrace 0, 0 \right\rbrace, \left\lbrace 0, 0 \right\rbrace \right\rbrace}$. On the right, the corresponding multiway causal graph for the same system (yielded by ``gluing'' the various causal graphs within the causal branchial graph together, in accordance with the tensor product structure of the corresponding Hilbert space).}
\label{fig:Figure29}
\end{figure}

\begin{figure}[ht]
\centering
\includegraphics[width=0.495\textwidth]{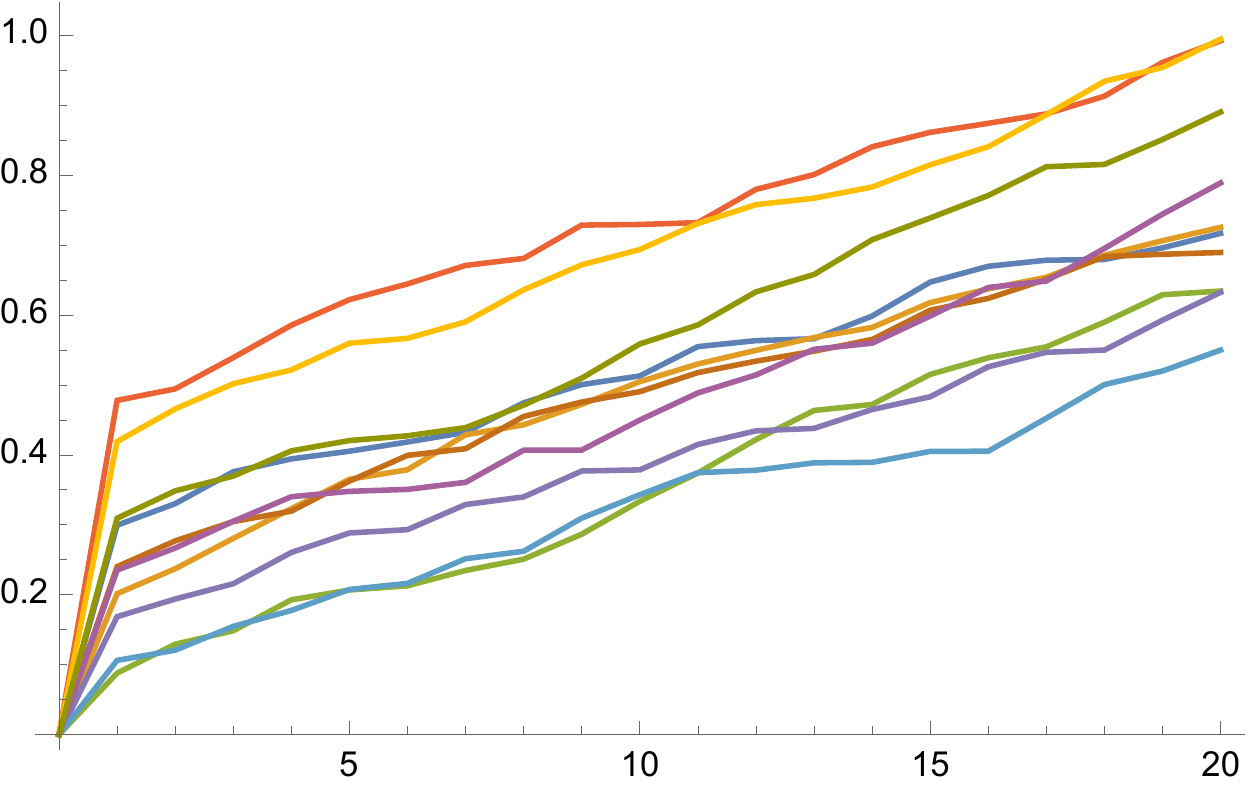}
\includegraphics[width=0.495\textwidth]{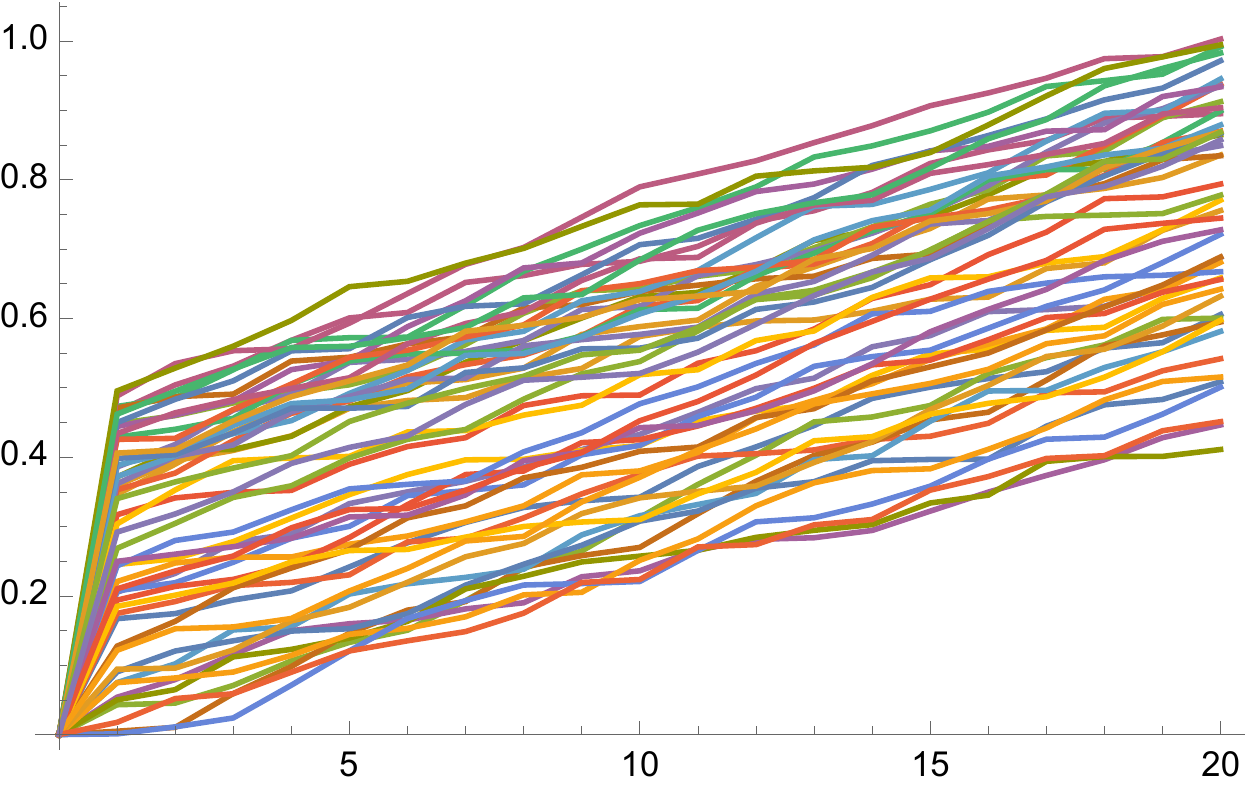}
\caption{Computations of the spacetime entanglement entropy $S$ for 10 (left) and 50 (right) randomly-selected Wolfram model evolutions of hypergraph rule signature ${2_2 \to 4_2}$; the vertical axes show the entanglement entropies as computed using the generalized eigenvalue approach detailed within the previous section, while the horizontal axes show the entanglement entropies as computed using the branchial graph/Fubini-Study metric distance, showing robust monotonic agreement between the two approaches.}
\label{fig:Figure30}
\end{figure}

However, in order to perform a systematic numerical comparison between entanglement entropies computed via branchial graphs and those computed via the Sorkin-Johnston construction, it is first necessary to reformulate the branchial graph procedure in a manifestly covariant form (at present, the decomposition of the Wolfram model multiway system into a sequence of branchial graphs, each consisting of a collection of discrete spacelike hypersurfaces, is equivalent to fixing a preferred spacetime gauge, and is therefore incompatible with the requisite covariance for the definition of a sensible notion of spacetime entropy). This may be achieved in a very natural way using the formalism \textit{causal} multiway systems, in which each vertex of a multiway evolution graph corresponds not to a particular hypergraph state (and therefore to a particular spacelike hypersurface), but rather to the complete causal history of the hypergraph rewriting system up to that point (and therefore to an extended region of discrete spacetime)\cite{gorard3}. In this way, causal multiway systems (which effectively represent the superposition of all possible causal histories for a given discrete spacetime) constitute a generalization of both classical and quantum sequential growth dynamics in causal set theory. An example of such a causal multiway system, generated by the non-deterministic evolution of a simple Wolfram model rule, is shown in Figure \ref{fig:Figure28}, effectively representing a superposition of several possible causal histories for a discrete (algorithmic) spacetime. The ``factorization'' or ``decomposition'' of those causal histories into a tensor product of classical discrete spacetimes (and thus of ``singleway'' causal graphs - causal graphs corresponding to a single choice of evolution path through the multiway system) can be represented using the corresponding \textit{causal branchial graphs}, as shown in Figure \ref{fig:Figure29}. Selecting a random sample of Wolfram model rules of a given hypergraph signature (in this case, the ${2_2 \to 4_2}$ signature, designating an input consisting of two hyperedges of arity-2 and an output consisting of four hyperedges of arity-2), we are able to evolve the selected rules non-deterministically before selecting a random pair of causal histories in the resulting causal branchial graphs. An entanglement entropy between the two discrete spacetimes (regarded as causal sets) can then be computed using the generalized eigenvalue algorithm derived within the preceding section from the Sorkin-Johnston construction; another entanglement entropy can also be computed by simply determining the branchial distance between the two discrete spacetimes within the corresponding causal branchial graph, in accordance with procedure outlined above. A numerical comparison of these two approaches is shown in Figure \ref{fig:Figure30}, robustly illustrating the expected monotonic relationship between the two notions of discrete spacetime entanglement entropy for algorithmically-generated causal sets, and confirming the preliminary numerical results previously obtained in \cite{dannemannfreitag}.

Note that, due to the considerable computational expense associated with evolving causal multiway systems for significant numbers of time steps, we have adopted a Monte Carlo sampling approach when generating these results, in order to make the computation more directly amenable to execution on massively-parallel supercomputer architectures. More precisely, each concurrent thread computes a different (randomly-selected) evolution history of the overall causal multiway system, with the assembly of the resulting threads being performed as a (relatively inexpensive) post-processing step. The sampling process is then continued until the assembled causal multiway system appears to reach a steady-state configuration.

\section{Concluding Remarks}
\label{sec:Section5}

Given that Wolfram model evolution effectively provides one with an explicit algorithmic dynamics by which causal sets may be naturally grown\cite{gorard3}, it is perhaps unsurprising that the two approaches to discrete quantum gravity share many formalistic features. However, the fact that two seemingly so radically different approaches to defining a quantum field theory over a discrete spacetime (one being essentially to equip a causal set with the complete apparatus of an algebraic field theory defined over its elements, the other being to use a directly combinatorial approach based on the non-deterministic dynamics of the hypergraph rewriting approach itself) should give equivalent results for such a fundamental calculation as the determination of spacetime entanglement entropies in generic algorithmically-generated spacetimes is both somewhat remarkable and rather encouraging. Moreover, these results may hint at a deeper underlying principle: as discussed previously, the standard formulation of quantum mechanics over general Wolfram model multiway systems effectively breaks covariance by assuming a decomposition into a time-ordered sequence of branchial graphs, each consisting of a collection of instantaneous spacelike hypersurfaces. As we have shown, once one reintroduces manifest covariance by promoting the ordinary multiway system to a \textit{causal} multiway system (in which the causal branchial graphs now consist of extended discrete spacetime regions), the same mathematical apparatus immediately yields an algebraic quantum field theory that is naturally free from ultraviolet divergences, and whose 2-point correlation functions are in both analytical and numerical agreement with those derived using the Sorkin-Johnston construction from causal set quantum field theory, without the need to impose a posteriori truncations on the spectra of the discrete operators. This leads to a highly tempting conjecture that the algebraic quantum field theory that one obtains over causal multiway evolution graphs is, in some sense, ``canonical'' (specifically, in the same sense that the categorical quantum mechanics model that one obtains over ordinary multiway evolution graphs is ``canonical'', in that it satisfies a certain universal property in the category-theoretic sense\cite{gorard5}\cite{gorard6}\cite{gorard7}). In this vein, it is notable that although the resulting quantum field theory in the Wolfram model case is free from ultraviolet divergences, and therefore does not require regularization or renormalization in order to yield finite observables, it nevertheless requires a procedure that is formally equivalent to dimensional regularization\cite{hooft} in order to ``analytically continue'' the requisite Green's functions/propagators from integer-dimensional spacetimes to the more general non-integer-dimensional case. This is perhaps indicative of a fairly general underlying consistency condition, wherein one cannot, in some sense, ``evade'' the need for regularization, even by passing to a discrete spacetime, because although one may achieve ultraviolet-finiteness by imposing a discrete cutoff/lattice regularization on spacetime, it may ultimately come at the expense of the guaranteed (homogeneous) integer-dimensionality of the limiting manifold. This would imply that the equivalence between the two analytic continuations is not merely mathematical, but also physical; in particular, it would indicate that one way to understand ``why'' dimensional regularization works for Feynman integrals is that it is somehow formally equivalent to passing from an ultraviolet-divergent quantum field theory in a continuum (integer-dimensional) spacetime to an ultraviolet-finite quantum field theory in a discrete (non-integer-dimensional) one. These conjectures are highly speculative, but also highly evocative, and seem worthy of further, more systematic, investigation via the methods developed within this article.

On a pragmatic level, although the quantum field theory derived via causal branchial graphs is arguably more conceptually elegant than the Sorkin-Johnston construction, at least for the case of spacetime entanglement entropy computations (in part because of the lack of need to impose eigenvalue truncations to obtain the desired area laws), it is also worth noting that it is, in general, significantly more computationally demanding, since it requires simulating the evolution of an entire causal multiway system (with potentially unbounded numbers of discrete parallel histories). It is also notably less general, since it requires complete knowledge of the full multiway/branchial structure of the Wolfram model evolution to calculate, making it inappropriate for the computation of entanglement entropies in (for instance) causal graphs obtained by sprinkling into predefined Lorentzian geometries, whereas the Sorkin-Johnston approach is general enough to apply to \textit{any} finite causal set (including those obtained algorithmically through Wolfram model evolution). However, since there exist algorithms by which Wolfram model evolution rules can be constructed that provably simulate the evolution of the Einstein field equations (for instance via the ADM, BSSN or CCZ4 initial-value formulations) from arbitrary initial Cauchy data\cite{gorard4}, most recently through the \href{https://github.com/JonathanGorard/Gravitas}{\textsc{Gravitas} framework}, this constitutes less of a restriction than it might initially appear to be.

Among the many avenues of potential future research opened up by the investigations presented within this article, perhaps the most immediate is the specialization of these methods to the case of known limiting spacetime geometries, and especially to black hole spacetimes, in which the connection between spacetime entanglement entropy and the Bekenstein-Hawking entropy law has already been studied extensively in holographic and string-theoretic contexts (for instance as a special case of the AdS/CFT correspondence)\cite{emparan}\cite{solodukhin}\cite{jacobson}. The eventual aim would be to determine whether the kinds of discrete spacetime approaches outlined within this article constitute a sufficiently complete description of spacetime microstates (and particularly black hole microstates) that semiclassical results such as the Bekenstein-Hawking law may be derived from first principles (for instance via a pure counting argument). Relatedly, by considering discrete models of maximally-extended anti-de Sitter Schwarzschild black holes (for instance in Kruskal-Szekeres coordinates, as studied in \cite{gorard4}), it may be possible to determine whether the kinds of correlations between entanglement entropies and limiting geometries of Einstein-Rosen wormholes predicted by the ER=EPR conjecture\cite{vanraamsdonk}\cite{maldacena} hold in the discrete case\cite{shah}. Since Wolfram model evolutions permit covariance-breaking foliations into discrete spacelike hypersurfaces (in a way which pure causal set models do not, or at least not as naturally), it may even be possible to perform the same restriction of quantum fields to spacelike hypersurfaces that occurs in non-manifestly covariant definitions of entanglement entropy within standard (continuum) quantum field theory descriptions, including semiclassical descriptions of quantum fields in curved spacetimes. On the more mathematical side, it is generally presumed that multiway systems are more naturally described through homotopic\cite{arsiwalla}\cite{arsiwalla2} or functorial\cite{gorard8} methods (specifically via the languages of homotopy type theory and functorial field theory) than through purely algebraic ones. Thus, an extension of the methods developed here from the Heisenberg (algebraic) picture to the Schr\"odinger (functorial) one would be very welcome; intuitively, this corresponds to transforming from a description of the resulting field theory in terms of branchial graphs (since these effectively encode categories of endomorphism algebras and their isomorphisms\cite{gorard2}\cite{gorard5}, as used in the Heisenberg/algebraic picture) to one in terms of multiway evolution graphs directly (since these effectively encode categories of vector spaces and their isomorphisms\cite{gorard5}\cite{gorard6}, as used in the Schr\"odinger/functorial picture). Amongst other things, such a reformulation might offer the opportunity to construct rigorous notions of topological and conformal field theories in the discrete spacetime case, for instance via the Atiyah-Segal axiomatization\cite{atiyah}\cite{atiyah2}\cite{segal}. Finally, for the sake of simplification of the analysis, we have restricted ourselves within this article purely to the case of massless, free scalar field theories. The extension to massive scalar field theories ought to be relatively straightforward, since we have already implicitly derived the relevant massive forms of the discrete propagators/Green's functions in the preceding sections. The extension to fully-interacting scalar field theories is more complicated, and will likely involve, at least in the first instance, using methods of perturbation theory to investigate the effects of quartic perturbations away from purely Gaussian states. The extension from symmetric to antisymmetric Fock spaces, and therefore from the description of bosonic to fermionic quantum fields (for instance via discrete formulations of spinor bundles), also remains an exciting open frontier of investigation.

\section*{Acknowledgments}

The authors would like to thank Fabrizio Genovese and Bob Coecke for useful early conversations regarding the extension of rigorous diagrammatic methods to relativistic quantum mechanics and quantum field theories, Shivaji Sondhi and the Rudolf Peierls Centre for Theoretical Physics at the University of Oxford for allowing us to present a preliminary version of these results and receive useful feedback, and Stephen Wolfram for his steadfast encouragement and for many clarifying conversations throughout.


\begin{thebibliography}{99}
\bibitem{popescu}
\textsc{S. Popescu and D. Rohrlich} (1997), ``Thermodynamics and the Measure of Entanglement'', \textit{Physical Review A} \textbf{56} (5): R3319(R). \url{https://arxiv.org/abs/quant-ph/9610044}.

\bibitem{vedral}
\textsc{V. Vedral, M. B. Plenio, M. A. Rippin and P. L. Knight} (1997), ``Quantifying Entanglement'', \textit{Physical Review Letters} \textbf{78} (12): 2275. \url{https://arxiv.org/abs/quant-ph/9702027}.

\bibitem{sorkin}
\textsc{R. D. Sorkin} (1983), ``On the Entropy of the Vacuum outside a Horizon'', \textit{Tenth International Conference on General Relativity and Gravitation Volume II}, B. Bertotti, F. de Felice and A. Pascolini (eds): 734--736. \url{https://arxiv.org/abs/1402.3589}.

\bibitem{bombelli}
\textsc{L. Bombelli, R. K. Koul, J. Lee and R. D. Sorkin} (1986), ``Quantum source of entropy for black holes'', \textit{Physical Review D} \textbf{34} (2): 373. \url{https://journals.aps.org/prd/abstract/10.1103/PhysRevD.34.373}.

\bibitem{ryu}
\textsc{S. Ryu, T. Takayanagi} (2006), ``Holographic Derivation of Entanglement Entropy from the anti-de Sitter Space/Conformal Field Theory Correspondence'', \textit{Physical Review Letters} \textbf{96} (18): 181602. \url{https://arxiv.org/abs/hep-th/0603001}.

\bibitem{bombelli2}
\textsc{L. Bombelli, J. Lee, D. A. Meyer and R. D Sorkin} (1987), ``Space-time as a causal set'', \textit{Physical Review Letters} \textbf{59} (5): 521. \url{https://journals.aps.org/prl/abstract/10.1103/PhysRevLett.59.521}.

\bibitem{bombelli3}
\textsc{L. Bombelli and D. A. Meyer} (1989), ``The origin of Lorentzian geometry'', \textit{Physics Letters A} \textbf{141} (5-6): 226--228. \url{https://www.sciencedirect.com/science/article/abs/pii/037596018990474X}.

\bibitem{sorkin2}
\textsc{R. D. Sorkin} (1997), ``Spacetime and Causal Sets'', \textit{Proceedings of the SILARG VII Conference on Relativity and Gravitation: Classical and Quantum}, J. C. D'Olivo, E. Nahmad-Achar, M. Rosenbaum, M. P. Ryan, L. F. Urrutia and F. Zertuche (eds): 150--173. World Scientific, Singapore. \url{https://www2.perimeterinstitute.ca/personal/rsorkin/some.papers/66.cocoyoc.pdf}.

\bibitem{sorkin3}
\textsc{R. D. Sorkin} (1997), ``Forks in the Road on the Way to Quantum Gravity'', \textit{International Journal of Theoretical Physics} \textbf{36}: 2759--2781. \url{https://arxiv.org/abs/gr-qc/9706002}.

\bibitem{hawking}
\textsc{S. W. Hawking, A. R. King and P. J. McCarthy} (1976), ``A new topology for curved space-time which incorporates the causal, differential and conformal structures'', \textit{Journal of Mathematical Physics} \textbf{17} (2): 174--181. \url{https://aip.scitation.org/doi/10.1063/1.522874}.

\bibitem{malament}
\textsc{D. B. Malament} (1977), ``The class of continuous timelike curves determines the topology of spacetime'', \textit{Journal of Mathematical Physics} \textbf{18} (7): 1399--1404. \url{https://aip.scitation.org/doi/10.1063/1.523436}.

\bibitem{wolfram}
\textsc{S. Wolfram} (2002), \textit{A New Kind of Science}. Champaign, Illinois: Wolfram Media, Inc. \url{https://www.wolframscience.com}

\bibitem{wolfram2}
\textsc{S. Wolfram} (2020), ``A Class of Models with the Potential to Represent Fundamental Physics'', \textit{Complex Systems} \textbf{29} (2): 107--536. \url{https://arxiv.org/abs/2004.08210}.

\bibitem{gorard}
\textsc{J. Gorard} (2020), ``Some Relativistic and Gravitational Properties of the Wolfram Model'', \textit{Complex Systems} \textbf{29} (2): 599--654. \url{https://arxiv.org/abs/2004.14810}.

\bibitem{gorard2}
\textsc{J. Gorard} (2020), ``Some Quantum Mechanical Properties of the Wolfram Model'', \textit{Complex Systems} \textbf{29} (2): 537--598. \url{https://www.complex-systems.com/abstracts/v29_i02_a02/}.

\bibitem{bolognesi}
\textsc{T. Bolognesi} (2010), ``Causal sets from simple models of computation'', \textit{International Journal of Unconventional Computing} \textbf{6} (6): 489--524. \url{https://arxiv.org/abs/1004.3128}.

\bibitem{bolognesi2}
\textsc{T. Bolognesi} (2012), ``Algorithmic Causal Sets for a Computational Spacetime'', \textit{A Computable Universe}, H. Zenil (ed): 451--477. World Scientific. \url{https://www.worldscientific.com/doi/10.1142/9789814374309_0024}.

\bibitem{gorard3}
\textsc{J. Gorard} (2020), ``Algorithmic Causal Sets and the Wolfram Model'', \textit{arXiv preprint}: \url{https://arxiv.org/abs/2011.12174}.

\bibitem{gorard4}
\textsc{J. Gorard} (2021), ``Hypergraph Discretization of the Cauchy Problem in General Relativity via Wolfram Model Evolution'', \textit{arXiv preprint}: \url{https://arxiv.org/abs/2102.09363}.

\bibitem{gorard5}
\textsc{J. Gorard, M. Namuduri and X. D. Arsiwalla} (2020), ``ZX-Calculus and Extended Hypergraph Rewriting Systems I: A Multiway Approach to Categorical Quantum Information Theory'', \textit{arXiv preprint}: \url{https://arxiv.org/abs/2010.02752}.

\bibitem{gorard6}
\textsc{J. Gorard, M. Namuduri and X. D. Arsiwalla} (2021), ``ZX-Calculus and Extended Wolfram Model Systems II: Fast Diagrammatic Reasoning with an Application to Quantum Circuit Simplification'', \textit{arXiv preprint}: \url{https://arxiv.org/abs/2103.15820}.

\bibitem{coecke}
\textsc{B. Coecke and R. Duncan} (2008), ``Interacting Quantum Observables'', \textit{International Colloquium on Automata, Languages and Programming}, Lecture Notes in Computer Science \textbf{5126}: 298--310. Springer-Verlag Berlin, Heidelberg. \url{https://link.springer.com/chapter/10.1007/978-3-540-70583-3_25}.

\bibitem{coecke2}
\textsc{B. Coecke and R. Duncan} (2011), ``Interacting Quantum Observables: Categorical Algebra and Diagrammatics'', \textit{New Journal of Physics} \textbf{13} (4): 043016. \url{https://arxiv.org/abs/0906.4725}.

\bibitem{gorard7}
\textsc{J. Gorard, M. Namuduri and X. D. Arsiwalla} (2021), ``Fast Automated Reasoning over String Diagrams using Multiway Causal Structure'', \textit{arXiv preprint}: \url{https://arxiv.org/abs/2105.04057}.

\bibitem{sorkin4}
\textsc{R. D. Sorkin} (2014), ``Expressing entropy globally in terms of (4D) field-correlations'', \textit{Journal of Physics: Conference Series Volume 484}. \url{https://arxiv.org/abs/1205.2953}.

\bibitem{peierls}
\textsc{R. E. Peierls} (1952), ``The commutation laws of relativistic field theory'', \textit{Proceedings of the Royal Society A} \textbf{214} (1117): 143--157. \url{https://royalsocietypublishing.org/doi/10.1098/rspa.1952.0158}.

\bibitem{jordan}
\textsc{P. Jordan and W. Pauli Jr.} (1928), ``Zur Quantenelektrodynamik ladungsfreier Felder'', \textit{Zeitschrift f\"ur Physik} \textbf{47}: 151--173. \url{https://link.springer.com/article/10.1007/BF02055793}.

\bibitem{dowker}
\textsc{F. Dowker and L. Glaser} (2013), ``Causal set d'Alembertians for various dimensions'', \textit{Classical and Quantum Gravity} \textbf{30} (19): 195016. \url{https://arxiv.org/abs/1305.2588}.

\bibitem{sorkin5}
\textsc{R. D. Sorkin} (2007), ``Does Locality Fail at Intermediate Length-Scales'', \textit{Towards Quantum Gravity}, D. Oriti (ed): 26--43. Cambridge University Press. \url{https://arxiv.org/abs/gr-qc/0703099}.

\bibitem{johnston}
\textsc{S. Johnston} (2008), ``Particle propagators on discrete spacetime'', \textit{Classical and Quantum Gravity} \textbf{25} (20): 202001. \url{https://arxiv.org/abs/0806.3083}.

\bibitem{johnston2}
\textsc{S. Johnston} (2009), ``Feynman Propagator for a Free Scalar Field on a Causal Set'', \textit{Physical Review Letters} \textbf{103} (18): 180401. \url{https://arxiv.org/abs/0909.0944}.

\bibitem{johnston3}
\textsc{S. P. Johnston} (2010), ``Quantum Fields on Causal Sets'', \textit{PhD Thesis, Imperial College London}. \url{https://arxiv.org/abs/1010.5514}.

\bibitem{sorkin6}
\textsc{R. D. Sorkin} (2011), ``Scalar Field Theory on a Causal Set in Histories Form'', \textit{Journal of Physics: Conference Series Volume 306}. \url{https://arxiv.org/abs/1107.0698}.

\bibitem{sorkin7}
\textsc{R. D. Sorkin and Y. K. Yazdi} (2018), ``Entanglement Entropy in Causal Set Theory'', \textit{Classical and Quantum Gravity} \textbf{35} (7): 074003. \url{https://arxiv.org/abs/1611.10281}.

\bibitem{marian}
\textsc{P. Marian and T. A. Marian} (2008), ``Bures distance as a measure of entanglement for symmetric two-mode Gaussian states'', \textit{Physical Review A} \textbf{77} (6): 062319. \url{https://arxiv.org/abs/0705.1138}.

\bibitem{ollivier}
\textsc{Y. Ollivier} (2007), ``Ricci curvature of metric spaces'', \textit{Comptes Rendus Math\'ematique de l'Acad\'emie des Sciences} \textbf{345} (11): 643--646. \url{https://www.sciencedirect.com/science/article/pii/S1631073X07004414}.

\bibitem{eidi}
\textsc{M. Eidi and J. Jost} (2020), ``Ollivier Ricci curvature of directed hypergraphs'', \textit{Scientific Reports} \textbf{10}: 12466 \url{https://arxiv.org/abs/1907.04727}.

\bibitem{hooft}
\textsc{G. 't Hooft and M. Veltman} (1972), ``Regularization and renormalization of gauge fields'', \textit{Nuclear Physics B} \textbf{44} (1): 189--213. \url{https://www.sciencedirect.com/science/article/abs/pii/0550321372902799}.

\bibitem{myrheim}
\textsc{J. Myrheim} (1978), ``Statistical geometry'', \textit{CERN Preprint TH-2538}. \url{https://cds.cern.ch/record/293594?ln=en}.

\bibitem{meyer}
\textsc{D. A. Meyer} (1989), ``The dimension of causal sets'', \textit{PhD Thesis, Massachusetts Institute of Technology}. \url{https://dspace.mit.edu/handle/1721.1/14328}.

\bibitem{cocchiarella}
\textsc{D. Cocchiarella, S. Scali, S. Ribisi, B. Nardi, G. Bel-Hadj-Aissa and R. Franzosi} (2020), ``Entanglement distance for arbitrary $M$-qudit hybrid systems'', \textit{Physical Review A} \textbf{101} (4): 042129. \url{https://arxiv.org/abs/2003.05771}.

\bibitem{bekenstein}
\textsc{J. D. Bekenstein} (1972), ``Black holes and the second law'', \textit{Lettere al Nuovo Cimento} \textbf{4} (15): 737--740. \url{https://link.springer.com/article/10.1007/BF02757029}.

\bibitem{hawking2}
\textsc{S. W. Hawking} (1975), ``Particle creation by black holes'', \textit{Communication in Mathematical Physics} \textbf{43} (3): 199--220. \url{https://link.springer.com/article/10.1007/BF02345020}.

\bibitem{vanraamsdonk}
\textsc{M. Van Raamsdonk} (2010), ``Building up spacetime with quantum entanglement'', \textit{General Relativity and Gravitation} \textbf{42} (14): 2323--2329. \url{https://arxiv.org/abs/1005.3035}.

\bibitem{maldacena}
\textsc{J. Maldacena and L. Susskind} (2013), ``Cool horizons for entangled black holes'', \textit{Fortschritte der Physik} \textbf{61} (9): 781--811. \url{https://arxiv.org/abs/1306.0533}.

\bibitem{shah}
\textsc{R. Shah, J. Gorard} (2019), ``Quantum Cellular Automata, Black Hole Thermodynamics and the Laws of Quantum Complexity'', \textit{Complex System} \textbf{28} (4): 393--410. \url{https://arxiv.org/abs/1910.00578}.

\bibitem{arsiwalla}
\textsc{X. D. Arsiwalla, J. Gorard and H. Elshatlawy} (2021), ``Homotopies in Multiway (Non-Deterministic) Rewriting Systems as $n$-Fold Categories'', \textit{arXiv preprint}: \url{https://arxiv.org/abs/2105.10822}.

\bibitem{arsiwalla2}
\textsc{X. D. Arsiwalla, J. Gorard} (2021), ``Pregeometric Spaces from Wolfram Model Rewriting Systems as Homotopy Types'', \textit{arXiv preprint}: \url{https://arxiv.org/abs/2111.03460}.

\bibitem{gorard8}
\textsc{J. Gorard} (2022), ``A Functorial Perspective on (Multi)computational Irreducibility'', \textit{arXiv preprint}: \url{https://arxiv.org/abs/2301.04690}.

\bibitem{henson}
\textsc{J. Henson} (2009), ``Discovering the Discrete Universe'', \textit{Proceedings of the Foundations of Space and Time Conference}. Cape Town. \url{https://arxiv.org/abs/1003.5890}.

\bibitem{benincasa}
\textsc{D. M. T. Benincasa and F. Dowker} (2010), ``Scalar Curvature of a Causal Set'', \textit{Physical Review Letters} \textbf{104} (18): 181301. \url{https://arxiv.org/abs/1001.2725}.

\bibitem{gelfand}
\textsc{I. Gel'fand and G. Shilov} (1964), \textit{Generalized Functions}. American Mathematical Society. ISBN: 978-1470426583.

\bibitem{egorov}
\textsc{Y. V. Egorov and M. A. Shubin} (1994), \textit{Partial Differential Equations II}. Springer-Verlag Berlin, Heidelberg. ISBN: 978-3-540-52001-6.

\bibitem{bogoliubov}
\textsc{N. N. Bogoliubov and D. V. Shirkov} (1959), \textit{Introduction to the Theory of Quantized Fields Volume III}. John Wiley \& Sons, Inc. ISBN: 978-0471042235.

\bibitem{dejager}
\textsc{E. M. de Jager} (1967), ``The Lorentz-Invariant Solutions of the Klein-Gordon Equation'', \textit{SIAM Journal on Applied Mathematics} \textbf{15} (4): 944--963. \url{https://www.jstor.org/stable/2099798}.

\bibitem{rideout}
\textsc{D. Rideout and S. Zohren} (2006), ``Evidence for an entropy bound from fundamentally discrete gravity'', \textit{Classical and Quantum Gravity} \textbf{23} (22): 6195--6213. \url{https://arxiv.org/abs/gr-qc/0606065}.

\bibitem{bombelli4}
\textsc{L. Bombelli} (1987), ``Space-time as a causal set'', \textit{PhD Thesis, Syracuse University}. \url{https://surface.syr.edu/phy_etd/88/}.

\bibitem{champeney}
\textsc{D. C. Champeney} (1987), \textit{A Handbook of Fourier Transforms}. Cambridge University Press. ISBN: 978-1139171823.

\bibitem{birrell}
\textsc{N. D. Birrell and P. C. W Davies} (1982), \textit{Quantum Fields in Curved Space}. Cambridge University Press. ISBN: 978-0511622632.

\bibitem{fulling}
\textsc{S. A. Fulling} (1989), \textit{Aspects of Quantum Field Theory in Curved Space-Time}. Cambridge University Press. ISBN: 978-1139172073.

\bibitem{forman}
\textsc{R. Forman} (2003), ``Bochner's Method for Cell Complexes and Combinatorial Ricci Curvature'', \textit{Discrete \& Computational Geometry} \textbf{29}: 323--374. \url{https://link.springer.com/article/10.1007/s00454-002-0743-x}.

\bibitem{ollivier2}
\textsc{Y. Ollivier} (2009), ``Ricci curvature of Markov chains on metric spaces'', \textit{Journal of Functional Analysis} \textbf{256} (3): 810--864. \url{https://arxiv.org/abs/math/0701886}.

\bibitem{ollivier3}
\textsc{Y. Ollivier} (2011), ``A visual introduction to Riemannian curvatures and some discrete generalizations'', \textit{Analysis and Geometry of Metric Measure Spaces: Lecture Notes of the 50th S'eminaire de Math\'ematiques Sup\'erieures (SMS)} \textbf{56}: 197--219. \url{https://hal.inria.fr/hal-00858008/en}.

\bibitem{roy}
\textsc{M. Roy, D. Sinha and S. Surya} (2013), ``Discrete geometry of a small causal diamond'', \textit{Physical Review D} \textbf{87} (4): 044046. \url{https://arxiv.org/abs/1212.0631}.

\bibitem{khetrapal}
\textsc{S. Khetrapal and S. Surya} (2013), ``Boundary term contributions to the volume of a small causal diamond'', \textit{Classical and Quantum Gravity} \textbf{30} (6): 065005. \url{https://arxiv.org/abs/1212.0629}.

\bibitem{jost}
\textsc{J. Jost} (2011), \textit{Riemannian Geometry and Geometric Analysis}. Springer-Verlag Berlin, Heidelberg. ISBN: 978-3-642-21298-7.

\bibitem{gray}
\textsc{A. Gray} (2004), \textit{Tubes}. Birkh\"auser Basel. ISBN: 978-3-7643-6907-1.

\bibitem{reid}
\textsc{D. D. Reid} (2003), ``Manifold dimension of a causal set: Tests in conformally flat spacetimes'', \textit{Physical Review D} \textbf{67} (2): 024034. \url{https://arxiv.org/abs/gr-qc/0207103}.

\bibitem{carlson}
\textsc{F. Carlson} (1914), ``Sur une classe de s\'eries de Taylor'', \textit{Doctoral Thesis, Uppsala University}.

\bibitem{haag}
\textsc{R. Haag and D. Kastler} (1964), ``An Algebraic Approach to Quantum Field Theory'', \textit{Journal of Mathematical Physics} \textbf{5} (7): 848. \url{https://aip.scitation.org/doi/10.1063/1.1704187}.

\bibitem{wightman}
\textsc{A. S. Wightman and L. Gardin} (1965), ``Fields as Operator-valued Distributions in Relativistic Quantum Theory'', \textit{Arkiv f\"or Fysik} \textbf{28} (13). \url{https://www.osti.gov/biblio/4606723}.

\bibitem{stone}
\textsc{M.  H. Stone} (1932), \textit{Linear Transformations in Hilbert Space}. American Mathematical Society. ISBN: 978-0821810156.

\bibitem{spiegel}
\textsc{M. R. Spiegel} (1953), ``The Summation of Series Involving Roots of Transcendental Equations and Related Applications'', \textit{Journal of Applied Physics} \textbf{24} (9): 1103--1106. \url{https://aip.scitation.org/doi/abs/10.1063/1.1721455}.

\bibitem{gorard9}
\textsc{J. Gorard} (2016), ``Uniqueness Trees: A Possible Polynomial Approach to the Graph Isomorphism Problem'', \textit{arXiv preprint}: \url{https://arxiv.org/abs/1606.06399}.

\bibitem{abramsky}
\textsc{S. Abramsky and B. Coecke} (2004), ``A categorical semantics of quantum protocols'', \textit{Proceedings of the 19th IEEE Symposium on Logic in Computer Science}. Turku, Finland: 415--425. \url{https://arxiv.org/abs/quant-ph/0402130}.

\bibitem{abramsky2}
\textsc{S. Abramsky and B. Coecke} (2008), ``Categorical quantum mechanics'', \textit{Handbook of Quantum Logic and Quantum Structures}, K. Engesser, D. M. Gabbay and D. Lehmann (eds): 261--323. Elsevier. \url{https://arxiv.org/abs/0808.1023}.

\bibitem{dannemannfreitag}
\textsc{J. Dannemann-Freitag} (2021), ``Comparing Wolfram Model and causal set entanglement entropies'', \textit{Wolfram Community}. \url{https://community.wolfram.com/groups/-/m/t/2312623}.

\bibitem{emparan}
\textsc{R. Emparan} (2006), ``Black hole entropy as entanglement entropy: a holographic derivation'', \textit{Journal of High Energy Physics} \textbf{2006}. \url{https://arxiv.org/abs/hep-th/0603081}.

\bibitem{solodukhin}
\textsc{S. N. Solodukhin} (2011), ``Entanglement Entropy of Black Holes'', \textit{Living Reviews in Relativity} \textbf{14} (8). \url{https://arxiv.org/abs/1104.3712}.

\bibitem{jacobson}
\textsc{T. Jacobson and A. Satz} (2013), ``Black hole entanglement entropy and the renormalization group'', \textit{Physical Review D} \textbf{87} (8): 084047. \url{https://arxiv.org/abs/1212.6824}.

\bibitem{atiyah}
\textsc{M. Atiyah} (1988), ``New Invariants of 3- and 4-Dimensional Manifolds'', \textit{The Mathematical Heritage of Hermann Weyl: Proceedings of Symposia in Pure Mathematics} \textbf{48}: 258--299. \url{https://archive.org/details/mathematicalheri0000symp/page/285/mode/2up}.

\bibitem{atiyah2}
\textsc{M. Atiyah} (1988), ``Topological quantum field theories'', \textit{Publications Math\'ematiques de l'Institut des Hautes \'Etudes Scientifiques} \textbf{68} (68): 175--186. \url{https://link.springer.com/article/10.1007/BF02698547}.

\bibitem{segal}
\textsc{G. B. Segal} (1988), ``The Definition of Conformal Field Theory'', \textit{Differential Geometrical Methods in Theoretical Physics}, K. Bleuler and M. Werner (eds) \textbf{250}: 165--171. \url{https://link.springer.com/chapter/10.1007/978-94-015-7809-7_9}.
\end{thebibliography}
\end{document}